\documentclass{article}   

\usepackage{graphicx}
\usepackage{hyperref}
\hypersetup{
    colorlinks=true,
    linkcolor=blue,
    urlcolor=blue,
    citecolor=blue,
}

\usepackage{subcaption}

\usepackage{float}
\usepackage{amssymb}
\usepackage{xfrac}
\usepackage{miller}

\usepackage{siunitx}
\usepackage{mathtools}
\usepackage{booktabs} 	
\usepackage[title]{appendix}
\DeclareMathAlphabet{\mathsfsl}{OT1}{cmss}{m}{sl}

\newcommand{\transp}{{\rm T}}

\newcommand{\fepx}{{\bfseries{\slshape{FEpX}}}}
\newcommand{\mechmet}{{\bfseries{\slshape{MechMet}}}}
\newcommand{\mechmonics}{{\bfseries{\slshape{MechMonics}}}}

\newcommand{\neper}{{\bfseries{\slshape{Neper}}}}
\newcommand{\paraview}{{\bfseries{\slshape{Paraview}}}}
\newcommand{\visit}{{\bfseries{\slshape{VisIt}}}}
\newcommand{\github}{{\bfseries{\slshape{GitHub}}}}
\newcommand{\matlab}{{\bfseries{\slshape{MATLAB}}}}

\newcommand{\vctr}[1]{\boldsymbol{#1}}

\newcommand{\tnsr}[1]{\mathsfsl{#1 }}

\newcommand{\matcapN}{\Big[ \,\mathsf{N}(\xi, \eta, \zeta) \,\Big]}

\newcommand{\Matstiff}{\Big[ \,\mathsf{K}\,\Big]} 
\newcommand{\Matmass}{\Big[ \,\mathsf{M}\,\Big]} 
  
\newcommand{\matstiff}{\Big[ \,\mathsf{k}^{\it e}\,\Big]} 
\newcommand{\matmass}{\Big[ \,\mathsf{m}^{\it e}\,\Big]}

\newcommand{\dee}{{\mathrm{d}}}

\begin{document}

\title{Mechanical Analyses using Discrete Harmonic Expansions\\ of Grain-Based Field Data}

\author{Paul R. Dawson \and Matthew P. Miller}

\date{Sibley School of Mechanical and Aerospace Engineering\\ Cornell University\\  Ithaca, NY 14850, USA \\ \today}

\maketitle

\begin{abstract}
A methodology for computing expansion basis functions  using discrete harmonic modes is presented.
The discrete harmonic modes are determined grain-by-grain for virtual polycrystals for which finite element meshes are available.  
The expansion weights associated with representing field variables over grain domains are determined by exploiting the orthogonality of the harmonic modes.  
The methodology is demonstrated with the representation of the axial stress distributions during tensile loading of a polycrystalline sample.  
An open source code, \mechmonics, is available to researchers wishing to use the methodology to analyze data.

\end{abstract}


\section{Introduction}

Structural alloys in general  are polycrystalline.  The individual crystals exhibit anisotropic mechanical behaviors, both for their elastic and plastic responses.  
 Because the orientations of the atomic lattices that define the individual crystals are not aligned with each other, deformations within polycrystals are heterogeneous at the scale of the crystals.  
 Likewise is the stress and strain.  
 A frequent goal in materials research is to better understand the nature of 
 spatial variations in the mechanical response 
 and to relate these to the microstructure and loading conditions.
 A more in-depth understanding opens the door to alloy improvement.
 
Development of experimental methods  and simulation capabilities have focused in recent years
on better resolving the distributions of stress and strain at the crystal scale, as well as
mapping features of the microstructure over grain domains. 
The improvements have been impressive, and have presented researchers with
a welcomed problem: difficulty in analyzing data sets that are spatially well-resolved and thus are large and complex.   To that end, automated methods have been invoked, such as machine-learning algorithms.
To most effectively use such methods, though, the data often need to be cast into a representation based on principal modes or vectors.  
Here we present a method to evaluate sets of discrete harmonic modes for polycrystals.
The modes span the physical domain of crystals -- namely their volumes.  
Using the modes, representation of field variables over grain domains are
readily constructed in terms of the modes and associated weights.
The method is implemented in \matlab\cite{matlab}, via a code named \mechmonics.

In this document we first present the underlying methods used by \mechmonics.
This includes both the method for determining the discrete harmonic modes and the
method for evaluating the expansion weights for a specified field variable.
We then demonstrate the methods by applying \mechmonics\, to simulation data
associated with the tensile loading of a virtual sample.
Through this example we  illustrate how grain morphology influences the 
attributes of the harmonic modes and the representation obtained using the modes.

\section{Methodology}
\label{sec:methodology}

\subsection{Representation of field data with an expansion}

As is routinely done in finite element methods, 
a smooth field can be represented over a domain, $V$, using piecewise polynomials
defined over subdomains of  $V$, referred to as elements.  
For a single element the representation has the form:
\begin{equation}
a({\bf{x}}) = [ N({\bf{x}})]^e \{ A \}^e 
\end{equation}
where $a({\bf{x}})$ is the smooth field, $[ N({\bf{x}})]^e $ are polynomial interpolation functions defined over an element, and $\{ A \}^e $ are the nodal point values where the superscript $e$ indicates a particular element~\cite{OCZ_6ed}.
The representation over the entire volume is the union of the elemental functions, and is written simply as a concatenation  of the elemental equations:
\begin{equation}
a({\bf{x}}) = [ N({\bf{x}})] \{ A\} 
\label{eq:fem_rep}
\end{equation}
where dropping the elemental superscript implies union over all elements in the mesh.
Continuity of $a({\bf{x}})$ over the full volume is  guaranteed as a consequence of the properties of the interpolation functions~\cite{thompson_book}.

For a scalar field, the size of $\{ A \} $, and thus the number of degrees of freedom in the representation of  $a({\bf{x}})$, coincides with the number of nodal points in the mesh. 
Thus, the representation of the field is done with a finite number of parameters. 
It is often the case that the field being represented is the solution  (or a component of the solution for vectors and tensors) of a differential equation and the implication of representing the
solution with a limited number of function with polynomials is that the solution is approximate.
For many physical systems, an attribute of the finite element method is that its solutions converge 
to the exact answer as the number of degrees of freedom  in the polynomial representation  ({\it e.g.} number of nodes in the mesh) increases.  The convergence rate depends on the type of 
element, but in any case this drives the practice to refine the mesh to obtain better results.    
So, while a better result is obtained, it is at the expense of an increased number of parameters employed in its representation. 

Often a goal of the analysis, though, is to extract basic trends from the results rather than to estimate the solution at specific points within the domain, such as extreme points of the field. 
Highly refined meshes  used to highlight the extremes can obscure such trends simply because of their complexity.  
In such cases a reduction in complexity can be useful.  
Expansion methods offer reduced-order descriptions in which systematic variations are captured with a relatively small number of parameters in comparison to the entire finite element representation. 
Fourier series are a well-known example.
For polycrystals in which field data are represented with finite element discretizations, such 
expansions can be written with respect to the nodal values of the finite element representation:
\begin{equation}
\{ A \} = \sum_k \beta^k \{ U^k \}
\label{eq:Aexpansion}
\end{equation}
Here,  $ \{ U^k \}$ constitutes one of $k$ modes that quantify systematic spatial variations ranging from a constant value over the volume to more complex linear or quadratic distributions.  
The $\beta^k $ are weighting coefficients.
The larger the relative value of one weight in comparison to others, the more significant
the functional dependence embodied in the corresponding mode.
The modes are determined {\it a priori}, so that when the expansion is substituted into 
Equation~\ref{eq:fem_rep}, the number of degrees of freedom in representing $a({\bf{x}})$ decreases from the 
number of finite element nodes to the number of spherical harmonic modes. 
This easily can be a factor greater than 10, and readily 100 or more.
The value of the expansion thus lies in its ability to efficiently identify dominant trends in the distributions, which were first represented using a detailed finite element representation, that 
may have a number of interesting features.

\subsection{Determination of the modes,  $\{ U^k \}$ }
\label{sec_modes}

To represent the field data with an expansion a set of modes are needed.  
One approach is to compute a set of discrete harmonic function from Laplace's equation:
\begin{equation}
	\nabla^2 u  = 0	
\end{equation}
subject to the boundary condition of zero fluxes\cite{harmonics,wiki:laplace}.  
In the absence of essential boundary conditions, this system of equation is singular and thus has no unique solution.
Instead, we seek the singular values for the system given by:
\begin{equation}
	\nabla^2 u  -  \lambda u  = 0
	\label{eq:singularform}
\end{equation}
By performing a singular value decomposition (SVD),  
a set of singular values and vectors, $(\lambda^i , \{ U^i \} )$, may be extracted~\cite{wiki:svd}.
The singular vectors are the harmonic modes, or bases, of the expansion and are orthonormal.
The number of modes equals the number of degrees of freedom, which here equals
the number of nodes in the mesh.
Clearly, no reduction-in-order is obtained if all the modes are computed and retained in the expansion.
Consequently, the number is restricted to a fraction of the number of nodes.  
Typically those modes with the smallest (magnitude) of the singular values are retained as they characteristically present the simpler distributions. 
Using only the lower modes  does not present an obstacle as 
linear algebra packages have SVD routines in which the number of modes 
being computed can be specified by the user.

A challenge does arise in computing the modes for grains with a polycrystal, however,
from the grain domains (volumes).  
The grains of a polycrystal typically have different sizes and shapes.  
In fact, it would be unusual to have two grains with exactly the same shape and size.
The harmonic modes depend on the domain size and shape, and differ from grain-to-grain.
The modes, therefore, must be computed specifically for each grain.
The methodology for this is presented in Section~\ref{sec:fem}.  
Gradients of modes are computed simply as $\nabla u$ using the derivatives of the interpolation functions.

\subsection{Finite element implementation}
\label{sec:fem}
\mechmonics\, is a finite element code that executes in a \matlab~\cite{matlab} environment.
It uses finite element meshes created by \neper~\cite{Quey_Neper_2011,Quey_Neper_2018} but
modified as described in Section~\ref{sec_modes}.
\mechmonics\, delivers files with the nodal point values of the discrete harmonic modes.
Modes are determined grain-by-grain.  
The finite element development begins by forming a residual from Equation~\ref{eq:singularform} 
independently over each grain:
\begin{equation}
 R = \int_V  \boldsymbol{\psi} \left(  \nabla^2 u  -  \lambda u  \right)\dee {V} 
\label{eq:weightedresidual}
\end{equation}
 where $\boldsymbol{\psi}$ are the weighting functions. 
The weak form of Equation~\ref{eq:singularform} is obtained by operating on the first term by integration by parts and requiring the result to equal zero:
\begin{equation}
 R = \int_V  { \nabla \boldsymbol{u} } \cdot {\nabla \boldsymbol{\psi}  } \dee {V} + \int_V \boldsymbol{\psi}  \lambda \boldsymbol{u} \dee {V} 
\label{eq:weakform}
\end{equation}
Zero-flux natural boundary conditions have been applied to eliminate the surface integral that results from using the 
divergence theorem to convert one of the volume integrals to a surface integral. 
Substitution of the trial and weight functions into Equation~\ref{eq:weakform} delivers a
matrix equation for the nodal displacements:
\begin{equation}
\left( \Matstiff  + \lambda\Matmass  \right)  \{ U \} = 0
\label{eq:globalmatrixequation}
\end{equation}
where  $\Matstiff $ is the assembled versions of the elemental coefficient  matrix:
\begin{equation}
\matstiff  =  \int_{V^{\it e}} [ \frac{dN}{dx}]^\transp  [ \frac{dN}{dx}]\dee{V}
\label{eq:femstiffnessmatrix}
\end{equation}
and  $\Matmass$ is the assembled versions of the elemental coefficient  matrix:
\begin{equation}
\matmass =  \int_{V^{\it e}} [N]^\transp  [N]\dee{V}
\label{eq:femmassmatrix}
\end{equation}
A simplification is possible at this point by replacing  the matrix $\Matmass$ with the identity matrix $[I]$.  In the mechanics literature for the dynamical behavior of solids, $\Matmass$ is scaled by the material density and is referred to as a consistent mass matrix.  Using the identity matrix scaled by the mass is referred to as a lumped mass matrix and is equivalent to lumping mass at the nodes rather than distributing it over the volume (which is the effect provided by $\Matmass$).  Both options are available in \mechmonics.  In the example presented later, this approximation was employed.  However, we expect that most users, including ourselves,  will opt for the full mass matrix as it is a more precise representation rendition of the residual. 

Equation~\ref{eq:globalmatrixequation} is computed every grain of a virtual polycrystal independently of all other grains.  
Equation~\ref{eq:globalmatrixequation} has the form of a generalized eigenvalue problem, which can be
analyzed directly with the \rm{eigs} function in \matlab\, to evaluate the eigenpairs, ($\lambda^i$,  $\{ U^i \}$).
The total number of eigenpairs equals the number of degrees of freedom of the mesh, but typically many fewer are employed in the expansions.  
The desired number of eigenpairs can be designated in  \rm{eigs}.  

\subsection{Evaluation of expansion weights}
\label{sec:expansion weights}
The reduced order representation of a field variable is given in Equation~\ref{eq:Aexpansion}.  The modes,   $\{ U^k \}$,  in this equation are the provided by the eigenvectors obtained from Equation~\ref{eq:globalmatrixequation}.
The weights, $\beta^k$, of the expansion must be determined from data, specifically the nodal point values of
a field variable on the same finite element mesh for which the eigenpairs are known. 
There is more than one way to evaluate the mode weights; one is to take advantage of the 
orthogonality properties of the eigenvectors. 
Orthogonality for $\Matstiff $ and $\Matmass$ may be stated as:
\begin{equation}
\{ U^i \}^T \Matstiff  \{ U^j\}   = \delta_{ij}\beta^j  
\label{eq:ortho4K}
\end{equation}
and
\begin{equation}
\{ U^i \}^T  \Matmass   \{ U^j \}   =  \delta_{ij}
\label{eq:ortho4M}
\end{equation}
Expansion coefficients are evaluated in the customary way for expansions with orthonormal modes.
To obtain $\beta^k$, the representation of $\{ A \} $ in Equation~\ref{eq:Aexpansion}  is multiplied by the transpose  $  \Matmass   \{ U^k \}$:
\begin{equation}
\{ U^j \}^T  \Matmass^T   \{ A \} = \{ U^j \}^T  \Matmass^T  \sum_k \beta^k \{ U^k \}
\end{equation}
Enforcing the orthogonality relation from Equation~\ref{eq:ortho4M}  isolates $\beta^k$:
\begin{equation}
\beta^k  = \{ U^k \}^T  \Matmass^T \{ A \}  
\end{equation}
This approach for determining the expansion weights presupposes that the field is known in terms of a piecewise polynomial representation given by Equation~\ref{eq:fem_rep}.   Further, if the 'lumped mass' approximation is used, $\Matmass$ is replaced by the identity matrix.  

\mechmonics\, evaluates the eigenpairs for all the grains in a polycrystal (plus their gradients and the principal axes of every grain).  It does not evaluate the weights for any field data expansions.  This can be done with the auxiliary code, {\bf FitFieldDataWithHarmonics}.  Note that {\bf FitFieldDataWithHarmonics}\, also has the capability to evaluate the expansion weights with an optimization method.  The results are comparable, but using orthogonality is computationally superior.

\subsection{Finite element interpolation functions}
\label{sec:interpolationfunctions}
\begin{figure}[h]
\begin{center}
\includegraphics*[width=6cm]{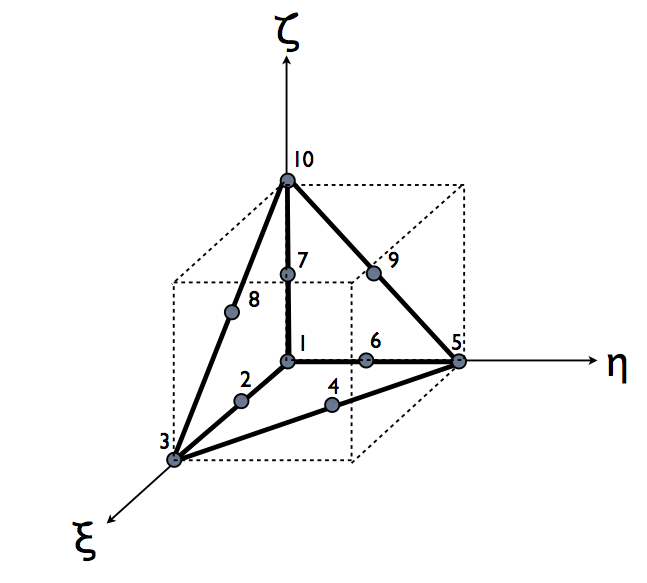}
\caption{10-node tetrahedral element with quadratic interpolation of the velocity, shown in the parent configuration and bounded by a unit cube.}
\label{fig:tet_element}
\end{center}
\end{figure}

For completeness, we again mention that \mechmonics\, utilizes isoparametric elements in the both the original methodology and in the alternative methodology.  
Using isoparametric elements, 
the mapping of the coordinates of points are specified by interpolation functions, $\matcapN$, and the coordinates of the nodal points, $\left\{ X \right\}$:
\begin{equation}
\left\{ x \right\}  = \matcapN  \left\{ X \right\}
\label{eq:coord_mapping}
\end{equation}
where $(\xi, \eta, \zeta)$ are local coordinates within an element. 
The same mapping functions are used for the solution (trial) functions and the weights (making this a Galerkin-type formulation):
\begin{equation}
a({\bf{x}}) = [ N({\bf{x}})] \{ A \} 
\end{equation}
and
\begin{equation}
\Big\{ \psi \Big\} =  \matcapN  \Big\{ {\it \Psi } \Big\}
\label{eq:trial_functions}
\end{equation}
Gradients of modes computed as:
\begin{equation}
  \frac{da}{dx}  =  [ \frac{dN}{dx}]  \{ A \} 
\end{equation}
For compatability with \neper, \fepx, and \mechmet, \mechmonics\, uses a 10-node, tetrahedral, serendipity element, as shown in Figure~\ref{fig:tet_element}. 
This $C^0$ element provides pure  quadratic interpolation of the trial and weight functions.  
The elements having $C^0$ continuity  means that the trial and weight functions are continuous across (and within) elements, but their spatial derivatives are not continuous across element boundaries.  

\subsection{Extraction of grain modes from Equation~\ref{eq:globalmatrixequation}.}
\label{sec:modeextraction}

The objectives of \mechmonics\, is to generate sets of discrete harmonic modes for every grain in a virtual polycrystal and to approximate field data using the modes in an expansion.
To accomplish this, the grains must be treated separately from one another. 
The mesh generated by \neper\, has nodal numbering globally over the entire mesh and enforces continuity of the trial and weights across the grain
boundaries by having grains share nodal points on the grain boundaries.
This mesh is modified In \mechmonics\, to introduce redundant nodes on the grain boundaries so that grains do not share nodes.   Supplementary meshes are then defined for each grain using only the nodal points for that grain.  This facilitates solving for the modes grain by grain.

The modes for a particular grain are the singular vectors of $\Matstiff$ evaluated for that grain. 
This is accomplished in \mechmonics\, with built-in functions available in  \matlab\,. 
Namely, either an eigenvalue function ({\it eigs})  or a singular value function ({\it svds})
is called depending on whether of not the mass matrix simplification is invoked 
(if the matrix $\Matmass$ has been replaced with the identity matrix $[I]$). 

\subsection{Supplemental grain  metrics}
\label{sec:grainaxes}
As previously discussed, the discrete harmonic modes are determined from Laplace's equation applied to the grain domains.  Only the grain geometry enters the formulation; there are no physical properties required.  Further, no boundary conditions are applied.  Consequently, the only relevant difference between the matrix equation used to determine modes for two grains within a polycrystal stems from the differences in their geometries.  Because the grains can have unique geometries, the sets of modes for any given grain can be different from every other  grain in the polycrystal.   
However, the modes do tend to share certain characteristics from grain-to-grain, as discussed in 
Section~\ref{sec:harmonicmodes}.
To quantify the possible correlations between the modes and the geometry of a grain, 
a grain shape tensor~\cite{bar_daw_01a} is computed in \mechmonics.  
To begin, the grain centroid, $\bar{\vctr{x}}$, is determined from the element centroids, $\hat{\vctr{x}} $.
Next, the relative position of the element centroid, $\tilde{\vctr{x}} $, to the grain centroid is defined as:
\begin{equation}
\tilde{\vctr{x}} = \hat{\vctr{x}}  - \bar{\vctr{x}}
\end{equation}
The  shape tensor is computed as the weighted sum of the diads formed using the relative 
centroid position vectors:
\begin{equation}
\tnsr{M} = \sum_{i=1}^{m} \varphi^{i} (\tilde{\vctr{x}}^{i} \times \tilde{\vctr{x}}^{i} )
\end{equation}
where the weights, $\varphi^{i} $, are the ratios of the element volumes to the grain volume. 
Summation is performed over the total number of elements in the grain, $m$.
The final step is to perform a singular value decompositon on $\tnsr{M}$. 
The singular vectors give the principal directions of the spatial orientation of the grain volume,
referred to here as the grain axes.  The singular values are the principal values, which are the squares of the grain diameters.
The shape tensor is an ellipsoid whose principal values are the grain diameters.  
\subsection{Computational platforms}
\label{sec:platforms}
The \mechmonics\, source code is available at the GitHub repository: \\
https://github.com/CHESSComputing/StructuralMaterialsRepository \\
in the MechToolBox folder.
In addition to the source code, the repository contains example files and input instructions.
The required input data consist solely of the output files from \neper, namely the microstructure and mesh files.
Output files consist of  mode data and visualization files in VTK format.
\mechmonics\, has been developed  as a \matlab\, program for use on laptop and desktop computers.  
Execution on other platforms with \matlab\, installed is also possible, of course.  The demonstration application presented in Section~\ref{sec:demo_application}, for example,  executed on a MacBook Pro with 6 cores and 64 GB memory.  (The simulations of elastoplastic responses of the sample used to generate data were performed on 64-core linux cluster using \fepx, which is a distinct code.).

\section{Demonstration -- Extension of FCC polycrystals}
\label{sec:demo_application}
To demonstrate the application of \mechmonics, the harmonic modes of three variants of one sample were computed and used to approximate the axial stress over a loading event (tensile extension to 1\% nominal strain).  The three variants of the sample were constructed to alter the grain morphologies, changing the uniformity of the grain size (diameter) and the uniformity of the grain sphericity.  The intent is to illustrate qualitatively the sensitivity of the harmonic mode expansions to the sample instantiation.  The sample variants are in some ways similar and other ways different owing to using the same starting points in their instantiations but afterward invoking constraints on grain shape and size.  

In this section we discuss the \neper\, instantiations of the three variants in terms of their grain diameter and grain sphericity statistics.  The grain structure is presented for each sample variant along with maps of the single-crystal directional stiffness and Schmid factor defined in \mechmet~\cite{mechmet_immi}.  
Mechanical attributes of the variants, evaluated using the sibling code, \mechmet~\cite{mechmet_immi}, are presented next for extensional loading.  
This is followed with presentation of the harmonic modes computed with \mechmonics and their associations with the grain axes. 
These modes are used to represent the axial stress generated with \fepx\, for extension to 1\% nominal strain.  
Attention is focused on 5 specific grains for the sample variants.
For these 5 grains, both the 'raw'  (\fepx \, data) and the harmonic mode expansion stress distributions are shown for several points in the loading sequence.
Finally, the evolution of the mode weights are given for the 5 grains to illustrate how the stress is re-distributing through the elastic-plastic transition.

\subsection{Sample instantiation and attributes}
The discrete harmonic modes that are defined by the Laplace's equation depend on the grain geometry.  
Samples instantiated by means of tessellations that are seeded with random points (irregularly positioned) will have dissimilar grains and thus dissimilar harmonic modes.  A question that naturally arises, then, is to what extent are the modes sensitive to differences in the instantiations.  The sample variants explored here were defined to exhibit distinctly different characteristics.  

\begin{figure}[htbp]
	\centering
	\begin{subfigure}[b]{.3\textwidth}
		\centering
		\includegraphics[width=1\linewidth]{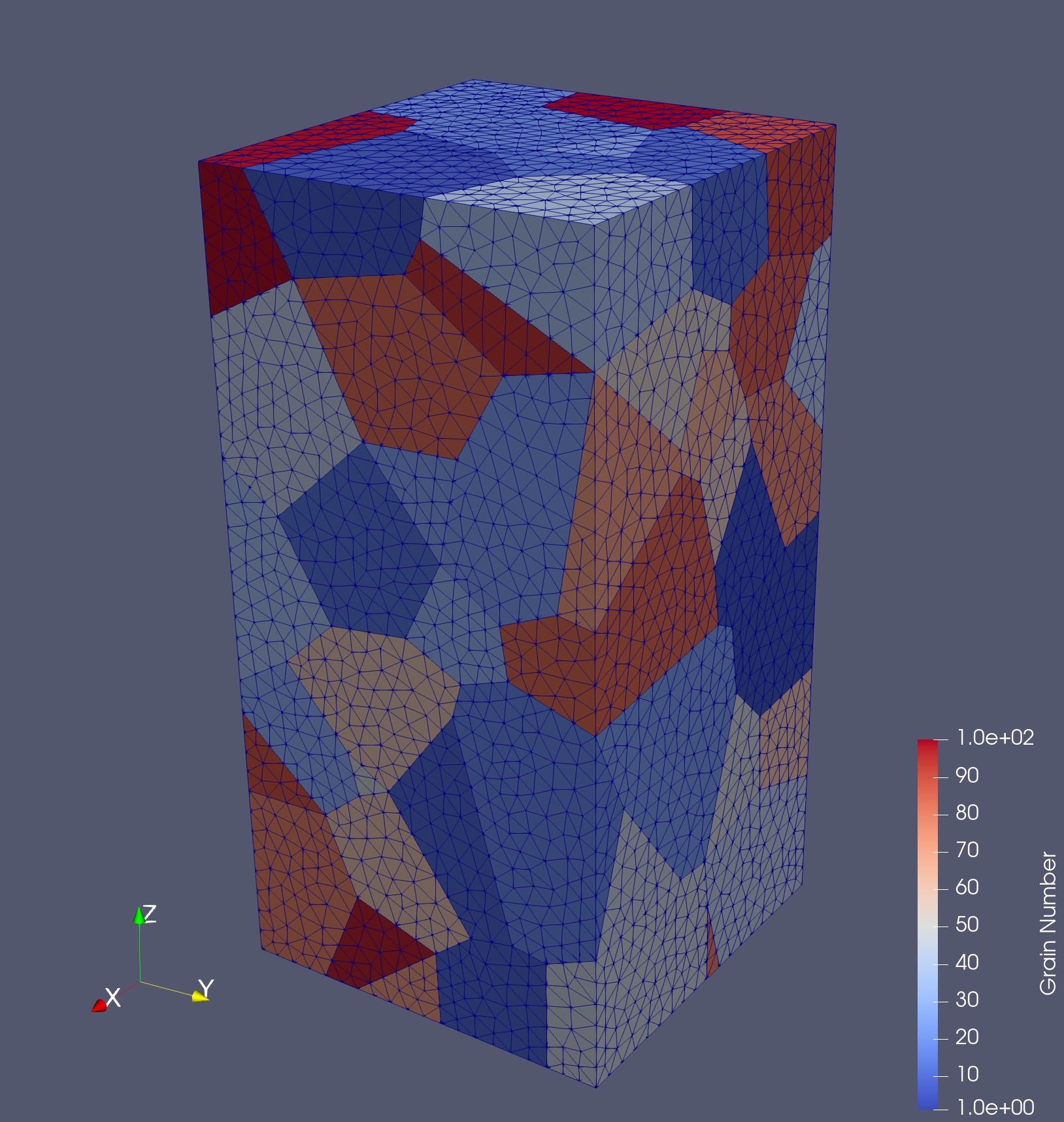}
		\caption{ }
		\label{fig:voronoi_mesh}
	\end{subfigure}%
	\quad
	\begin{subfigure}[b]{.3\textwidth}
		\centering
		\includegraphics[width=1\linewidth]{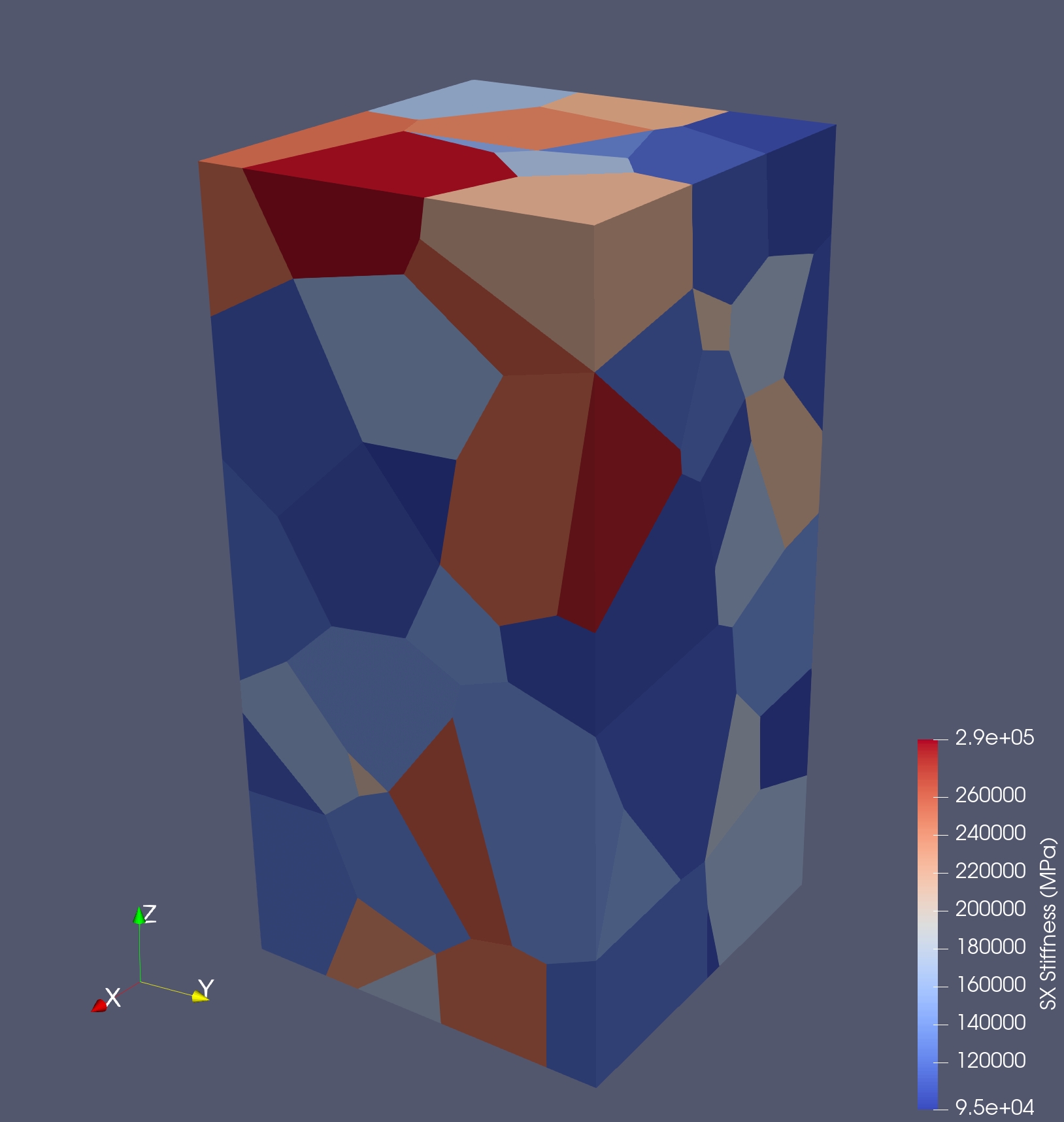}
		\caption{ }
		\label{fig:voronoi_sxstiff}
	\end{subfigure}%
	\quad
	\begin{subfigure}[b]{.3\textwidth}
		\centering
		\includegraphics[width=1\linewidth]{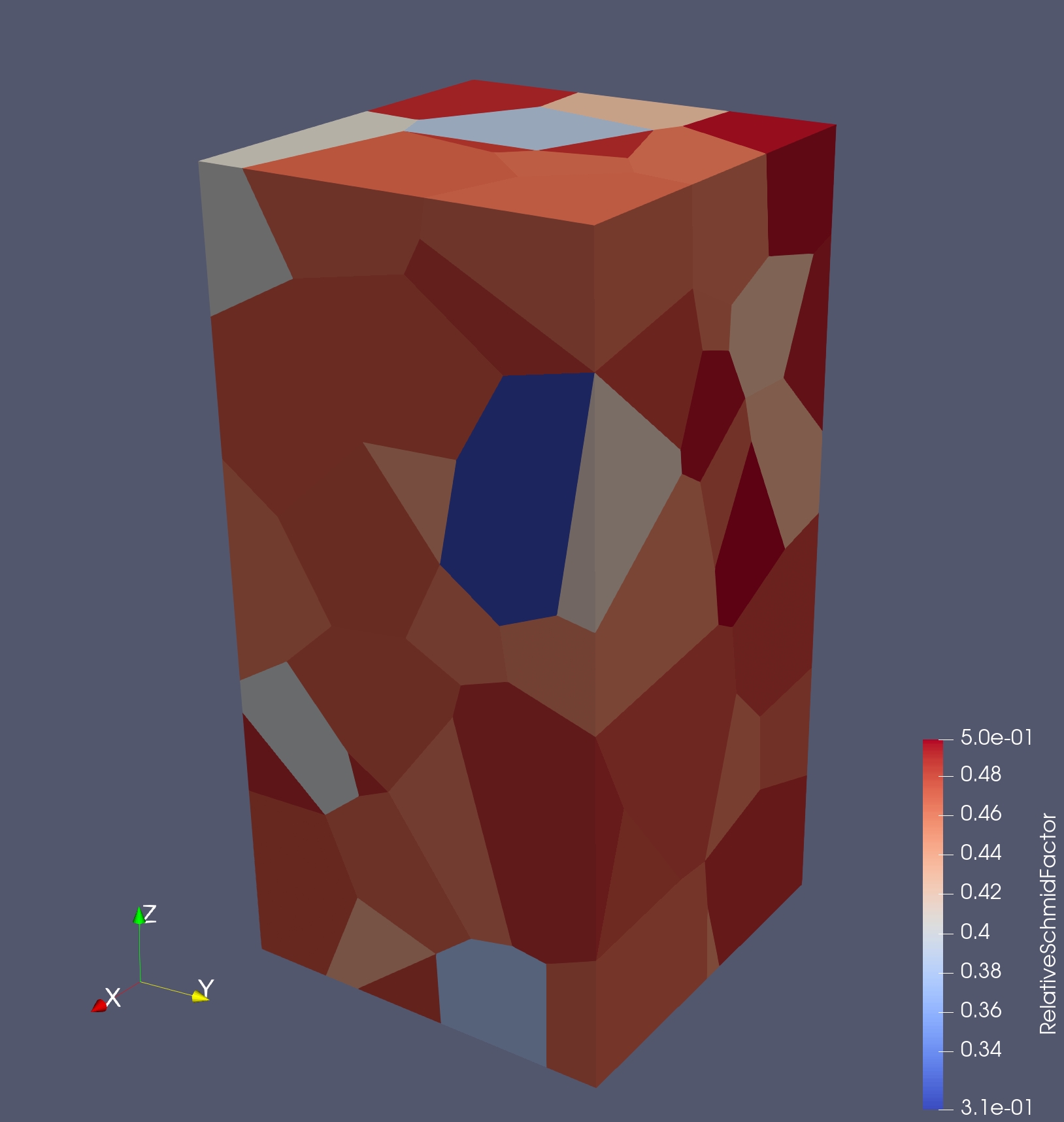}
		\caption{ }
		\label{fig:voronoi_relschmid}
	\end{subfigure}%
	\caption{Attributes of the Voronoi sample.  (a) mesh over grains; (b) single crystal stiffness; (c) relative Schmid factor. }
		\label{fig:sample_voronoi}
\end{figure}

\begin{figure}[htbp]
	\centering
	\begin{subfigure}{.3\textwidth}
		\centering
		\includegraphics[width=1\linewidth]{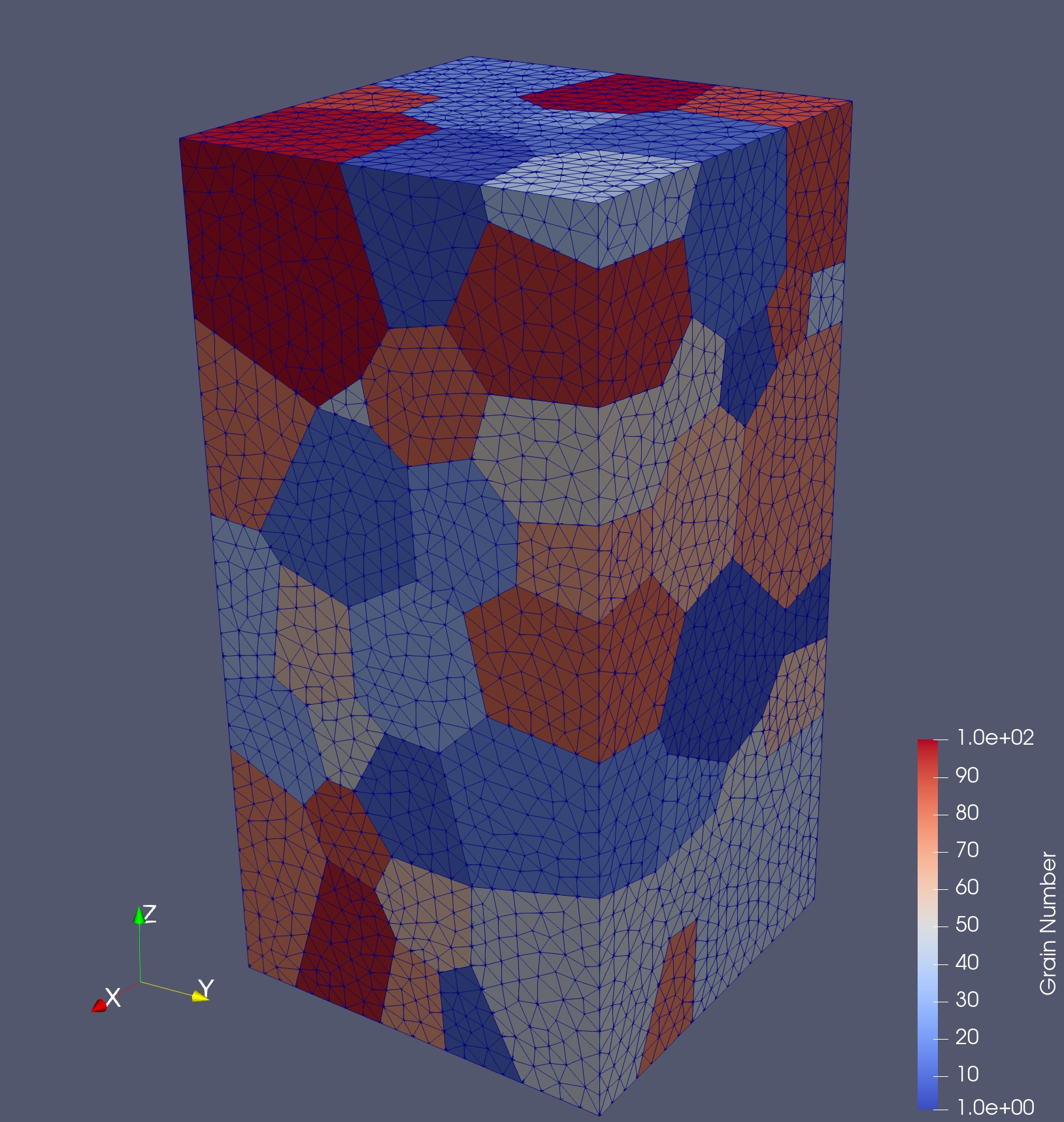}
		\caption{ }
		\label{fig:dia0p35_sph0p06_mesh}
	\end{subfigure}%
	\quad
	\begin{subfigure}{.3\textwidth}
		\centering
		\includegraphics[width=1\linewidth]{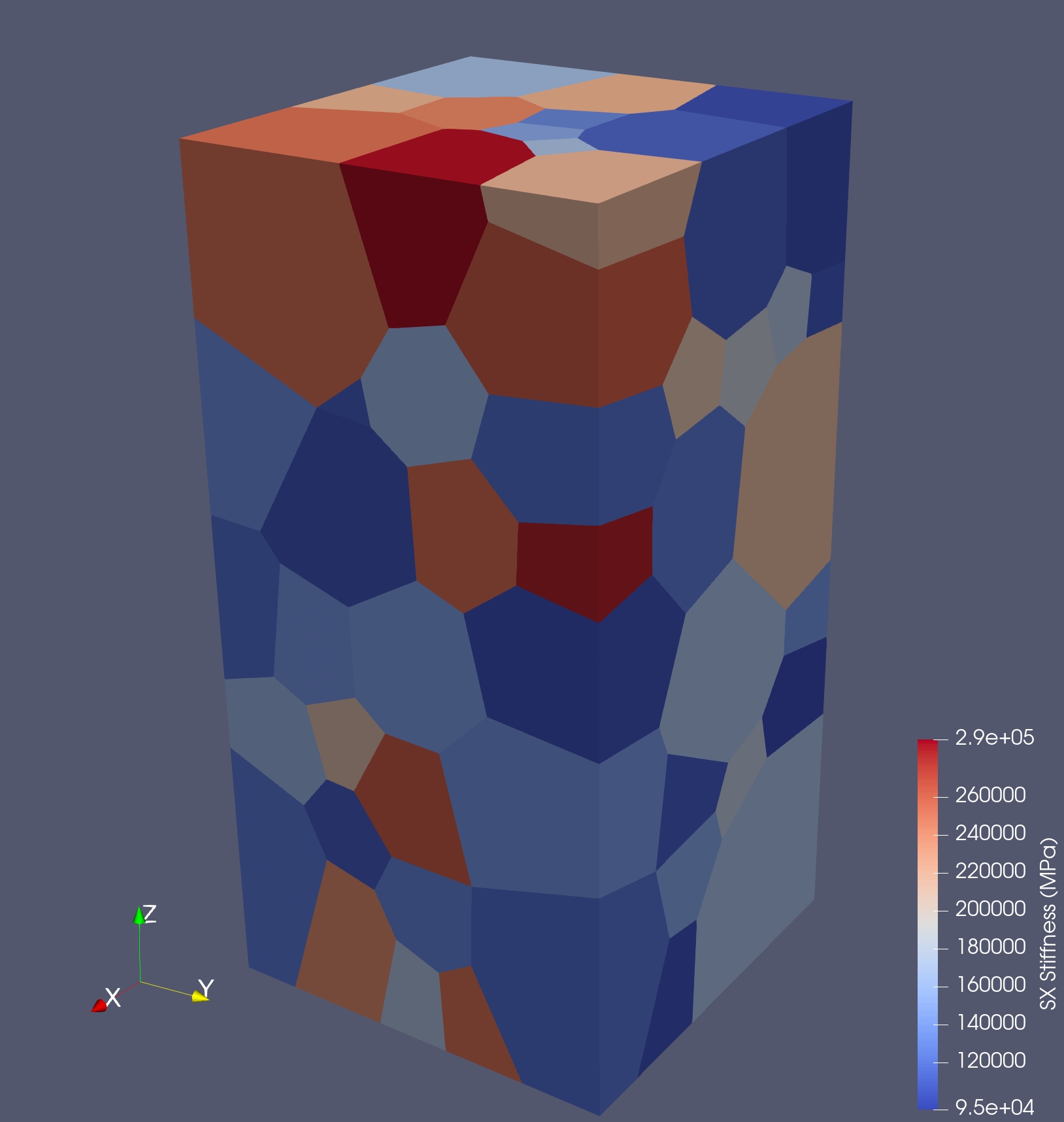}
		\caption{ }
		\label{fig:dia0p35_sph0p06_sxstiff}
	\end{subfigure}%
	\quad
	\begin{subfigure}{.3\textwidth}
		\centering
		\includegraphics[width=1\linewidth]{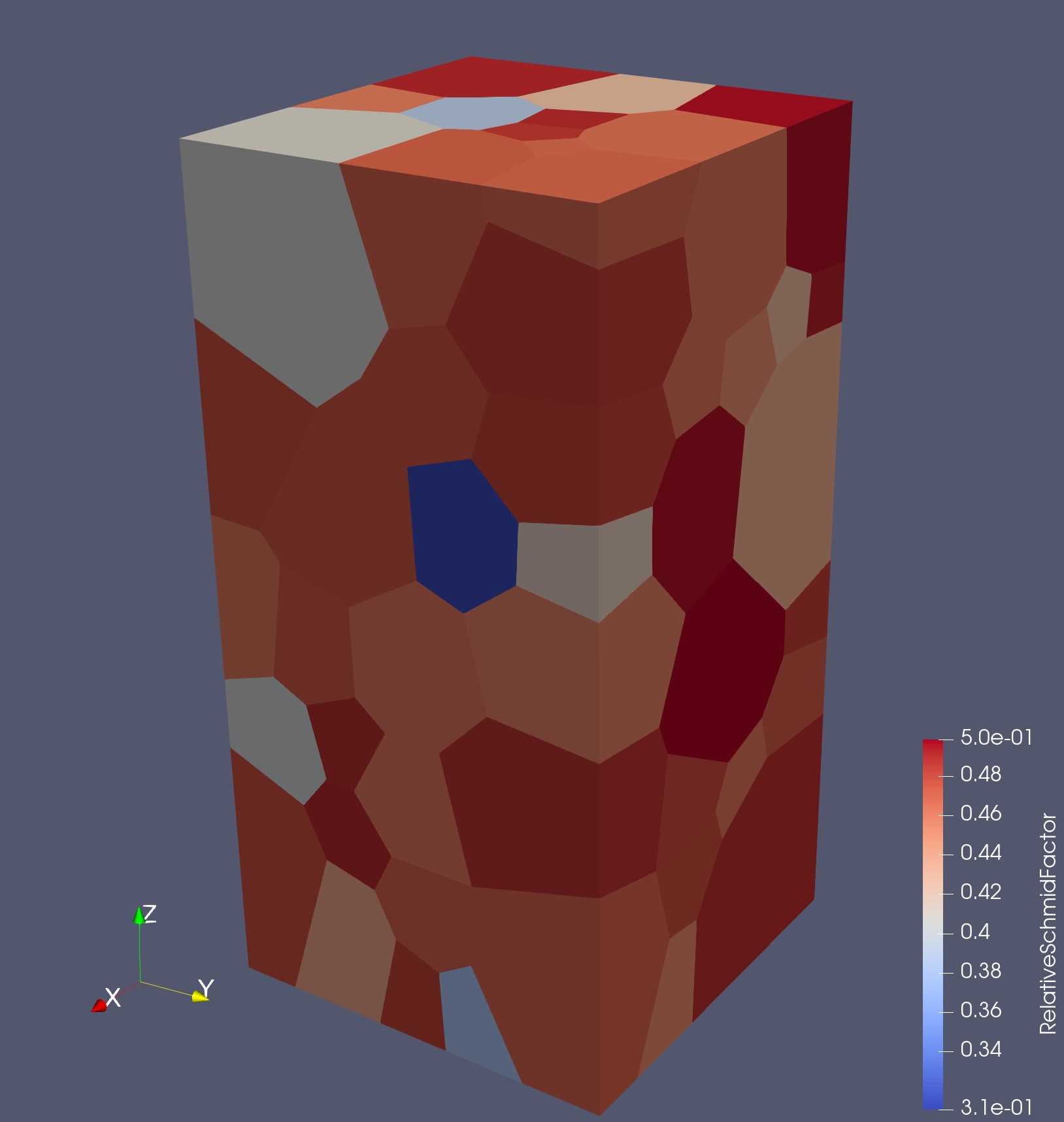}
		\caption{ }
		\label{fig:dia0p35_sph0p06_relschmid}
	\end{subfigure}%
	\caption{Attributes of the LULS sample.  (a) mesh over grains; (b) single crystal stiffness; (c) relative Schmid factor.}
		\label{fig:sample_dia0p35_sph0p06}
\end{figure}

\begin{figure}[htbp]
	\centering
	\begin{subfigure}{.3\textwidth}
		\centering
		\includegraphics[width=1\linewidth]{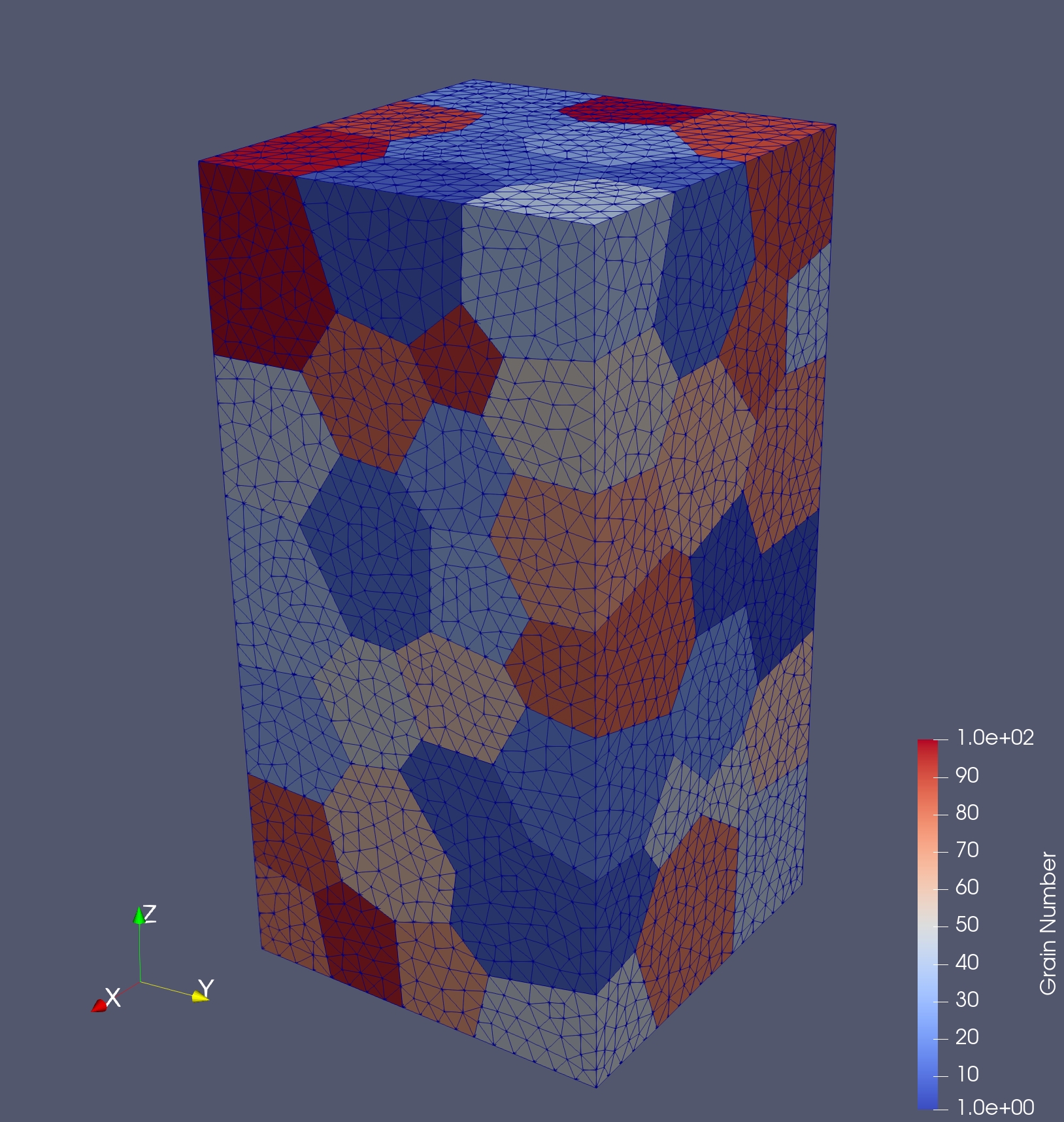}
		\caption{ }
		\label{fig:dia0p15_sph0p03_mesh}
	\end{subfigure}%
	\quad
	\begin{subfigure}{.3\textwidth}
		\centering
		\includegraphics[width=1\linewidth]{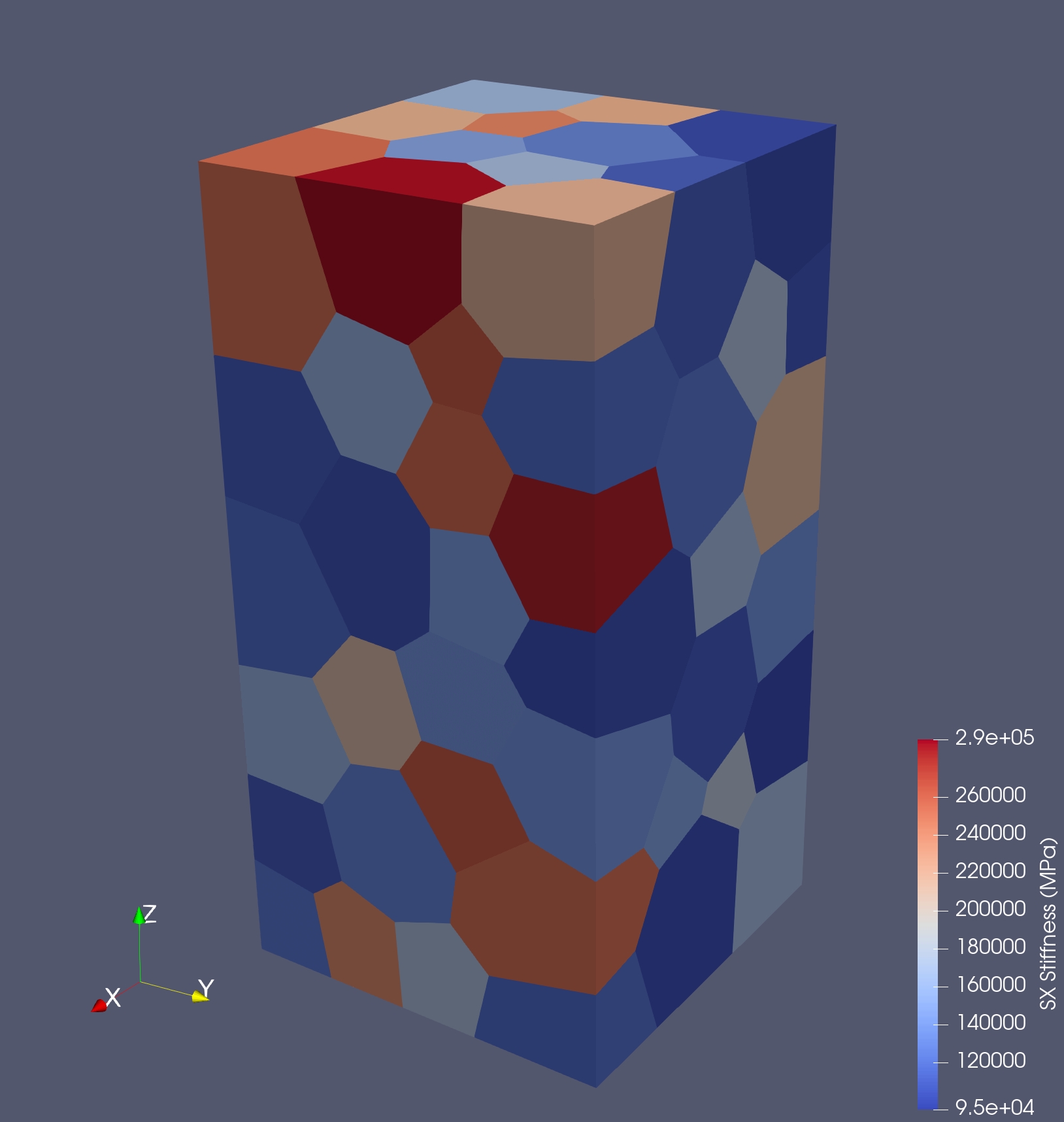}
		\caption{ }
		\label{fig:dia0p15_sph0p03_sxstiff}
	\end{subfigure}%
	\quad
	\begin{subfigure}{.3\textwidth}
		\centering
		\includegraphics[width=1\linewidth]{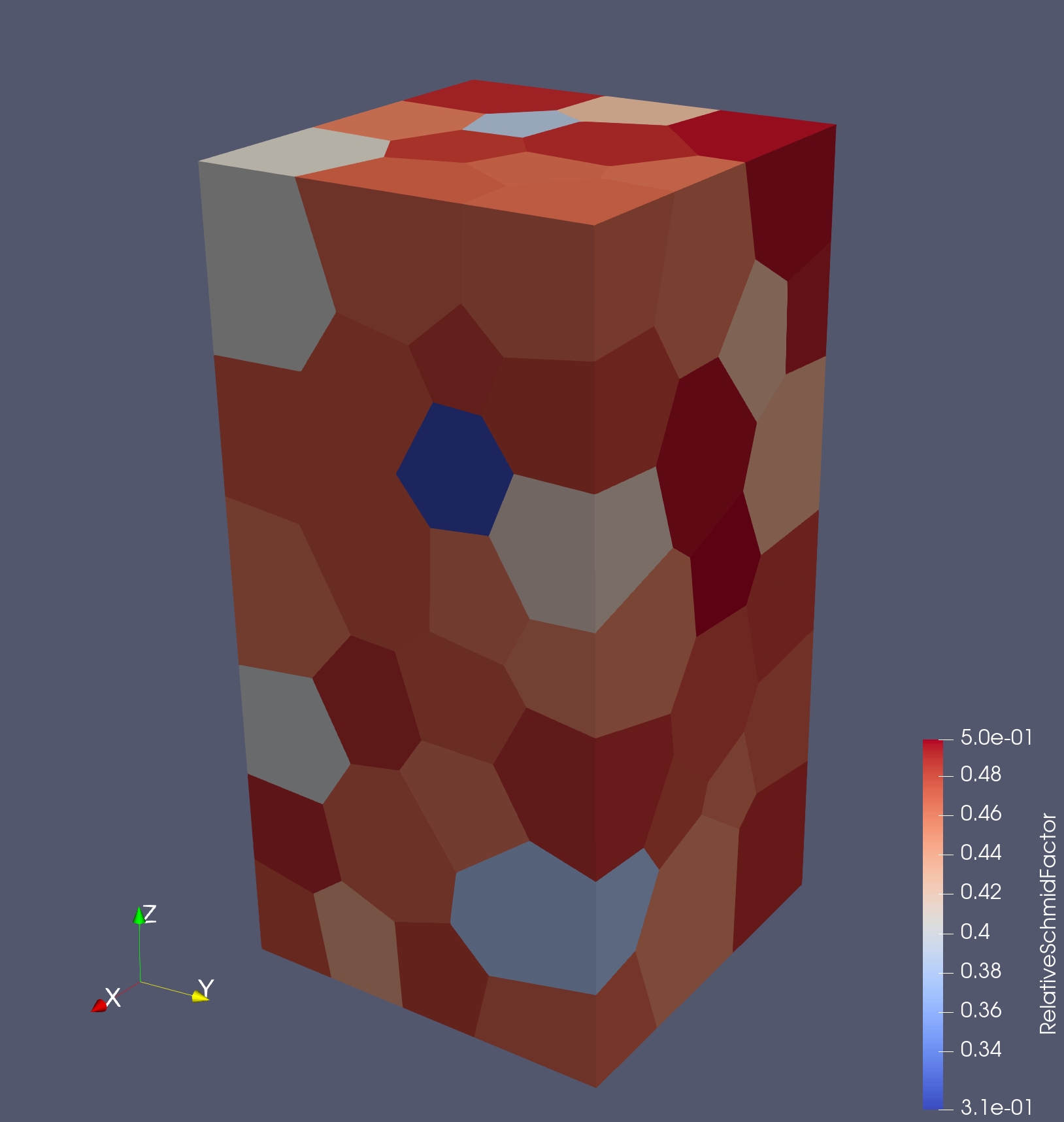}
		\caption{ }
		\label{fig:dia0p15_sph0p03_relschmid}
	\end{subfigure}%
	\caption{Attributes of the HUHS sample.  (a) mesh over grains; (b) single crystal stiffness; (c) relative Schmid factor. }
		\label{fig:sample_dia0p15_sph0p03}
\end{figure}

All variants were generated with \neper\, and have 100 grains.  
All variants are Laguerre tessellations, but have been generated to have different grain morphological characteristics. 
 In particular, tessellations were created with different grain diameter distributions and different grain sphericity distributions by altering the seed attributes (as accommodated in the \neper\, input data).  
The three variants are shown in  Figures~\ref{fig:sample_voronoi}, \ref{fig:sample_dia0p35_sph0p06} and \ref{fig:sample_dia0p15_sph0p03}.  
Respectively, the variants are designated as (a) Voronoi, (b) Low grain size uniformity/low sphericity (LULS) and (c) high grain size uniformity/high sphericity (HUHS).
The  Voronoi tessellation exhibits the most irregular grain shapes.
 The LULS variant has more spherical grains than the Voronoi tessellation, but also has larger variations in the diameter.
  The HUHS variant has greater sphericity and greater size uniformity than either of the two other variants.
  Mean and standard deviation (SD) statistics are listed in Table~\ref{tab:sample_stats}.
\begin{table}[ht]
\small
\centering
\begin{tabular}{c | c c c }
\hline
  &  Voronoi	& LULS & HUHS	\\ 
\hline
 Volume Mean & 0.020 & 0.020 & 0.020 	\\
 Volume SD& 0.0092& 0.0217 & 0.0093 	\\
 Sphericity Mean & 0.761 & 0.837 & 0.862 	\\
 Sphericity SD  & 0.067 & 0.057& 0.0287 	\\
 \hline
\end{tabular}
\caption{Diameter and sphericity statistical metrics for the three sample variants. }
\label{tab:sample_stats}
\end{table}
 
\begin{figure}[htbp]
	\centering		
	\begin{subfigure}{.3\textwidth}
		\centering
		\includegraphics[width=1\linewidth]{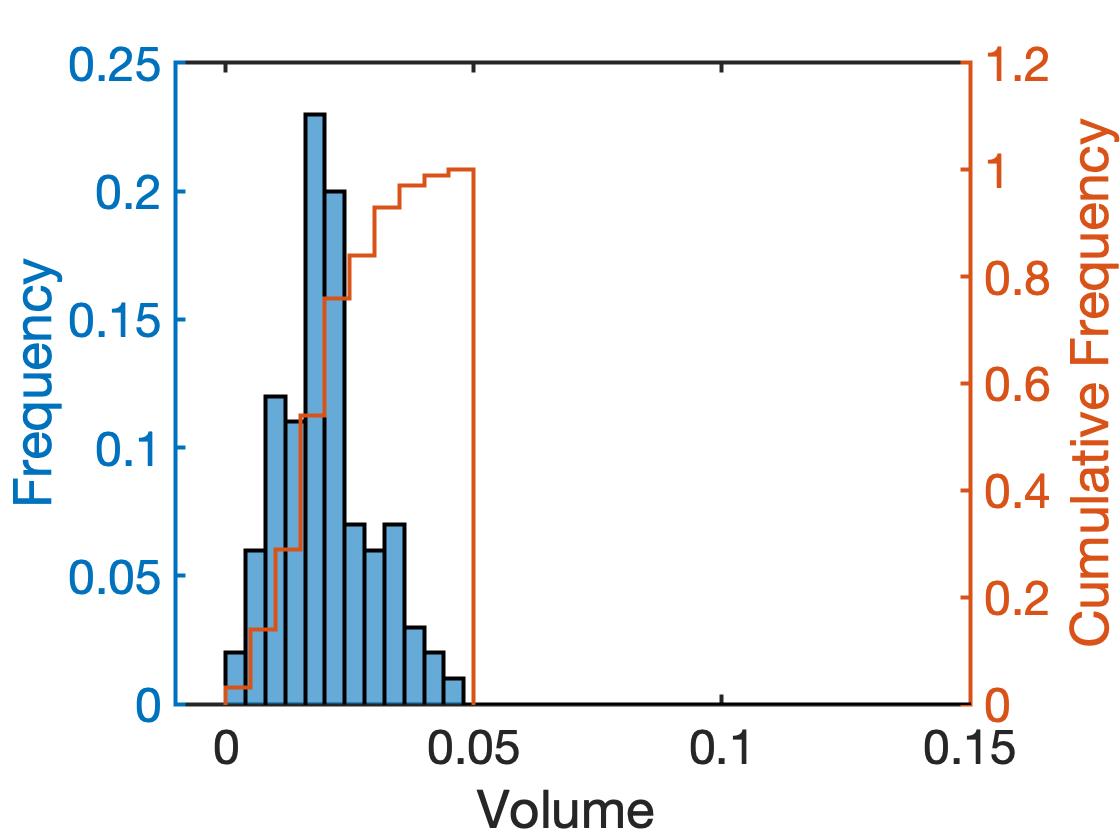}
		\caption{ }
		\label{fig:voronoi_voldist}
	\end{subfigure}%
	\quad
	\begin{subfigure}{.3\textwidth}
		\centering
		\includegraphics[width=1\linewidth]{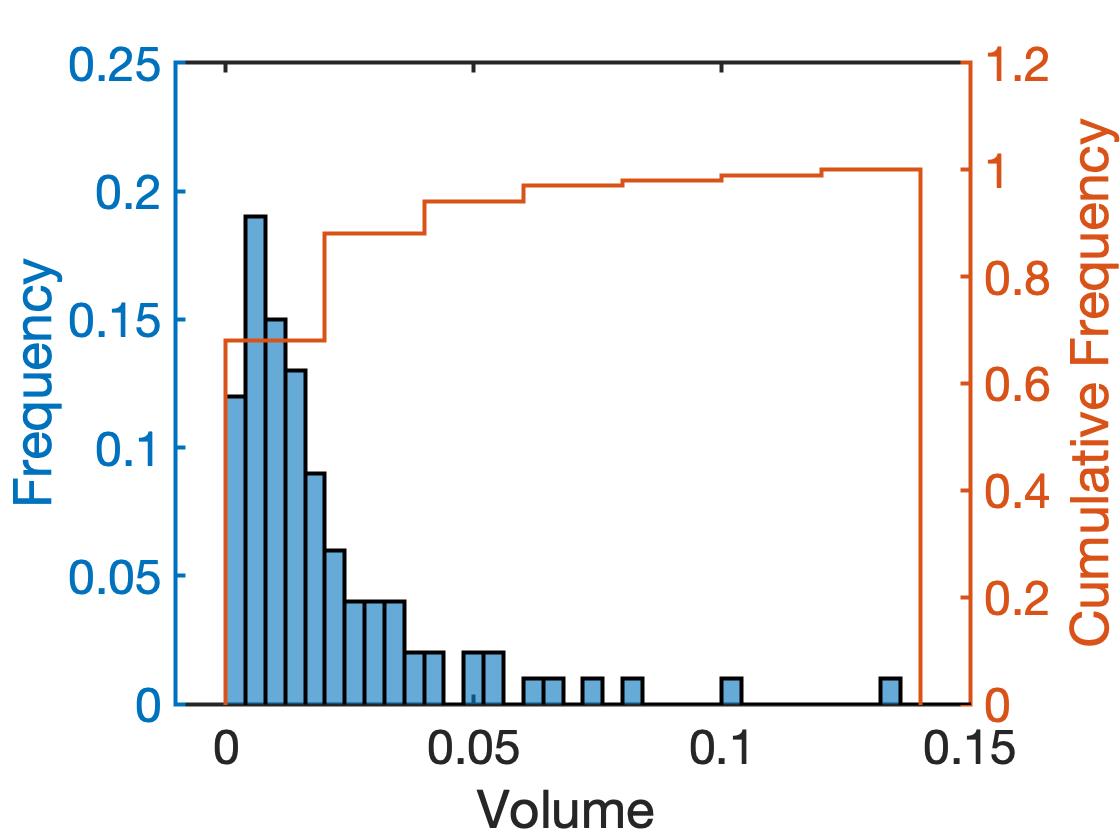}
		\caption{ }
		\label{fig:dia0p15_sph0p03_voldist}
	\end{subfigure}%
		\quad
	\begin{subfigure}{.3\textwidth}
		\centering
		\includegraphics[width=1\linewidth]{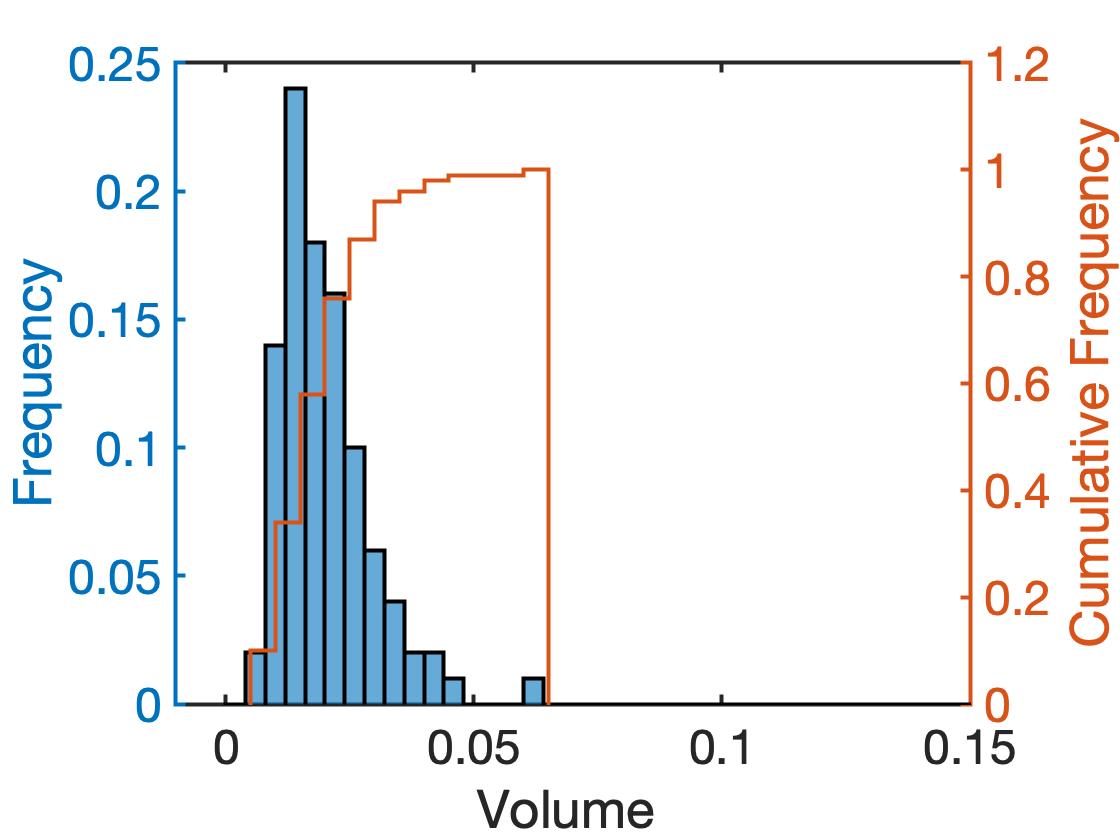}
		\caption{ }
		\label{fig:vdia0p35_sph0p06_voldist}
	\end{subfigure}%
		\caption{Diameter frequency distributions for the sample variants  (a) Voronoi; (b) LULS (c) HUHS. }
		\label{fig:sample_volume_frequency}
\end{figure}

\begin{figure}[htbp]
	\centering		
		\begin{subfigure}{.3\textwidth}
		\centering
		\includegraphics[width=1\linewidth]{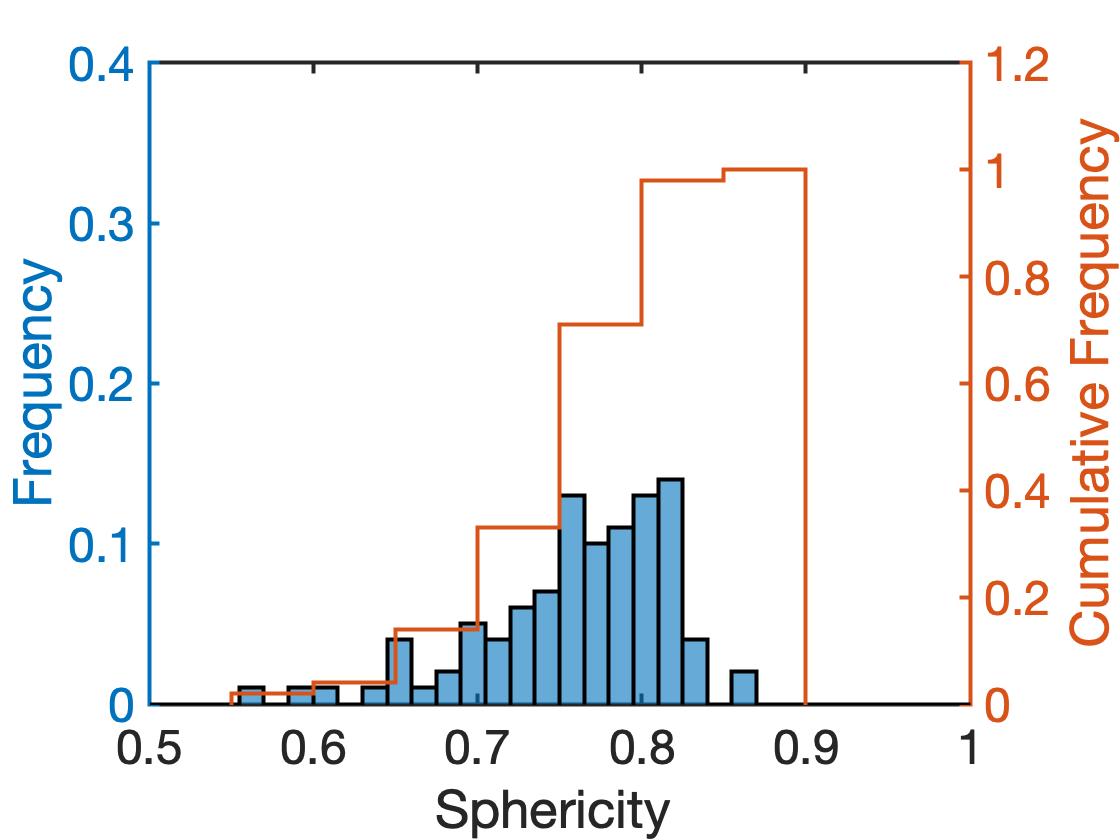}
		\caption{ }
		\label{fig:voronoi_sphdist}
	\end{subfigure}
	\quad
	\begin{subfigure}{.3\textwidth}
		\centering
		\includegraphics[width=1\linewidth]{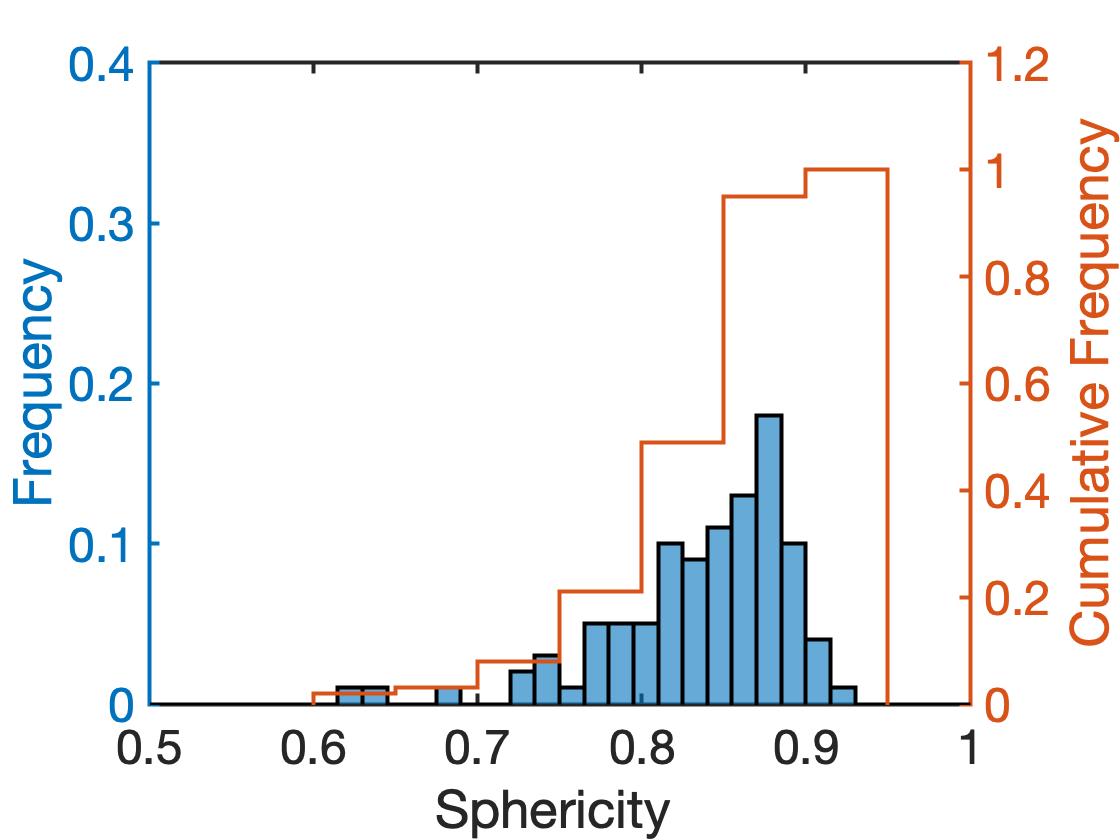}
		\caption{ }
		\label{fig:dia0p35_sph0p06_sphdist}
	\end{subfigure}
	\quad
	\begin{subfigure}{.3\textwidth}
		\centering
		\includegraphics[width=1\linewidth]{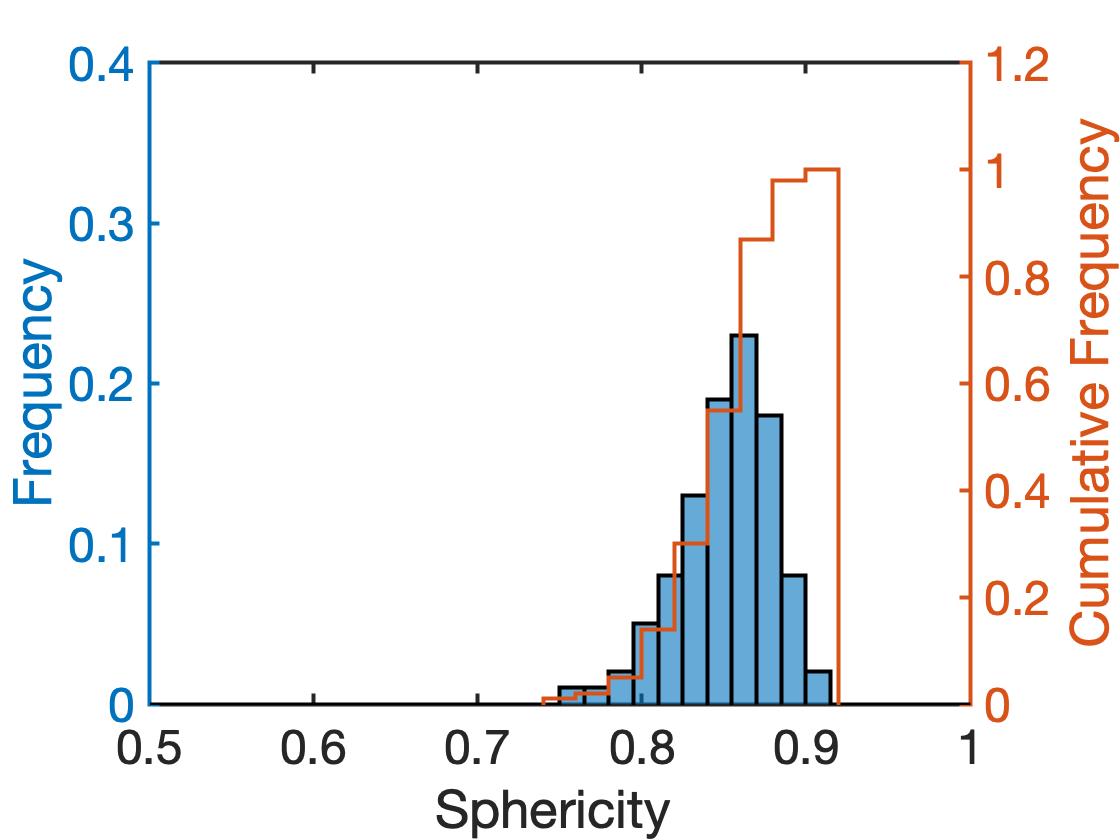}
		\caption{ }
		\label{fig:dia0p15_sph0p03_sphdist}
	\end{subfigure}
		\caption{Sphericity frequency distributions for the sample variants  (a) Voronoi; (b) LULS (c) HUHS.}
		\label{fig:sample_sphericity_frequency}
\end{figure}

  Figures~\ref{fig:sample_voronoi}, \ref{fig:sample_dia0p35_sph0p06} and \ref{fig:sample_dia0p15_sph0p03} show several aspects of the microstructures.
  Part(a) shows the grains with the finite element mesh.  For all variants, the meshing controlled to produce virtual polycrystals with about 100K elements.  
  Lattice orientations were assigned to the grains randomly from a uniform distribution and are   
  uniform over the grains initially.  
  Two single-crystal properties are shown:  the z-direction (axial) directional elastic modulus (Part(b))  and the Schmid factor for z-direction (axial) tension (Part(c)), both evaluated for AL6XN~\cite{pos_daw_mmta_2019a}.
  In Figures~\ref{fig:sample_volume_frequency} and \ref{fig:sample_sphericity_frequency},
  Two frequency distributions are shown: the grain diameter and the grain sphericity.
  Comparisons of these plots provide confirmation of the variant characteristics mentioned above.

\subsection{Sample mechanical response and mechanical metrics}
\label{sec:mechmet_metric}
\begin{figure}[b!]
	\centering
	\begin{subfigure}{.3\textwidth}
		\centering
		\includegraphics[width=1\linewidth]{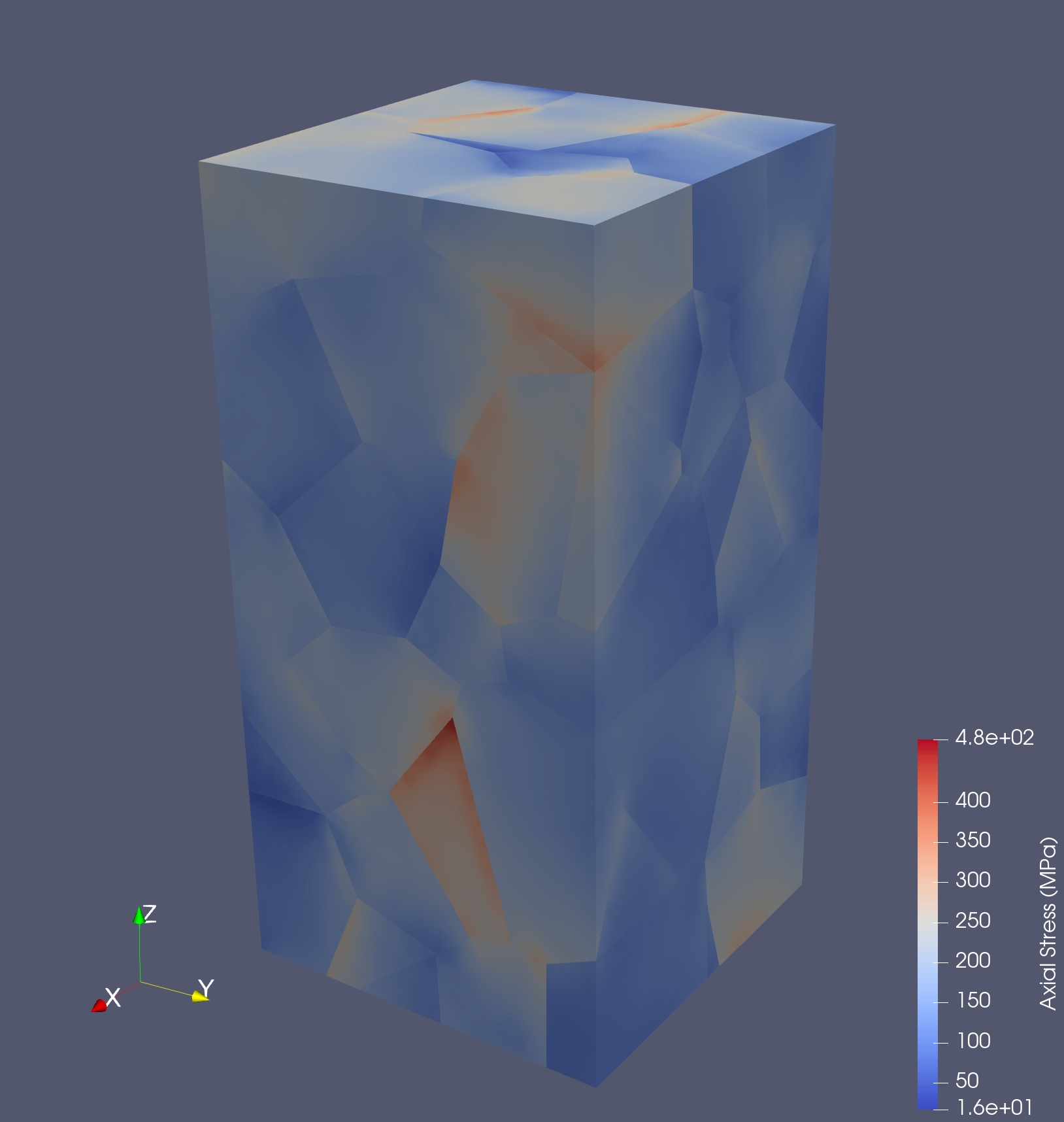}
		\caption{ }
		\label{fig:voronoi_sigzz}
	\end{subfigure}%
	\quad
	\begin{subfigure}{.3\textwidth}
		\centering
		\includegraphics[width=1\linewidth]{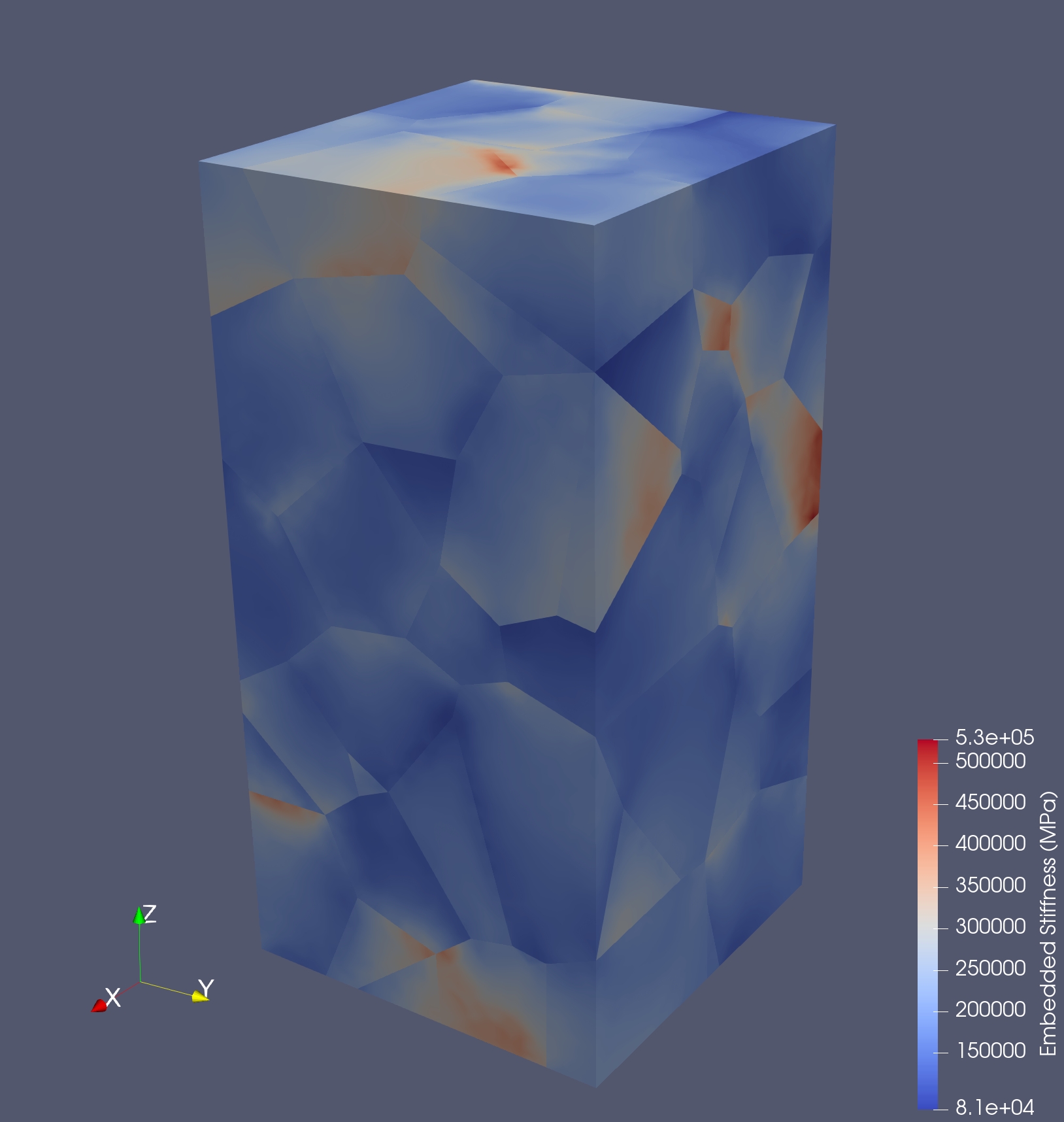}
		\caption{ }
		\label{fig:voronoi_embstiff}
	\end{subfigure}%
	\quad
	\begin{subfigure}{.3\textwidth}
		\centering
		\includegraphics[width=1\linewidth]{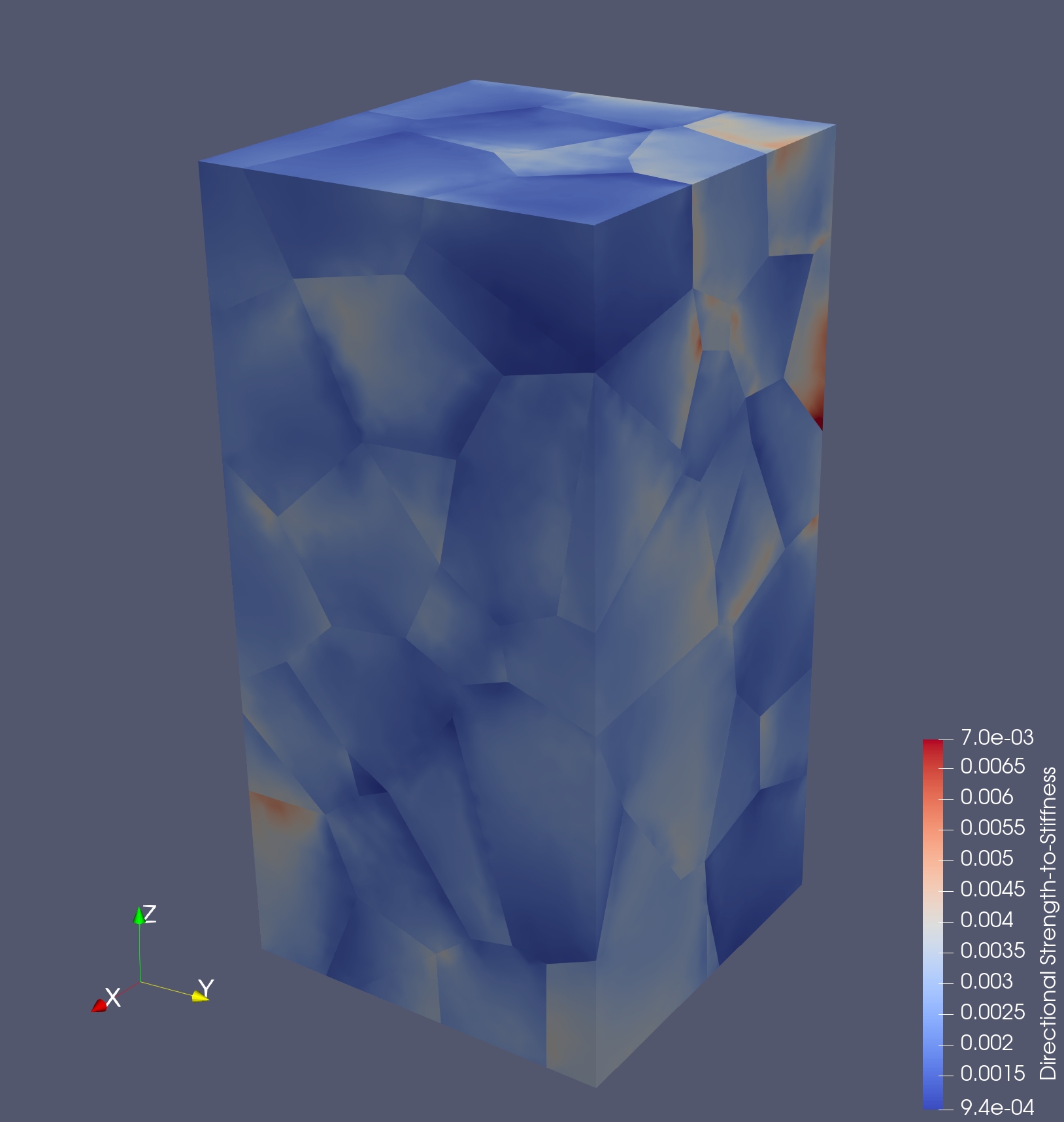}
		\caption{ }
		\label{fig:voronoi_y2e}
	\end{subfigure}%
	\caption{Mechanical metrics of the Voronoi sample.  (a) axial stress; (b) embedded stiffness; (c) directional strength-to-stiffness ratio. }
		\label{fig:metrics_voronoi}
\end{figure}

\begin{figure}[h!]
	\centering
	\begin{subfigure}{.3\textwidth}
		\centering
		\includegraphics[width=1\linewidth]{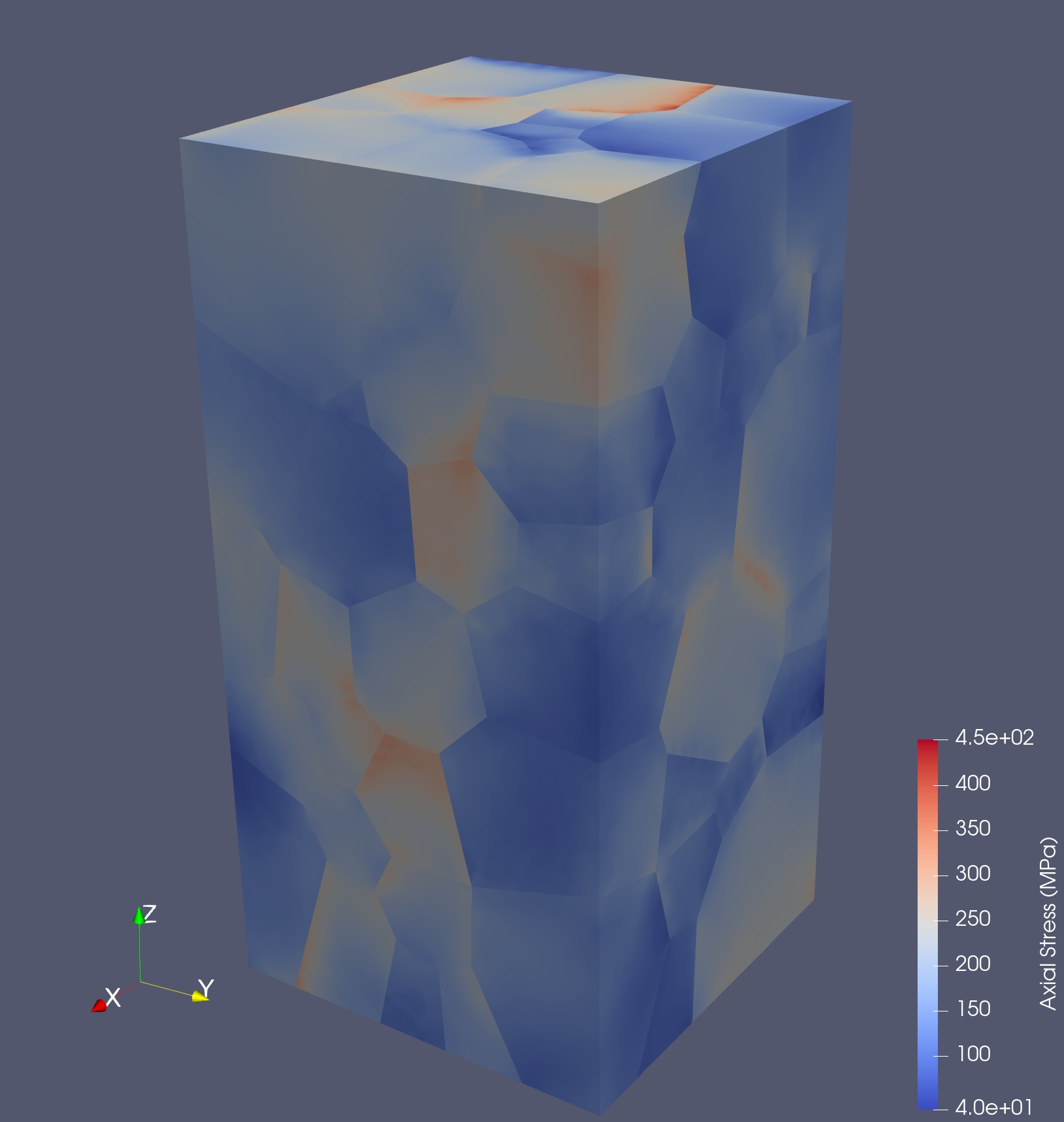}
		\caption{ }
		\label{fig:dia0p35_sph0p06_sigzz}
	\end{subfigure}%
	\quad
	\begin{subfigure}{.3\textwidth}
		\centering
		\includegraphics[width=1\linewidth]{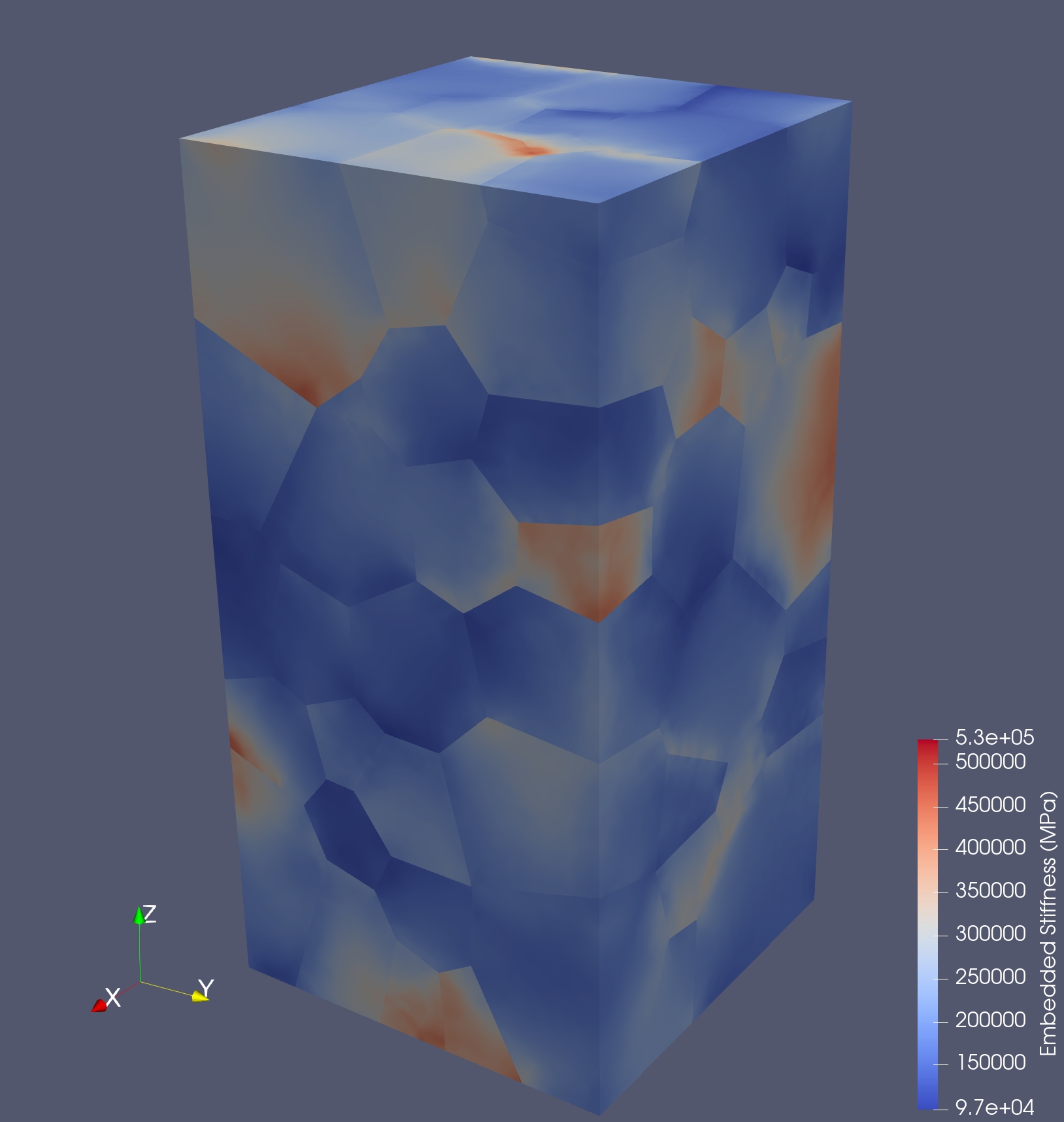}
		\caption{ }
		\label{fig:dia0p35_sph0p06_embstiff}
	\end{subfigure}%
	\quad
	\begin{subfigure}{.3\textwidth}
		\centering
		\includegraphics[width=1\linewidth]{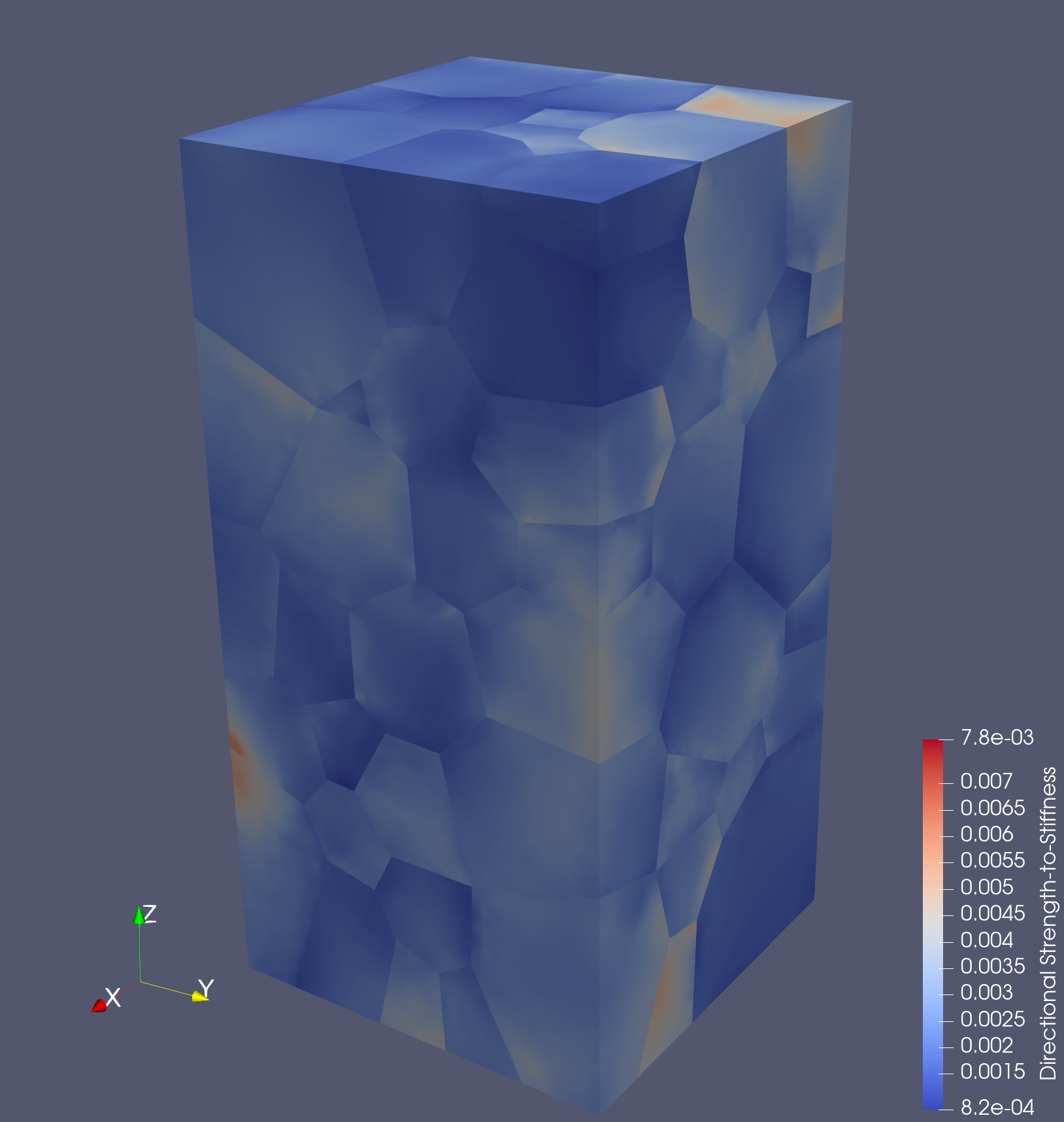}
		\caption{ }
		\label{fig:dia0p35_sph0p06_y2e}
	\end{subfigure}%
	\caption{Mechanical metrics of the LULS sample.  (a) axial stress; (b) embedded stiffness; (c) directional strength-to-stiffness ratio.}
		\label{fig:metrics_dia0p35_sph0p06}
\end{figure}

\begin{figure}[h!]
	\centering
	\begin{subfigure}{.3\textwidth}
		\centering
		\includegraphics[width=1\linewidth]{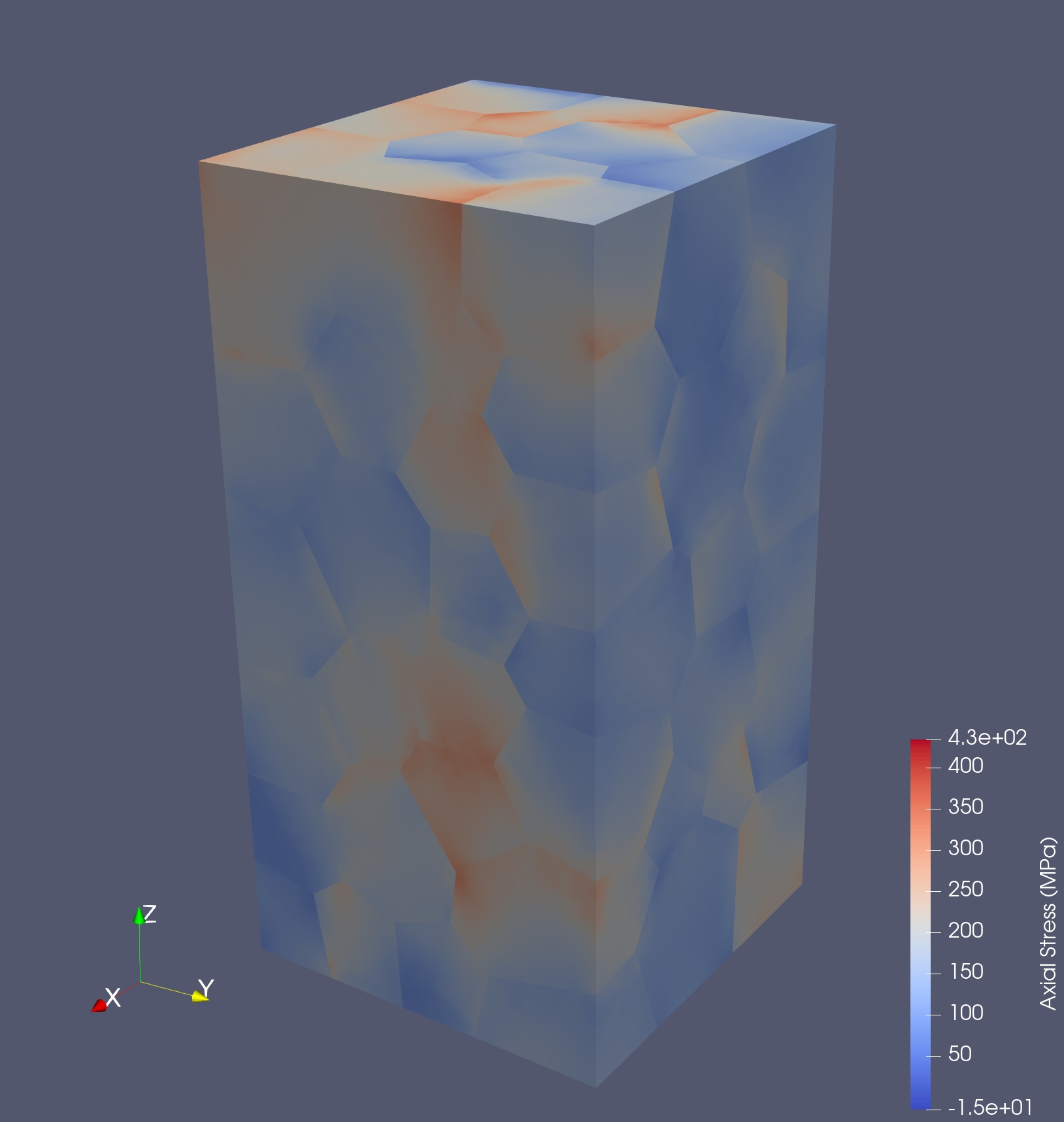}
		\caption{ }
		\label{fig:dia0p15_sph0p03_sigzz}
	\end{subfigure}%
	\quad
	\begin{subfigure}{.3\textwidth}
		\centering
		\includegraphics[width=1\linewidth]{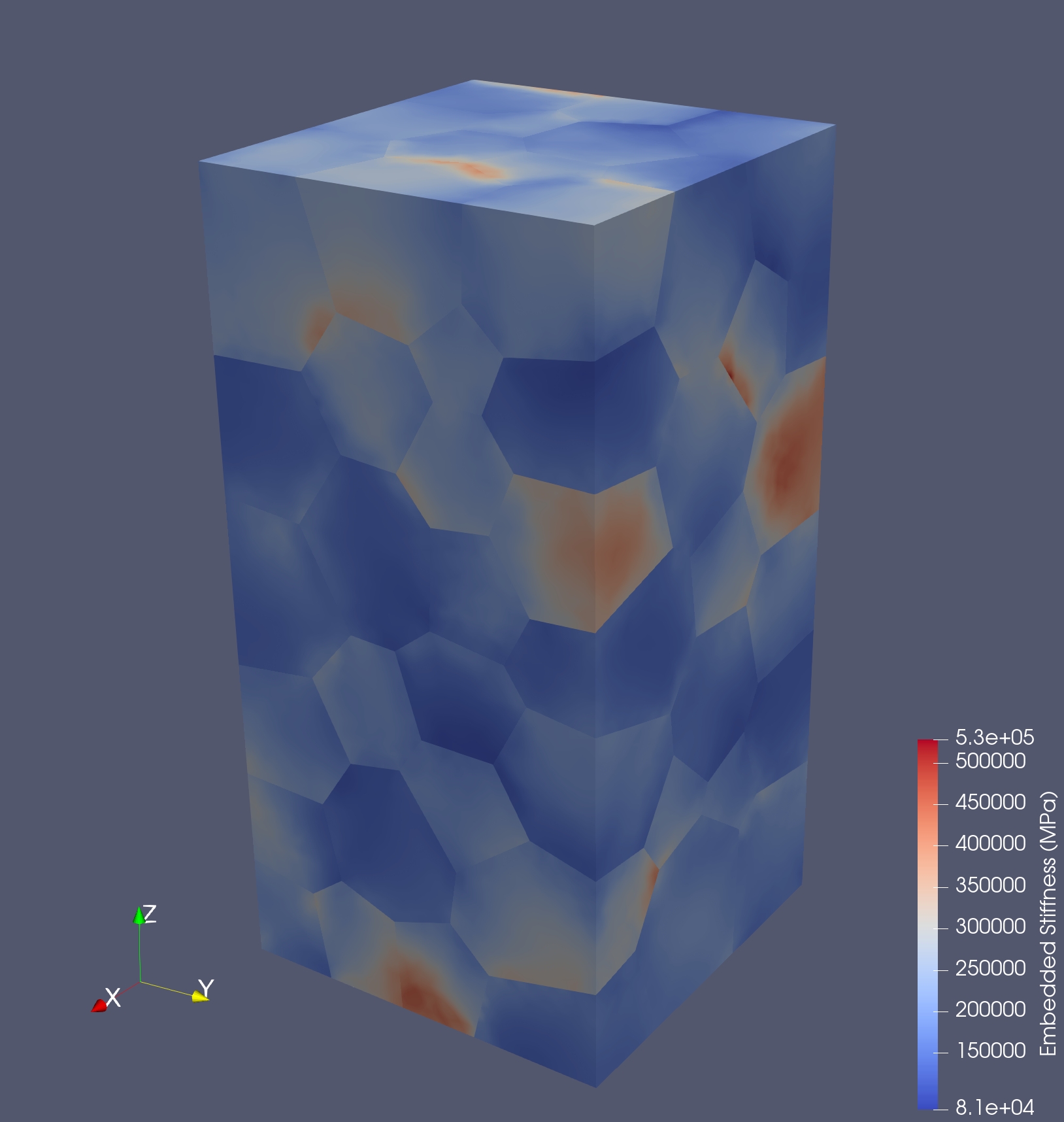}
		\caption{ }
		\label{fig:dia0p15_sph0p03_embstiff}
	\end{subfigure}%
	\quad
	\begin{subfigure}{.3\textwidth}
		\centering
		\includegraphics[width=1\linewidth]{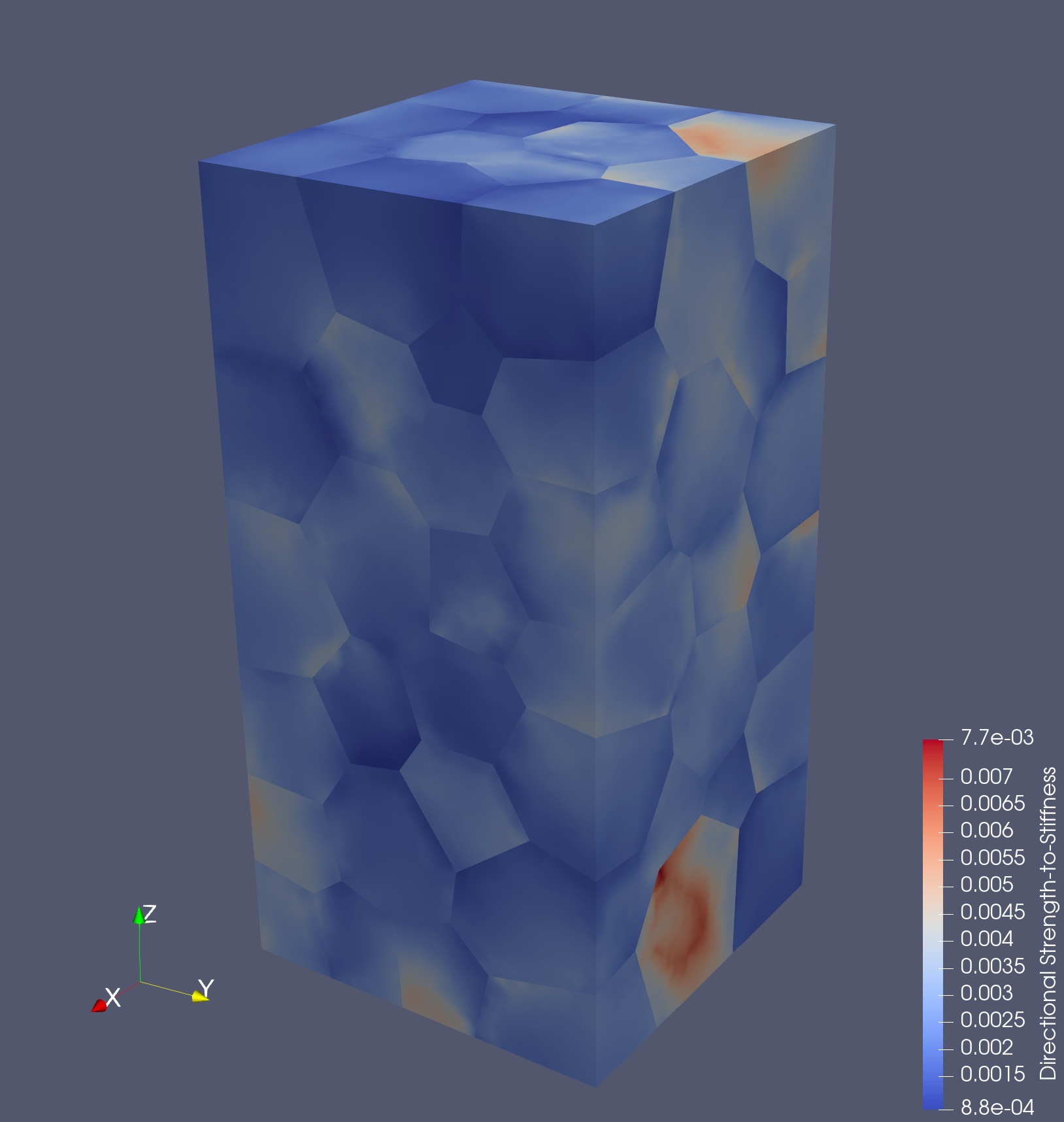}
		\caption{ }
		\label{fig:dia0p15_sph0p03_y2e}
	\end{subfigure}%
	\caption{Mechanical metrics of the HUHS sample.  (a) axial stress; (b) embedded stiffness; (c) directional strength-to-stiffness ratio.}
		\label{fig:metrics_dia0p15_sph0p03}
\end{figure}

The sample variants have relatively few grains and thus might be expected to show spread in
their mechanical behaviors.  The mechanical characteristics of the sample variants were examined 
with \mechmet~\cite{mechmet_immi}  to provide insight in this regard.  
The single-crystal elastic moduli correspond to AL6XN~\cite{pos_daw_mmta_2019a}.
Figures~\ref{fig:metrics_voronoi}, \ref{fig:metrics_dia0p35_sph0p06} and \ref{fig:metrics_dia0p15_sph0p03}  give information for the Voronoi, LULS and HUHS sample variants, respectively.  
Part (a) of each figure shows the axial stress distribution for 0.1\% extensional strain in the $z$ coordinate direction.   Spatial variations in the stress are evident; these arise in a grain within an anisotropic polycrystals from the influence of its neighbors on the loading.
Embedded (apparent) stiffness is shown in Part (b), and can be compared to the single-crystal directional  stiffness distributions shown in Part (b) of   Figures~\ref{fig:sample_voronoi}, \ref{fig:sample_dia0p35_sph0p06} and \ref{fig:sample_dia0p15_sph0p03}.
 Variations within grains exist due to intra-grain deformation heterogeneities associated with 
 neighborhood effects.
 The directional strength-to-stiffness ratios quantify the influence of stiffness on onset of yielding as well as neighborhood affects.  Distributions of strength-to-stiffness are provided in Part (c).
 One can say that there appear to exist trends in all these (related) fields that are shared
 across the sample variants.  However, 
 the grain morphologies do play a substantial role in the location of extreme values of the fields.
 This has a bearing on the activation of modes, particularly higher modes, in the representation of data. 
 \clearpage
 
\clearpage

\subsection{Harmonic modes}
\label{sec:harmonicmodes}
\begin{figure}[htbp]
	\centering
	\begin{subfigure}{.3\textwidth}
		\centering 
		\includegraphics[width=1\linewidth]{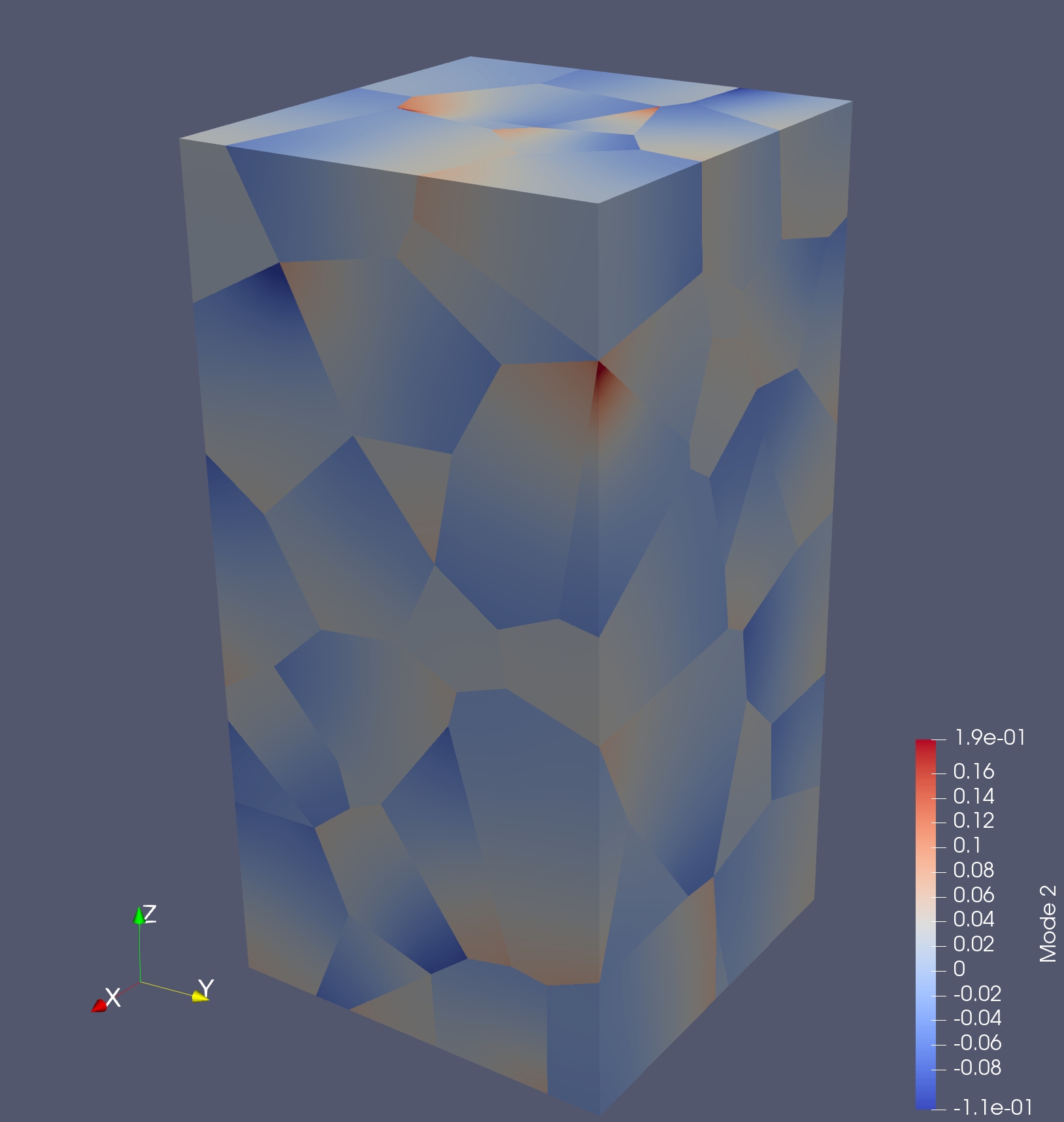}
		\caption{ }
		\label{fig:voronoi_mode1}
	\end{subfigure}%
	\quad
	\begin{subfigure}{.3\textwidth}
		\centering
		\includegraphics[width=1\linewidth]{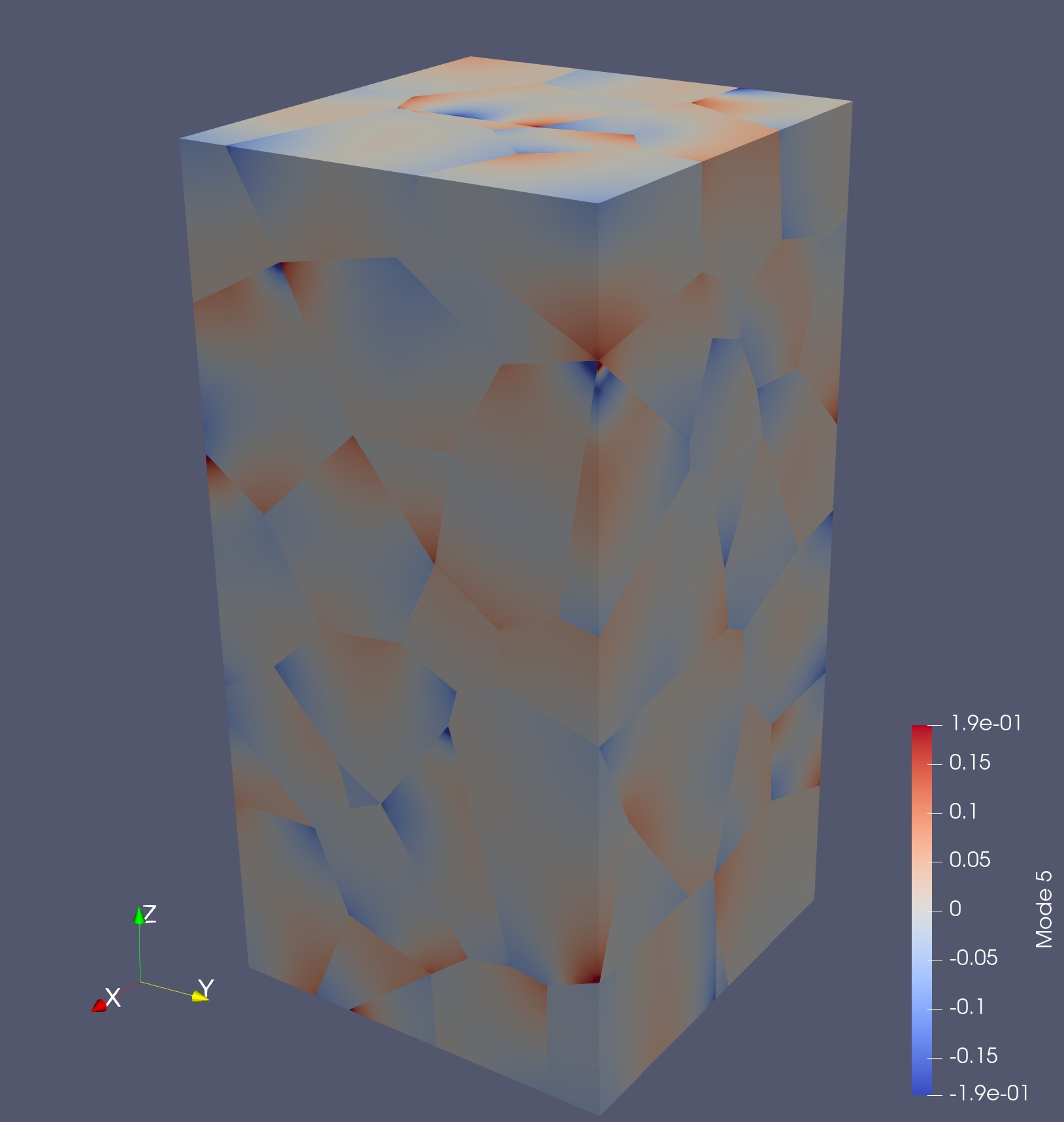}
		\caption{ }
		\label{fig:voronoi_mode4}
	\end{subfigure}%
	\quad
	\begin{subfigure}{.3\textwidth}
		\centering
		\includegraphics[width=1\linewidth]{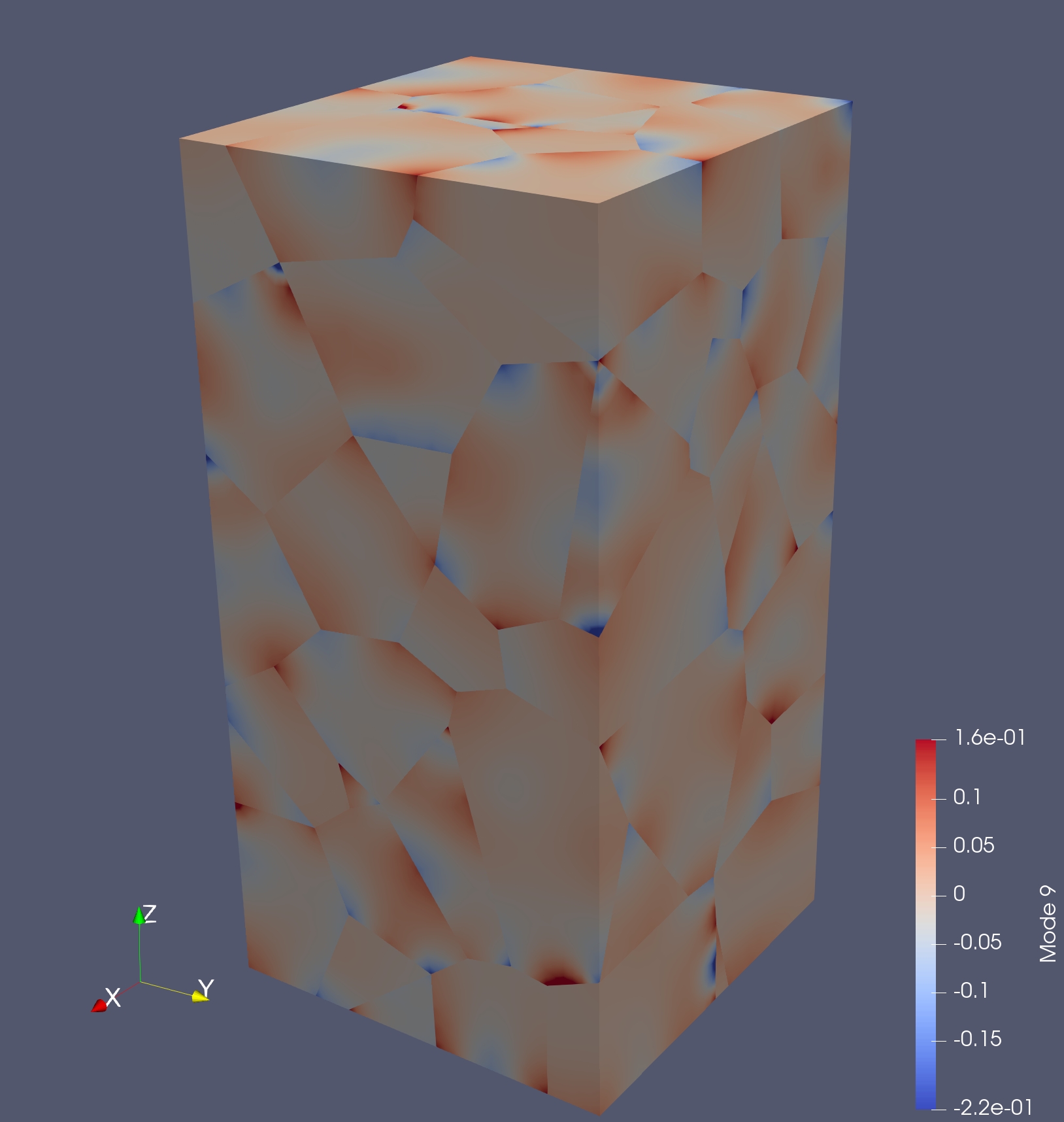}
		\caption{ }
		\label{fig:voronoi_mode8}
	\end{subfigure}%
	\caption{Harmonic modes of the Voronoi sample.  (a) Mode 2; (b) Mode 5; (c) Mode 9.}
		\label{fig:harmonicmodes_voronoi}
\end{figure}

\begin{figure}[h!]
	\begin{subfigure}{.3\textwidth}
		\centering
		\includegraphics[width=1\linewidth]{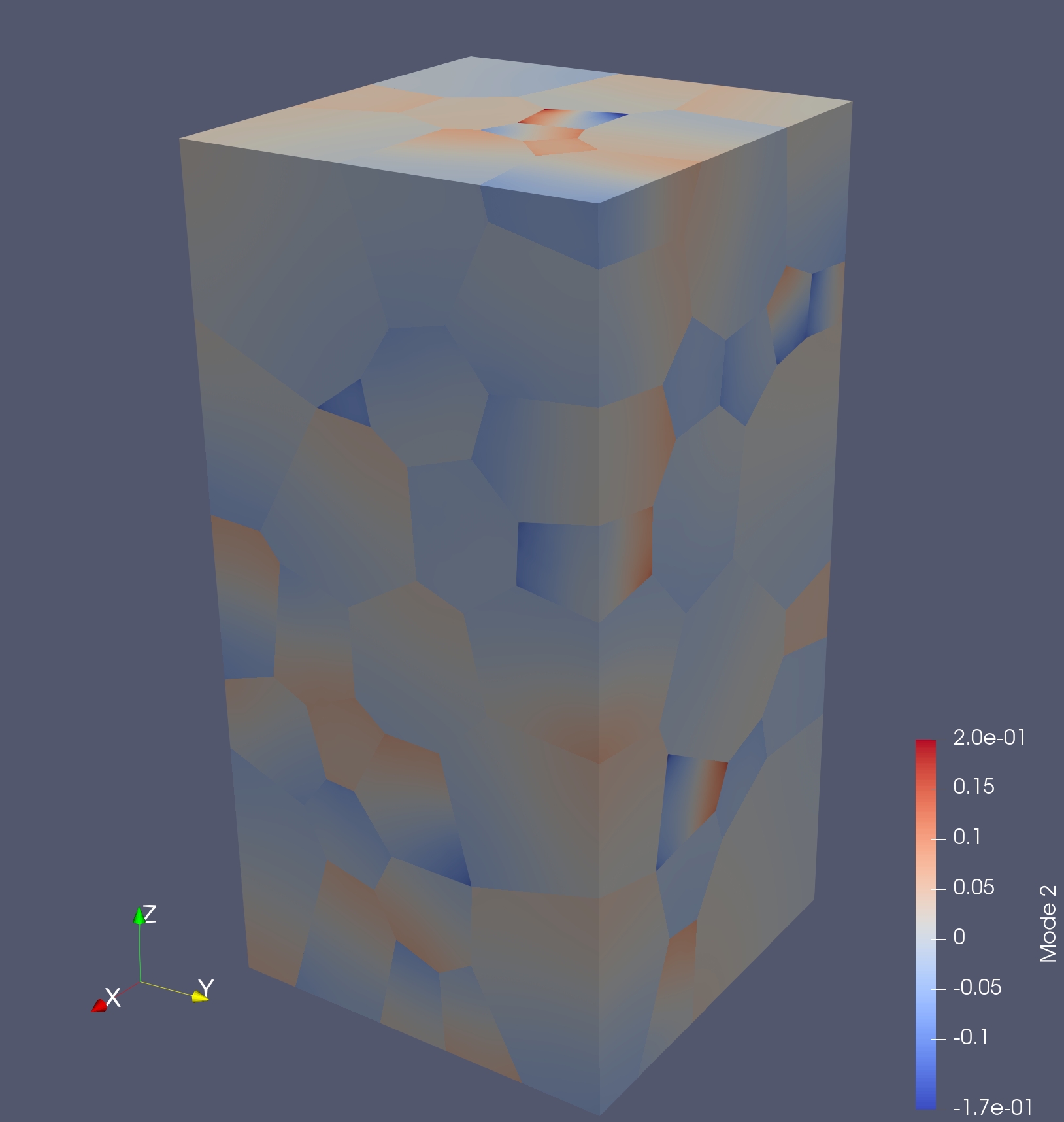}
		\caption{ }
		\label{fig:dia0p35_sph0p06_mode4}
	\end{subfigure}%
		\quad
	\begin{subfigure}{.3\textwidth}
		\centering
		\includegraphics[width=1\linewidth]{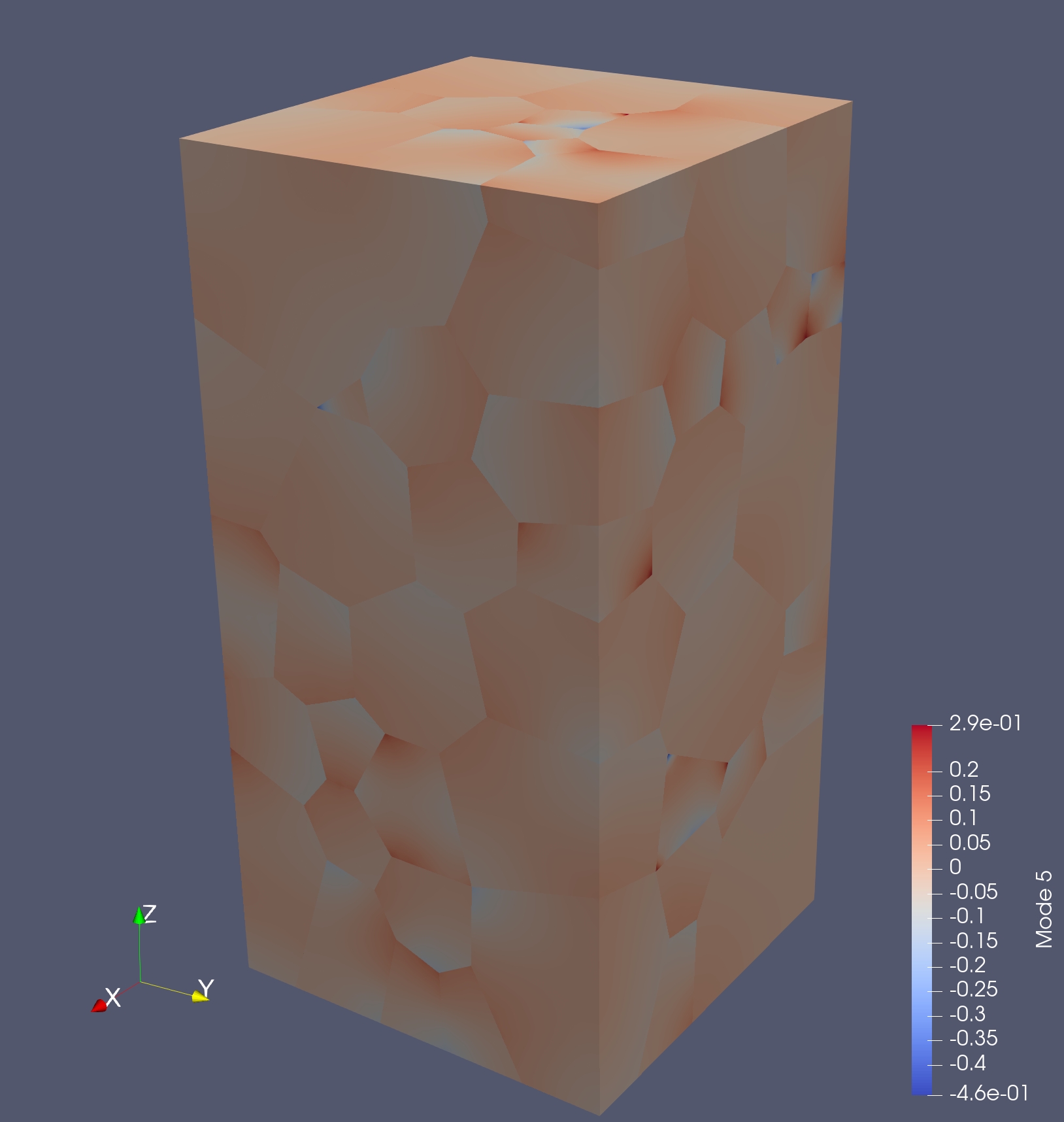}
		\caption{ }
		\label{fig:dia0p35_sph0p06_mode1}
	\end{subfigure}%
	\quad
	\begin{subfigure}{.3\textwidth}
		\centering
		\includegraphics[width=1\linewidth]{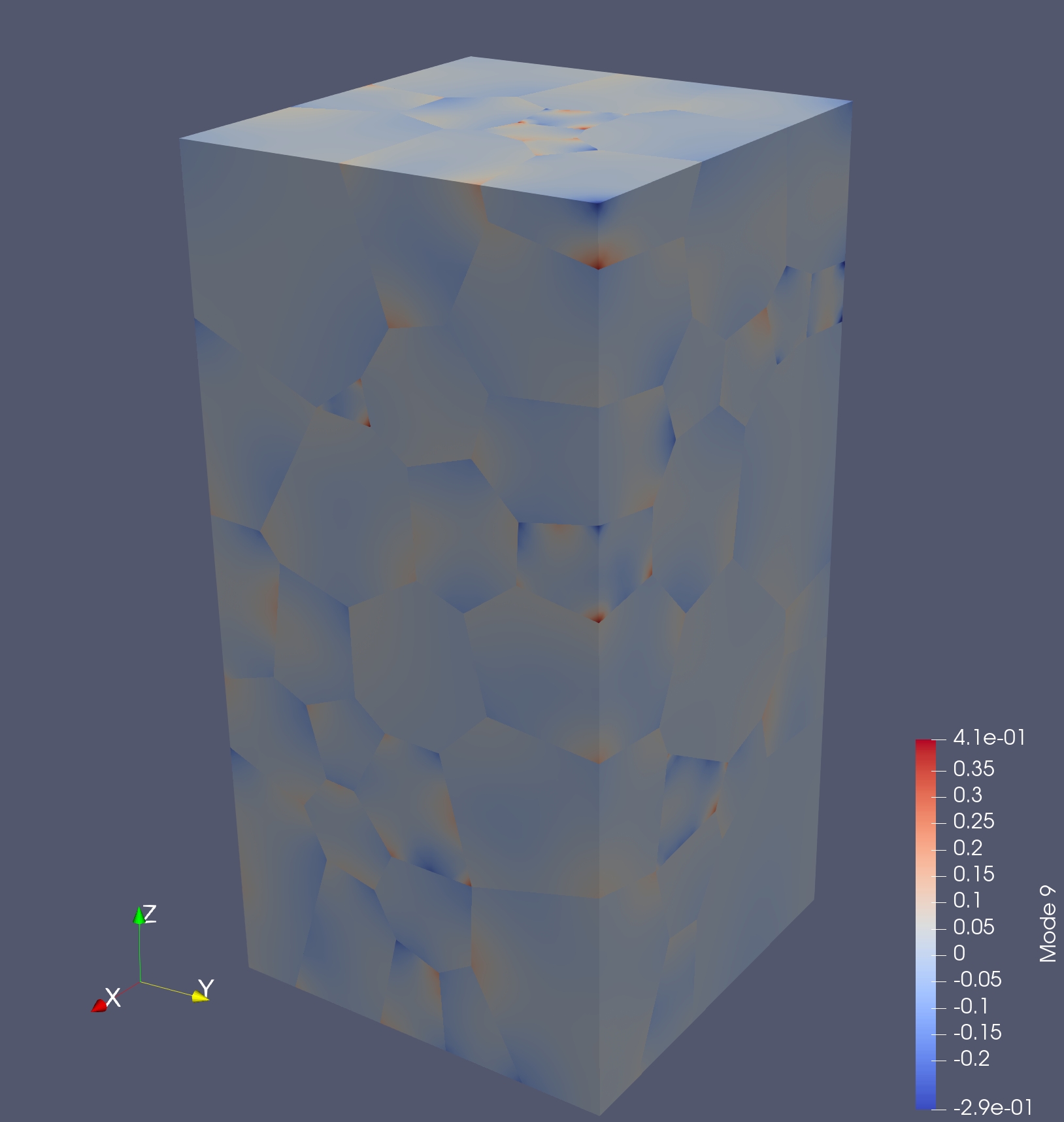}
		\caption{ }
		\label{fig:dia0p35_sph0p06_mode8}
	\end{subfigure}%
	\caption{Harmonic modes  of the LULS sample.  (a) Mode 2; (b) Mode 5; (c) Mode 9.}
		\label{fig:harmonicmodes_dia0p35_sph0p06}
\end{figure}

\begin{figure}[h!]
	\centering
	\begin{subfigure}{.3\textwidth}
		\centering
		\includegraphics[width=1\linewidth]{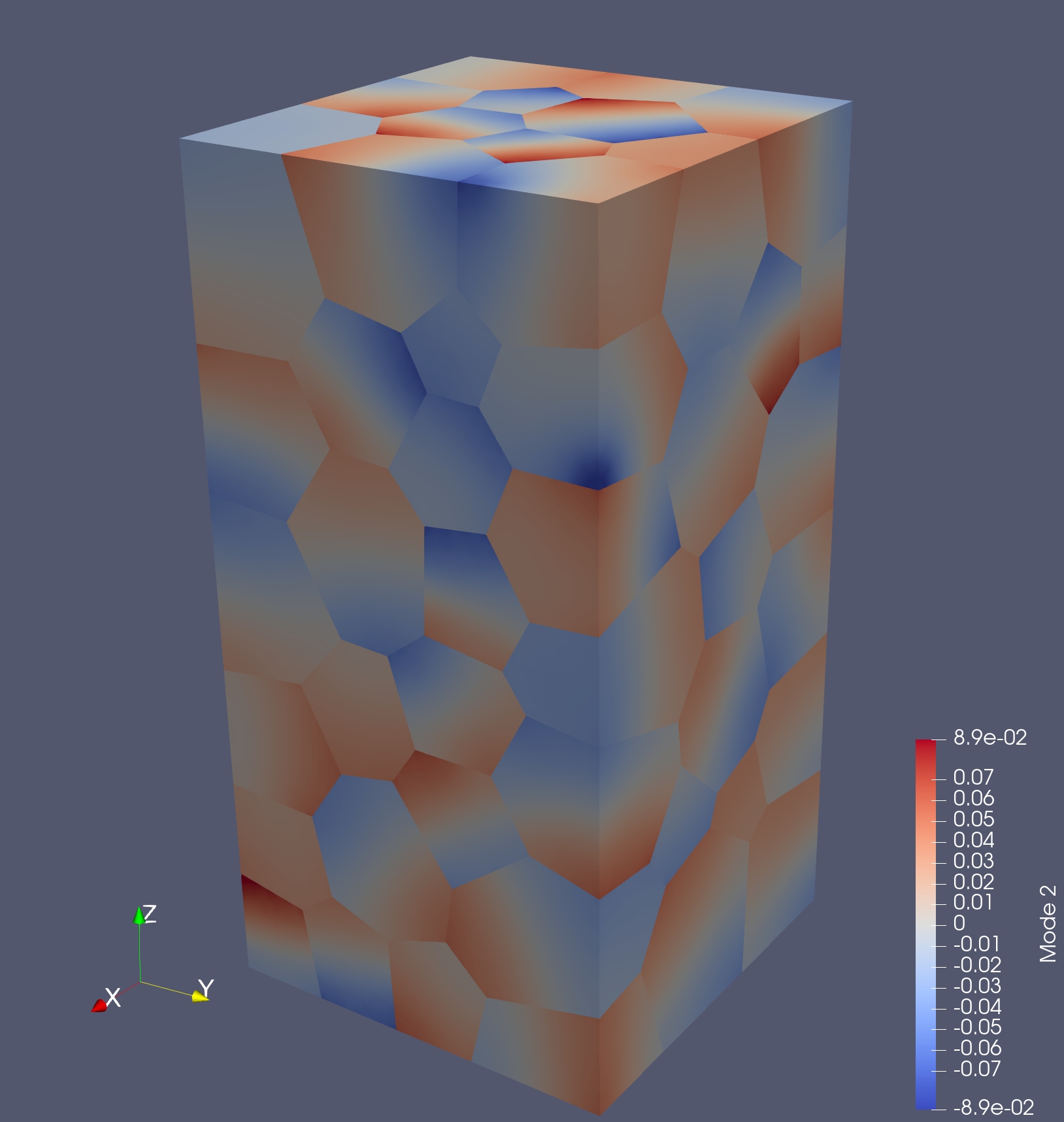}
		\caption{ }
		\label{fig:dia0p15_sph0p03_mode1}
	\end{subfigure}%
	\quad
	\begin{subfigure}{.3\textwidth}
		\centering
		\includegraphics[width=1\linewidth]{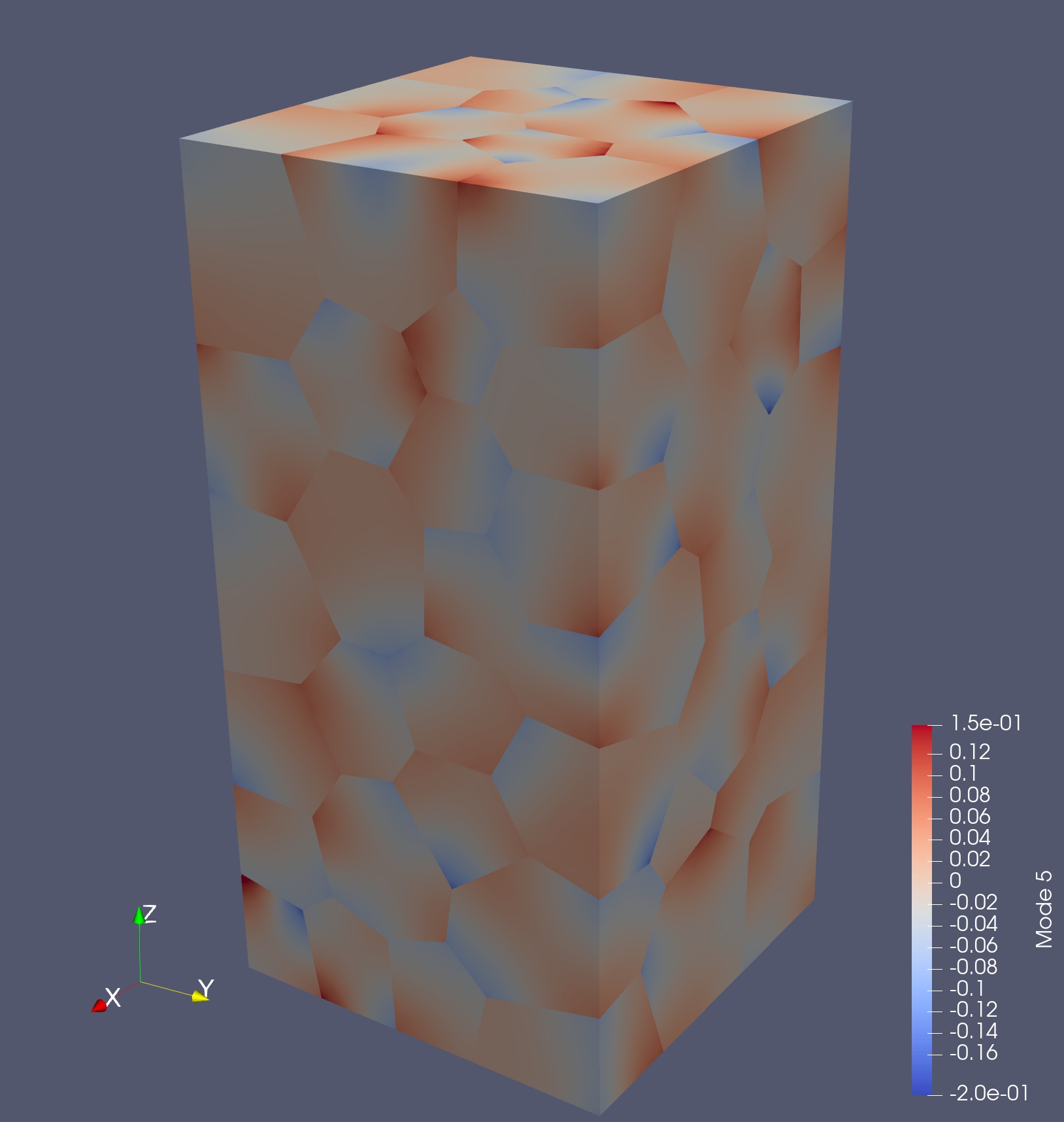}
		\caption{ }
		\label{fig:dia0p15_sph0p03_mode4}
	\end{subfigure}%
        \quad
	\begin{subfigure}{.3\textwidth}
		\centering
		\includegraphics[width=1\linewidth]{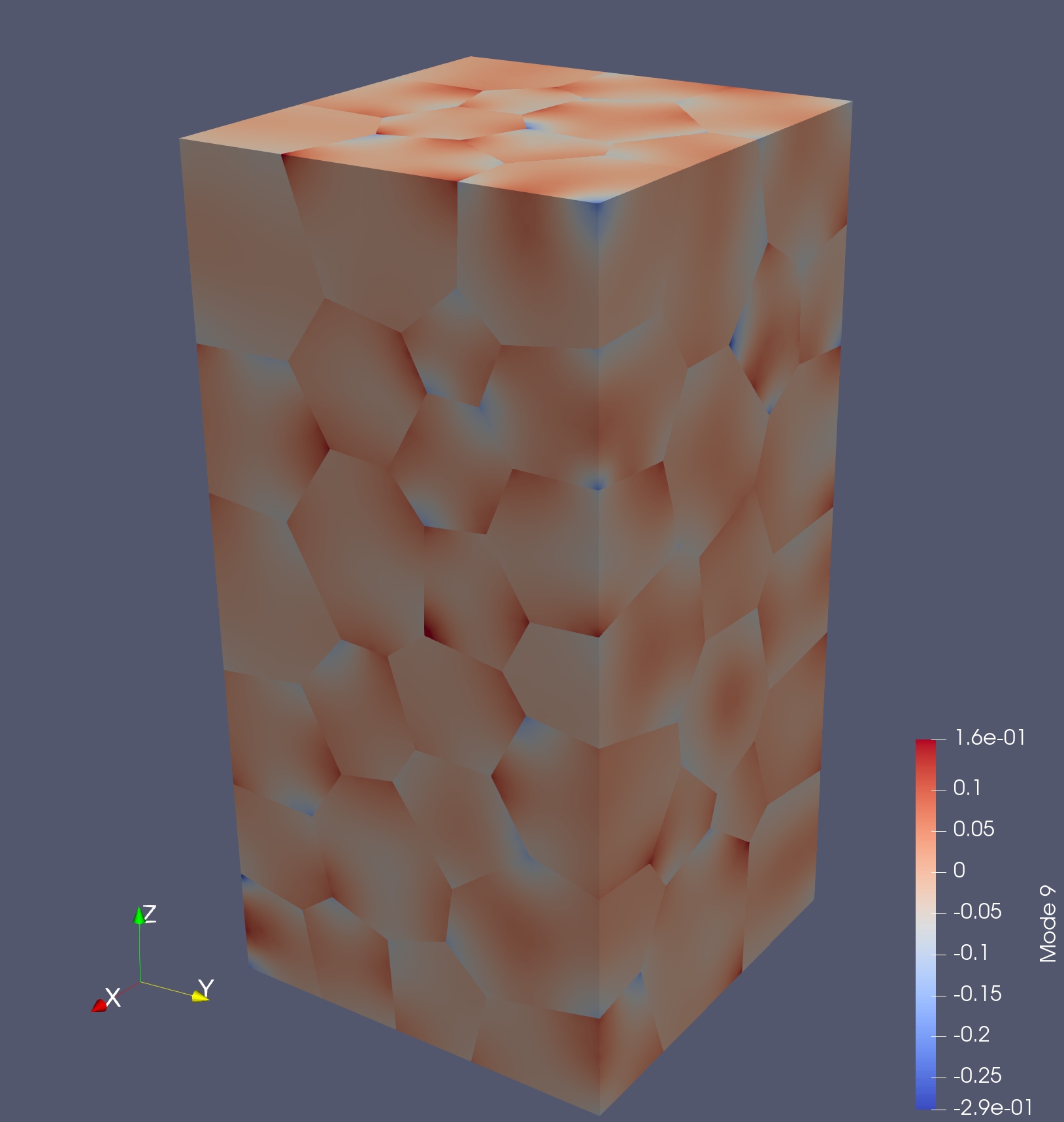}
		\caption{ }
		\label{fig:dia0p15_sph0p03_mode8}
	\end{subfigure}%
	\caption{Harmonic modes of the HUHS sample.  (a) Mode 2; (b) Mode 5; (c) Mode 9.}
		\label{fig:harmonicmodes_dia0p15_sph0p03}
\end{figure}

For the three sample variants, discrete harmonic modes were determined with \mechmonics\, according to the approach laid out in Section~\ref{sec:methodology}.
For the demonstration objectives here, only the first 10 modes are discussed.   
To represent the stress distribution computed in \fepx\, with error levels below 1\% 
would require 20-30 modes based on experience to date.  
This is a reflection of the real complexity of stress distributions in polycrystals that have 
complicated grain morphologies and anisotropic properties.  
Modes 2, 5 and 9 are shown for each sample in Figures~\ref{fig:harmonicmodes_voronoi}, \ref{fig:harmonicmodes_dia0p35_sph0p06} and \ref{fig:harmonicmodes_dia0p15_sph0p03}.
\clearpage

First, Mode 1  (not shown) is the constant mode for each set of harmonic modes. 
Modes 2 to 10 are not constant, but rather vary over the grain with increased complexity
as the mode number increases.
Mode 2 has monotonically increasing or decreasing distributions. 
However, its gradients are not constants (so the mode is not linear).
Generally speaking, the values of the gradient of Mode 2 are lower than corresponding values for the other modes.
Modes 5 and 9 have more complex distributions and definitely are not monotonic over grains.
 Note that in general the mode patterns do not  repeat the same pattern from grain to grain.
 Rather,  each grain has its own unique set of modes owing to its unique size and shape.  
Similarities among the modes appears to be the greatest for HUHS and the lowest for LULS, but this observation is made without quantitative analysis.

\begin{figure}[htbp]
	\centering
	\begin{subfigure}{.3\textwidth}
		\centering 
		\includegraphics[width=1\linewidth]{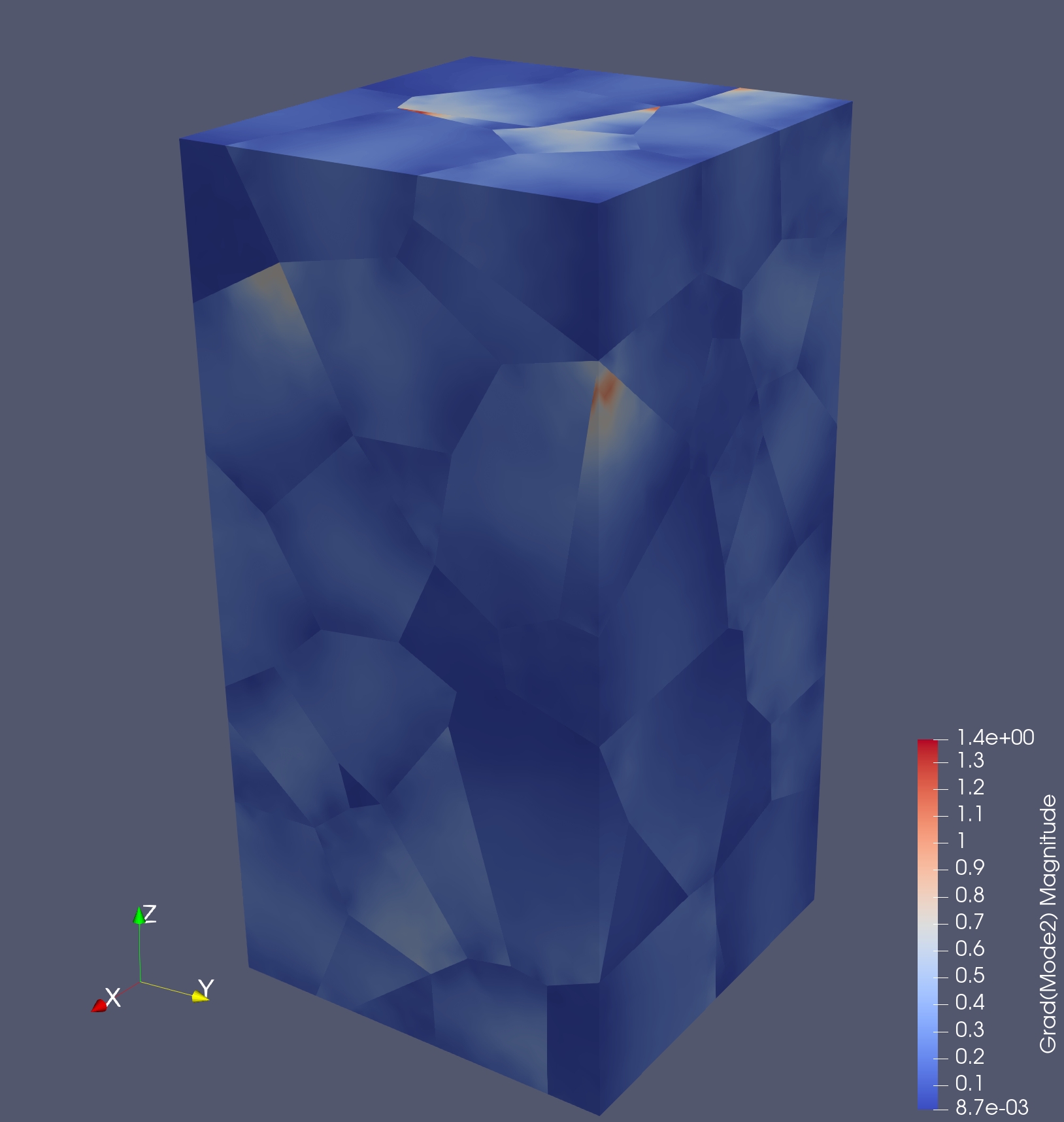}
		\caption{ }
		\label{fig:voronoi_gradmode1}
	\end{subfigure}%
	\quad
	\begin{subfigure}{.3\textwidth}
		\centering
		\includegraphics[width=1\linewidth]{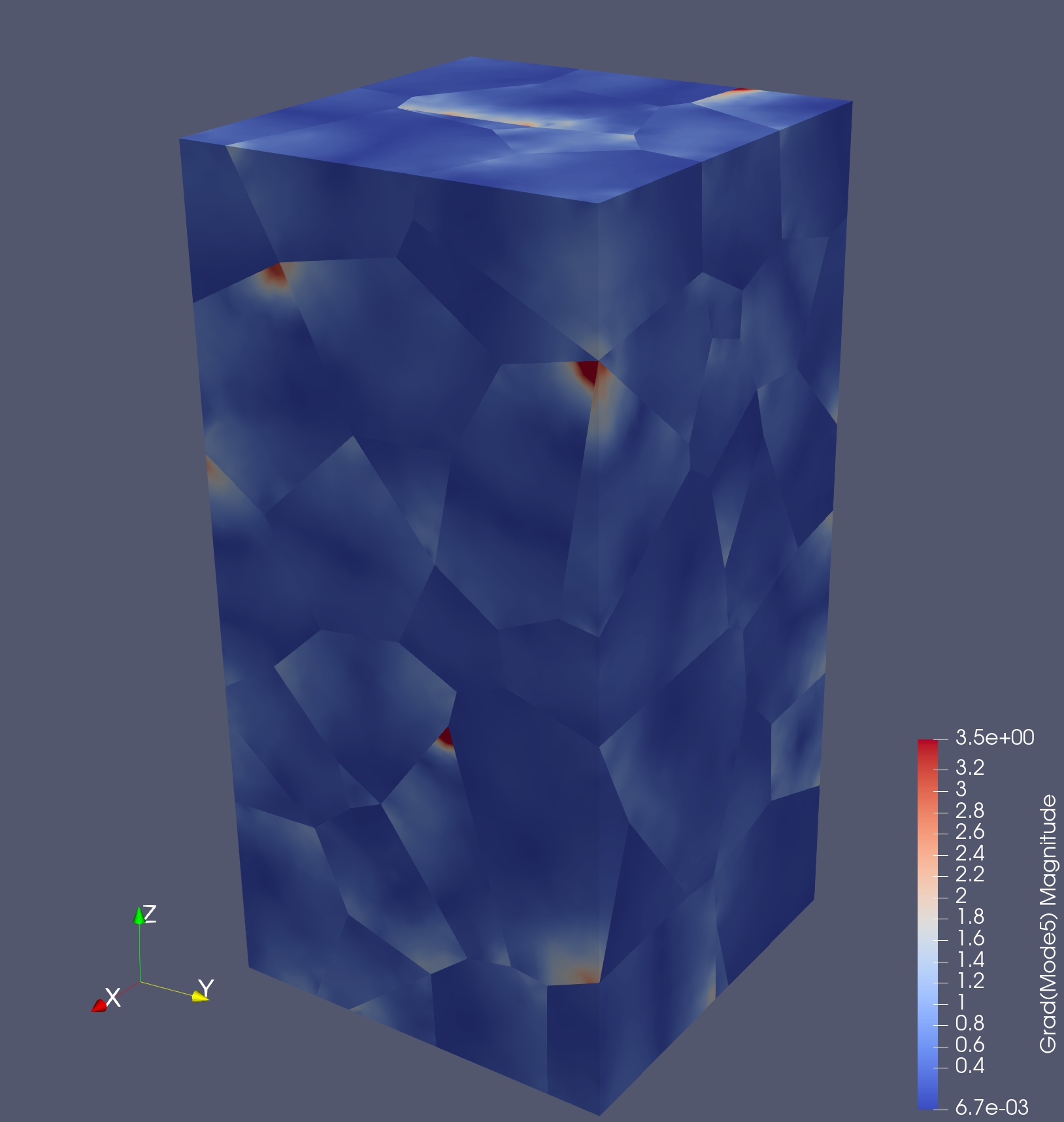}
		\caption{ }
		\label{fig:voronoi_gradmode4}
	\end{subfigure}%
	\quad
	\begin{subfigure}{.3\textwidth}
		\centering
		\includegraphics[width=1\linewidth]{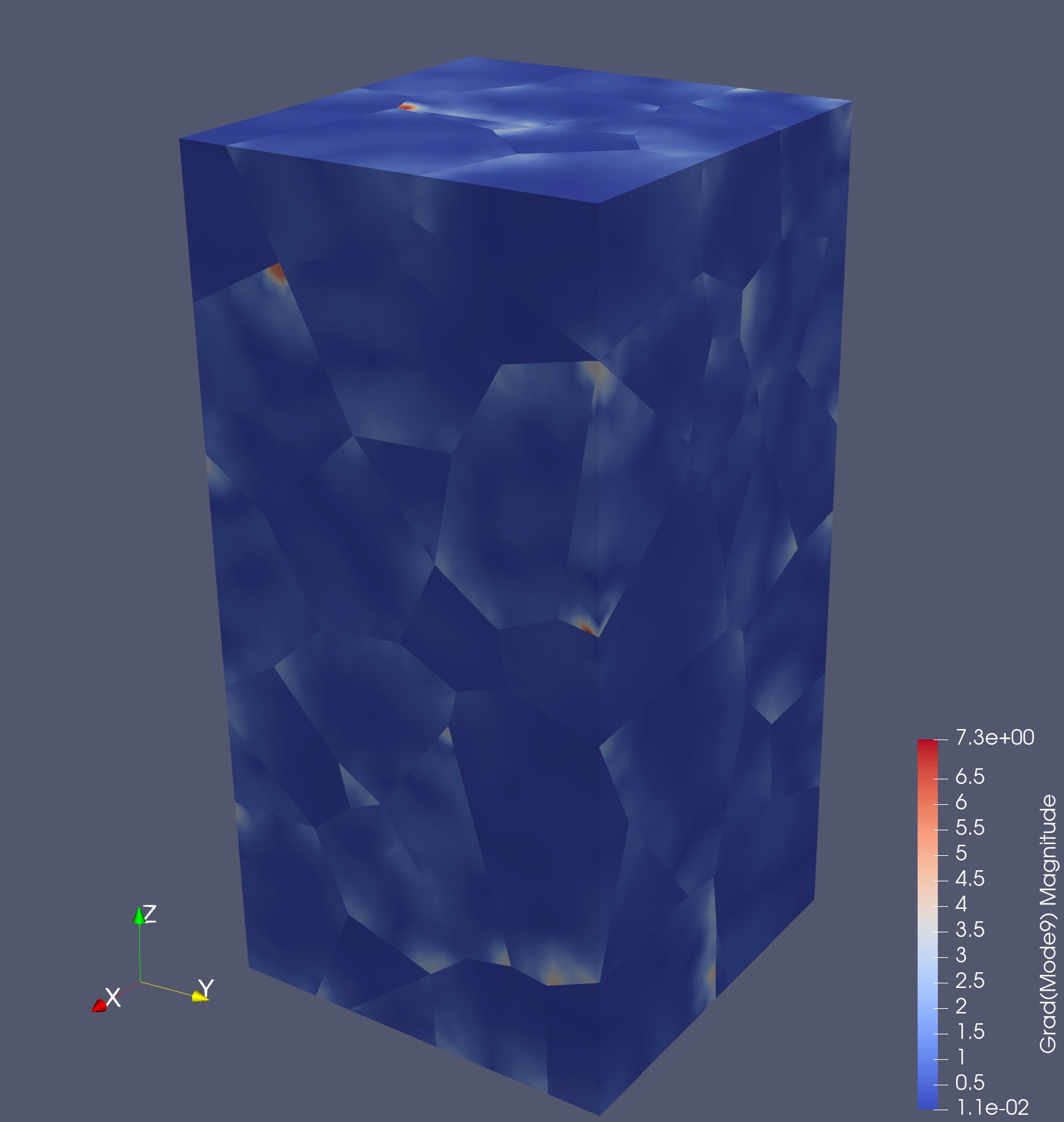}
		\caption{ }
		\label{fig:voronoi_gradmode8}
	\end{subfigure}%
	\caption{Harmonic mode gradients of the Voronoi sample.  (a) Mode 2; (b) Mode 5; (c) Mode 9.}
		\label{fig:harmonicmodesgrad_voronoi}
\end{figure}

\begin{figure}[h!]
	\begin{subfigure}{.3\textwidth}
		\centering
		\includegraphics[width=1\linewidth]{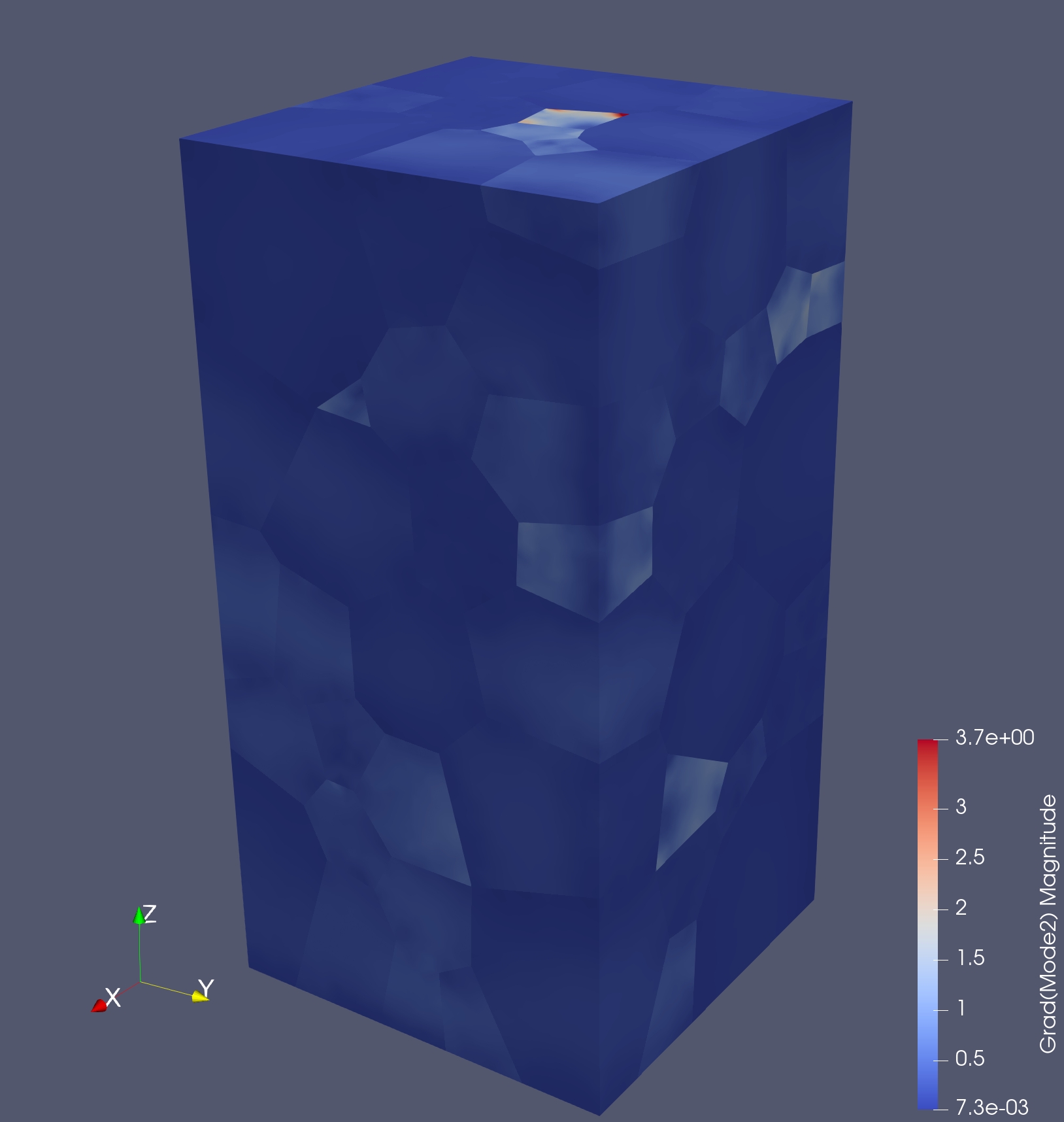}
		\caption{ }
		\label{fig:dia0p35_sph0p06_gradmode4}
	\end{subfigure}%
		\quad
	\begin{subfigure}{.3\textwidth}
		\centering
		\includegraphics[width=1\linewidth]{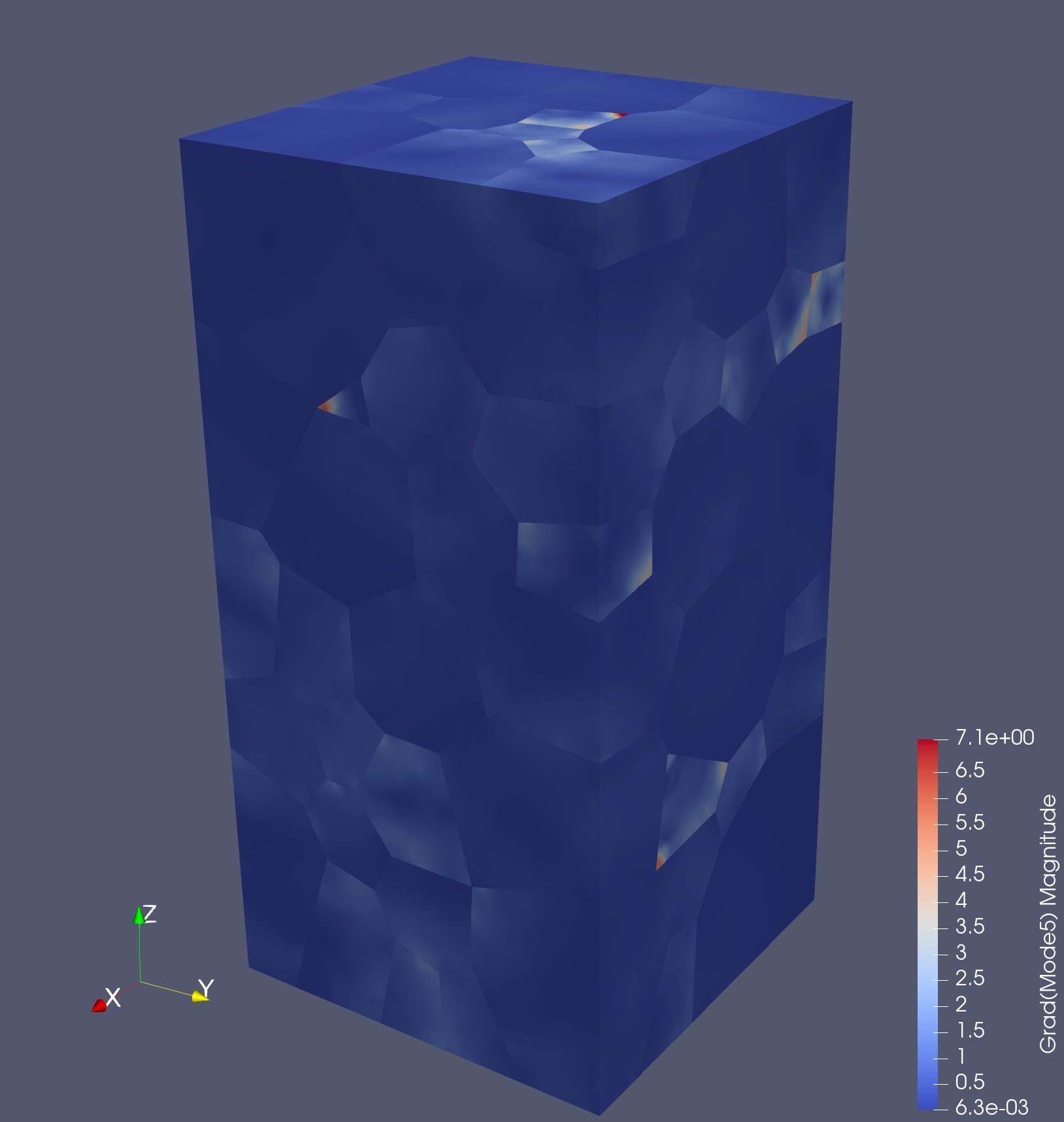}
		\caption{ }
		\label{fig:dia0p35_sph0p06_gradmode1}
	\end{subfigure}%
	\quad
	\begin{subfigure}{.3\textwidth}
		\centering
		\includegraphics[width=1\linewidth]{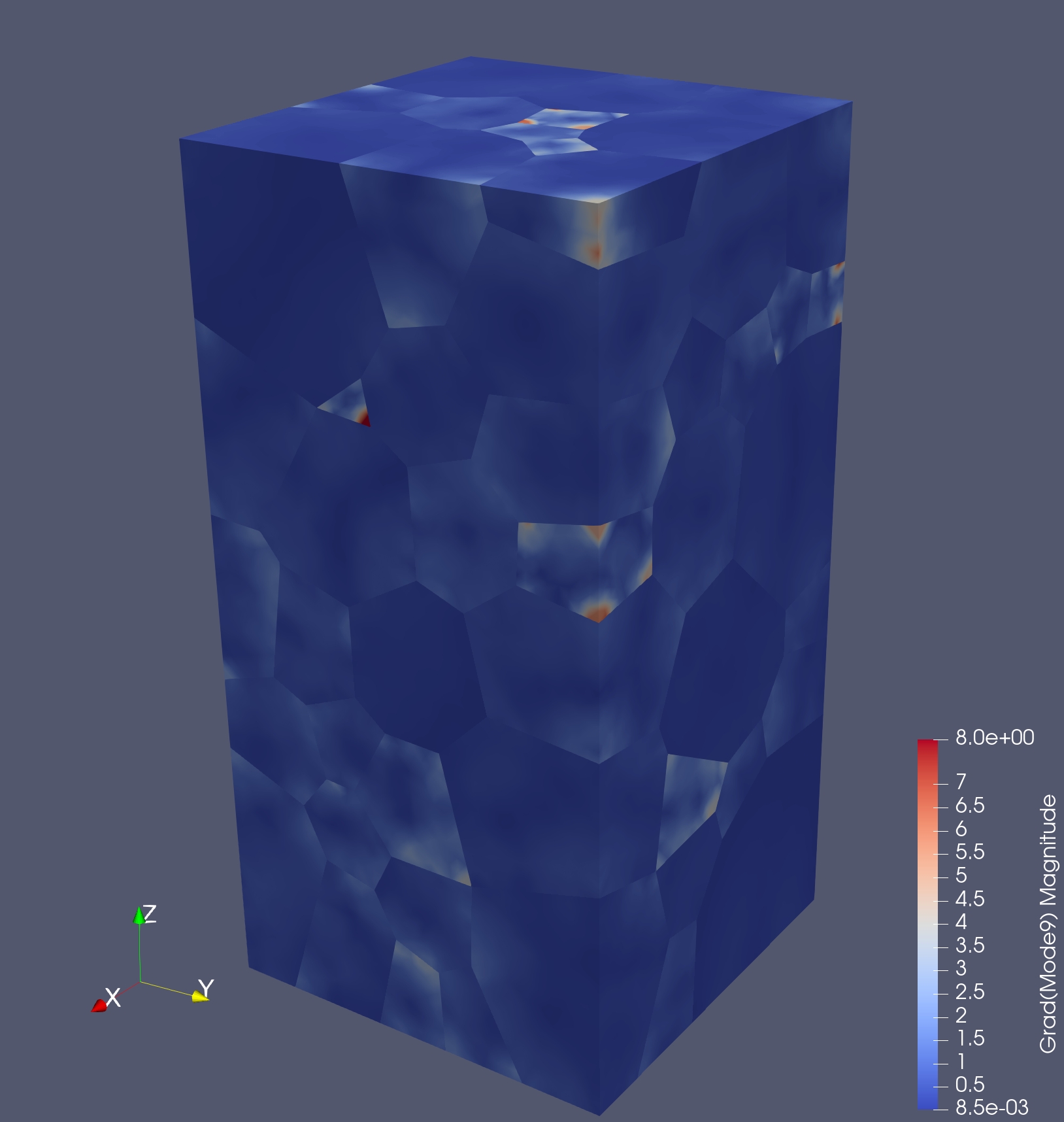}
		\caption{ }
		\label{fig:dia0p35_sph0p06_gradmode8}
	\end{subfigure}%
	\caption{Harmonic mode gradients of the LULS sample.  (a) Mode 2; (b) Mode 5; (c) Mode 9.}
		\label{fig:harmonicmodesgrad_dia0p35_sph0p06}
\end{figure}

\begin{figure}[h!]
	\centering
	\begin{subfigure}{.3\textwidth}
		\centering
		\includegraphics[width=1\linewidth]{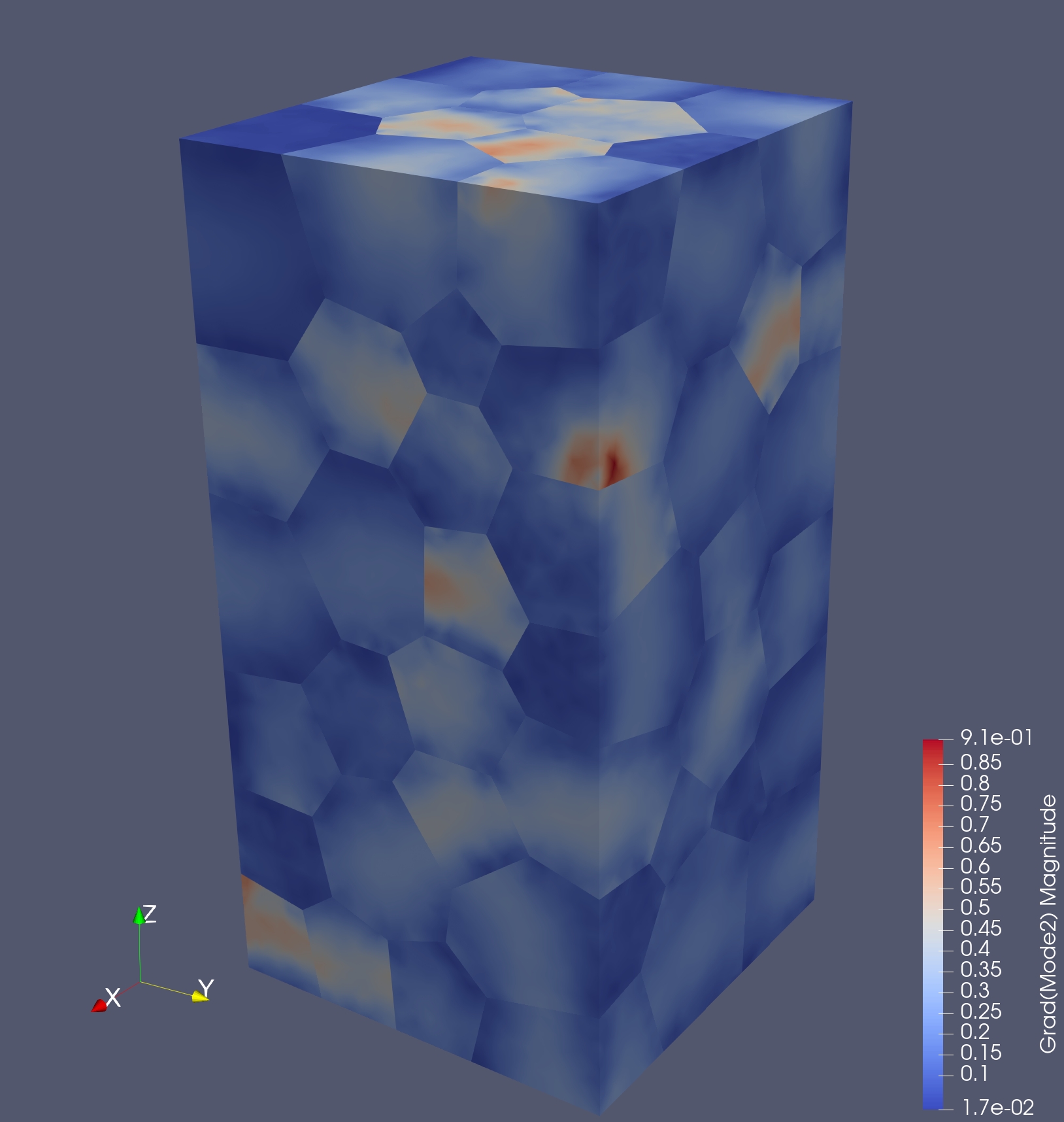}
		\caption{ }
		\label{fig:dia0p15_sph0p03_gradmode1}
	\end{subfigure}%
		\quad
	\begin{subfigure}{.3\textwidth}
		\centering
		\includegraphics[width=1\linewidth]{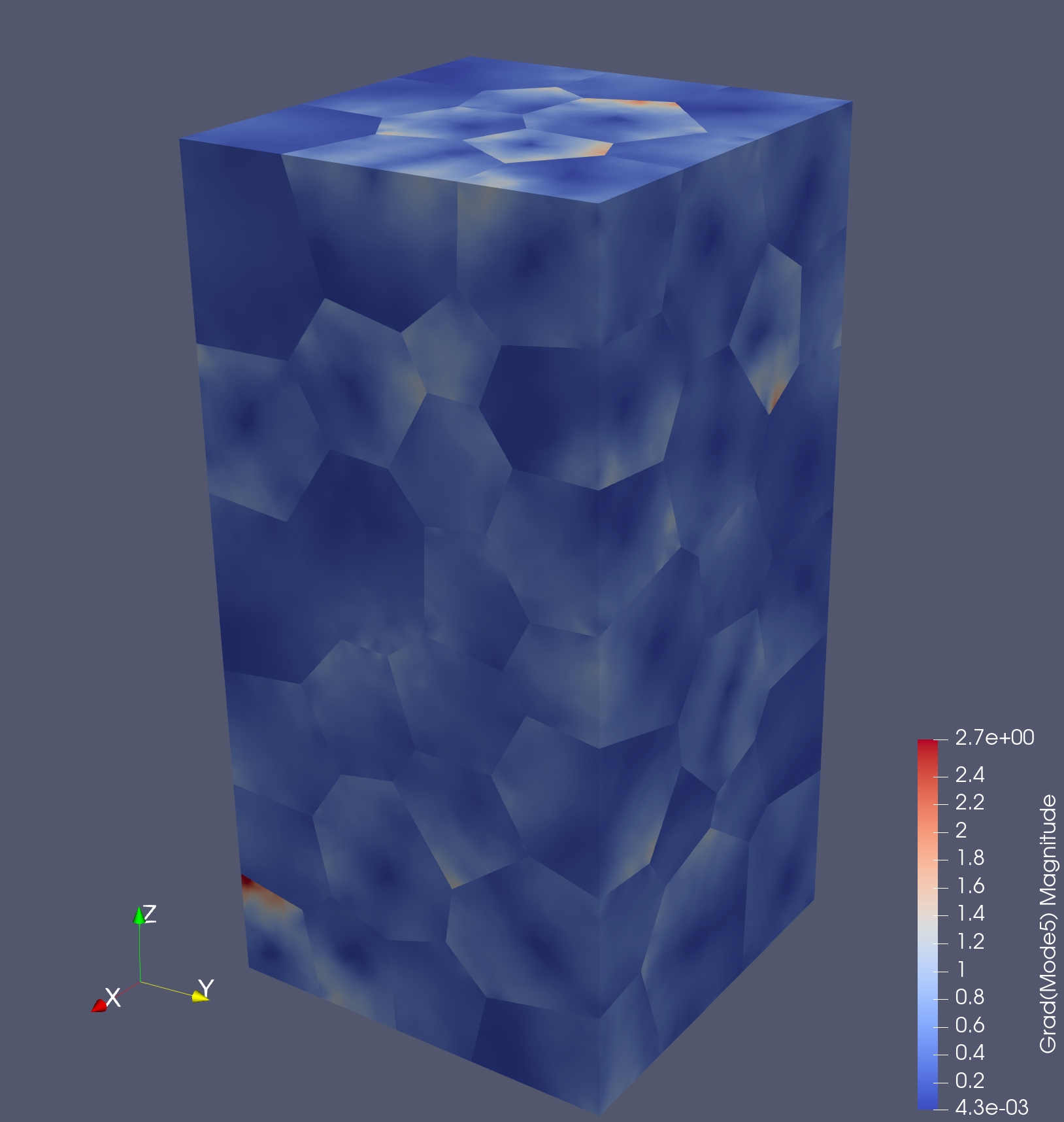}
		\caption{ }
		\label{fig:dia0p15_sph0p03_gradmode4}
	\end{subfigure}%
	\quad
	\begin{subfigure}{.3\textwidth}
		\centering
		\includegraphics[width=1\linewidth]{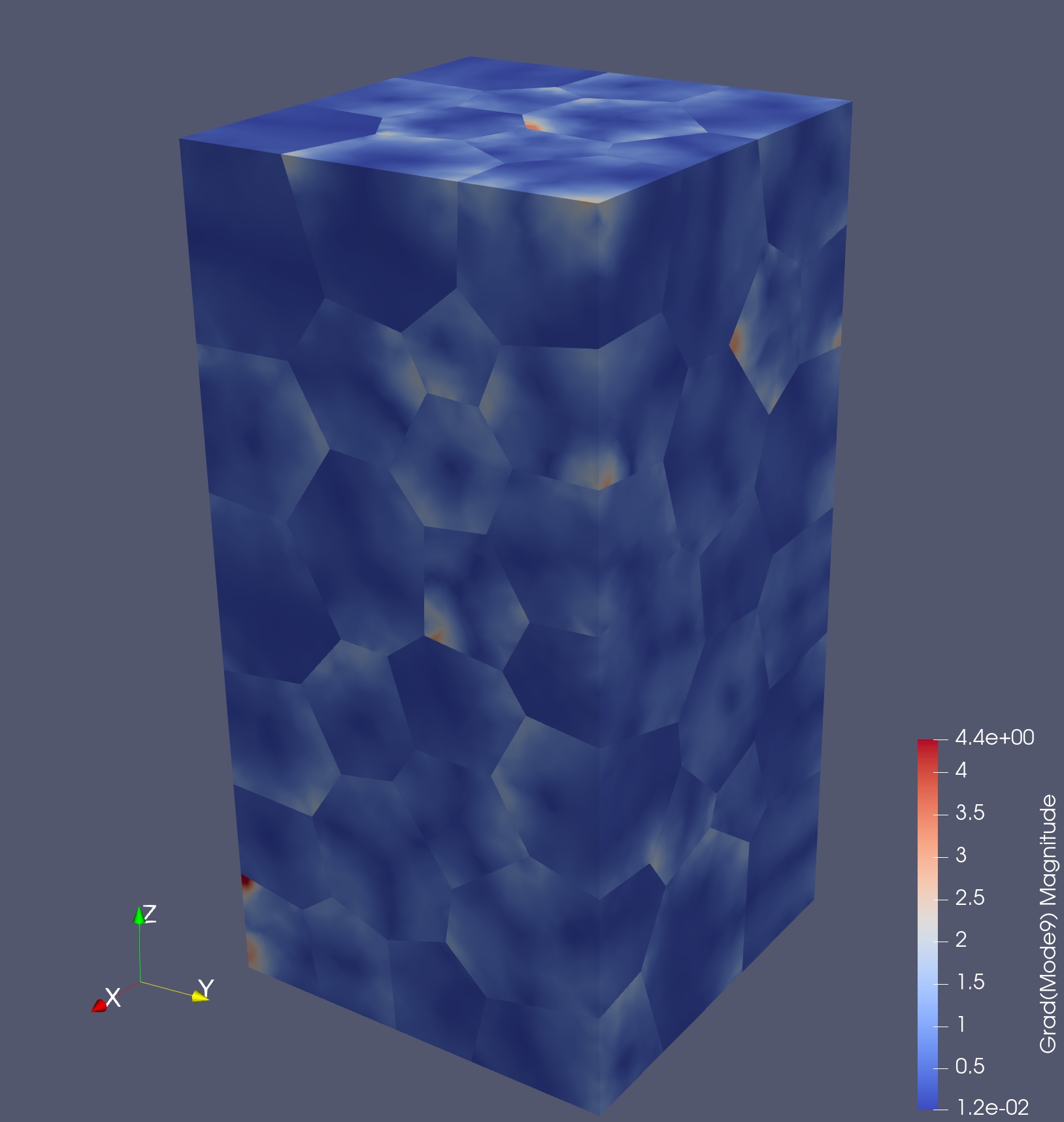}
		\caption{ }
		\label{fig:dia0p15_sph0p03_gradmode8}
	\end{subfigure}%
	\caption{Harmonic modes gradient of the HUHS sample.  (a) Mode 2; (b) Mode 5; (c) Mode 9.}
		\label{fig:harmonicmodesgrad_dia0p15_sph0p03}
\end{figure}

\begin{figure}[h!]
	\centering
	\begin{subfigure}{.3\textwidth}
		\centering
		\includegraphics[width=1\linewidth]{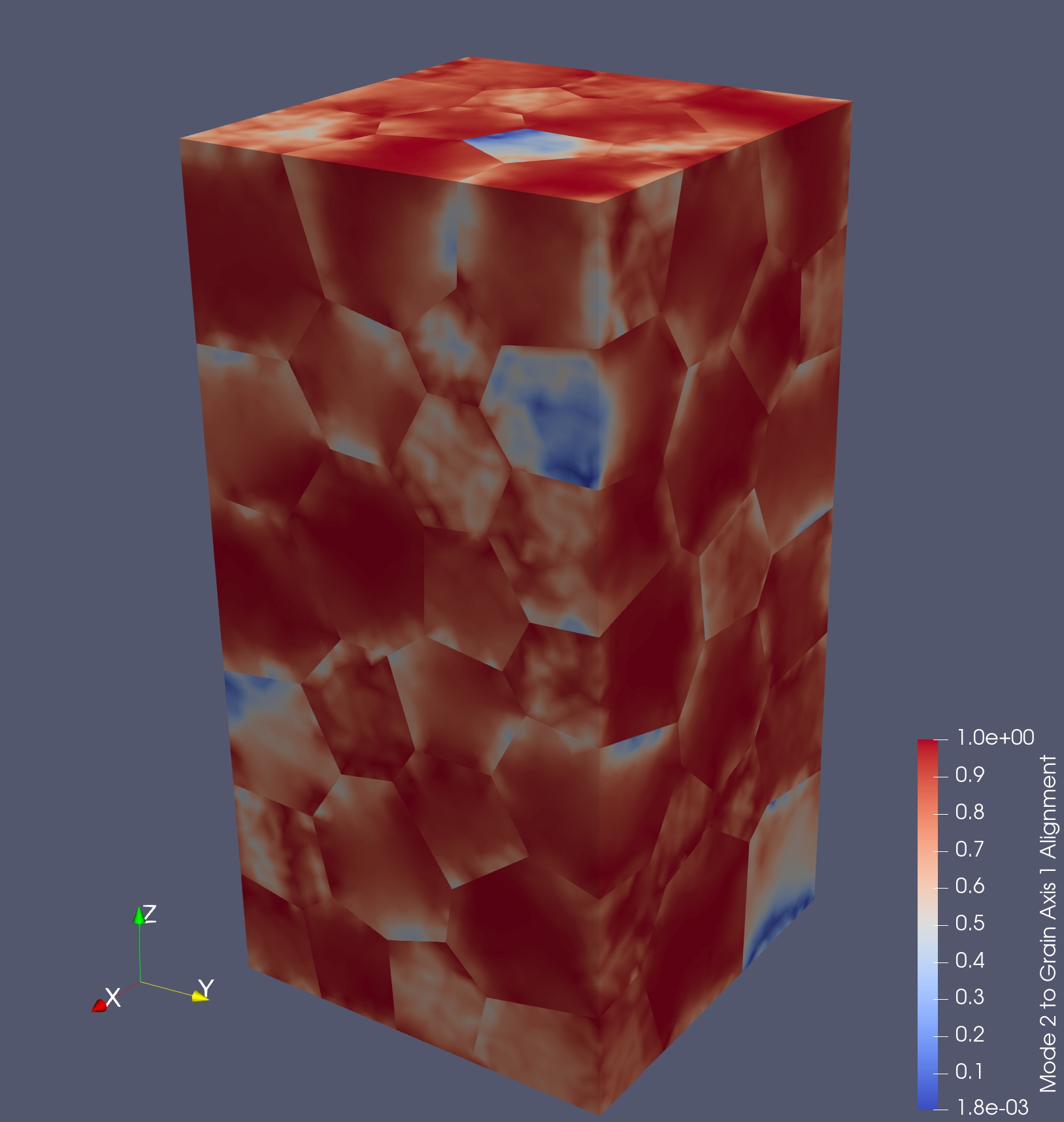}
		\caption{ }
		\label{fig:sample_a_m12e1}
	\end{subfigure}%
	\quad
	\begin{subfigure}{.3\textwidth}
		\centering
		\includegraphics[width=1\linewidth]{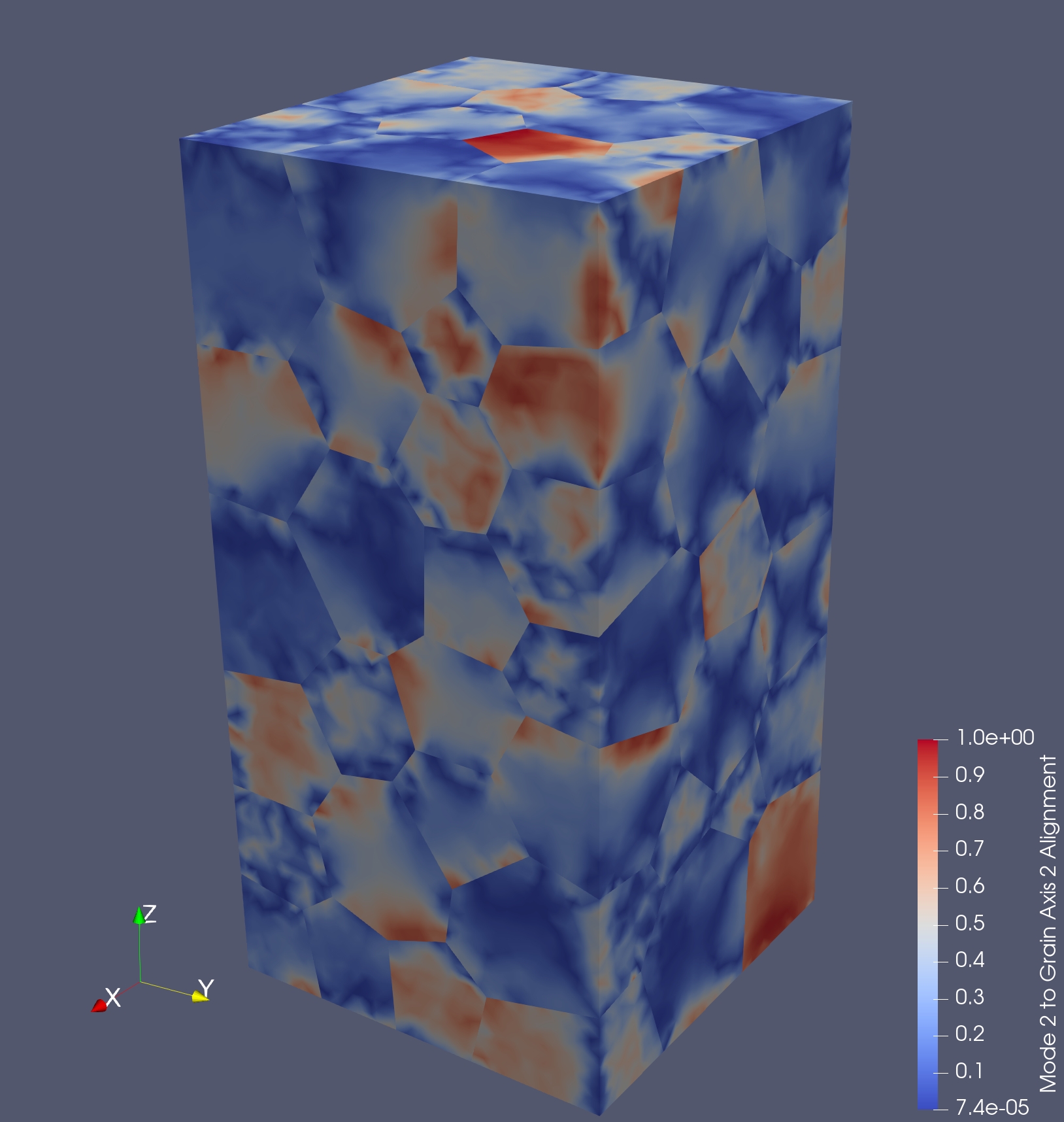}
		\caption{ }
		\label{fig:sample_a_m12e2}
	\end{subfigure}%
	\quad
	\begin{subfigure}{.3\textwidth}
		\centering
		\includegraphics[width=1\linewidth]{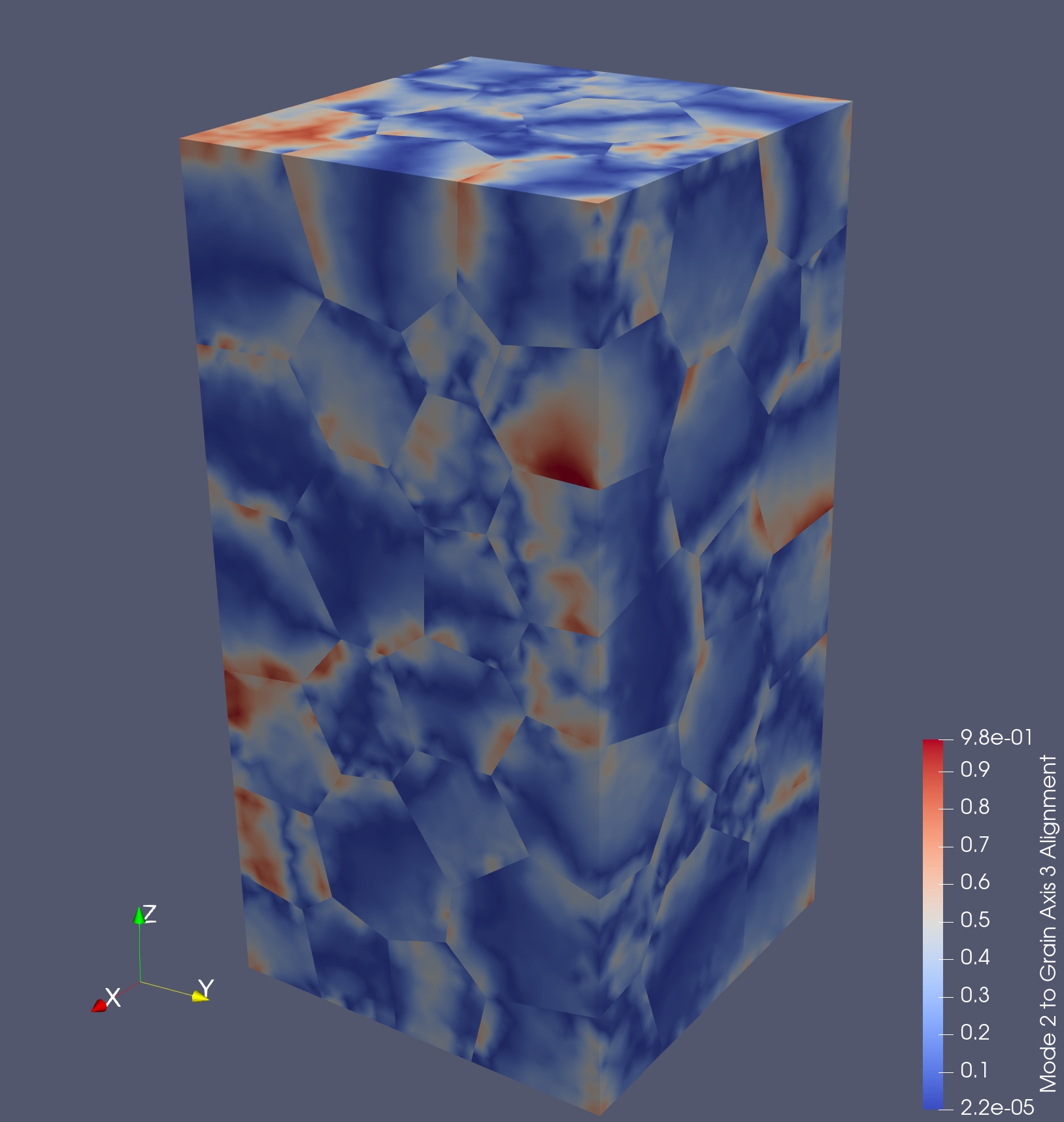}
		\caption{ }
		\label{fig:sample_a_m12e3}
	\end{subfigure}%
	\caption{Alignment of the gradient vector in the HUHS sample for Mode 2 with the grain axes: (a) grain axis 1; (b) grain axis 2; (c) grain axis 3. }
		\label{fig:HUHS_grnaxes2mode1}
\end{figure}

\begin{figure}[h!]
	\centering
	\begin{subfigure}{.3\textwidth}
		\centering
		\includegraphics[width=1\linewidth]{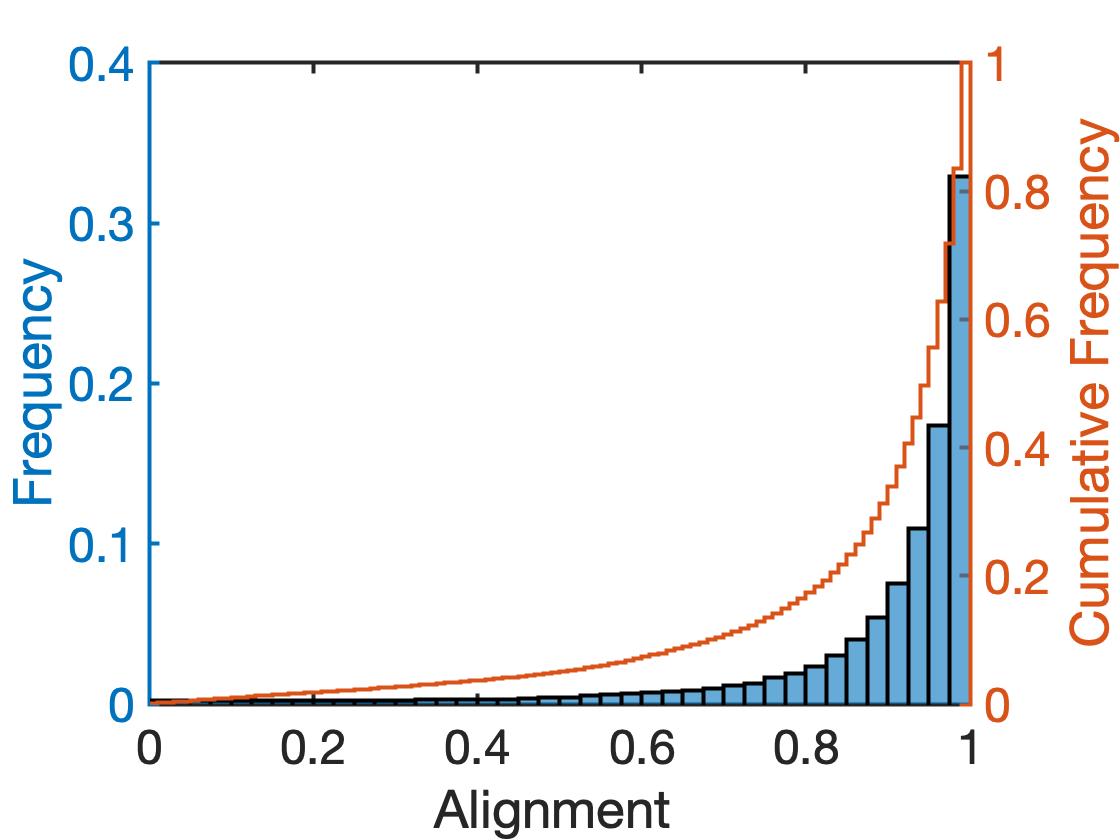}
		\caption{ }
		\label{fig:sample_a_m12e1_freq}
	\end{subfigure}%
	\quad
	\begin{subfigure}{.3\textwidth}
		\centering
		\includegraphics[width=1\linewidth]{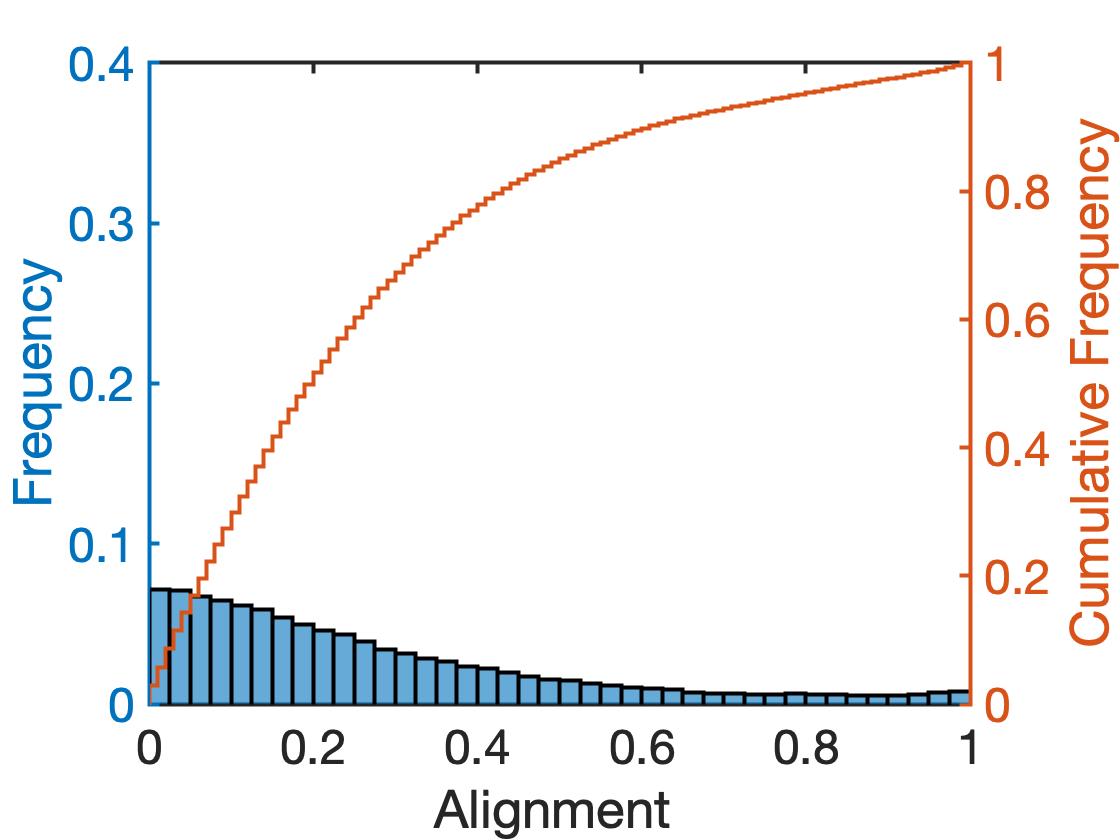}
		\caption{ }
		\label{fig:sample_a_m12e2_freq}
	\end{subfigure}%
	\quad
	\begin{subfigure}{.3\textwidth}
		\centering
		\includegraphics[width=1\linewidth]{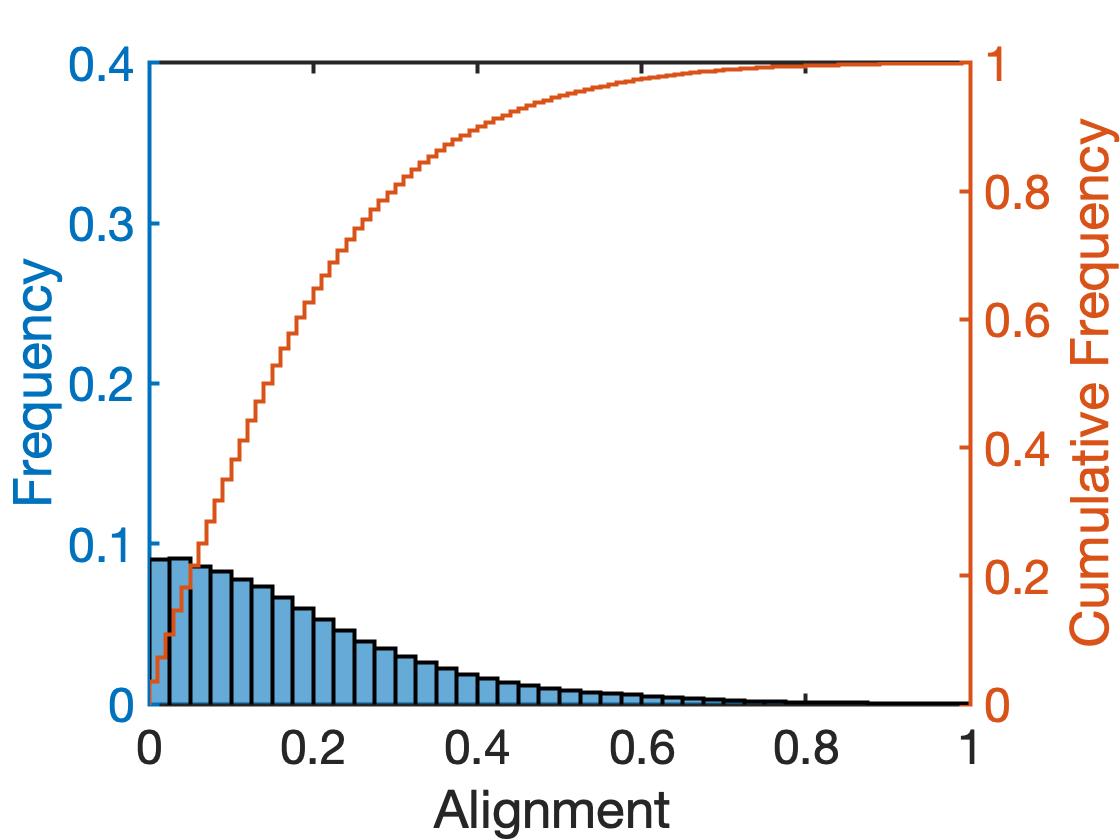}
		\caption{ }
		\label{fig:sample_a_m12e3_freq}
	\end{subfigure}%
	\caption{Frequency distribution for the alignment of the gradient vector in the HUHS sample for Mode 2 with the grain axes: (a) grain axis 1; (b) grain axis 2; (c) grain axis 3. }
		\label{fig:HUHS_grnaxes2mode1_frequency}
\end{figure}

\begin{figure}[h!]
	\centering
	\begin{subfigure}{.3\textwidth}
		\centering
		\includegraphics[width=1\linewidth]{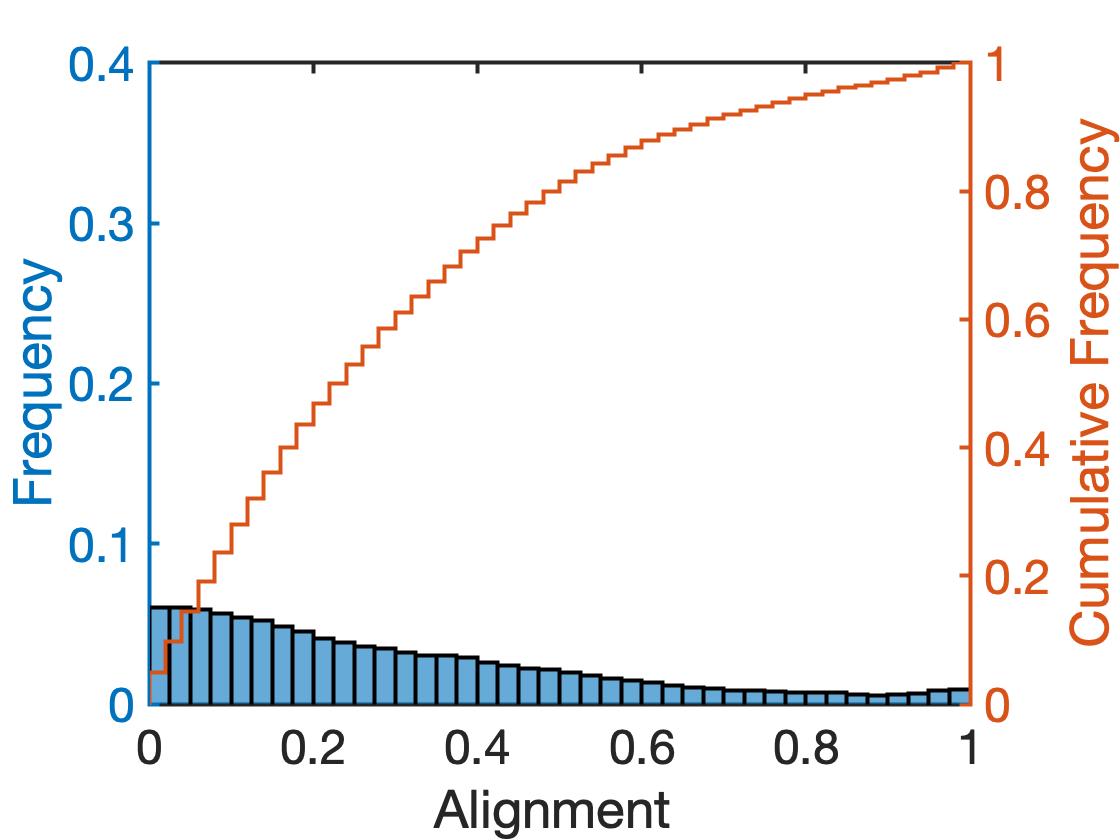}
		\caption{ }
		\label{fig:sample_a_m22e1_freq}
	\end{subfigure}%
	\quad
	\begin{subfigure}{.3\textwidth}
		\centering
		\includegraphics[width=1\linewidth]{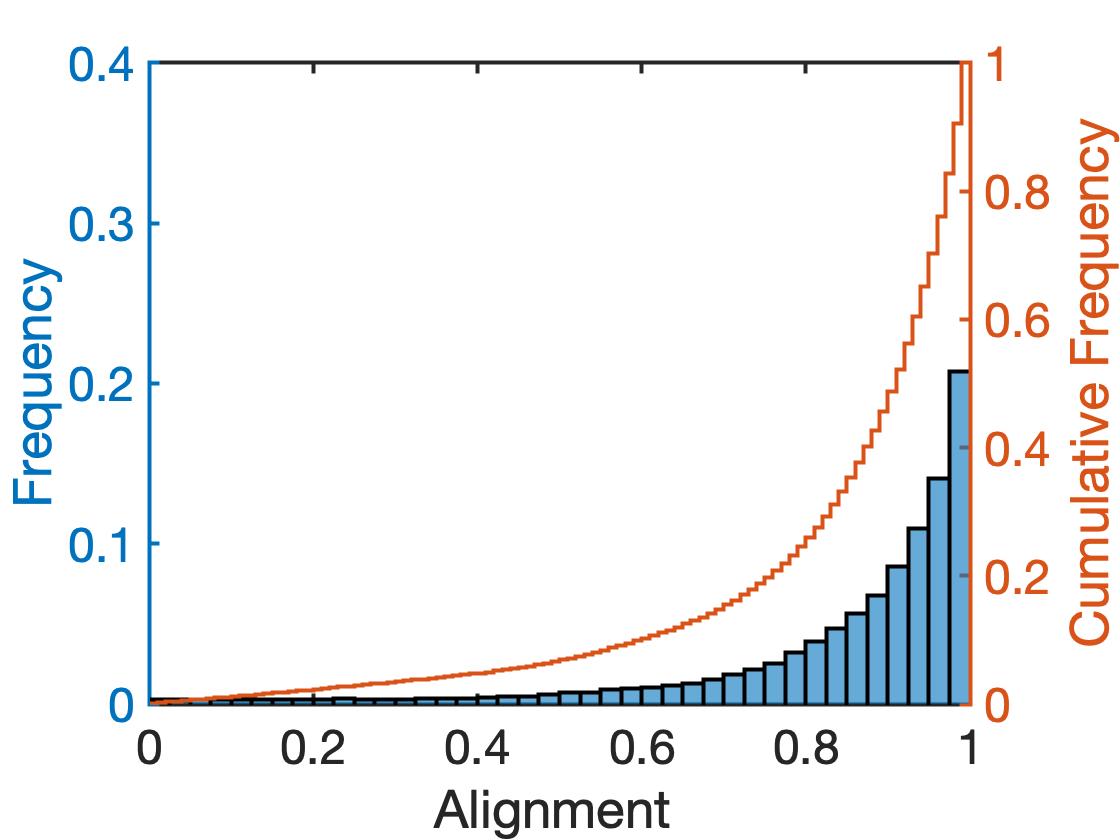}
		\caption{ }
		\label{fig:sample_a_m22e2_freq}
	\end{subfigure}%
	\quad
	\begin{subfigure}{.3\textwidth}
		\centering
		\includegraphics[width=1\linewidth]{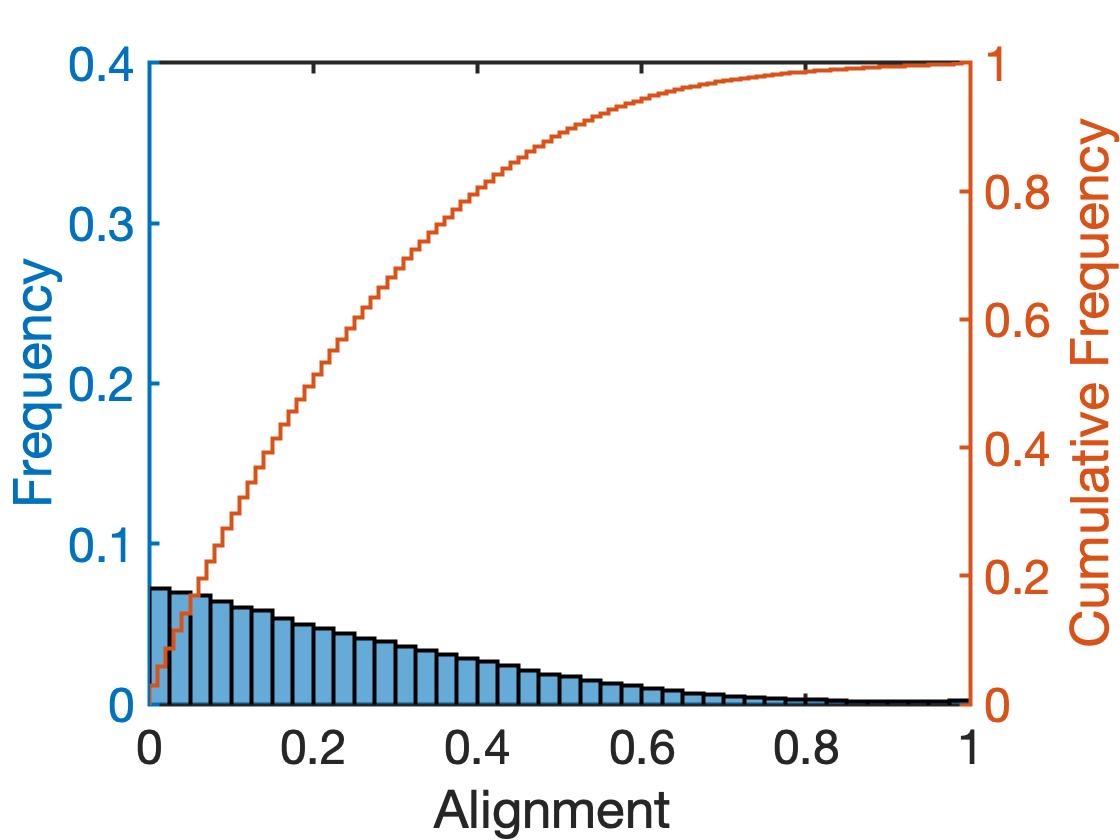}
		\caption{ }
		\label{fig:2}
	\end{subfigure}%
	\caption{Frequency distribution for the alignment of the gradient vector in the HUHS sample for Mode 2 with the grain axes: (a) grain axis 1; (b) grain axis 2; (c) grain axis 3. }
		\label{fig:HUHS_grnaxes2mode2_frequency}
\end{figure}

\begin{figure}[h!]
	\centering
	\begin{subfigure}{.3\textwidth}
		\centering
		\includegraphics[width=1\linewidth]{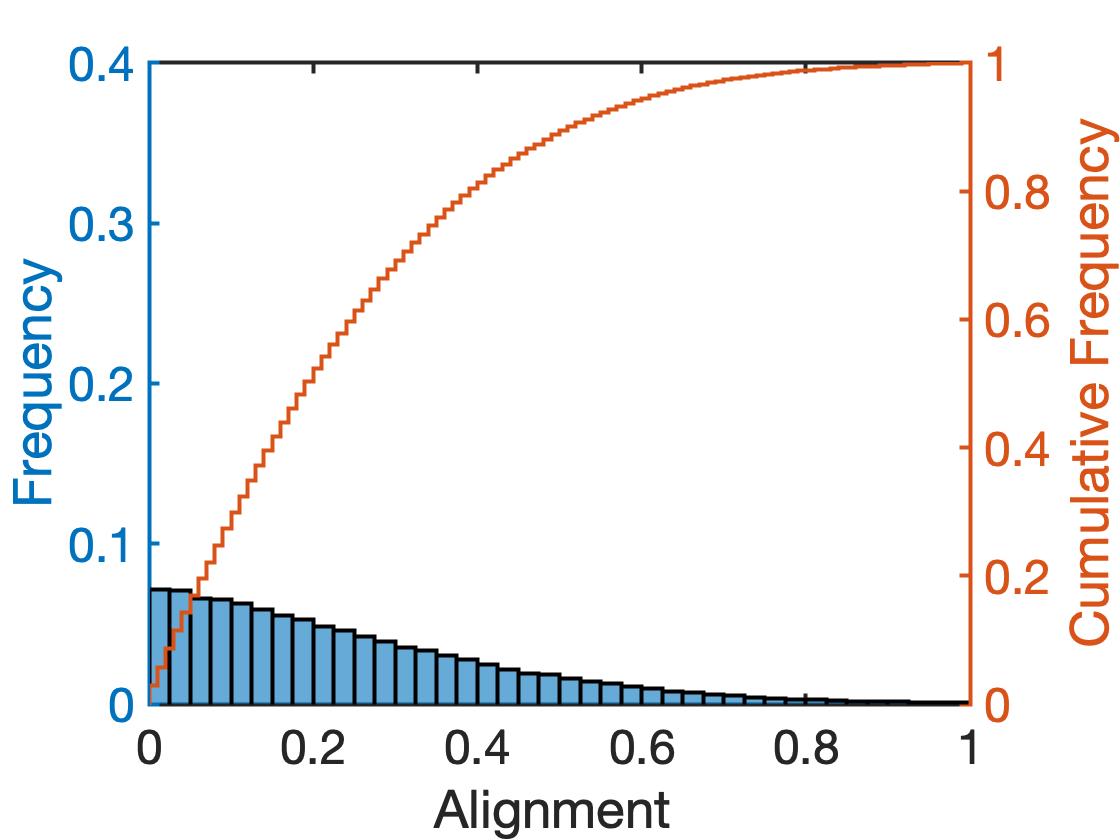}
		\caption{ }
		\label{fig:sample_a_m32e1_freq}
	\end{subfigure}%
	\quad
	\begin{subfigure}{.3\textwidth}
		\centering
		\includegraphics[width=1\linewidth]{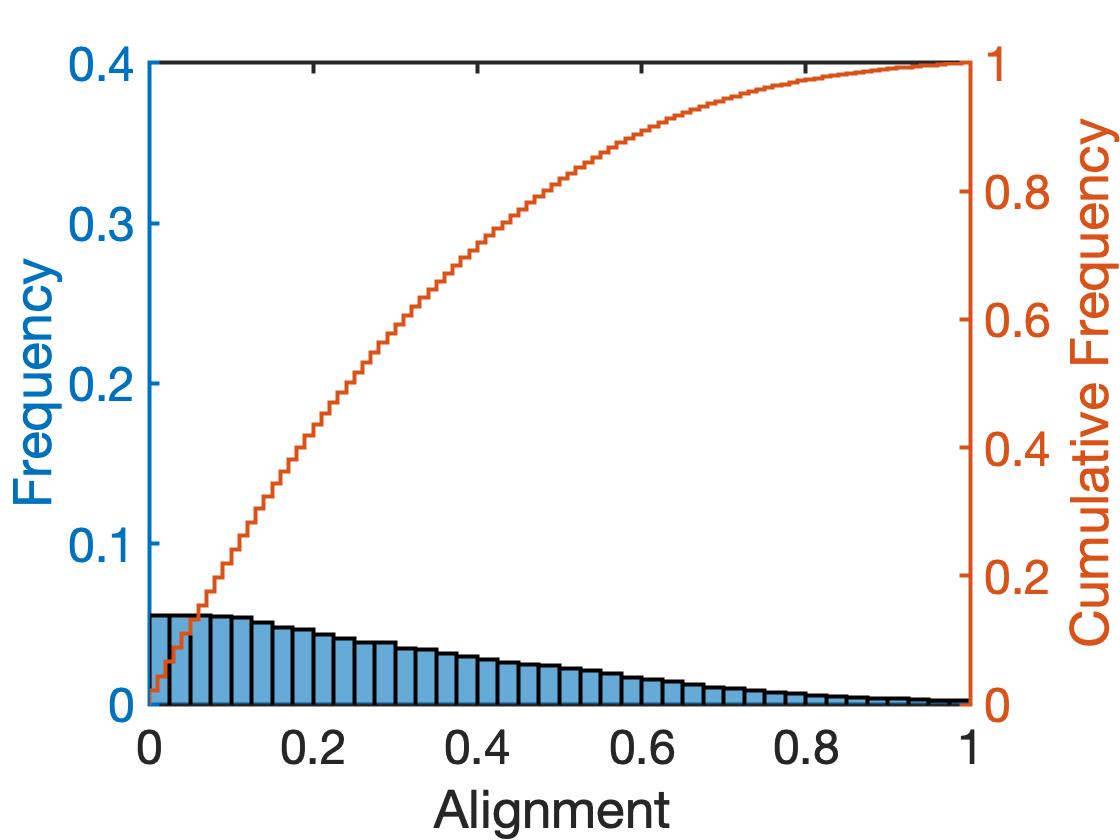}
		\caption{ }
		\label{fig:sample_a_m32e2_freq}
	\end{subfigure}%
	\quad
	\begin{subfigure}{.3\textwidth}
		\centering
		\includegraphics[width=1\linewidth]{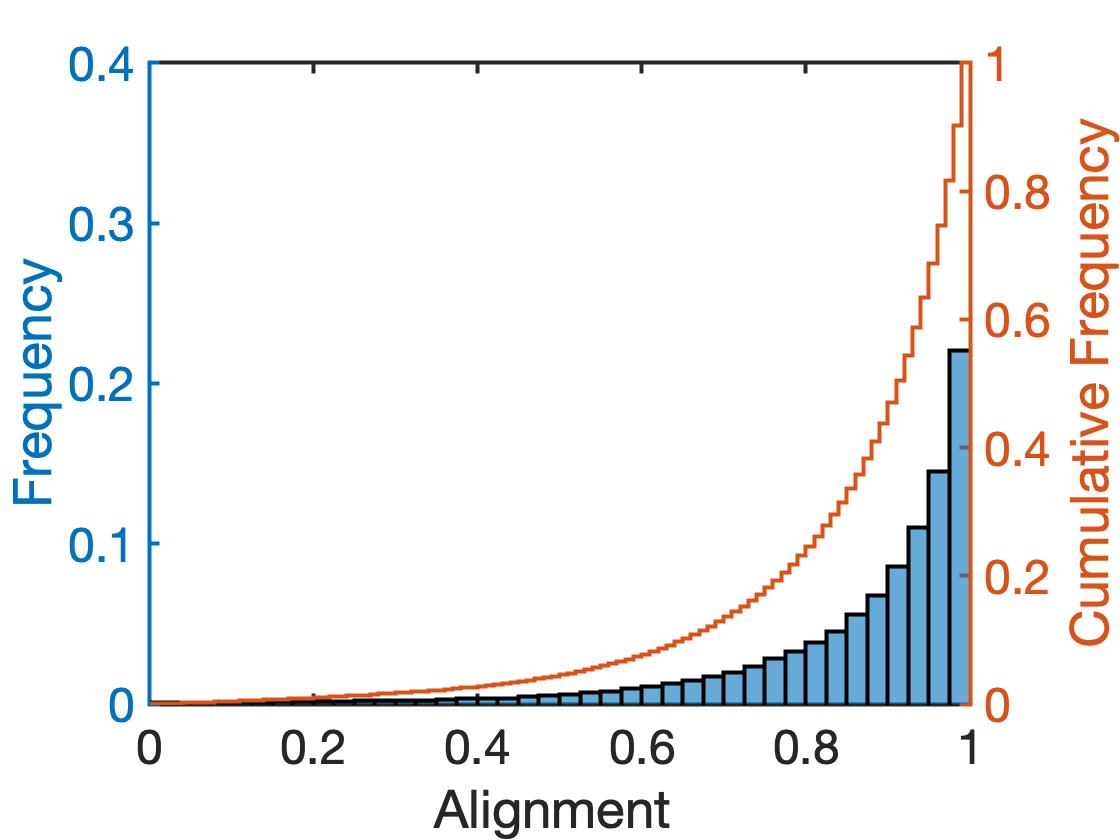}
		\caption{ }
		\label{fig:sample_a_m32e3_freq}
	\end{subfigure}%
	\caption{Frequency distribution for the alignment of the gradient vector in the HUHS sample for Mode 3 with the grain axes: (a) grain axis 1; (b) grain axis 2; (c) grain axis 3. }
		\label{fig:HUHS_grnaxes2mode3_frequency}
\end{figure}

Figures~\ref{fig:harmonicmodesgrad_voronoi}, \ref{fig:harmonicmodesgrad_dia0p35_sph0p06} and \ref{fig:harmonicmodesgrad_dia0p15_sph0p03}
show the gradients for Modes 2, 5 and  9 for each of the sample variants.
Here we show the magnitudes of the vector gradients. 
Although the direction of the modes (the gradient vector) appear to be haphazard, there is a correlation with the grain axes.  Figure~\ref{fig:HUHS_grnaxes2mode1} displays the alignment of the gradient vector for Mode 2 with each of the three grain axis vectors obtained from the dot product of the two vectors for the HUHS sample variant.  
Visually this plot suggest a strong correlation.
Figures~\ref{fig:HUHS_grnaxes2mode1_frequency}, \ref{fig:HUHS_grnaxes2mode2_frequency} and \ref{fig:HUHS_grnaxes2mode3_frequency} show the frequency distributions for the first three modes
of the HUHS sample variant with respect to the three grain axes.  
What can be inferred from the visual display in  Figure~\ref{fig:HUHS_grnaxes2mode1}  is confirmed statistically in these plots for the first three (non-constant) modes: the gradient vector of Mode 2 aligns with the major diameter axis while the gradient vectors of Modes 3 and 4 align with the two minor axes.
Qualitatively similar distributions exist for the Voronoi and LULS sample variants.
We have not explored the correlations between higher modes and the grain axes, but suspect correlations do exist.
\clearpage
\clearpage


\newpage
\subsection{Simulated responses of samples using \fepx }
Data for this demonstration application were generated with the finite element code, \fepx.  
\fepx\, simulates the elasto-viscoplastic response of polycrystals over  loading sequences that can 
induce large strain deformations.  The constitutive behaviors are defined at the crystal scale: anisotropic elasticity given by Hooke's law and anisotropic plasticity resulting from slip on a limited number of slip systems (referred to as restricted slip).  Slip is rate dependent.  The nonlinear solution algorithms facilitate the specification of low rate sensitivity as is typically the case of metallic alloys at low homologous temperatures.    For the simulation of the three sample variants, the single crystal properties of a FCC stainless steel (AL6XN)  were chosen.  The determination of the needed parameters is documented in \cite{pos_daw_mmta_2019a}.  

\begin{figure}[h]
\begin{center}
\includegraphics*[width=10cm]{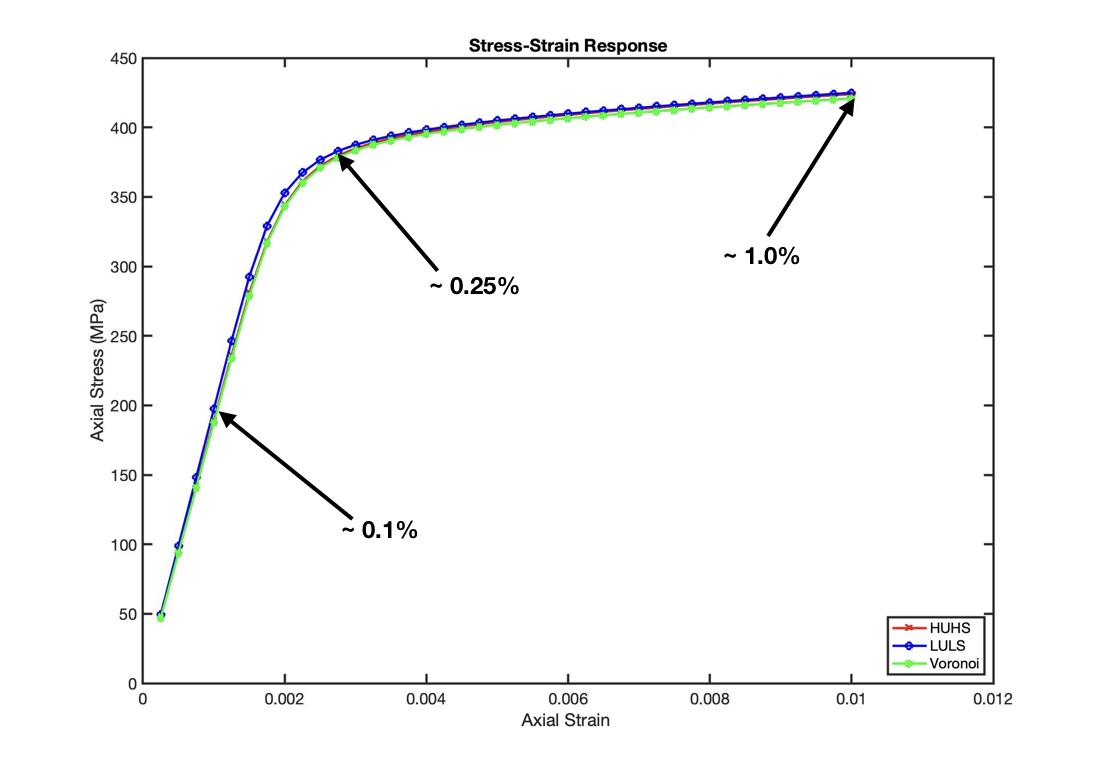}
\caption{Stress-strain curves for the Voronoi, LULS and HUHS samples for axial extension to 1\% nominal strain in the $z$ direction.}
\label{fig:sig_eps_allsamples}
\end{center}
\end{figure}

The samples variants all were subjected to the same loading:  extension in the $z$ coordinate at fixed 'cross-head' velocity to a nominal strain of 1\%.   40 equal time steps were used to reach the nominal strain. The lateral surfaces of the sample variants were traction-free.  The tractions on the exterior surfaces were integrated over those surfaces to generate nominal stress-strain curves, as shown in Figure~\ref{fig:sig_eps_allsamples}.  Small variations in the stress can be observed, particularly through the elastic-plastic transition (the knee of the curve). 

\begin{figure}[htbp]
	\centering
	\begin{subfigure}{.3\textwidth}
		\centering
		\includegraphics[width=1\linewidth]{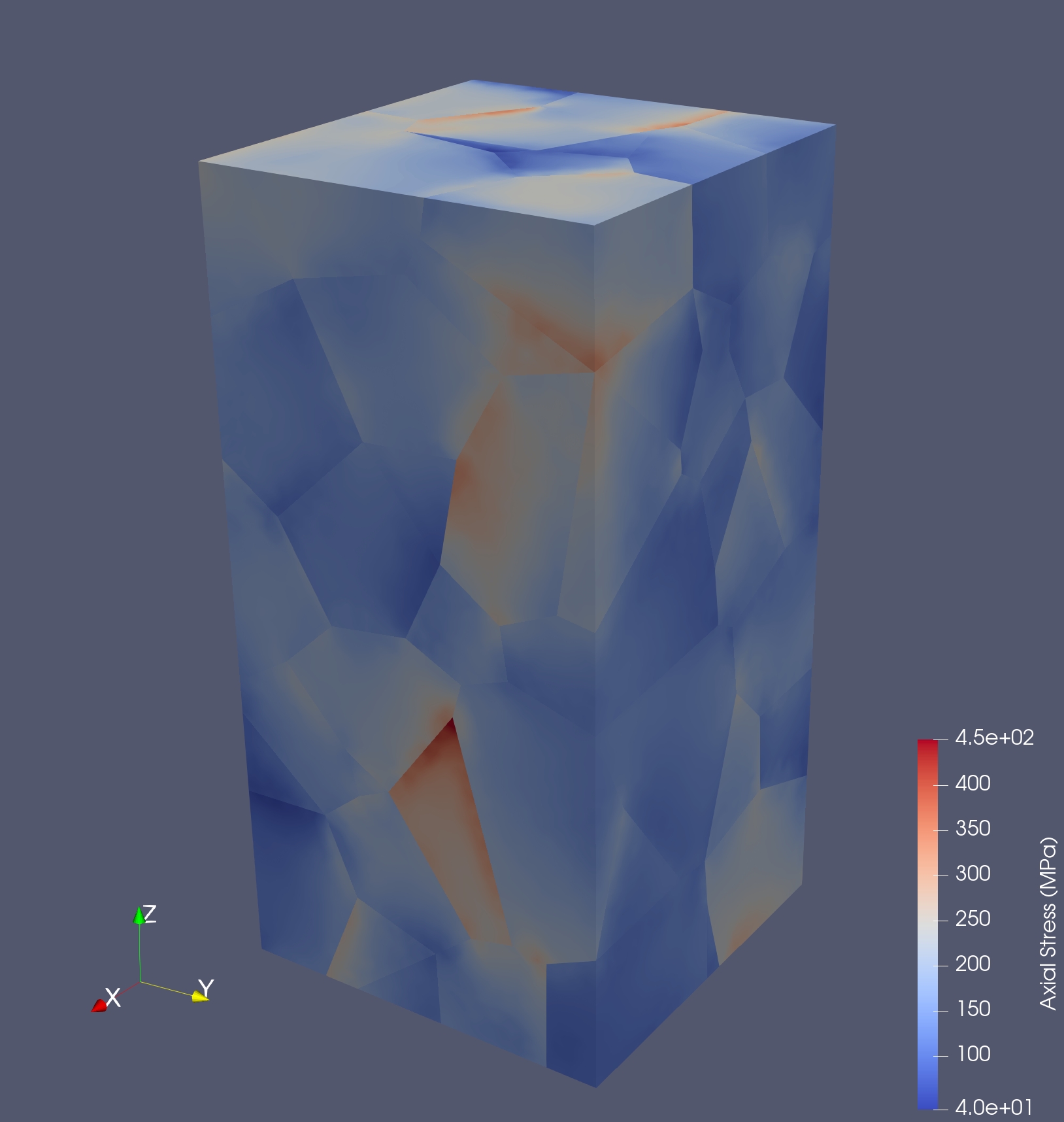}
		\caption{ }
		\label{fig:Voronoi_eps0p001}
	\end{subfigure}%
	\quad
	\begin{subfigure}{.3\textwidth}
		\centering
		\includegraphics[width=1\linewidth]{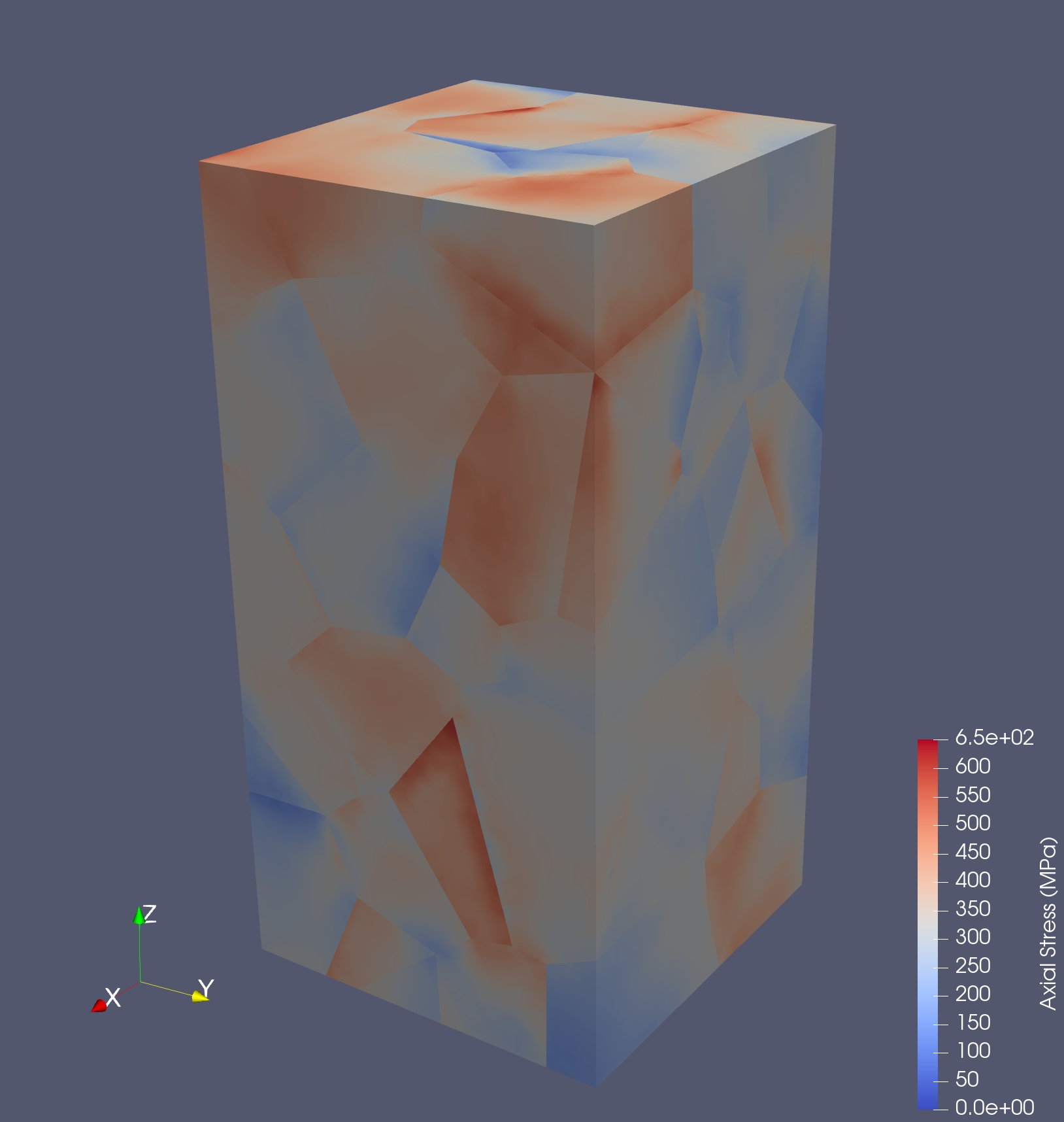}
		\caption{ }
		\label{fig:Voronoi_eps0p0025}
	\end{subfigure}%
	\quad
	\begin{subfigure}{.3\textwidth}
		\centering
		\includegraphics[width=1\linewidth]{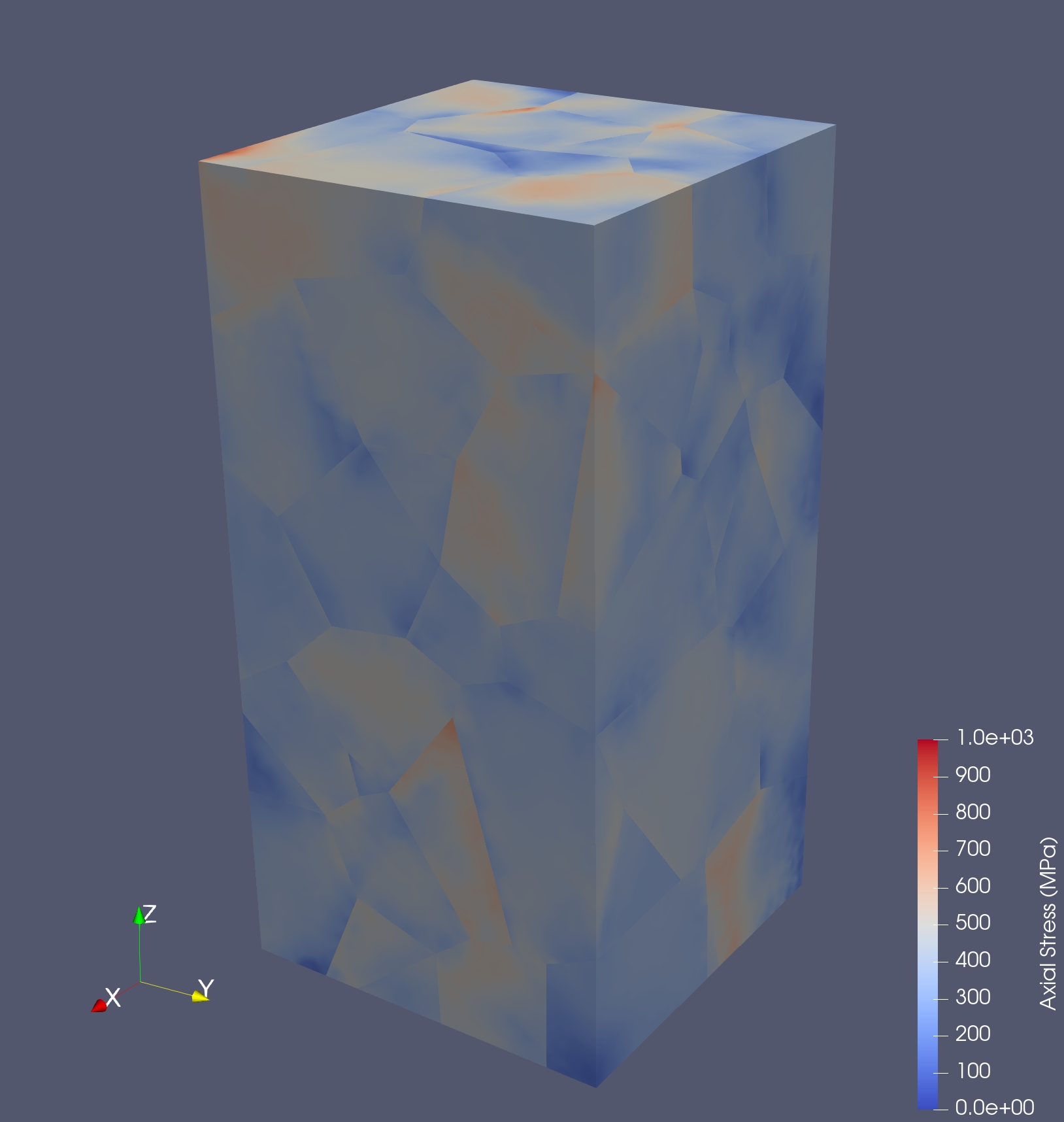}
		\caption{ }
		\label{fig:Voronoi_eps0p01}
	\end{subfigure}%
	\caption{Axial stress distributions over the Voronoi sample using \fepx ~ at nominal axial strains of  (a) 0.1\% ; (b) 0.25\%;  and (c) 1.0\%. Note that each distribution has its own scale.}
		\label{fig:stress_distributions_Voronoi}
\end{figure}

\begin{figure}[htbp]
	\centering
	\begin{subfigure}{.3\textwidth}
		\centering
		\includegraphics[width=1\linewidth]{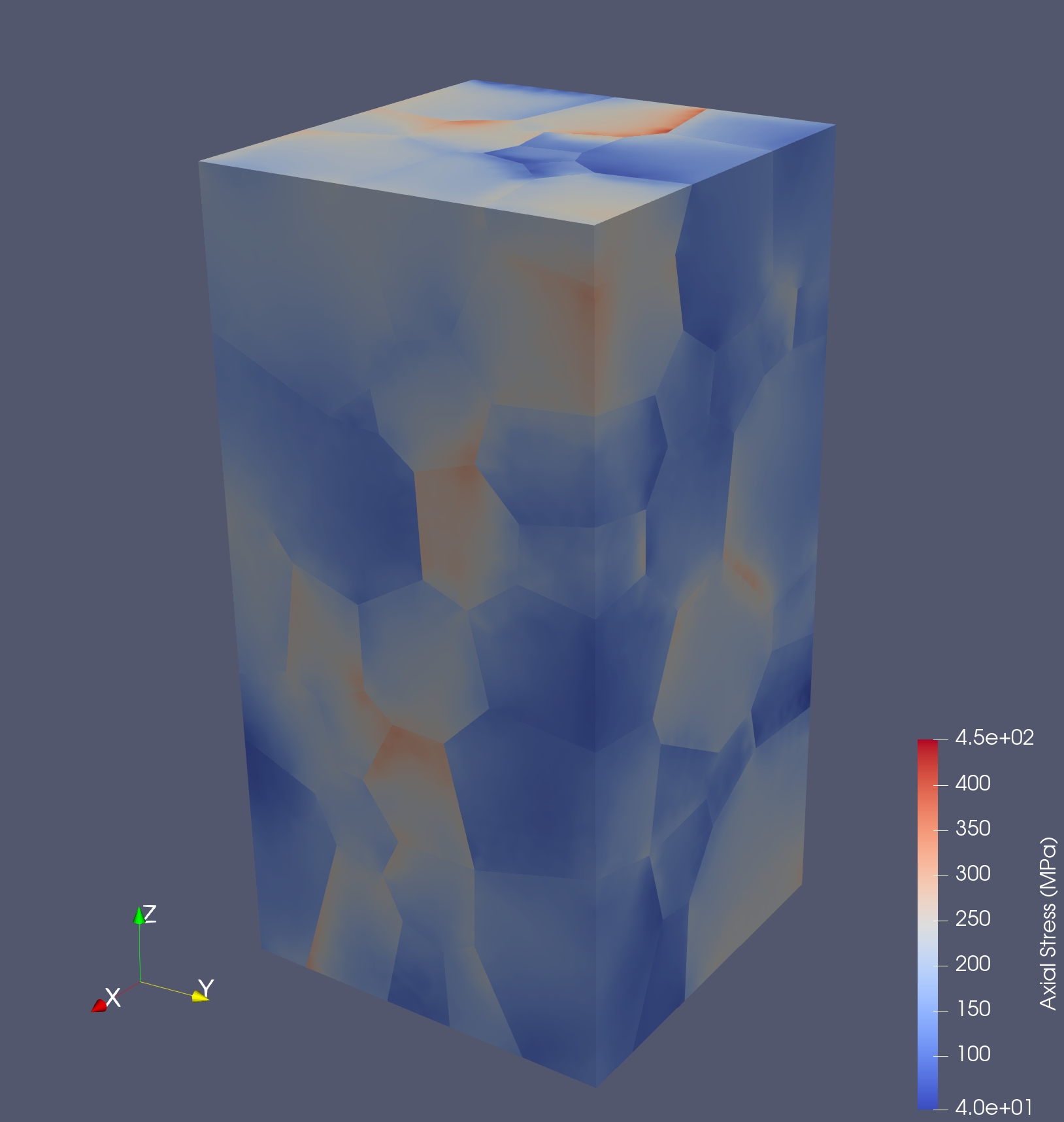}
		\caption{ }
		\label{fig:dia0p35_sph0p06_eps0p001}
	\end{subfigure}%
	\quad
	\begin{subfigure}{.3\textwidth}
		\centering
		\includegraphics[width=1\linewidth]{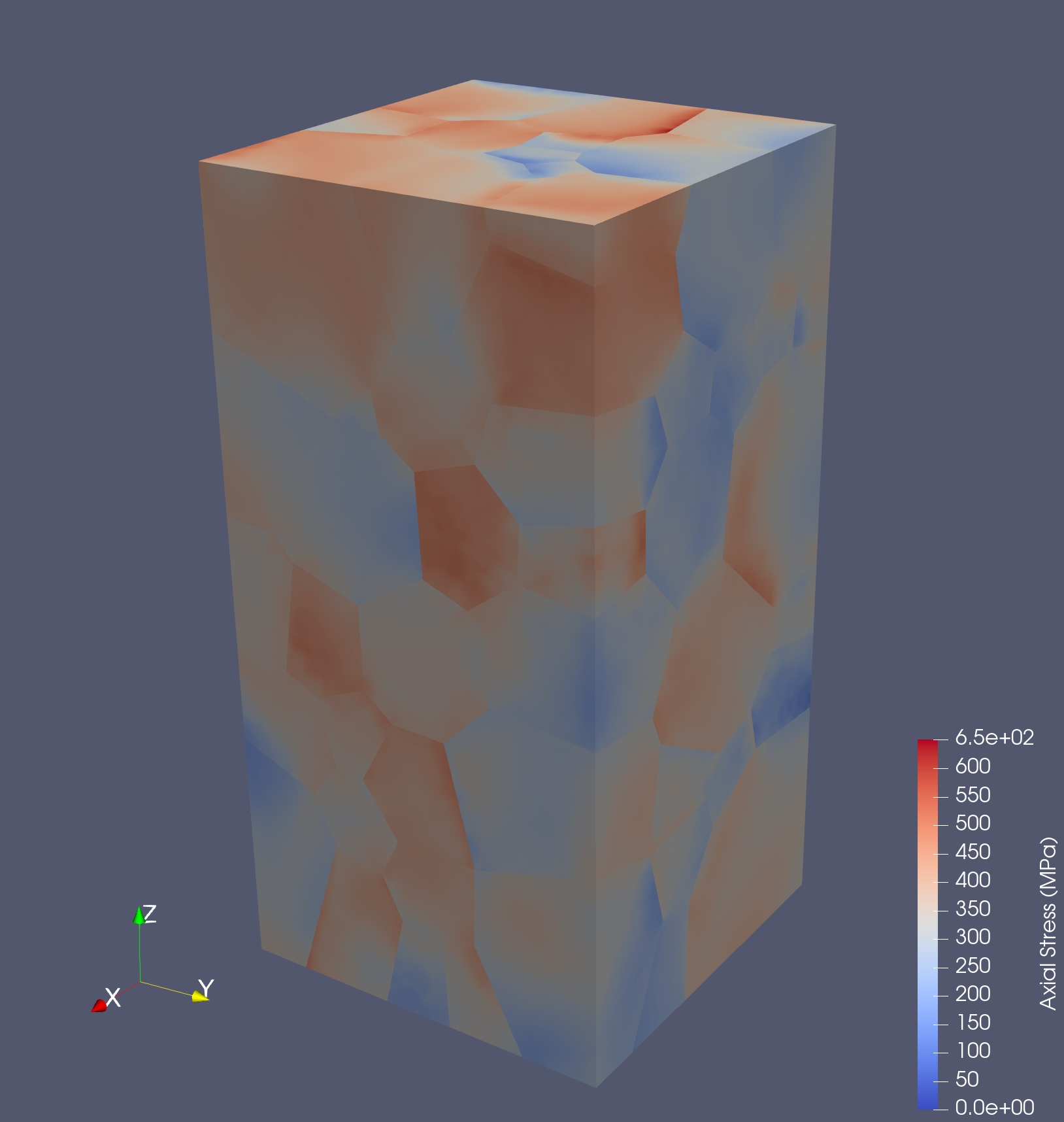}
		\caption{ }
		\label{fig:dia0p35_sph0p06_eps0p0025}
	\end{subfigure}%
	\quad
	\begin{subfigure}{.3\textwidth}
		\centering
		\includegraphics[width=1\linewidth]{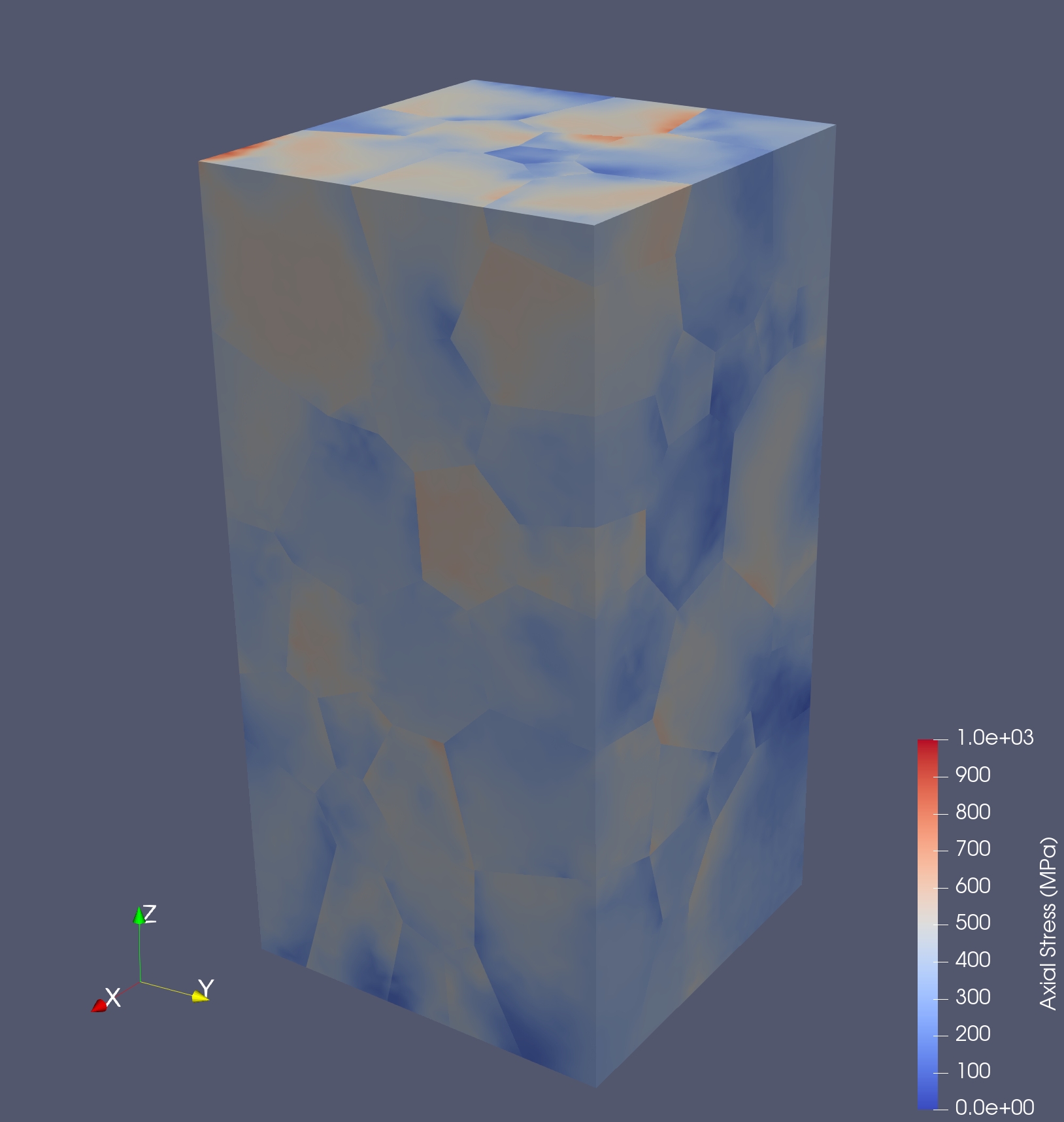}
		\caption{ }
		\label{fig:dia0p35_sph0p06_eps0p01}
	\end{subfigure}%
	\caption{Axial stress distributions over the LULS sample using \fepx ~ at nominal axial strains of  (a) 0.1\% ; (b) 0.25\%;  and (c) 1.0\%. Note that each distribution has its own scale.}
		\label{fig:stress_distributions_dia0p35_sph0p06}
\end{figure}

\begin{figure}[htbp]
	\centering
	\begin{subfigure}{.3\textwidth}
		\centering
		\includegraphics[width=1\linewidth]{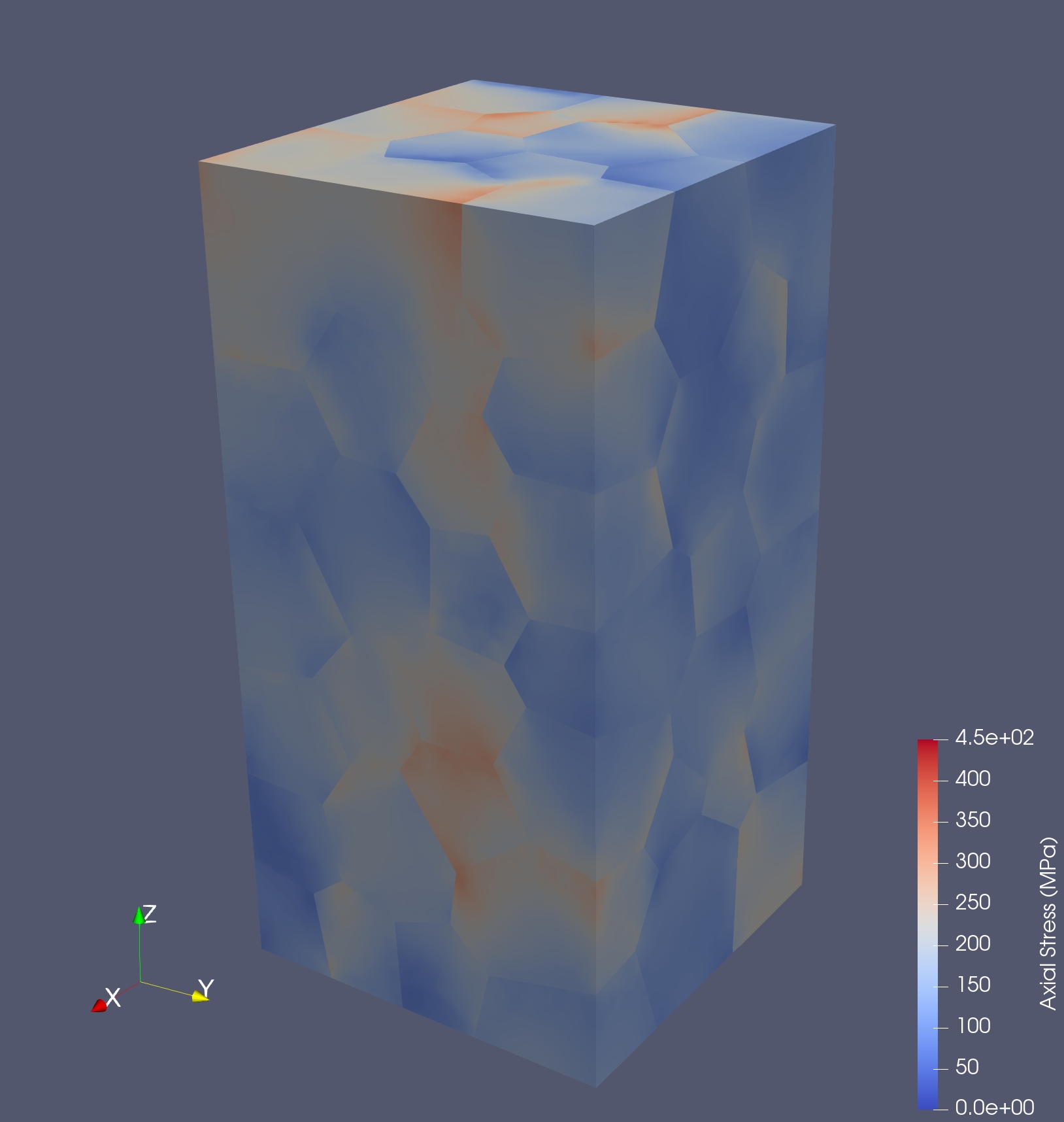}
		\caption{ }
		\label{fig:dia0p15_sph0p03_eps0p001}
	\end{subfigure}%
	\quad
	\begin{subfigure}{.3\textwidth}
		\centering
		\includegraphics[width=1\linewidth]{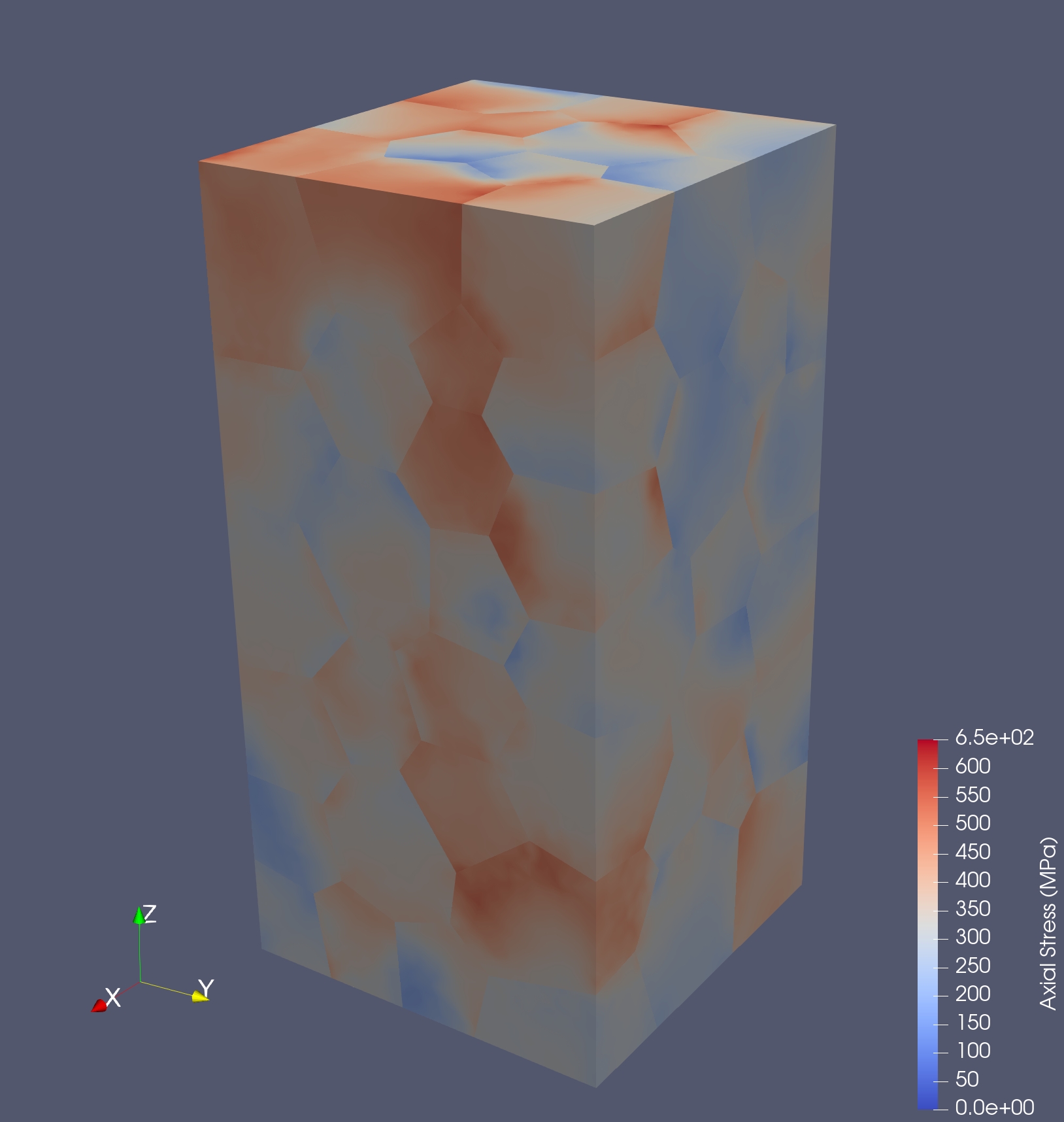}
		\caption{ }
		\label{fig:dia0p15_sph0p03_eps0p0025}
	\end{subfigure}%
	\quad
	\begin{subfigure}{.3\textwidth}
		\centering
		\includegraphics[width=1\linewidth]{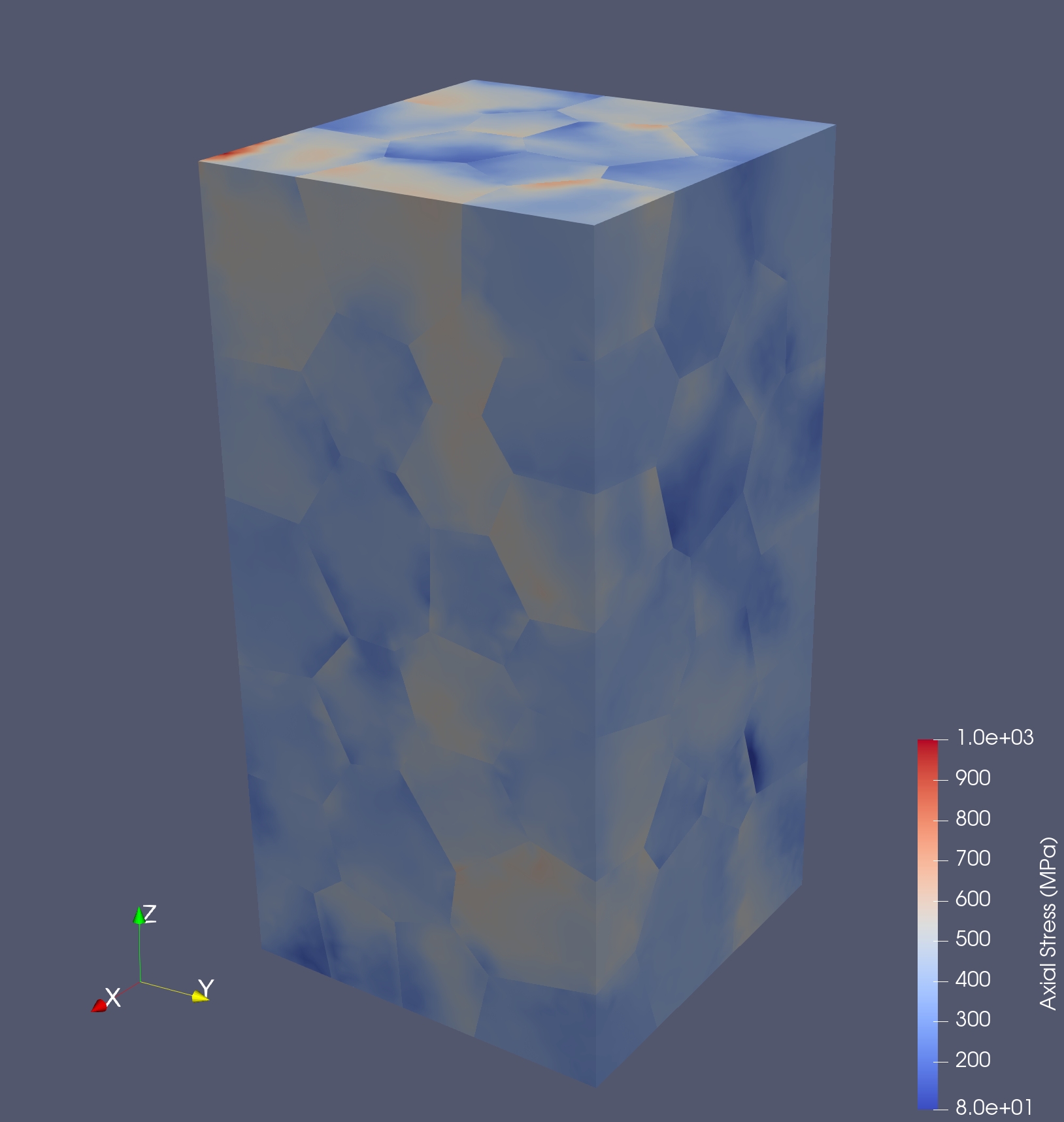}
		\caption{ }
		\label{fig:dia0p15_sph0p03_eps0p01}
	\end{subfigure}%
	\caption{Axial stress distributions over the HUHS sample using \fepx ~ at nominal axial strains of  (a) 0.1\% ; (b) 0.25\%;  and (c) 1.0\%. Note that each distribution has its own scale.}
		\label{fig:stress_distributions_dia0p15_sph0p03}
\end{figure}

Stress distributions on full sample at three levels of nominal strain (0.1\%, 0.25\% and 1\%) are shown in Figures~\ref{fig:stress_distributions_Voronoi}, \ref{fig:stress_distributions_dia0p35_sph0p06} and \ref{fig:stress_distributions_dia0p15_sph0p03}.  
The scales are different for the three strain levels but the same across the sample variants.
Note that the first strain level, 0.1\%, is the same as that specified for the elastic \mechmet\, analyses presented in Section~\ref{sec:mechmet_metric}.
Similarities of the stress distributions do appear to exist across the sample variants.
However, the most noteworthy feature is the changes in the patterns of stress that
occur as loading proceeds through the elastic-plastic transition, indicating that the stress re-distributes with the onset of yielding. 

\subsection{Harmonic representation of stress in 5 grains}
\begin{figure}[htbp]
	\centering
	\begin{subfigure}{0.30\textwidth}
		\centering
		\includegraphics[width=1\linewidth]{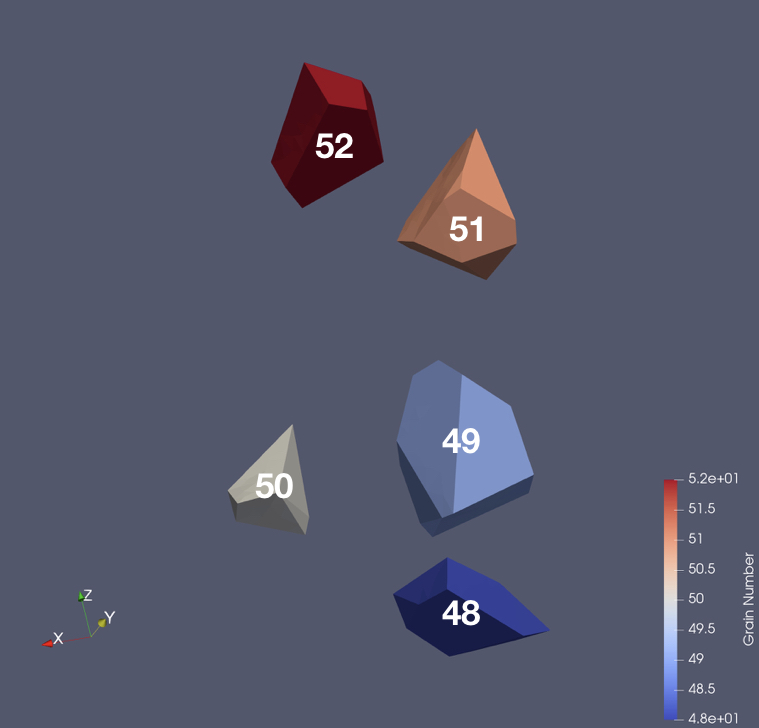}
		\caption{ }
		\label{fig:voronoi_5grains}
	\end{subfigure}%
	\quad
	\begin{subfigure}{0.30\textwidth}
		\centering
		\includegraphics[width=1\linewidth]{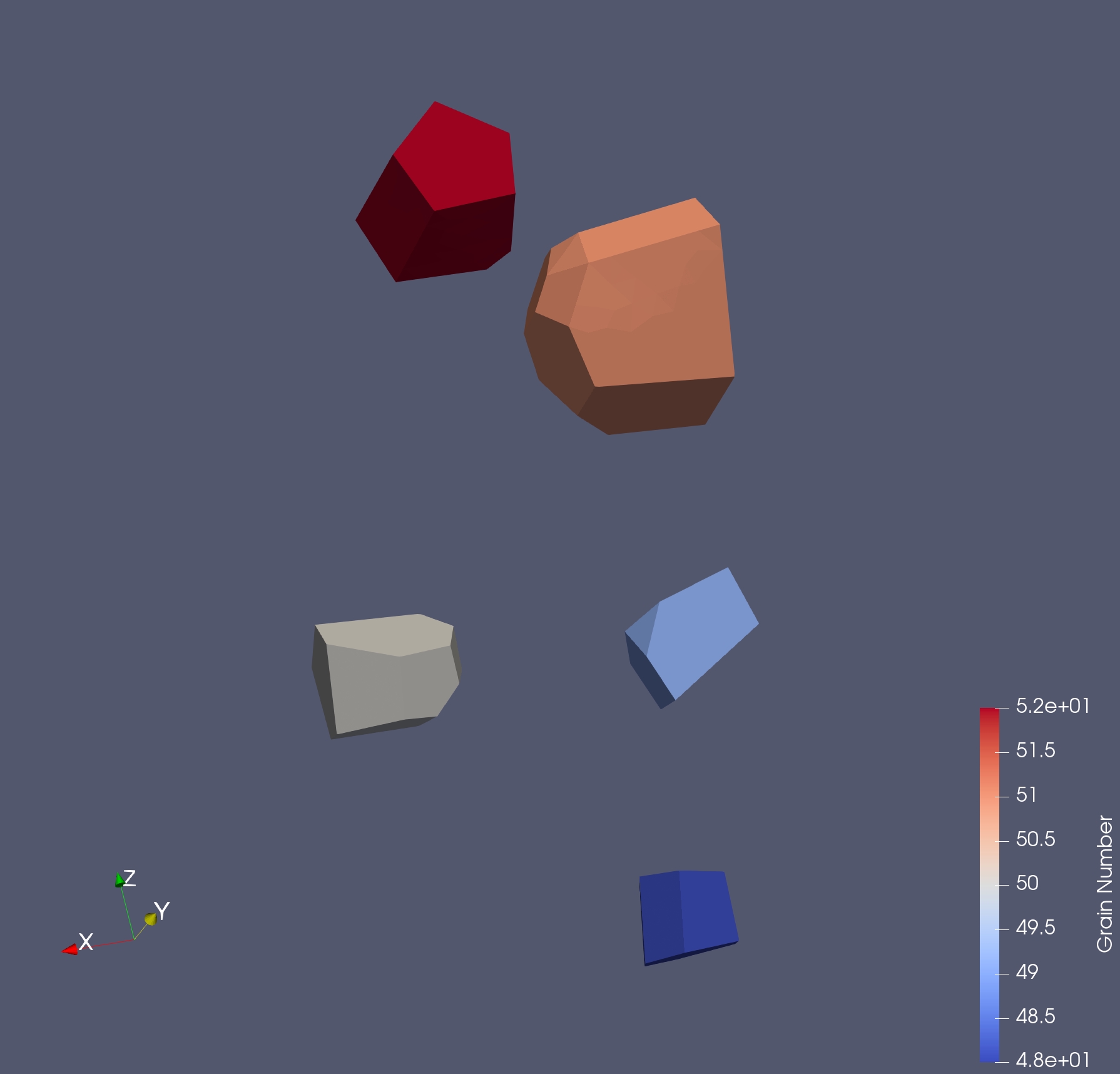}
		\caption{ }
		\label{fig:dia0p35_sph0p06_5grains}
	\end{subfigure}%
	\quad
	\begin{subfigure}{0.30\textwidth}
		\centering
		\includegraphics[width=1\linewidth]{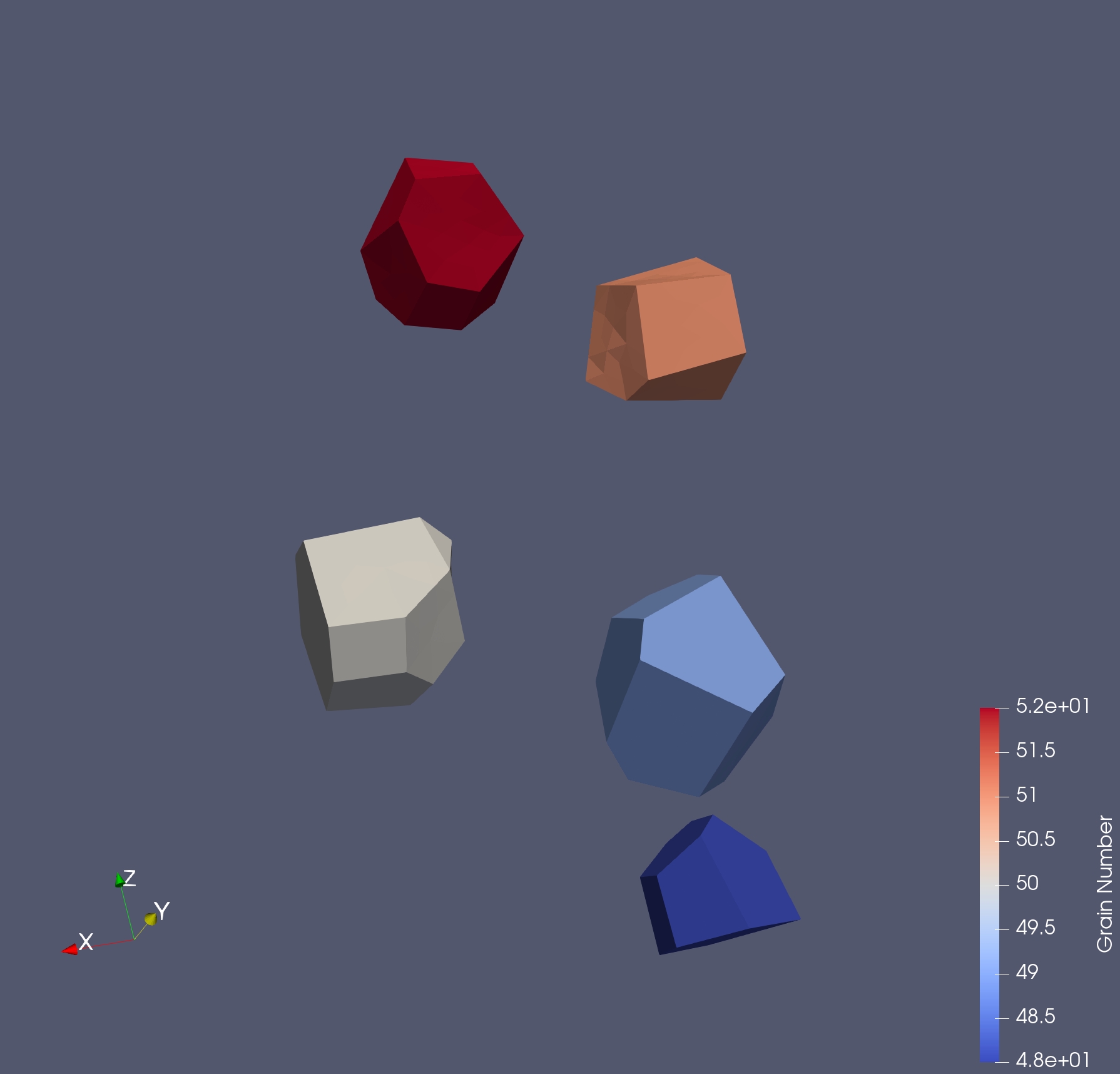}
		\caption{ }
		\label{fig:dia0p15_sph0p03_5grains}
	\end{subfigure}%
	\caption{Grains 48, 49, 50, 51 and 52 in:  (a) Voronoi sample ; (b) LULS sample;  and (c) HUHS sample. Note that grains with the same number have the same initial lattice orientation.}
		\label{fig:5gr_grainnumbers}
\end{figure}

We have chosen 5 grains to examine individually: Grains 48, 49, 50, 51, and 52.  These are the 5 grains in the middle of the numbered list and were chosen simply for convenience and because they were not end grains (having a facet on the loaded surface).  
 The same five grains are examined for all three sample variants. 
 Figure~\ref{fig:5gr_grainnumbers} shows the sets of grains.
 The morphological attributes built into the variants are evident: grains of the Voronoi sample have more irregular shapes; grains of the LULS sample have greater size variations; grains of the HUHS sample have more most regular shapes and most similar sizes.
 
\begin{figure}[htbp]
	\centering
	\begin{subfigure}{0.30\textwidth}
		\centering
		\includegraphics[width=1\linewidth]{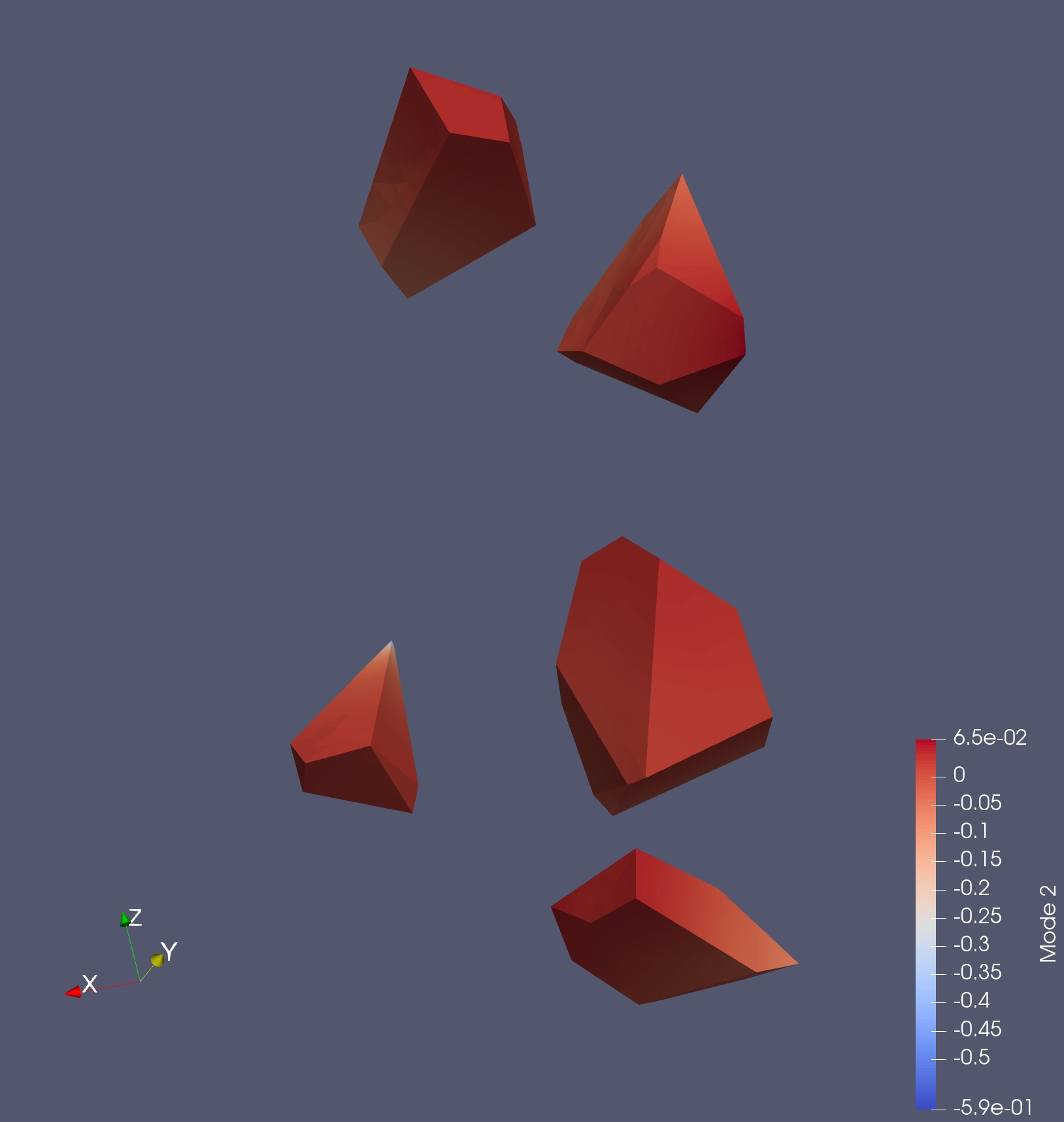}
		\caption{ }
		\label{fig:voronoi_mode1_5gr}
	\end{subfigure}%
	\quad
	\begin{subfigure}{0.30\textwidth}
		\centering
		\includegraphics[width=1\linewidth]{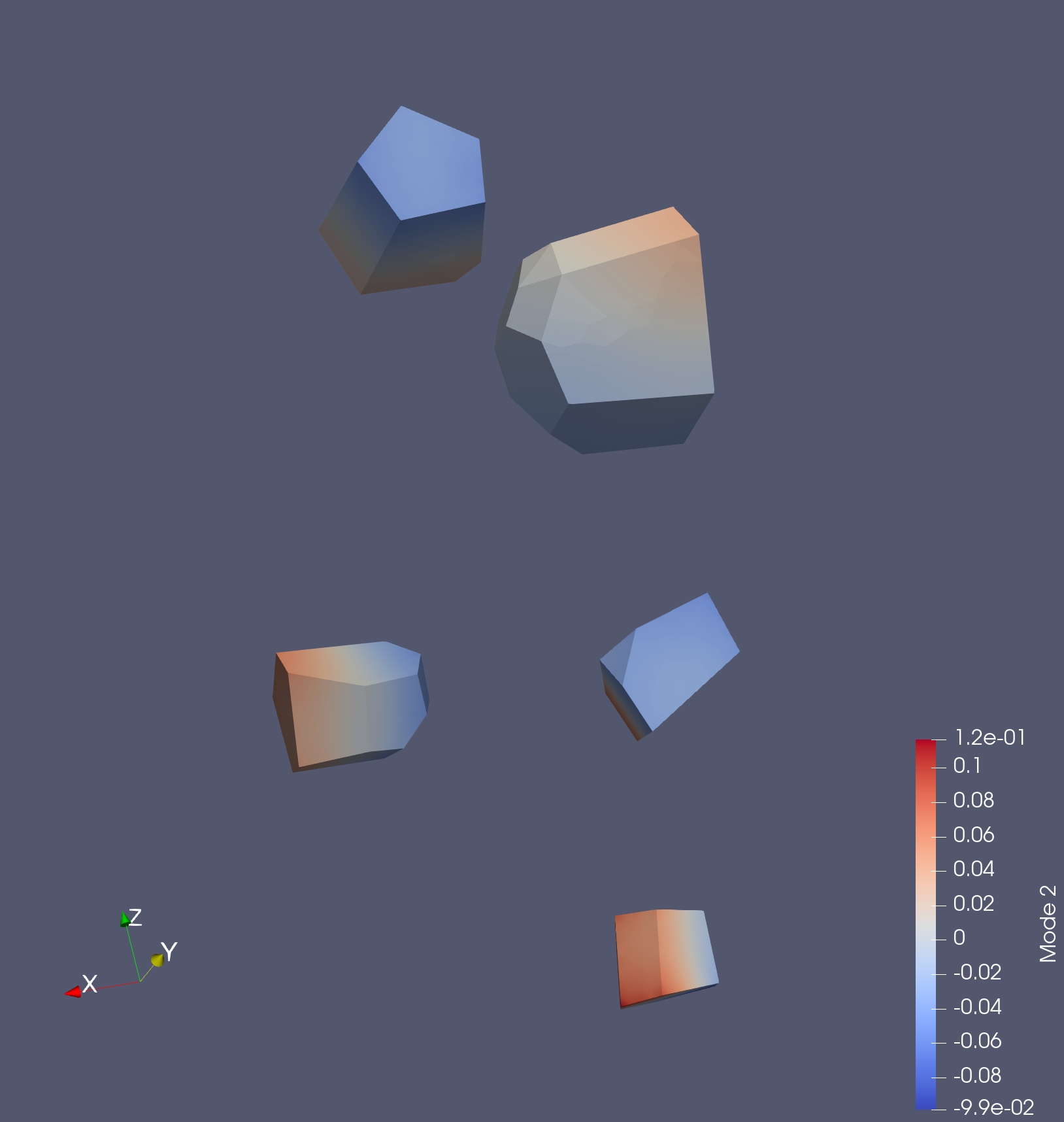}
		\caption{ }
		\label{fig:dia0p35_sph0p06_mode1_5gr}
	\end{subfigure}%
	\quad
	\begin{subfigure}{0.30\textwidth}
		\centering
		\includegraphics[width=1\linewidth]{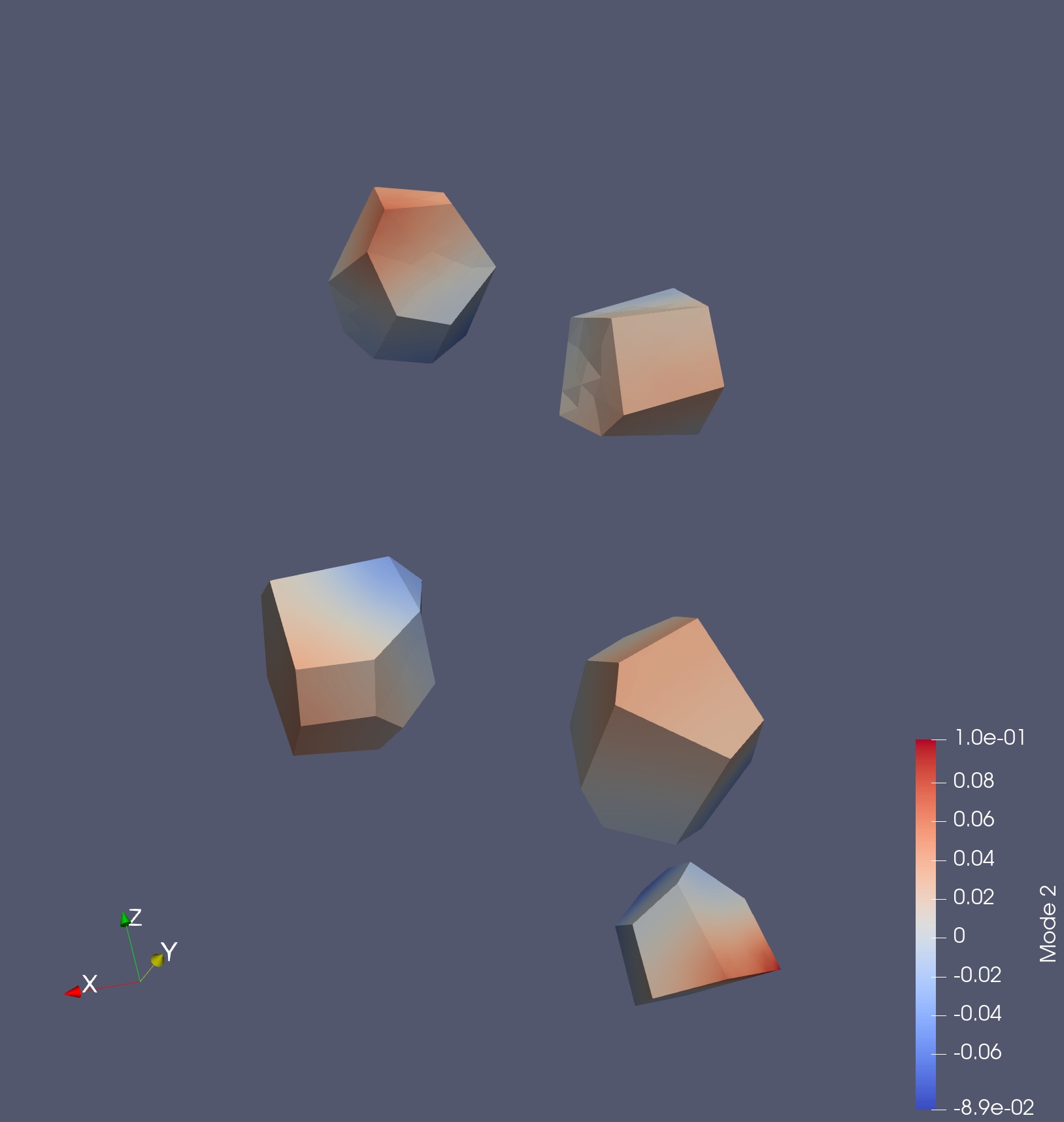}
		\caption{ }
		\label{fig:dia0p15_sph0p03_mode1_5gr}
	\end{subfigure}%
	\caption{Mode 2 for Grains 48, 49, 50, 51 and 52 in:  (a) Voronoi sample ; (b) LULS sample;  and (c) HUHS sample.  Note that the plots for the Voronoi sample are biased by an extreme value in Grain 50.}
		\label{fig:5gr_mode1}
\end{figure}

\begin{figure}[htbp]
	\centering
	\begin{subfigure}{0.30\textwidth}
		\centering
		\includegraphics[width=1\linewidth]{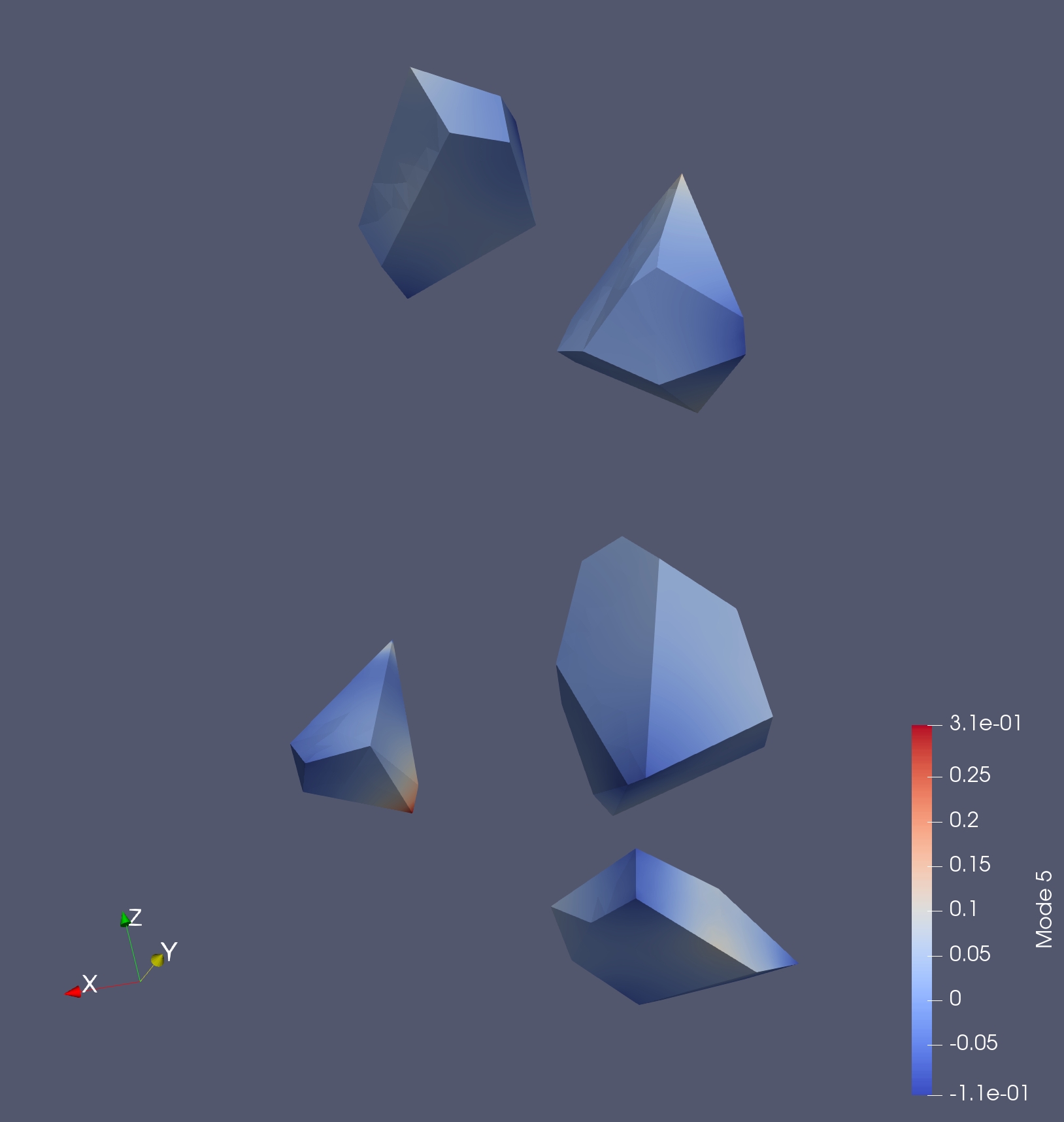}
		\caption{ }
		\label{fig:voronoi_mode4_5gr}
	\end{subfigure}%
	\quad
	\begin{subfigure}{0.30\textwidth}
		\centering
		\includegraphics[width=1\linewidth]{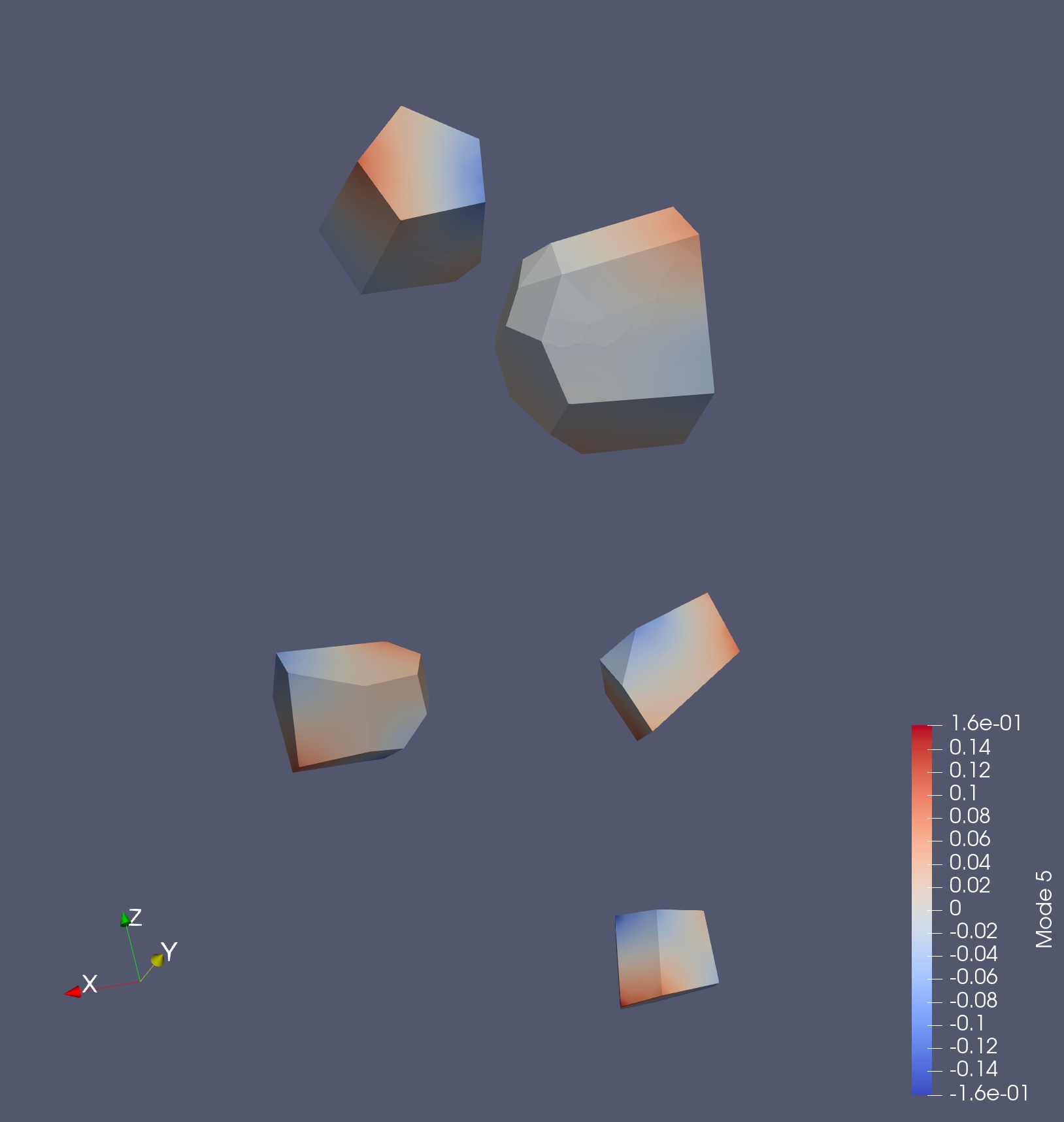}
		\caption{ }
		\label{fig:dia0p35_sph0p06_mode4_5gr}
	\end{subfigure}%
	\quad
	\begin{subfigure}{0.30\textwidth}
		\centering
		\includegraphics[width=1\linewidth]{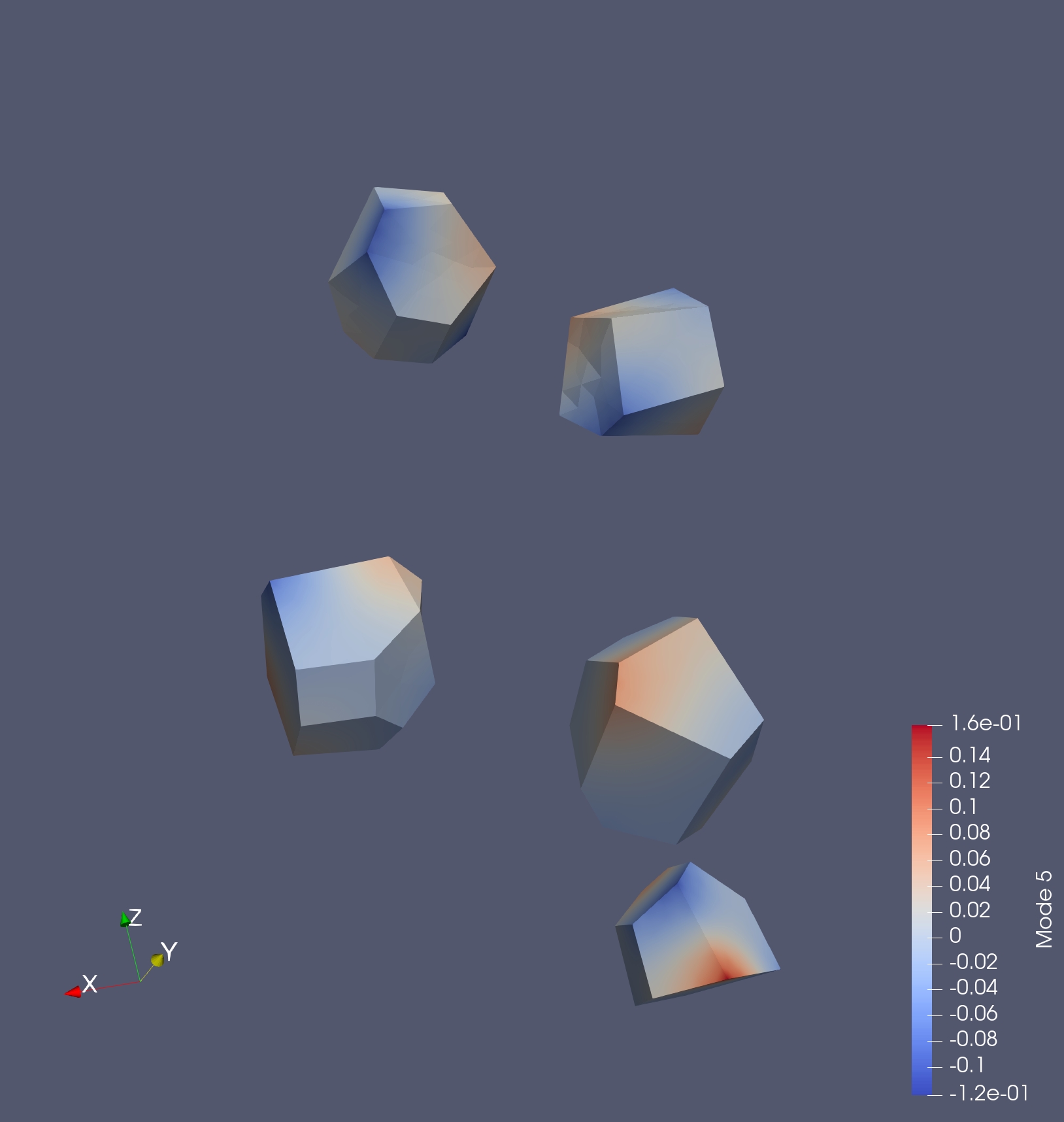}
		\caption{ }
		\label{fig:dia0p15_sph0p03_mode4_5gr}
	\end{subfigure}%
	\caption{Mode 5 for Grains 48, 49, 50, 51 and 52 in:  (a) Voronoi sample ; (b) LULS sample;  and (c) HUHS sample.  }
		\label{fig:5gr_mode4}
\end{figure}

\begin{figure}[htbp]
	\centering
	\begin{subfigure}{0.30\textwidth}
		\centering
		\includegraphics[width=1\linewidth]{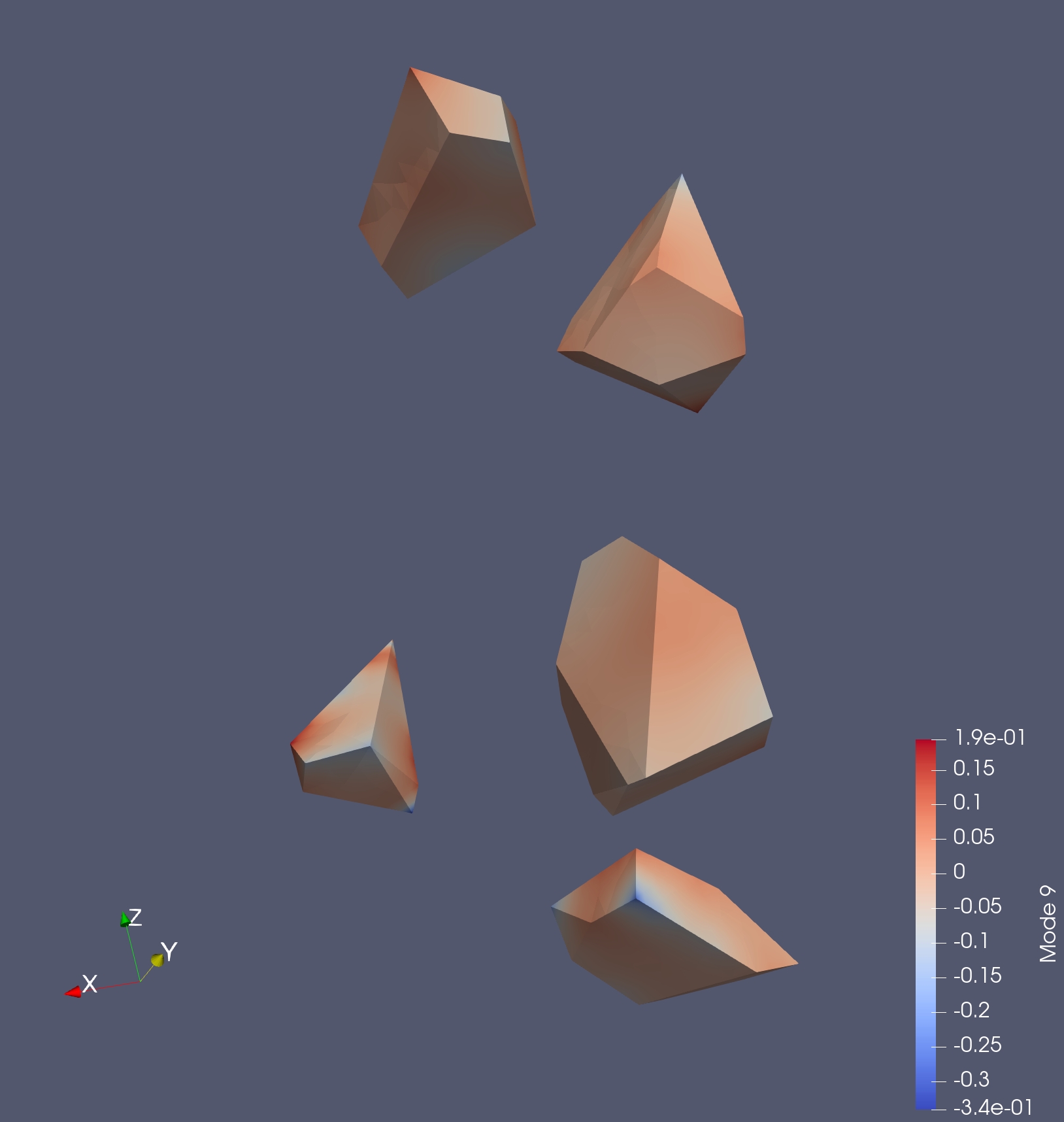}
		\caption{ }
		\label{fig:voronoi_mode8_5gr}
	\end{subfigure}%
	\quad
	\begin{subfigure}{0.30\textwidth}
		\centering
		\includegraphics[width=1\linewidth]{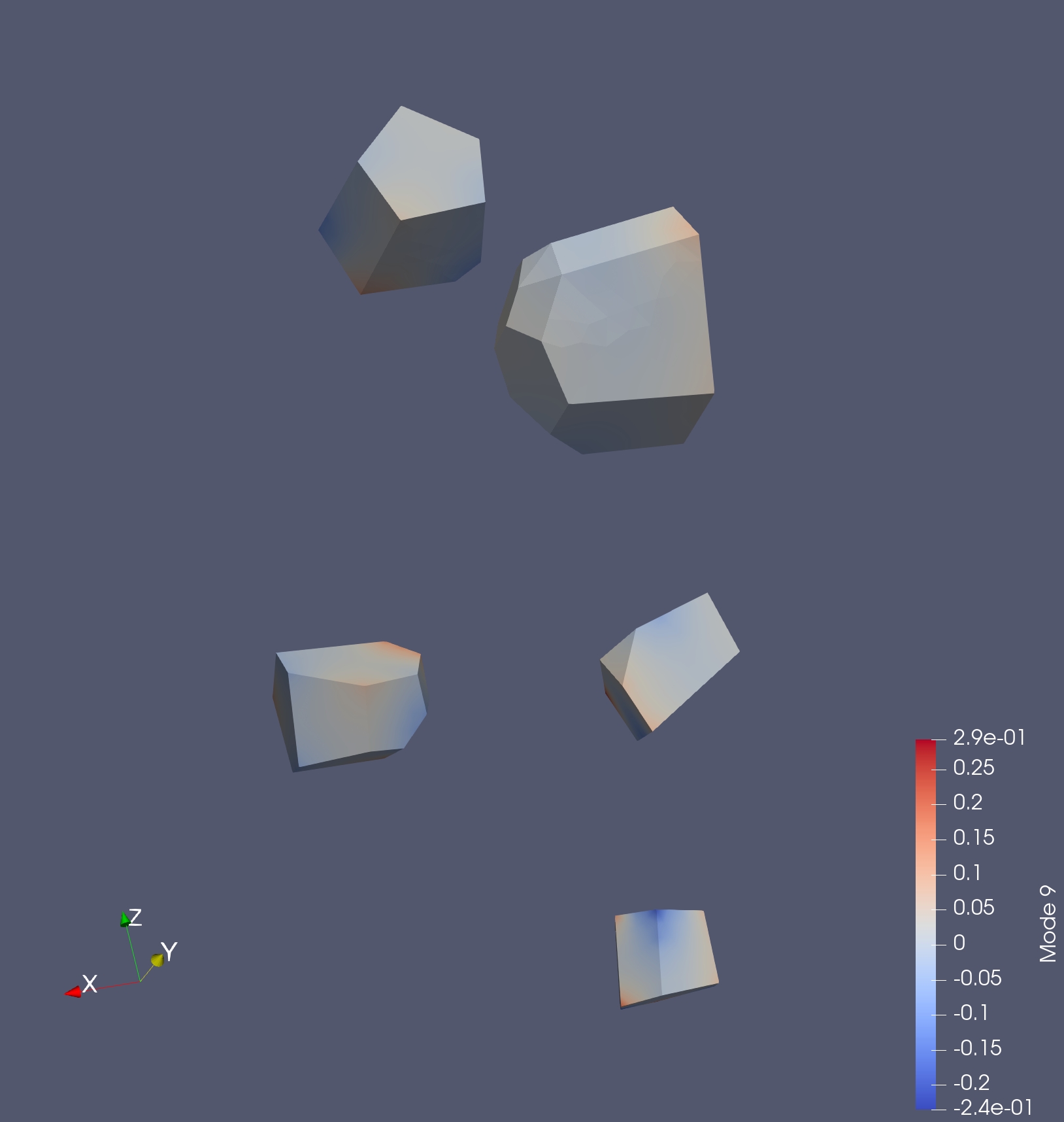}
		\caption{ }
		\label{fig:dia0p35_sph0p06_mode8_5gr}
	\end{subfigure}%
	\quad
	\begin{subfigure}{0.30\textwidth}
		\centering
		\includegraphics[width=1\linewidth]{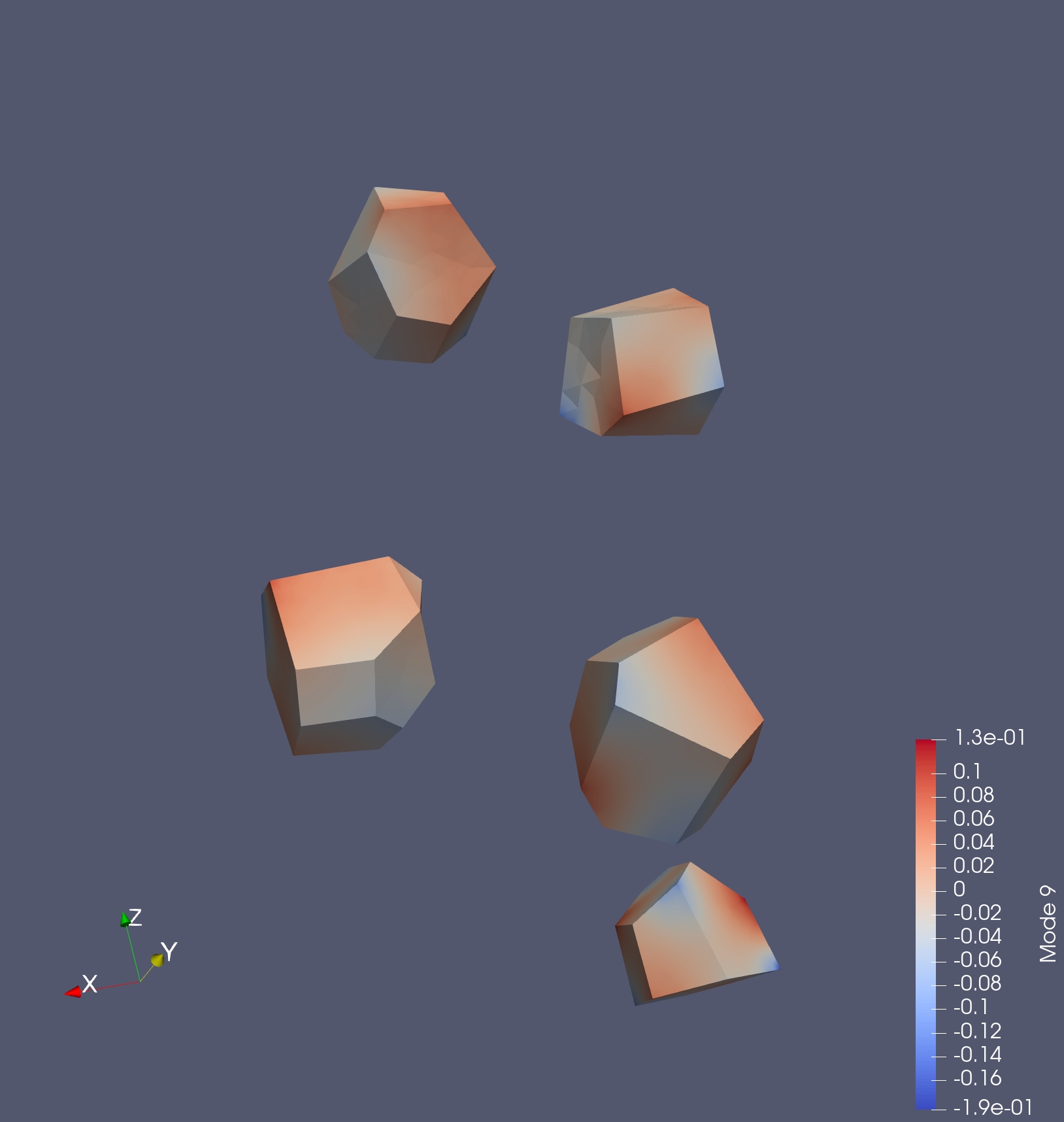}
		\caption{ }
		\label{fig:dia0p15_sph0p03_mode8_5gr}
	\end{subfigure}%
	\caption{Mode 9 for Grains 48, 49, 50, 51 and 52 in:  (a) Voronoi sample ; (b) LULS sample;  and (c) HUHS sample.  }
		\label{fig:5gr_mode8}
\end{figure}

 Modes 2, 5 and 9 for the 5 grains for all three sample variants are shown in  Figures~\ref{fig:5gr_mode1}, \ref{fig:5gr_mode4},  and \ref{fig:5gr_mode8}, respectively.
 As is the case for the entire sample, presented in Section~\ref{sec:harmonicmodes}, 
 the higher the mode number the more complex the mode distribution.  
 The Mode 2 distributions appear to be monotonic, whereas the other modes definitely are not.
  (Note that the Voronoi Mode 2 distribution in Figure~\ref{fig:voronoi_mode1_5gr} appears to be nearly constant, but actually is not; it appear so because of the
 extreme values of the harmonic function at one node. )

\begin{figure}[htbp]
	\centering
		\includegraphics[width=1\linewidth]{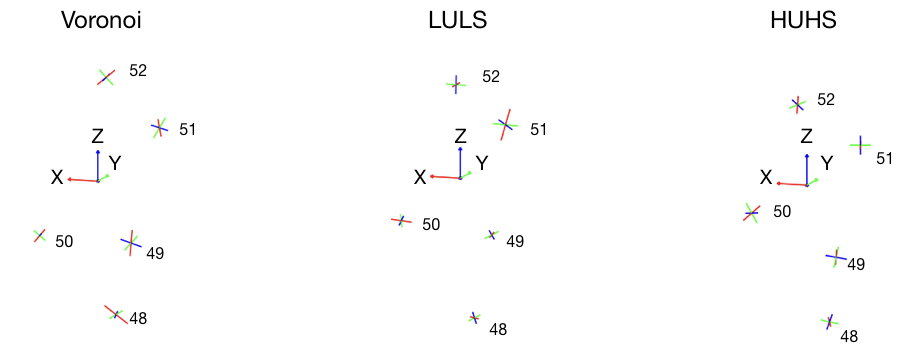}
	\caption{Grain axes for Grains 48, 49, 50, 51 and 52 in: Left: Voronoi sample ; Center LULS sample;  and Right HUHS sample.  Axis 1: red; Axis 2: green; Axis 3: blue. }
		\label{fig:grainaxes_3samples}
\end{figure}

\begin{figure}[htbp]
	\centering
	\begin{subfigure}{0.30\textwidth}
		\centering
		\includegraphics[width=1\linewidth]{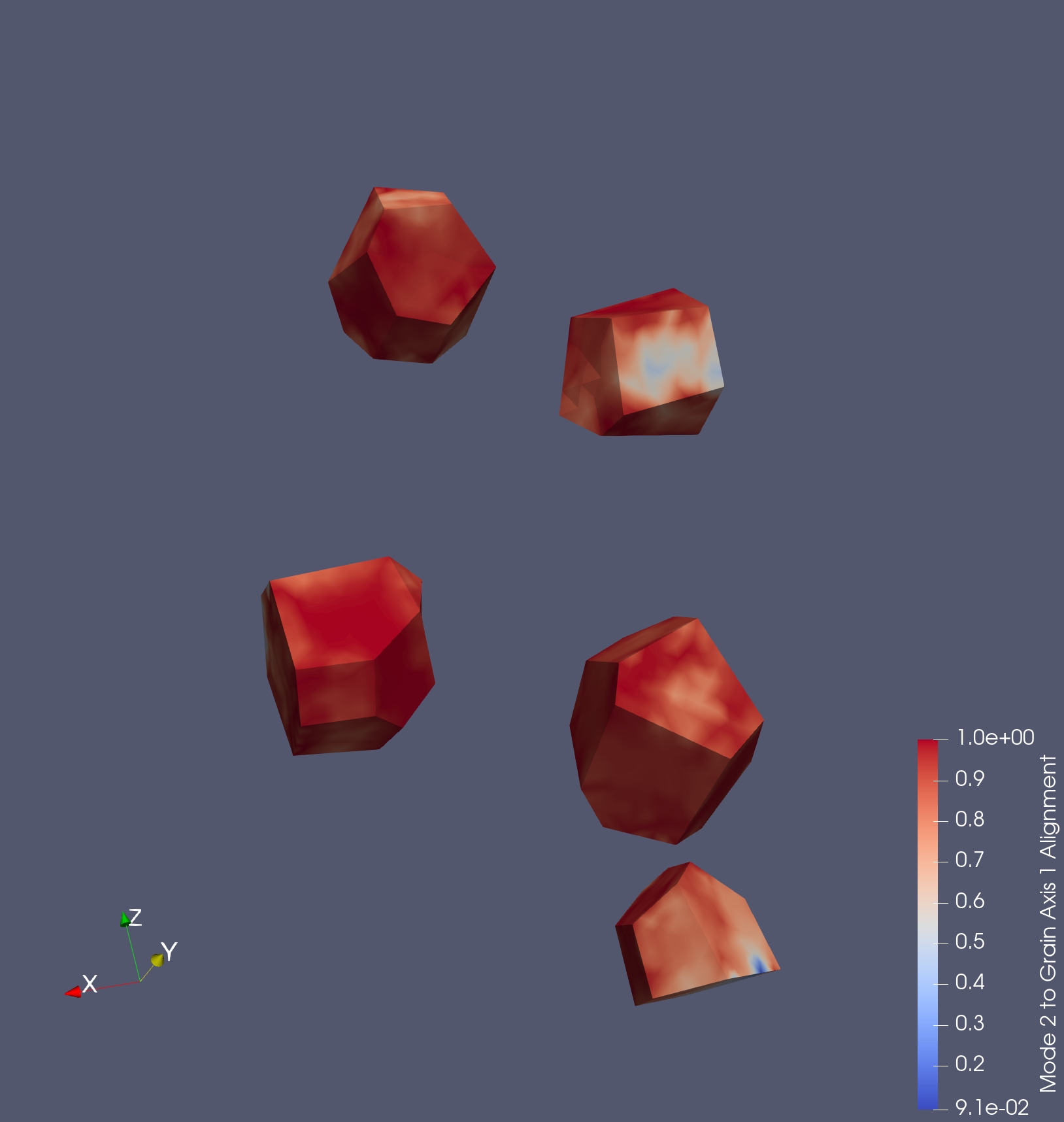}
		\caption{ }
		\label{fig:voronoi_5grains_m1e1}
	\end{subfigure}%
	\quad
	\begin{subfigure}{0.30\textwidth}
		\centering
		\includegraphics[width=1\linewidth]{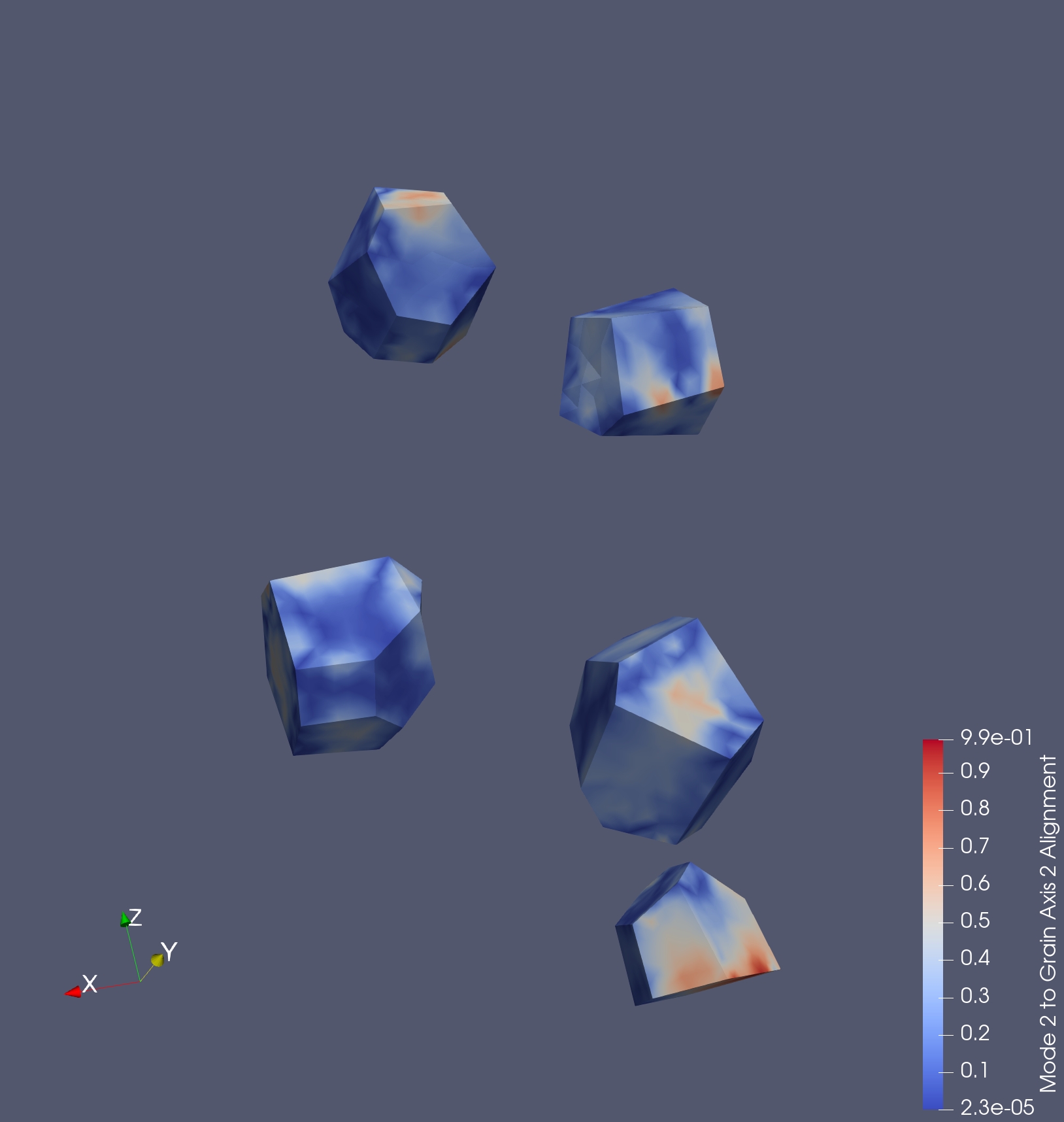}
		\caption{ }
		\label{fig:dia0p35_sph0p06_5grains_m1e2}
	\end{subfigure}%
	\quad
	\begin{subfigure}{0.30\textwidth}
		\centering
		\includegraphics[width=1\linewidth]{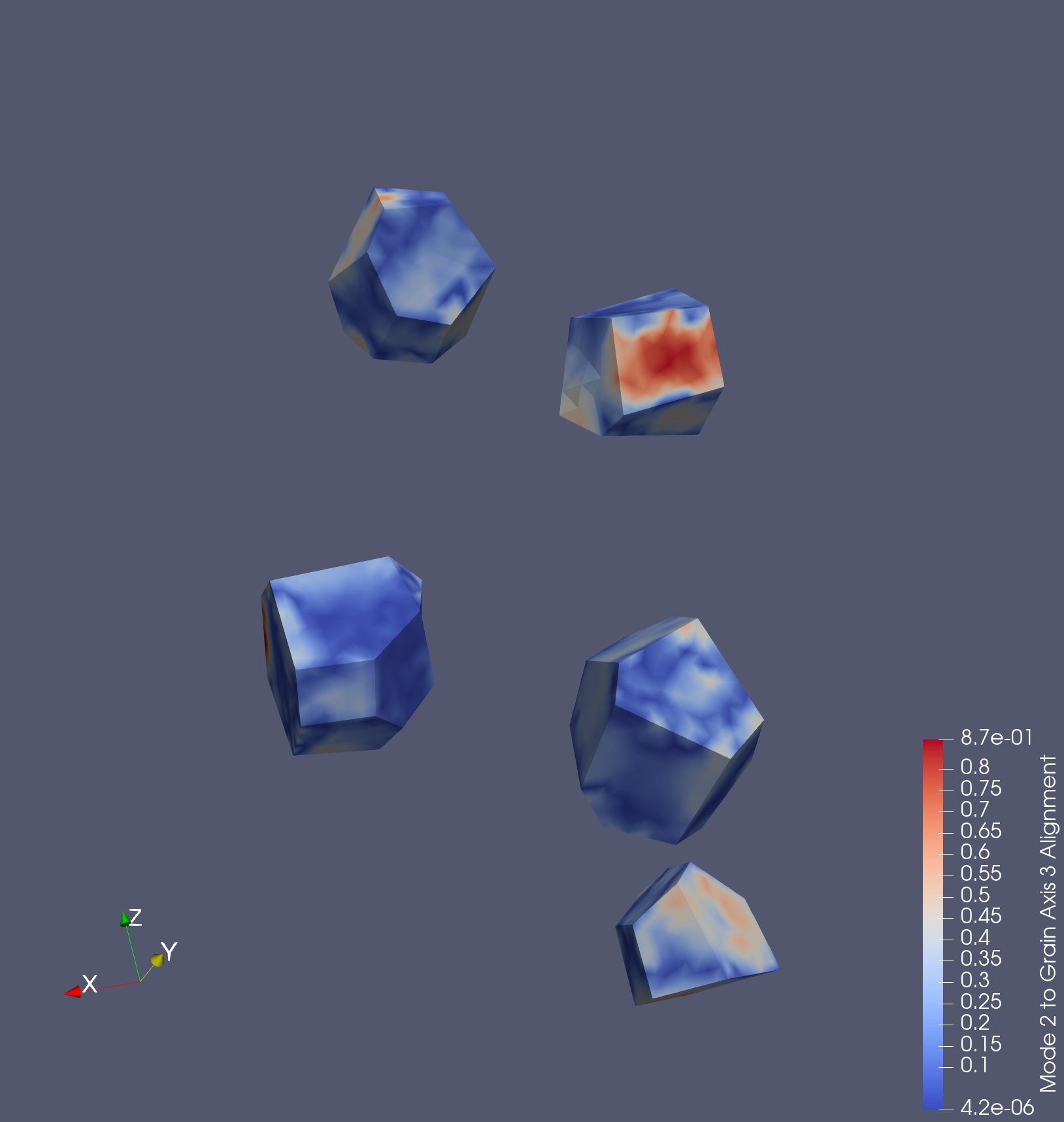}
		\caption{ }
		\label{fig:dia0p15_sph0p03_5grains_m1e3}
	\end{subfigure}%
	\caption{Alignment of the grain axes with the gradient vector of Mode 2 for the HUHS sample variant  for Grains 48, 49, 50, 51 and 52 in:  (a) Grain Axis 1 ; (b) Grain Axis 2;  and (c) Grain Axis 3.}
		\label{fig:5gr_grnaxes2mode1}
\end{figure}

 Figure~\ref{fig:grainaxes_3samples} shows the corresponding grain axes.  
 The alignment of Mode 2 with the Axis 1 of the grain axes is shown in Figure~\ref{fig:5gr_grnaxes2mode1} for the HUHS sample variant. 
 The five selected grains display the same alignment trend as the sample at-large.
Figures~\ref{fig:5gr_grnaxes2mode1} and \ref{fig:5gr_grnaxes2mode1}
show equivalent distributions for Modes 3 and 4.  Note that the for Mode 3 the alignment is strongest for Grain Axis 2 and for Mode 4 the alignment is strongest for Grain Axis 3. 
  The other sample variants show similar trends.
\newpage  
\begin{figure}[h!]
	\centering
	\begin{subfigure}{0.30\textwidth}
		\centering
		\includegraphics[width=1\linewidth]{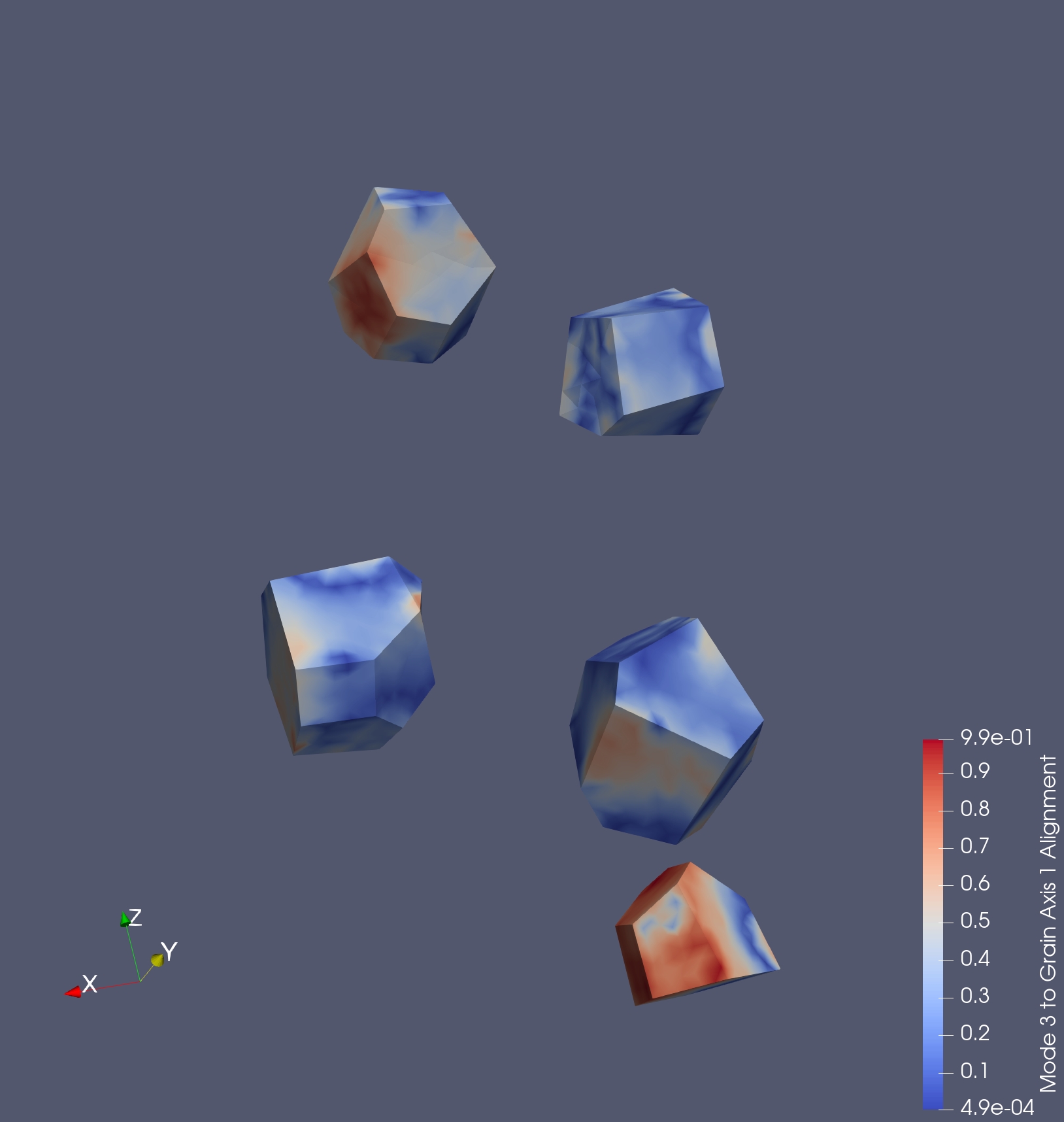}
		\caption{ }
		\label{fig:voronoi_5grains_m2e1}
	\end{subfigure}%
	\quad
	\begin{subfigure}{0.30\textwidth}
		\centering
		\includegraphics[width=1\linewidth]{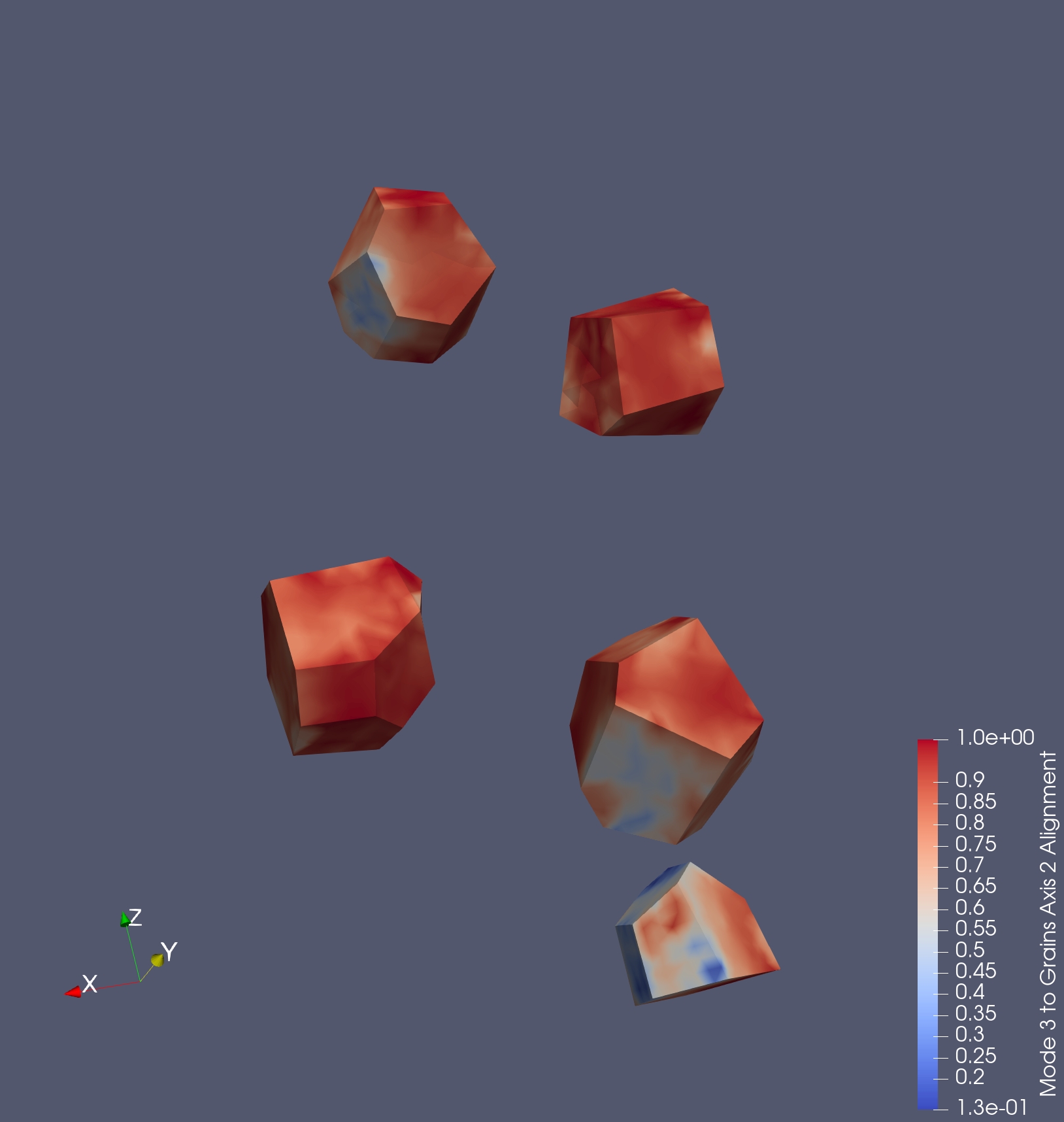}
		\caption{ }
		\label{fig:dia0p35_sph0p06_5grains_m2e2}
	\end{subfigure}%
	\quad
	\begin{subfigure}{0.30\textwidth}
		\centering
		\includegraphics[width=1\linewidth]{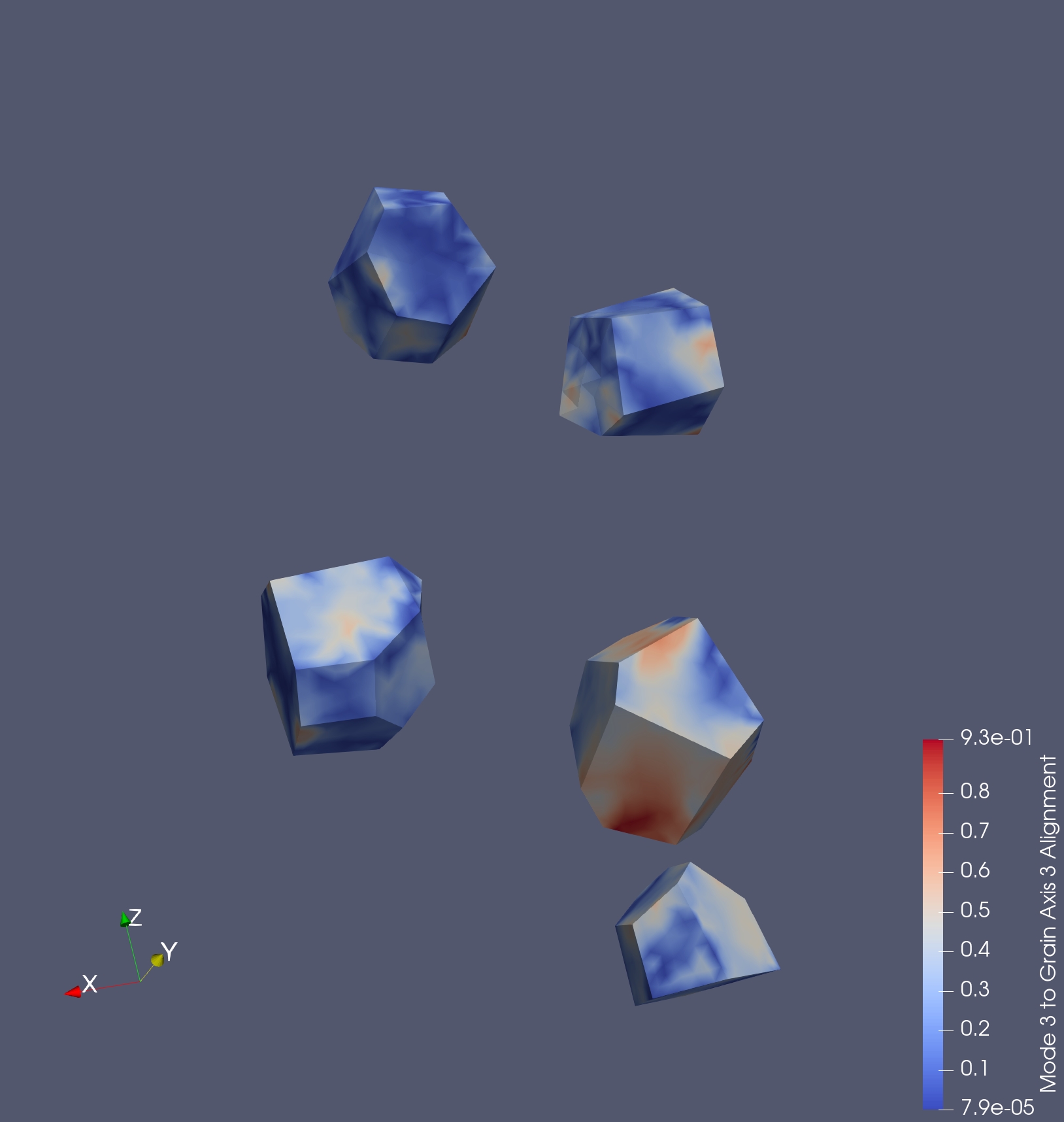}
		\caption{ }
		\label{fig:dia0p15_sph0p03_5grains_m2e3}
	\end{subfigure}%
	\caption{Alignment of the grain axes with the gradient vector of Mode 3 for the HUHS sample variant  for Grains 48, 49, 50, 51 and 52 in:  (a) Grain Axis 1 ; (b) Grain Axis 2;  and (c) Grain Axis 3.}
		\label{fig:5gr_grnaxes2mode2}
\end{figure}
\begin{figure}[h!]
	\centering
	\begin{subfigure}{0.30\textwidth}
		\centering
		\includegraphics[width=1\linewidth]{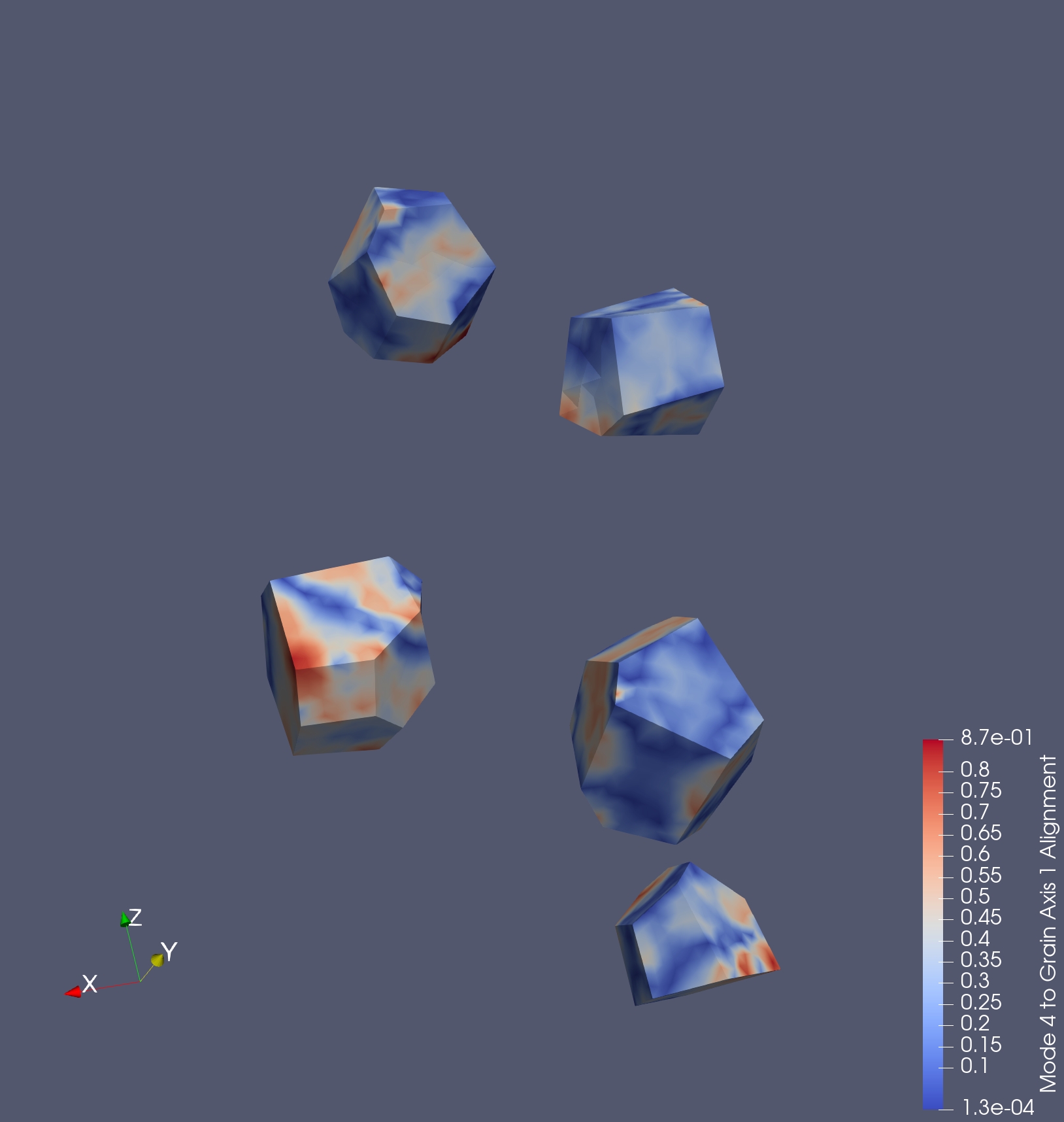}
		\caption{ }
		\label{fig:voronoi_5grains_m3e1}
	\end{subfigure}%
	\quad
	\begin{subfigure}{0.30\textwidth}
		\centering
		\includegraphics[width=1\linewidth]{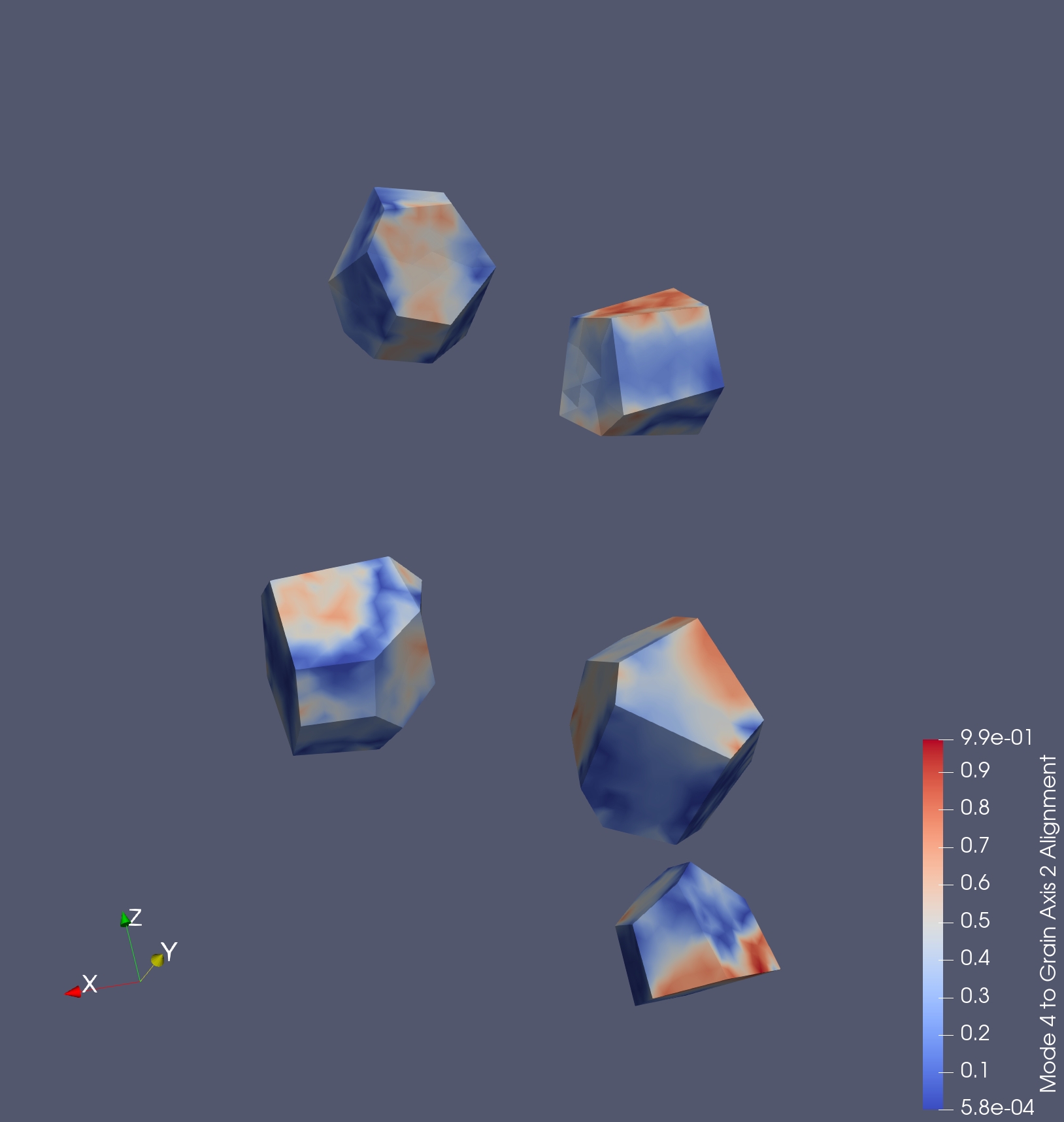}
		\caption{ }
		\label{fig:dia0p35_sph0p06_5grains_m3e2}
	\end{subfigure}%
	\quad
	\begin{subfigure}{0.30\textwidth}
		\centering
		\includegraphics[width=1\linewidth]{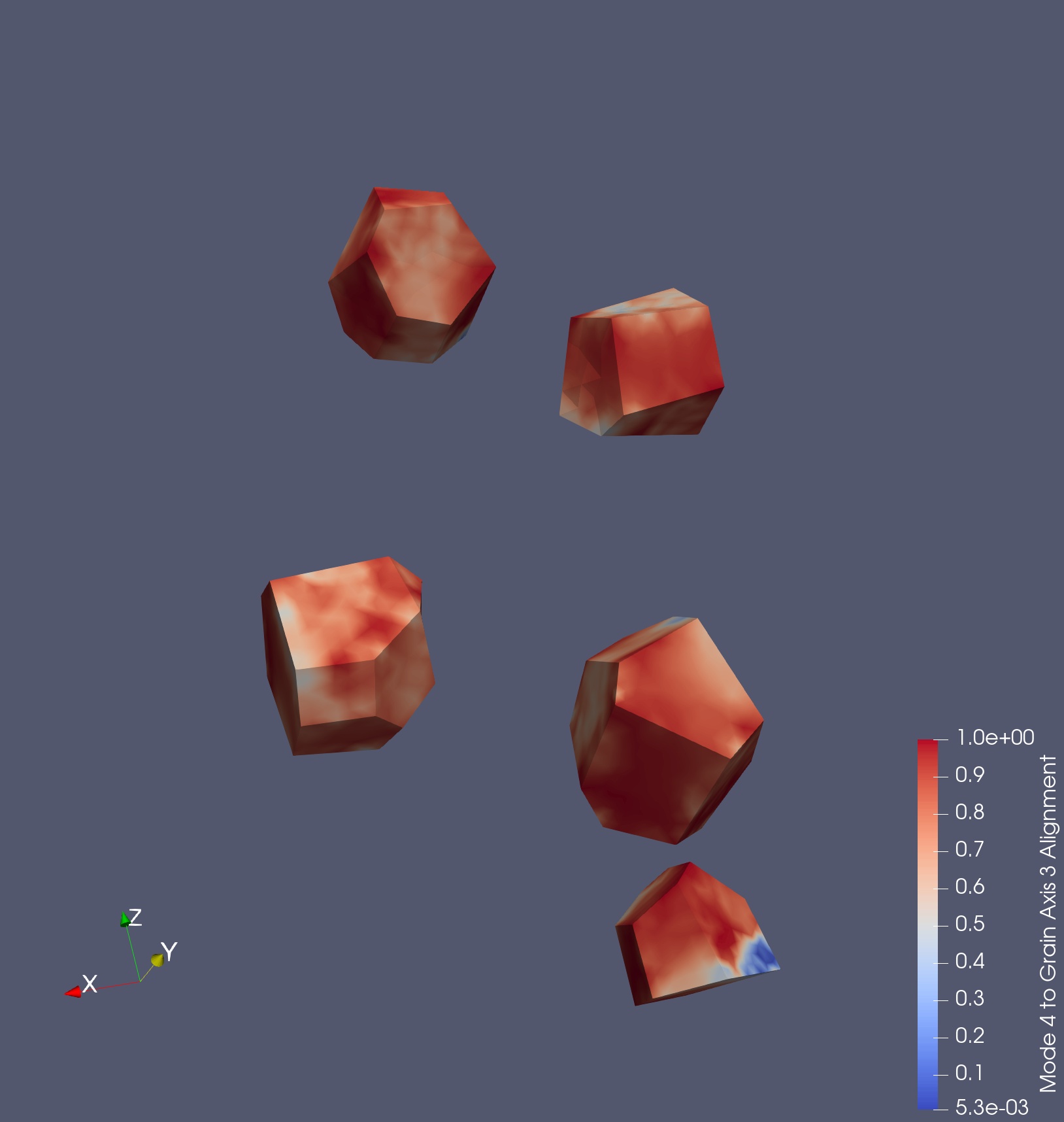}
		\caption{ }
		\label{fig:dia0p15_sph0p03_5grains_m3e3}
	\end{subfigure}%
	\caption{Alignment of the grain axes with the gradient vector of Mode 4 for the HUHS sample variant  for Grains 48, 49, 50, 51 and 52 in:  (a) Grain Axis 1 ; (b) Grain Axis 2;  and (c) Grain Axis 3.}
		\label{fig:5gr_grnaxes2mode3}
\end{figure}
\clearpage

\begin{figure}[htbp]
	\centering
	\begin{subfigure}{.25\textwidth}
		\centering
		\includegraphics[width=1\linewidth]{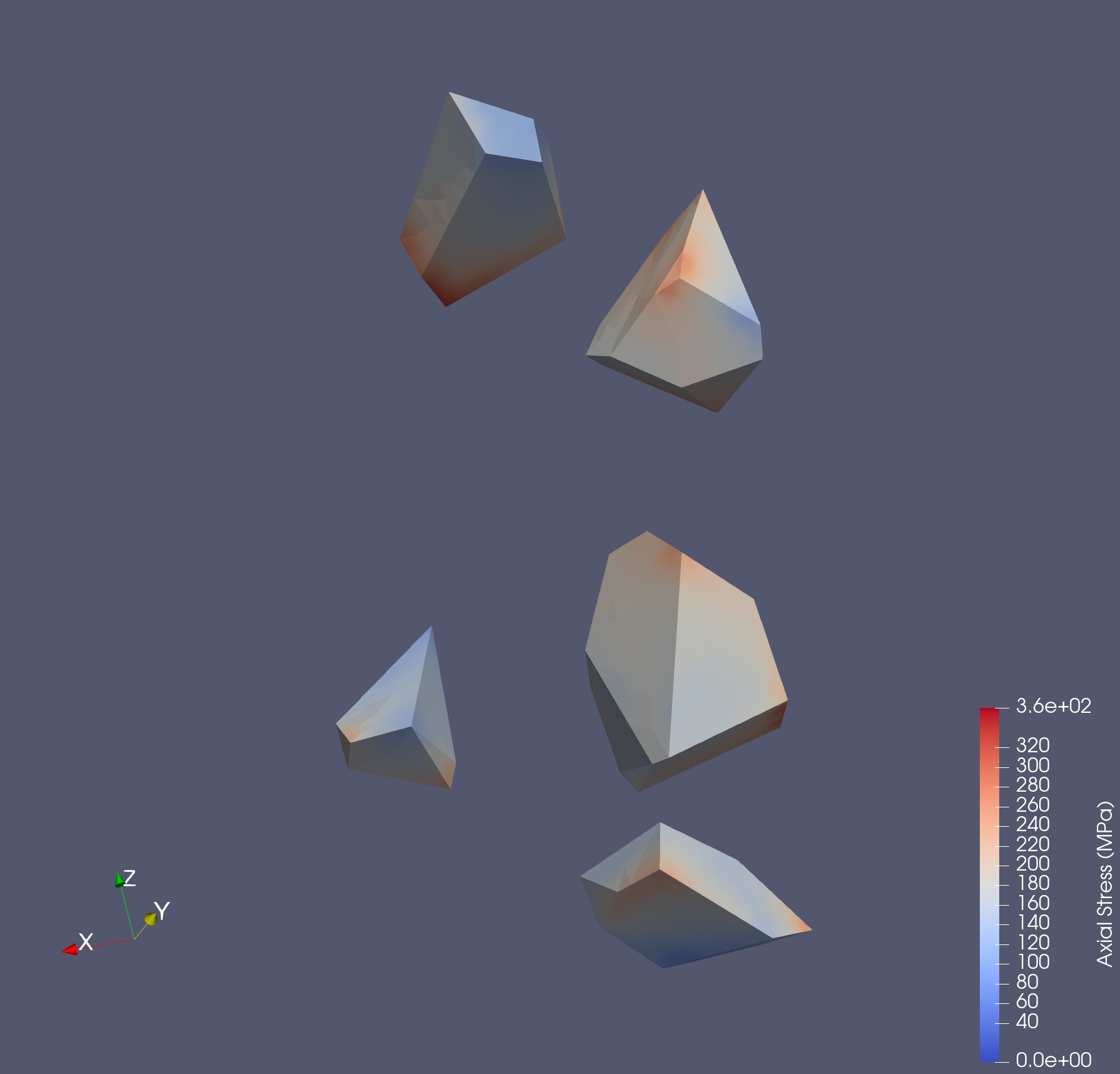}
		\caption{ }
		\label{fig:voronoi_sigzz_eps0p001_5gr_fepx}
	\end{subfigure}%
	\quad
	\begin{subfigure}{.25\textwidth}
		\centering
		\includegraphics[width=1\linewidth]{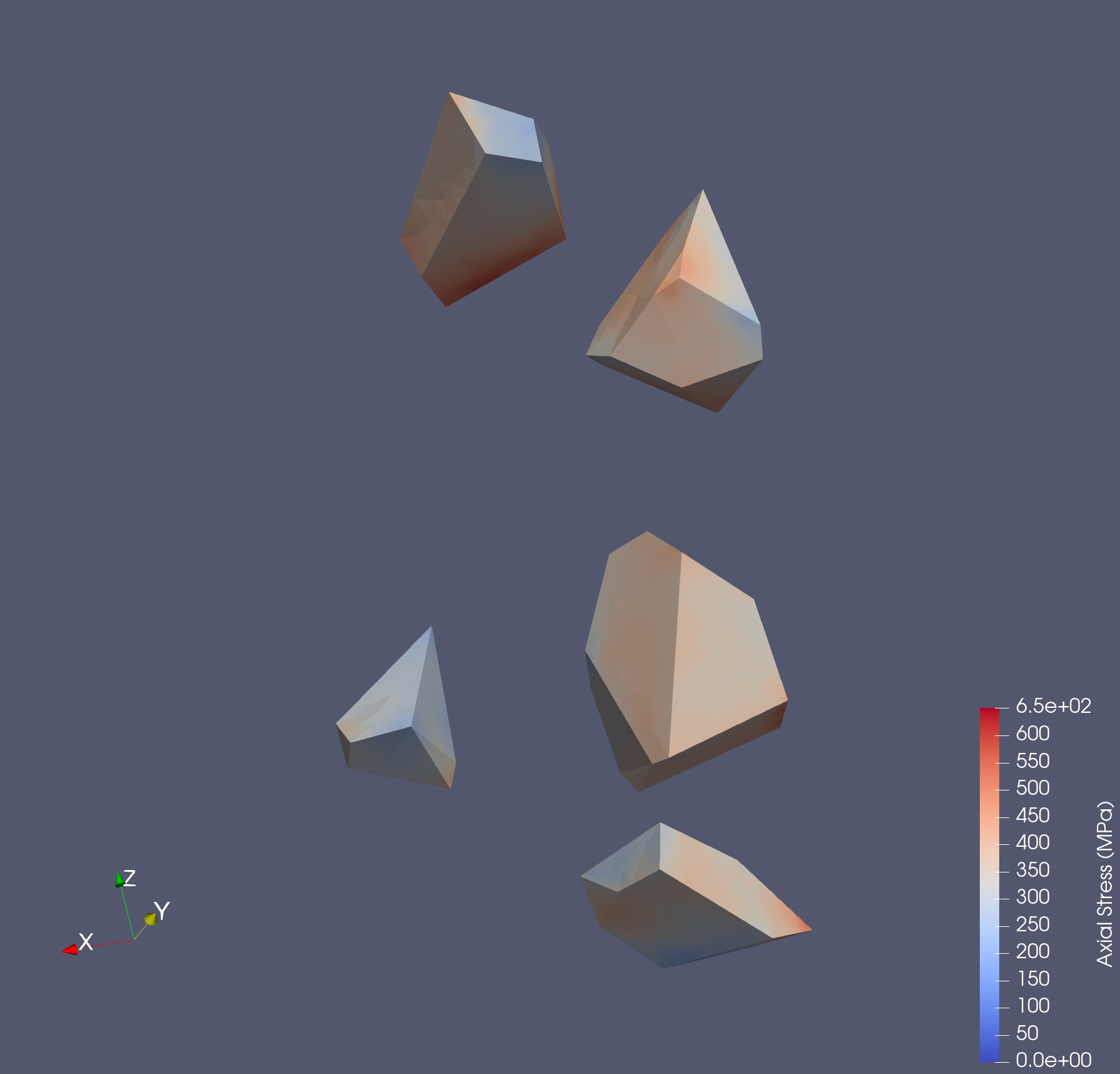}
		\caption{ }
		\label{fig:voronoi_sigzz_eps0p0025_5gr_fepx}
	\end{subfigure}%
	\quad
	\begin{subfigure}{.25\textwidth}
		\centering
		\includegraphics[width=1\linewidth]{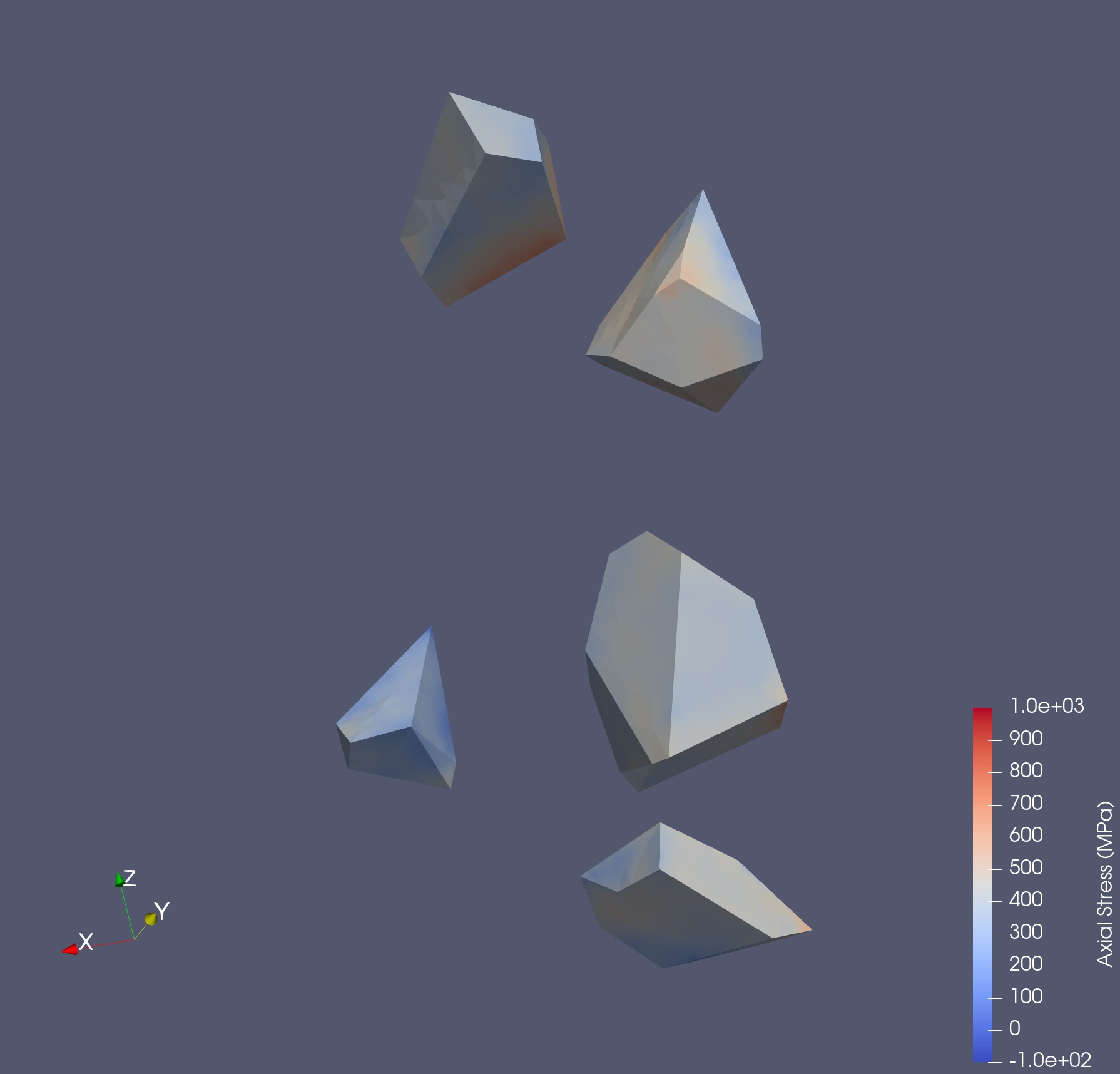}
		\caption{ }
		\label{fig:voronoi_sigzz_eps0p01_5gr_fepx}
	\end{subfigure}%
	\\
	\begin{subfigure}{.25\textwidth}
		\centering
		\includegraphics[width=1\linewidth]{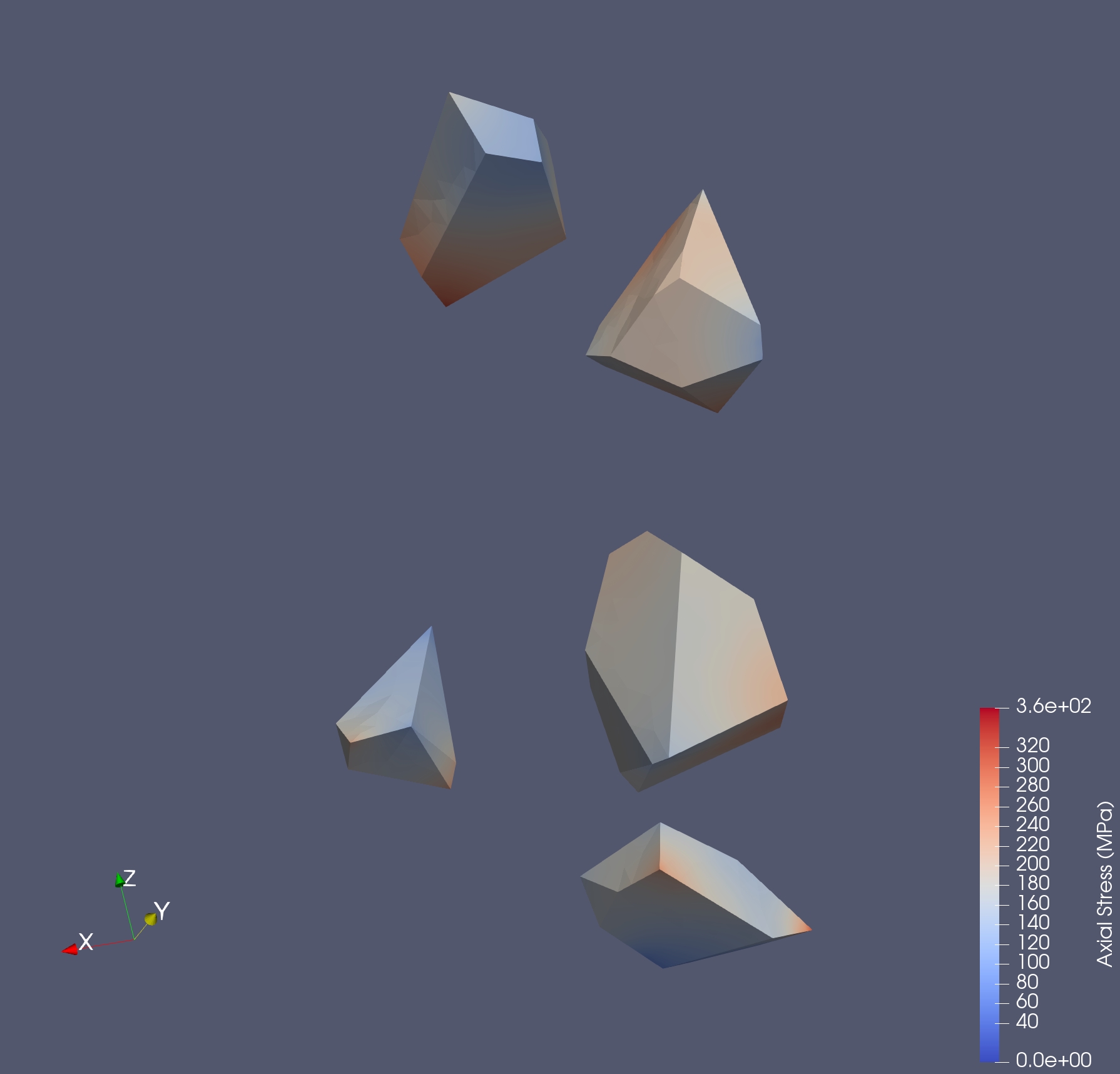}
		\caption{ }
		\label{fig:voronoi_sigzz_eps0p001_5gr_10m}
	\end{subfigure}%
	\quad
	\begin{subfigure}{.25\textwidth}
		\centering
		\includegraphics[width=1\linewidth]{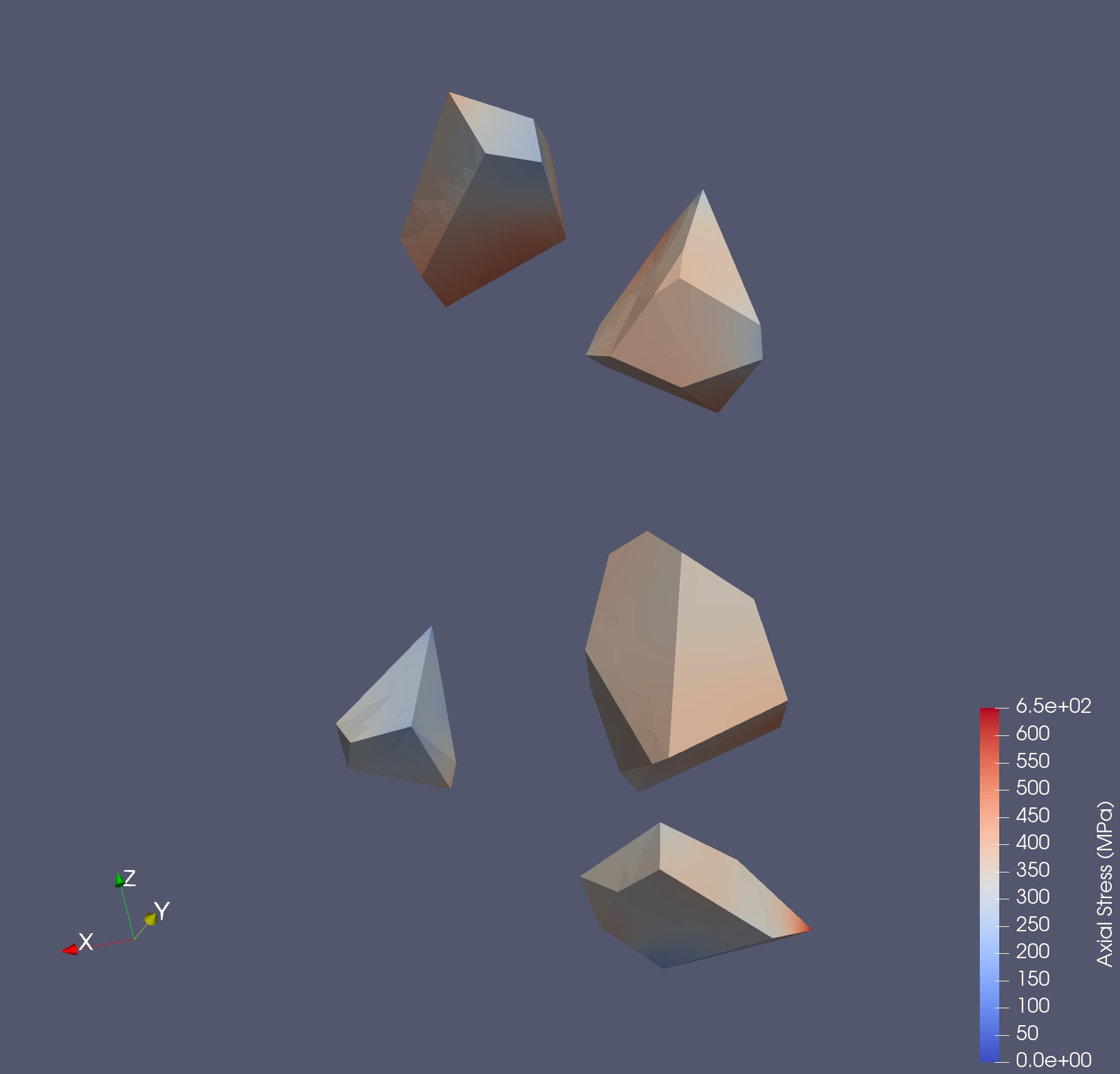}
		\caption{ }
		\label{fig:voronoi_sigzz_eps0p0025_5gr_10m}
	\end{subfigure}%
	\quad
	\begin{subfigure}{.25\textwidth}
		\centering
		\includegraphics[width=1\linewidth]{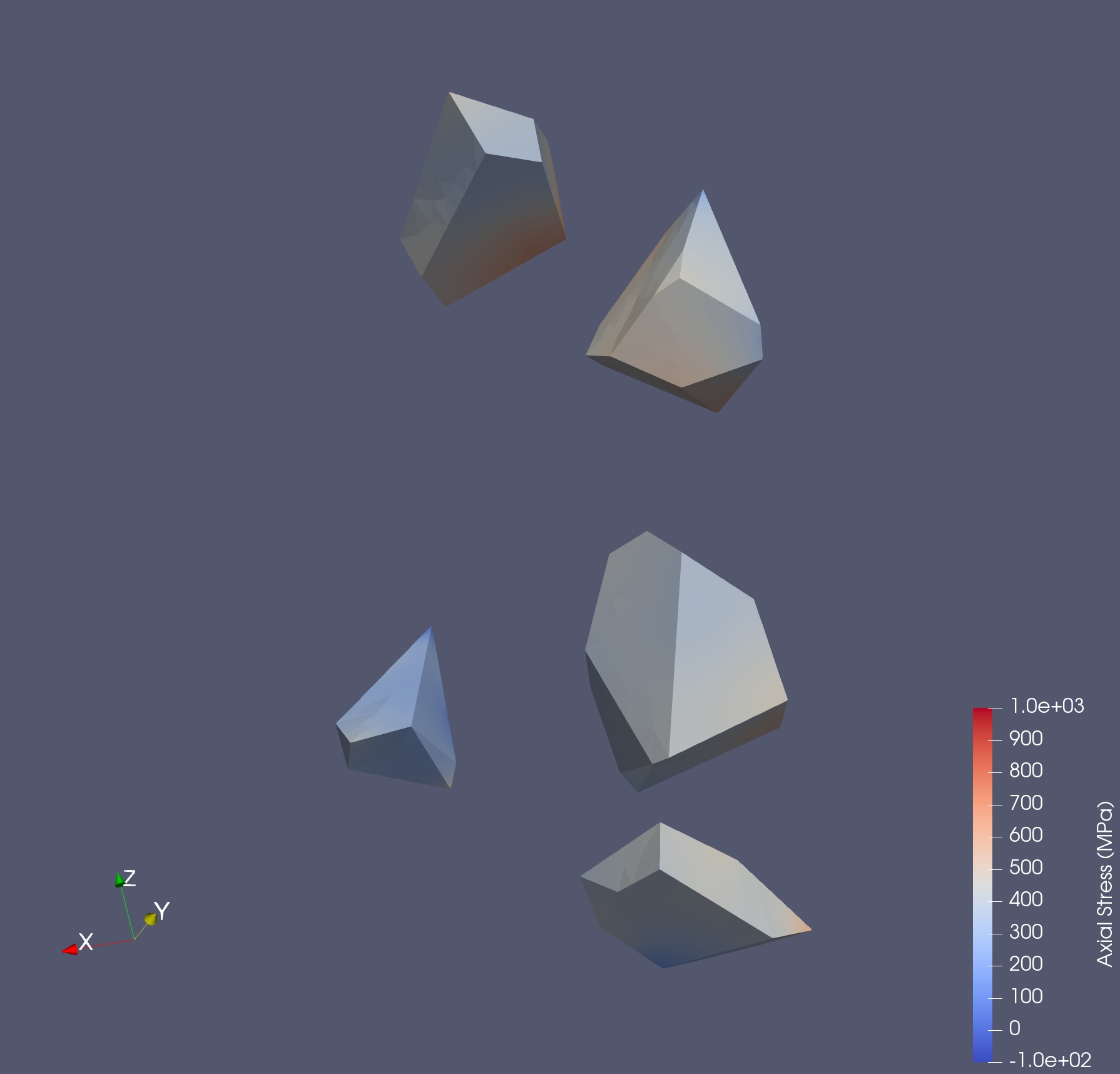}
		\caption{ }
		\label{fig:voronoi_sigzz_eps0p01_5gr_10m}
	\end{subfigure}%
	\caption{Simulated axial stress distributions over Grains 48, 49, 50, 51, and 52 in the Voronoi sample at nominal strains of  (a) 0.001 ; (b) 0.0025;  and (c) 0.01; 10 mode fits to the simulated stress distributions at nominal strains of (d) 0.001 ; (e) 0.0025;  and (f) 0.01.  Note that the scales are different for the different strain levels.}
		\label{fig:5gr_stress_distributions_voronoi}
\end{figure}

\begin{figure}[htbp]
	\centering
	\begin{subfigure}{.25\textwidth}
		\centering
		\includegraphics[width=1\linewidth]{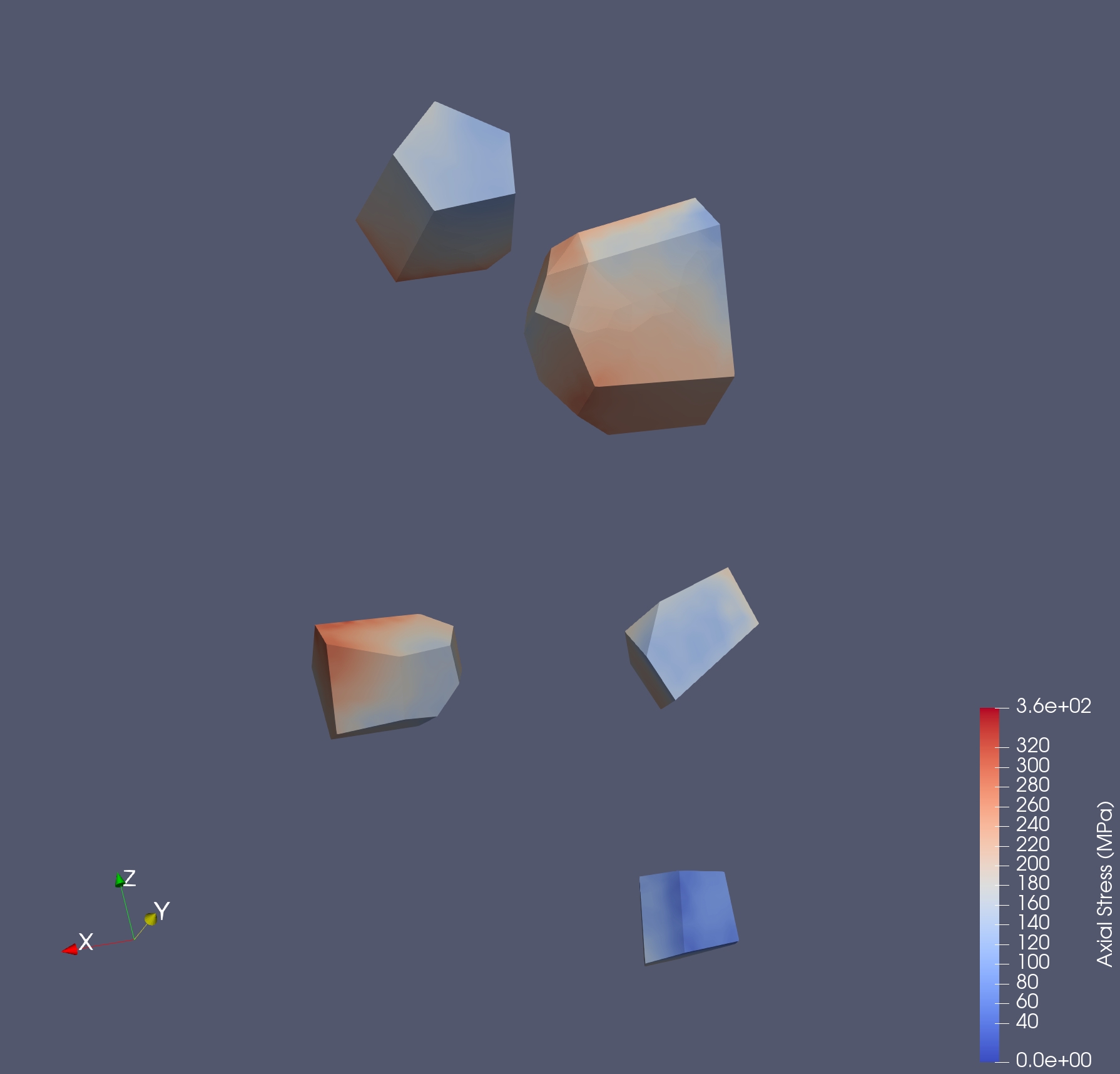}
		\caption{ }
		\label{fig:dia0p35_sph0p06_sigzz_eps0p001_5gr_fepx}
	\end{subfigure}%
	\quad
	\begin{subfigure}{.25\textwidth}
		\centering
		\includegraphics[width=1\linewidth]{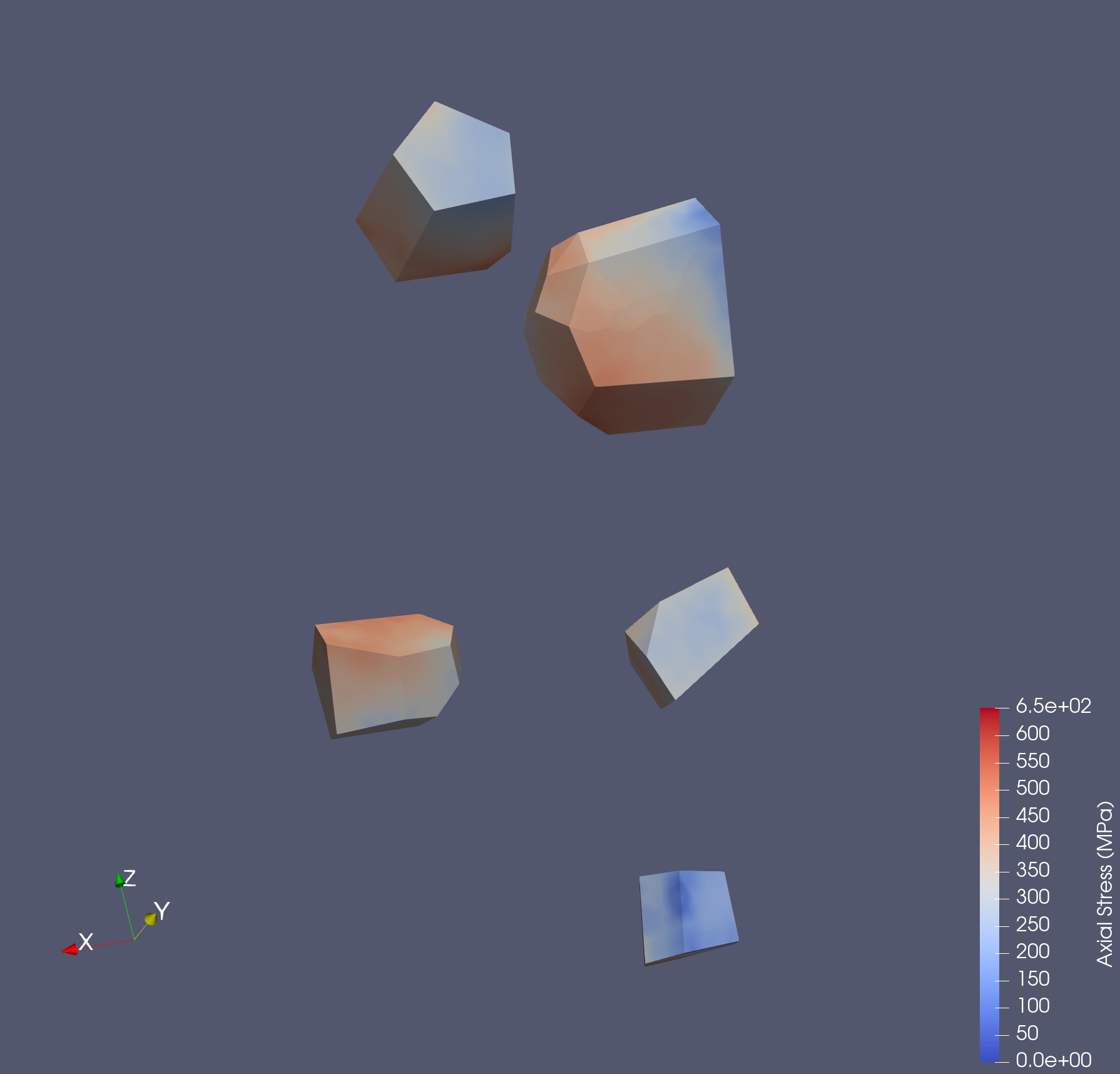}
		\caption{ }
		\label{fig:dia0p35_sph0p06_sigzz_eps0p0025_5gr_fepx}
	\end{subfigure}%
	\quad
	\begin{subfigure}{.25\textwidth}
		\centering
		\includegraphics[width=1\linewidth]{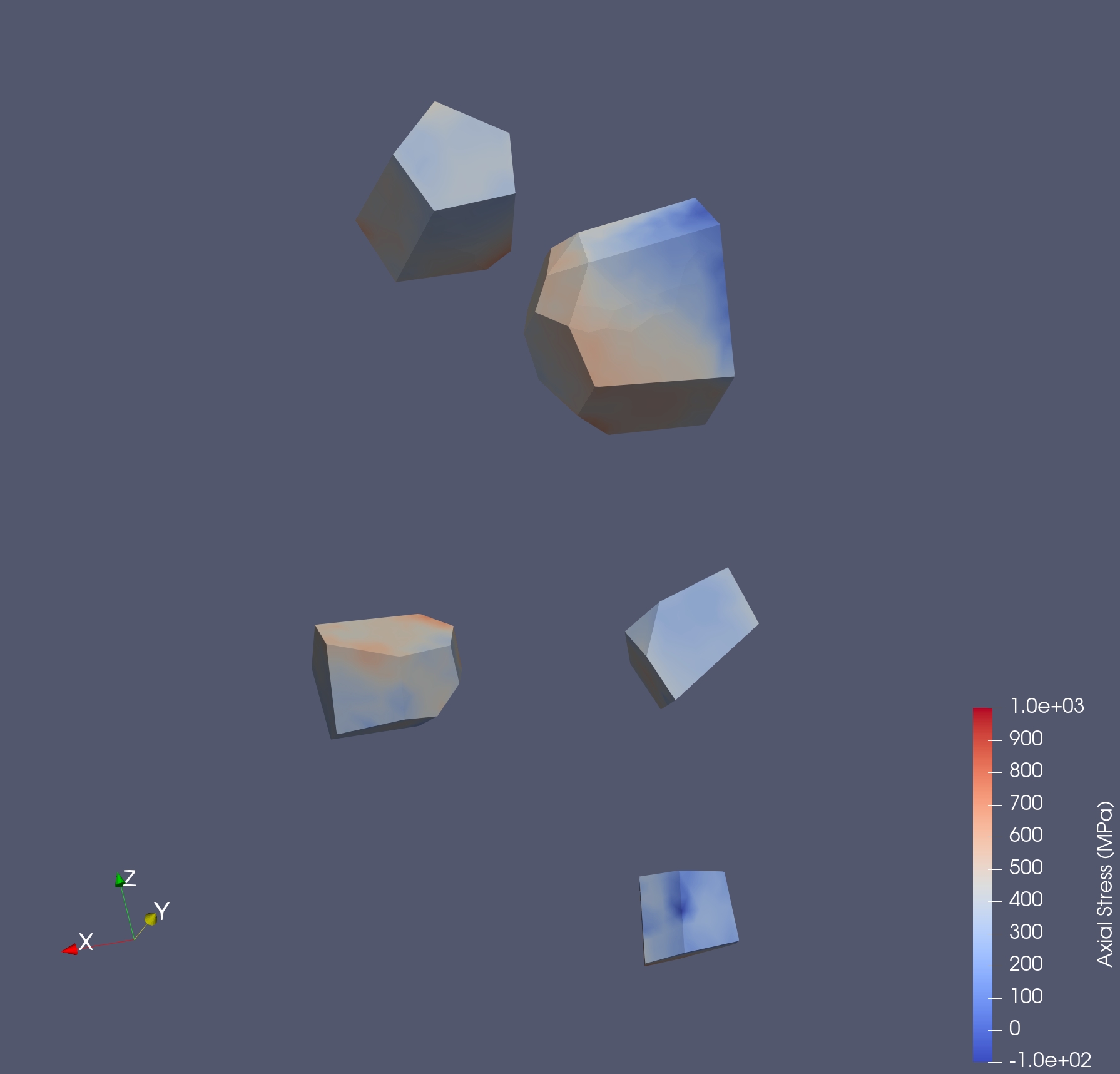}
		\caption{ }
		\label{fig:dia0p35_sph0p06_sigzz_eps0p01_5gr_fepx}
	\end{subfigure}%
	\\
	\begin{subfigure}{.25\textwidth}
		\centering
		\includegraphics[width=1\linewidth]{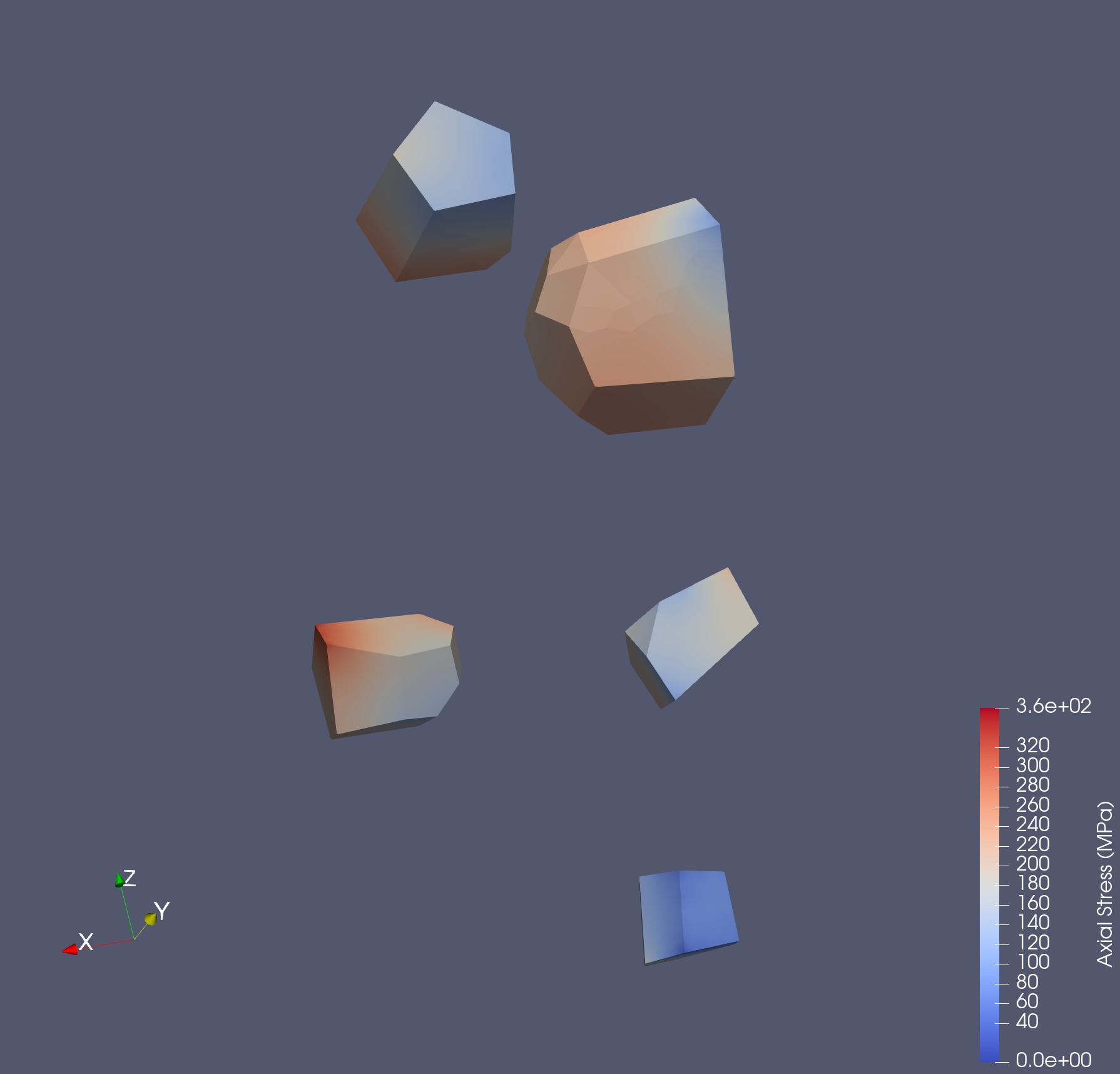}
		\caption{ }
		\label{fig:dia0p35_sph0p06_sigzz_eps0p001_5gr_10m}
	\end{subfigure}%
	\quad
	\begin{subfigure}{.25\textwidth}
		\centering
		\includegraphics[width=1\linewidth]{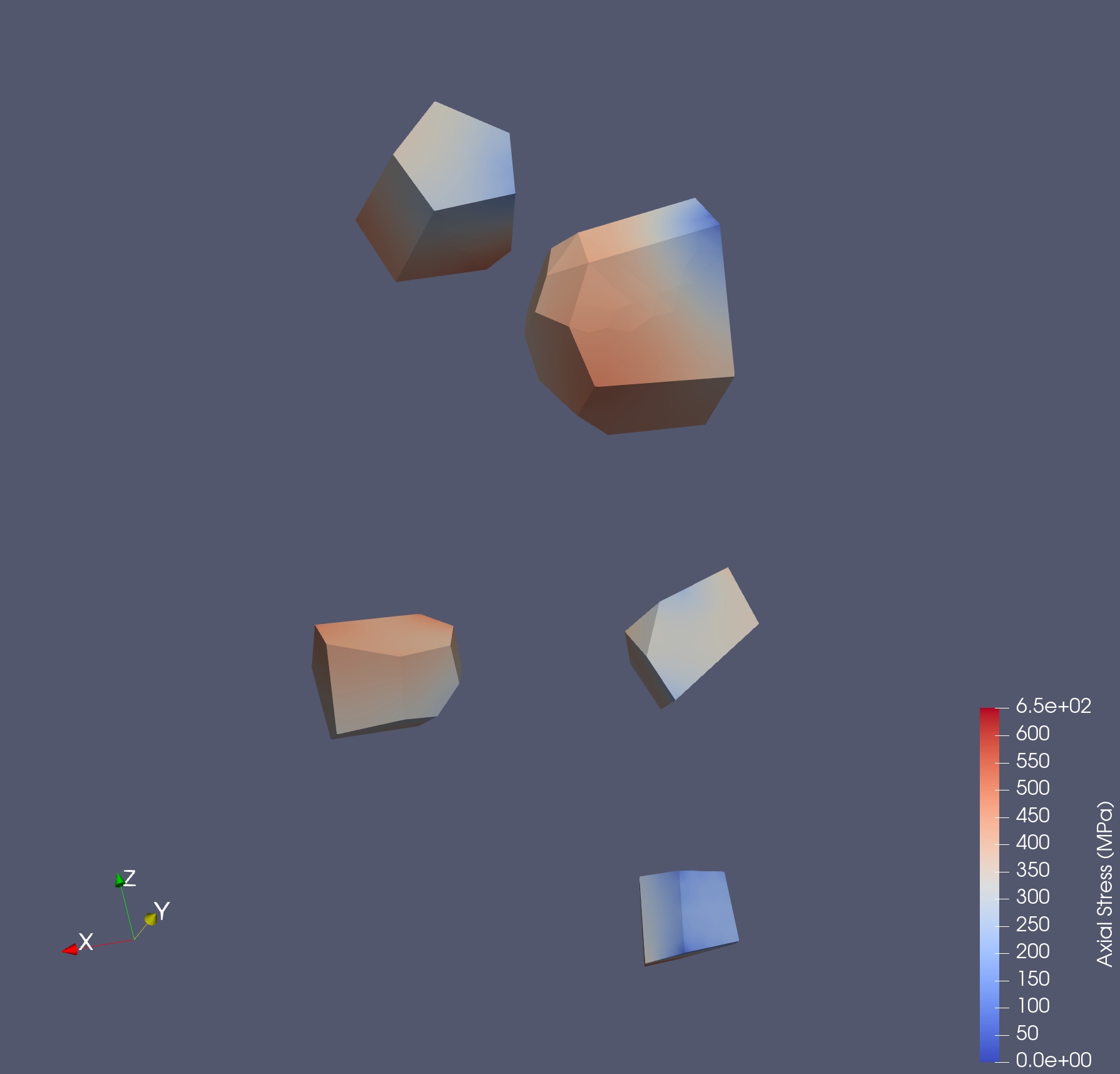}
		\caption{ }
		\label{fig:dia0p35_sph0p06_sigzz_eps0p0025_5gr_10m}
	\end{subfigure}
	\quad
	\begin{subfigure}{.25\textwidth}
		\centering
		\includegraphics[width=1\linewidth]{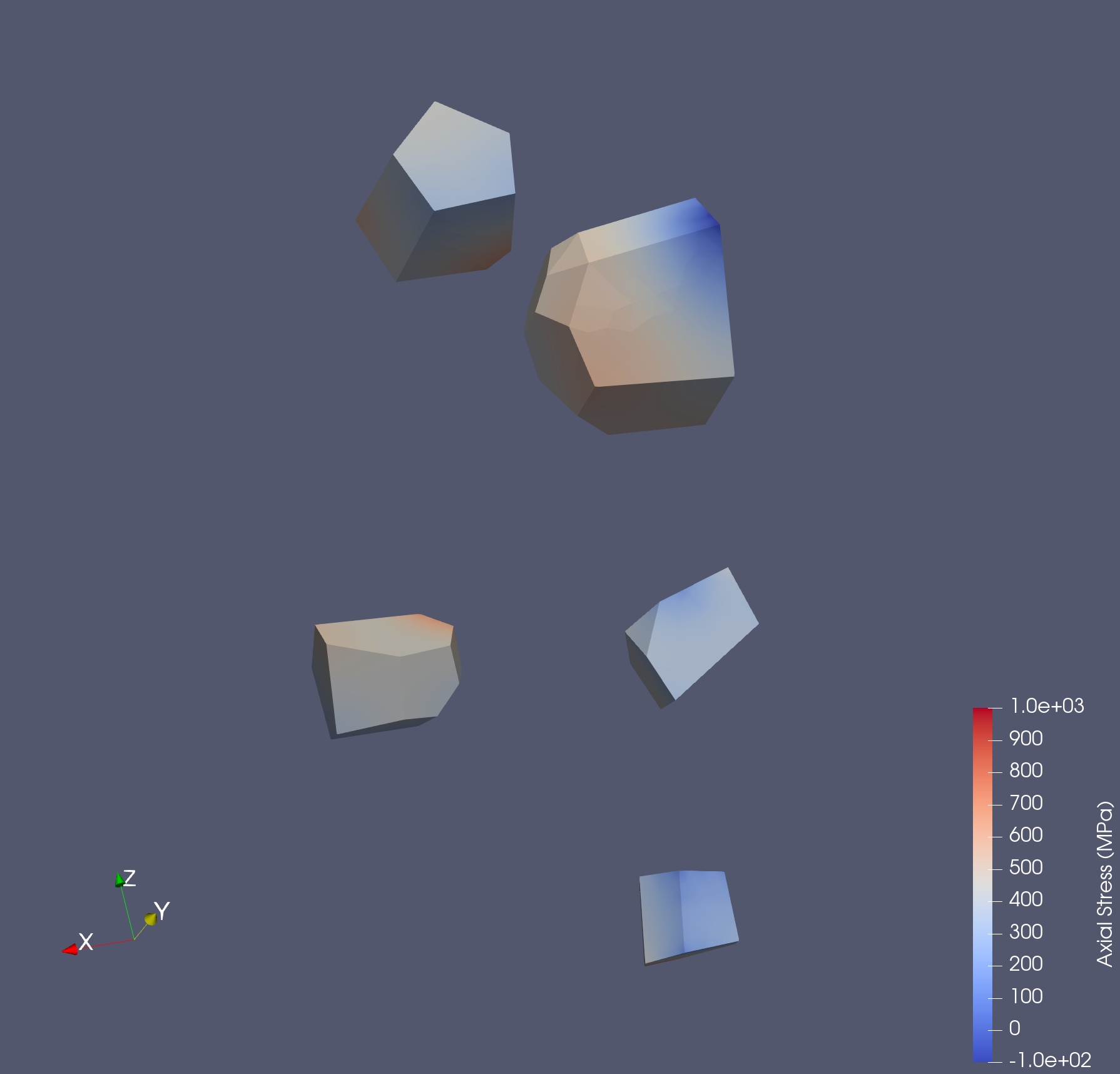}
		\caption{ }
		\label{fig:dia0p35_sph0p06_sigzz_eps0p01_5gr_10m}
	\end{subfigure}%
	\caption{Simulated axial stress distributions over Grains 48, 49, 50, 51, and 52 in the LULS sample at nominal strains of  (a) 0.001 ; (b) 0.0025;  and (c) 0.01; 10 mode fits to the simulated stress distributions at nominal strains of (d) 0.001 ; (e) 0.0025;  and (f) 0.01.  Note that the scales are different for the different strain levels.}
		\label{fig:5gr_stress_distributions_dia0p35_sph0p06}
\end{figure}

\begin{figure}[htbp]
	\centering
	\begin{subfigure}{.25\textwidth}
		\centering
		\includegraphics[width=1\linewidth]{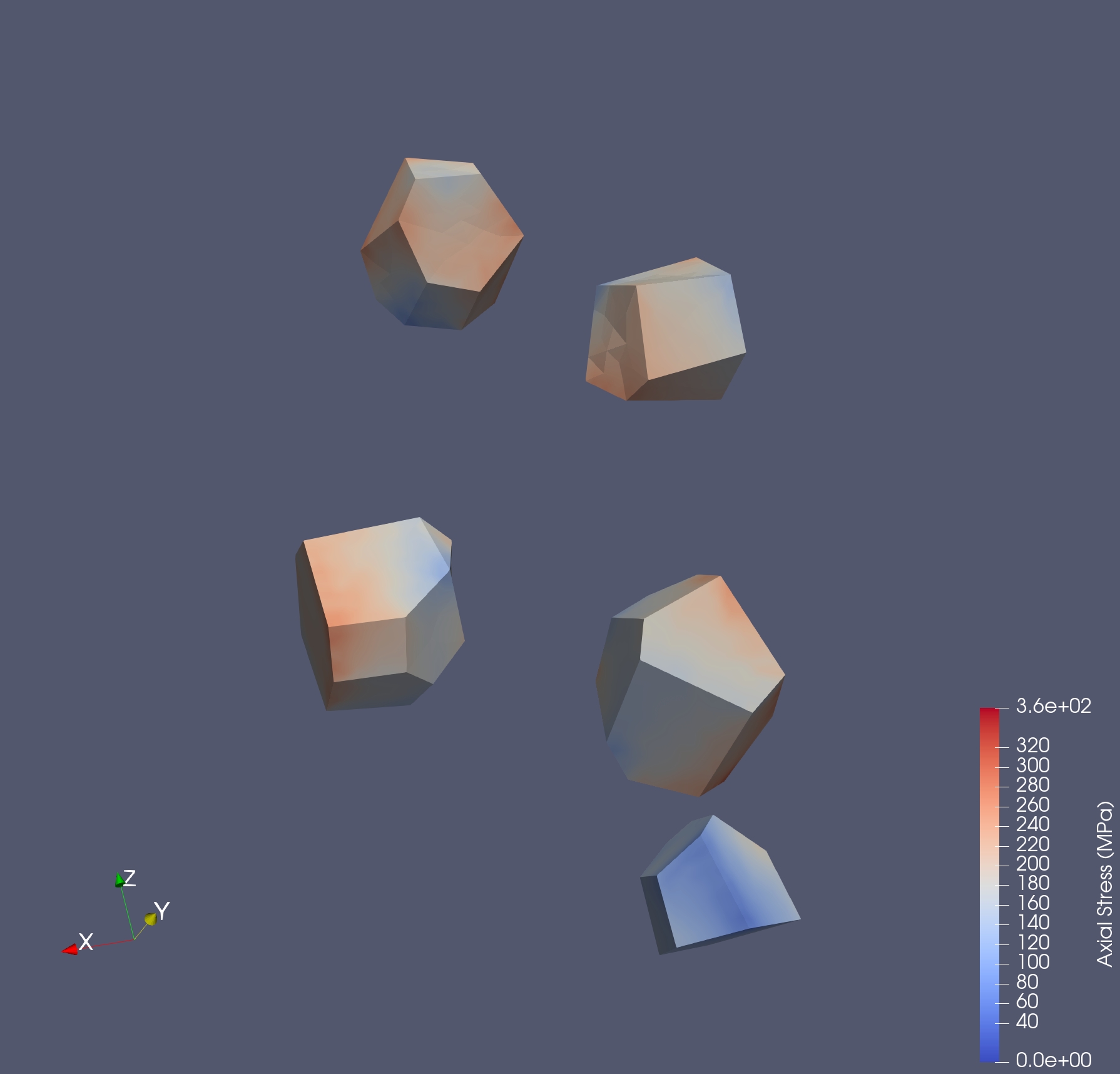}
		\caption{ }
		\label{fig:dia0p15_sph0p03_sigzz_eps0p001_5gr_fepx}
	\end{subfigure}%
	\quad
	\begin{subfigure}{.25\textwidth}
		\centering
		\includegraphics[width=1\linewidth]{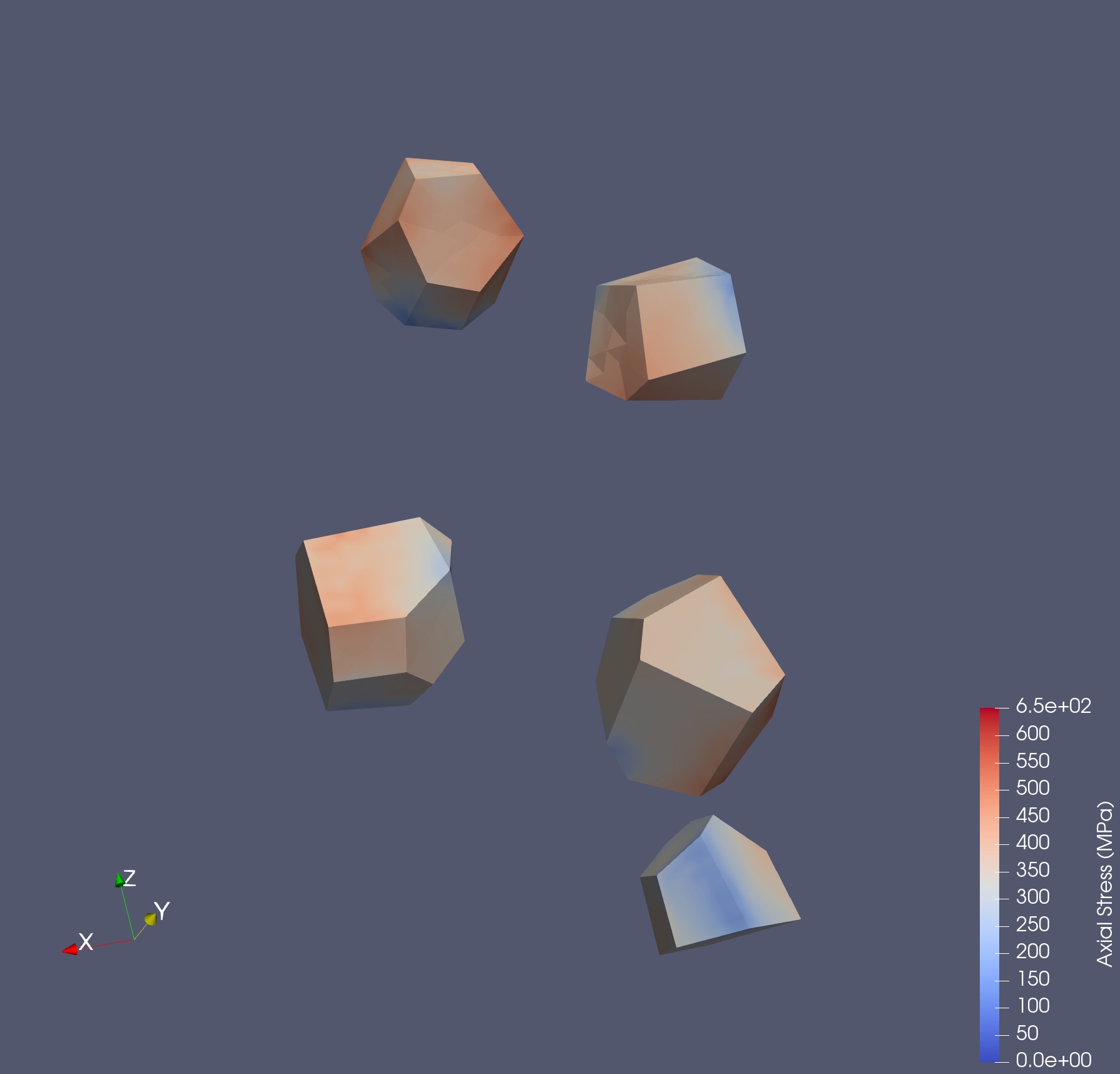}
		\caption{ }
		\label{fig:dia0p15_sph0p03_sigzz_eps0p0025_5gr_fepx}
	\end{subfigure}%
	\quad
	\begin{subfigure}{.25\textwidth}
		\centering
		\includegraphics[width=1\linewidth]{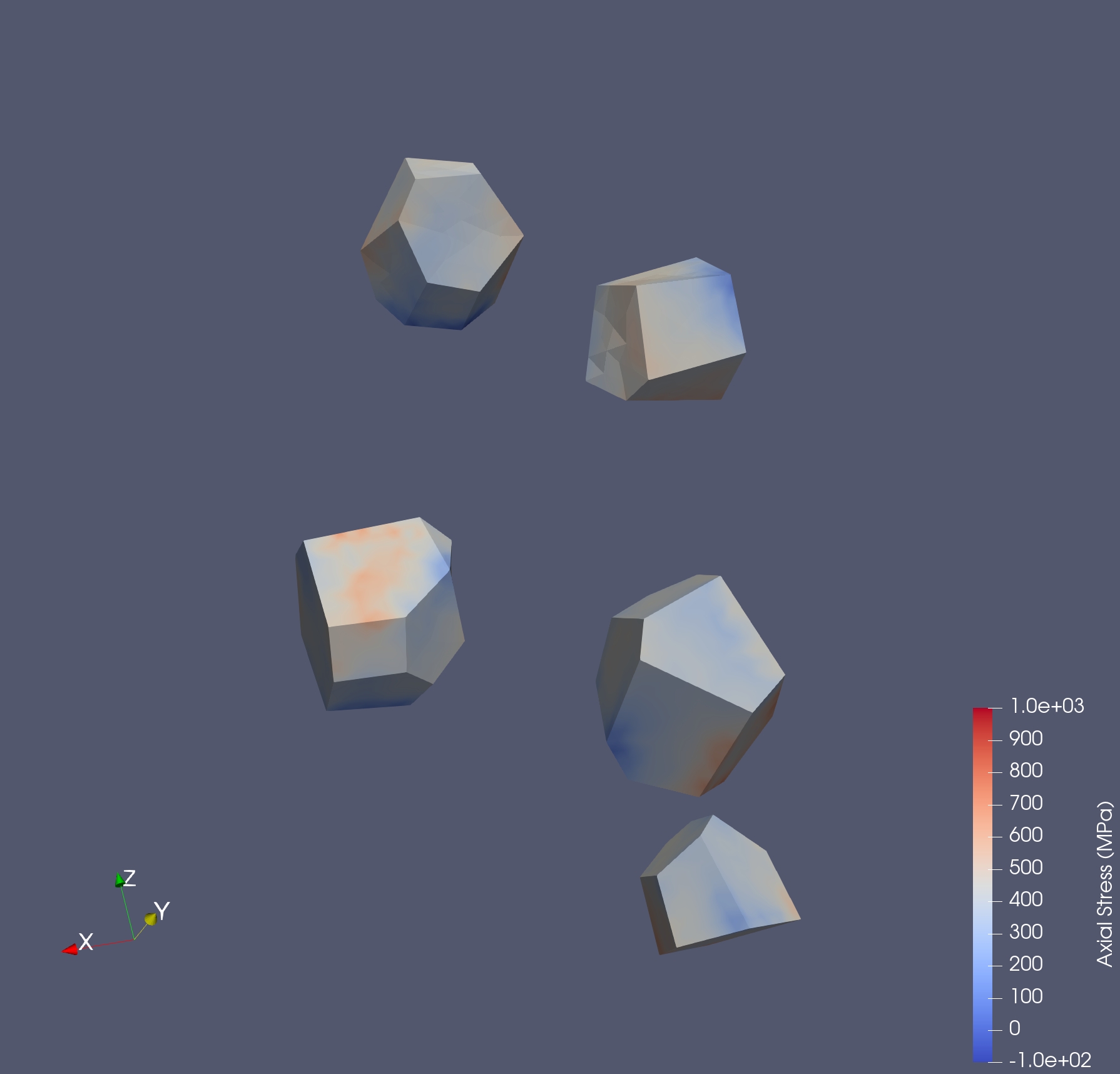}
		\caption{ }
		\label{fig:dia0p15_sph0p03_sigzz_eps0p01_5gr_fepx}
	\end{subfigure}%
	\\
	\begin{subfigure}{.25\textwidth}
		\centering
		\includegraphics[width=1\linewidth]{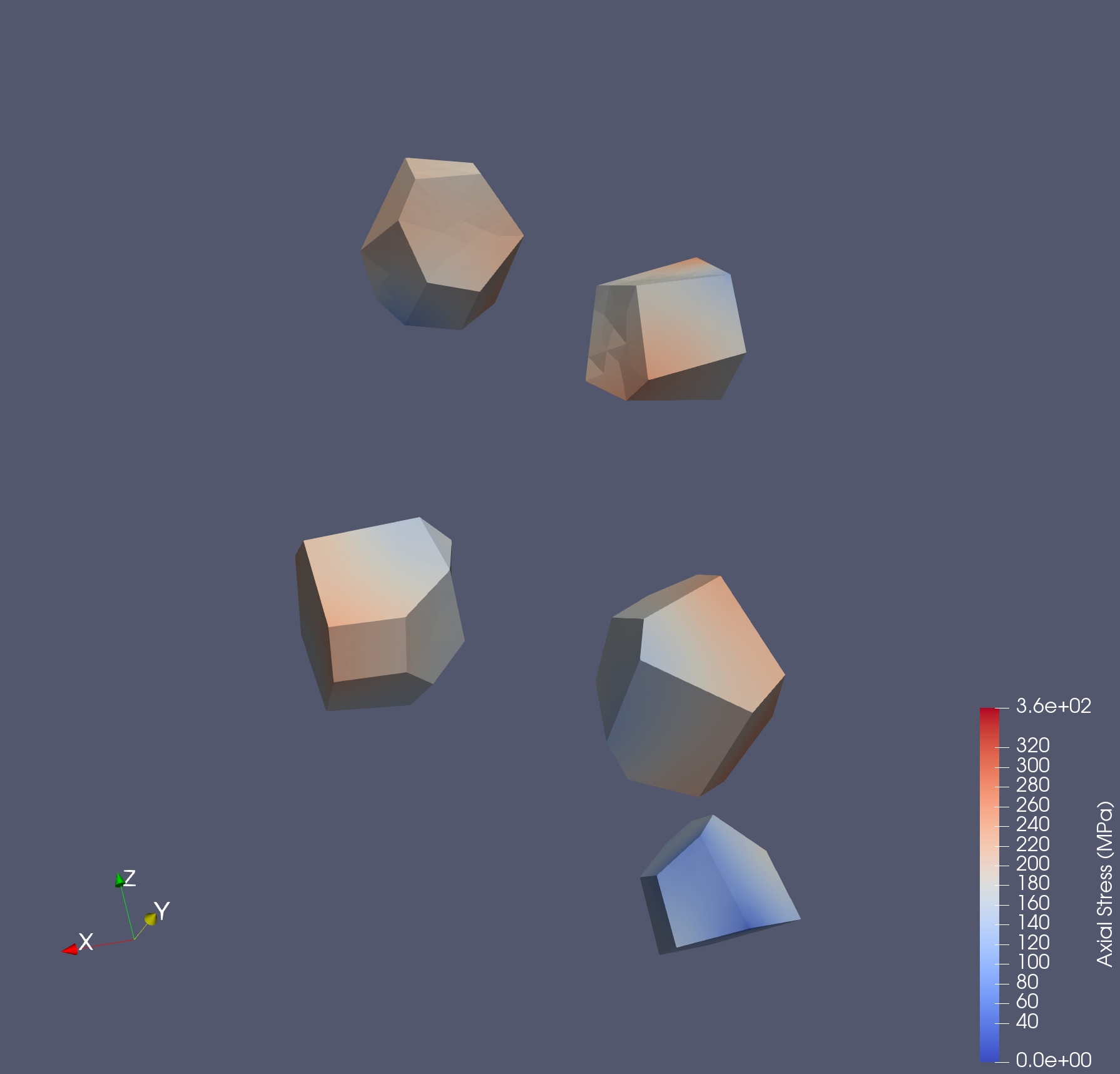}
		\caption{ }
		\label{fig:dia0p15_sph0p03_sigzz_eps0p001_5gr_10m}
	\end{subfigure}%
	\quad
	\begin{subfigure}{.25\textwidth}
		\centering
		\includegraphics[width=1\linewidth]{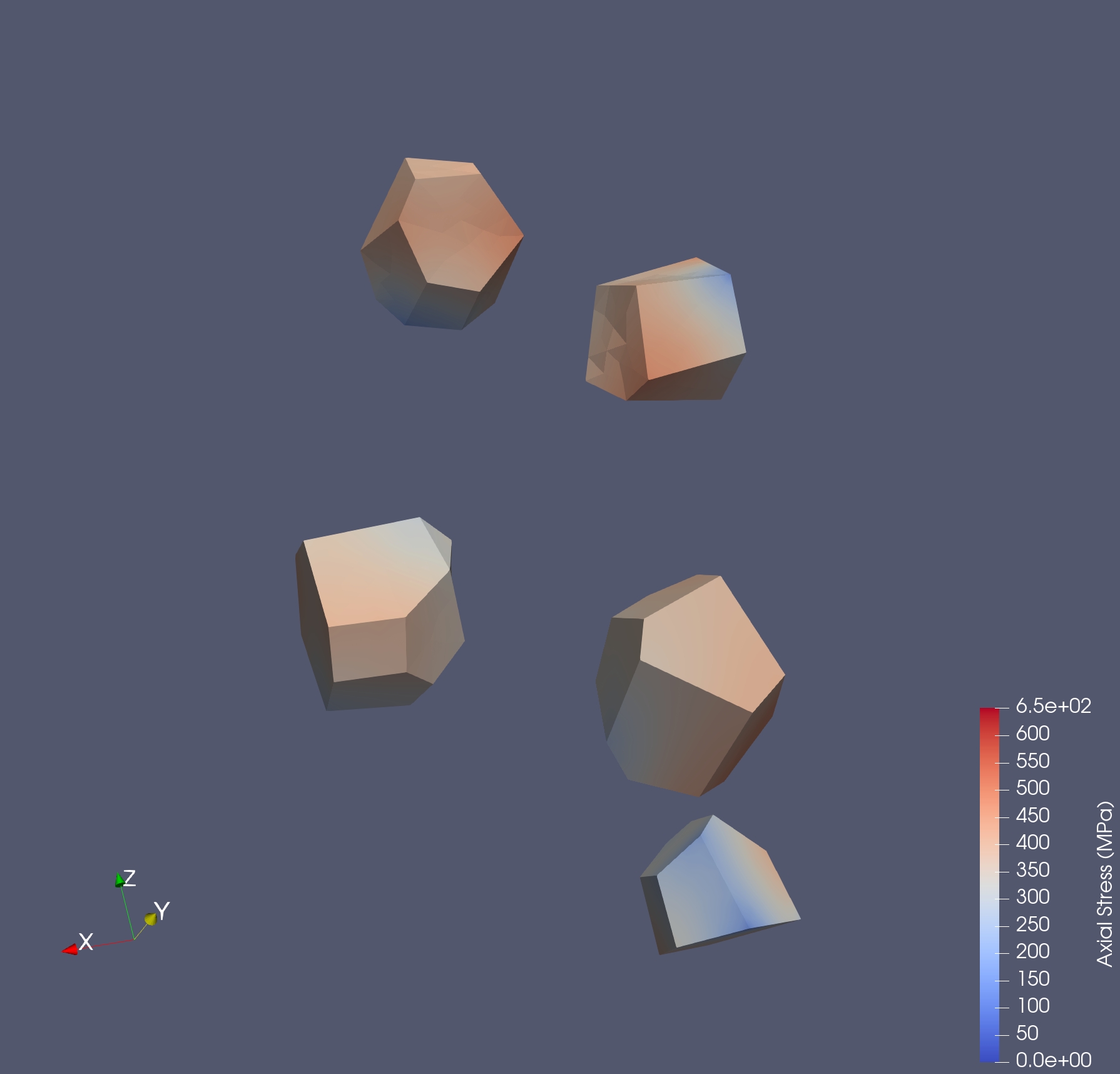}
		\caption{ }
		\label{fig:dia0p15_sph0p03_sigzz_eps0p0025_5gr_10m}
	\end{subfigure}%
	\quad
	\begin{subfigure}{.25\textwidth}
		\centering
		\includegraphics[width=1\linewidth]{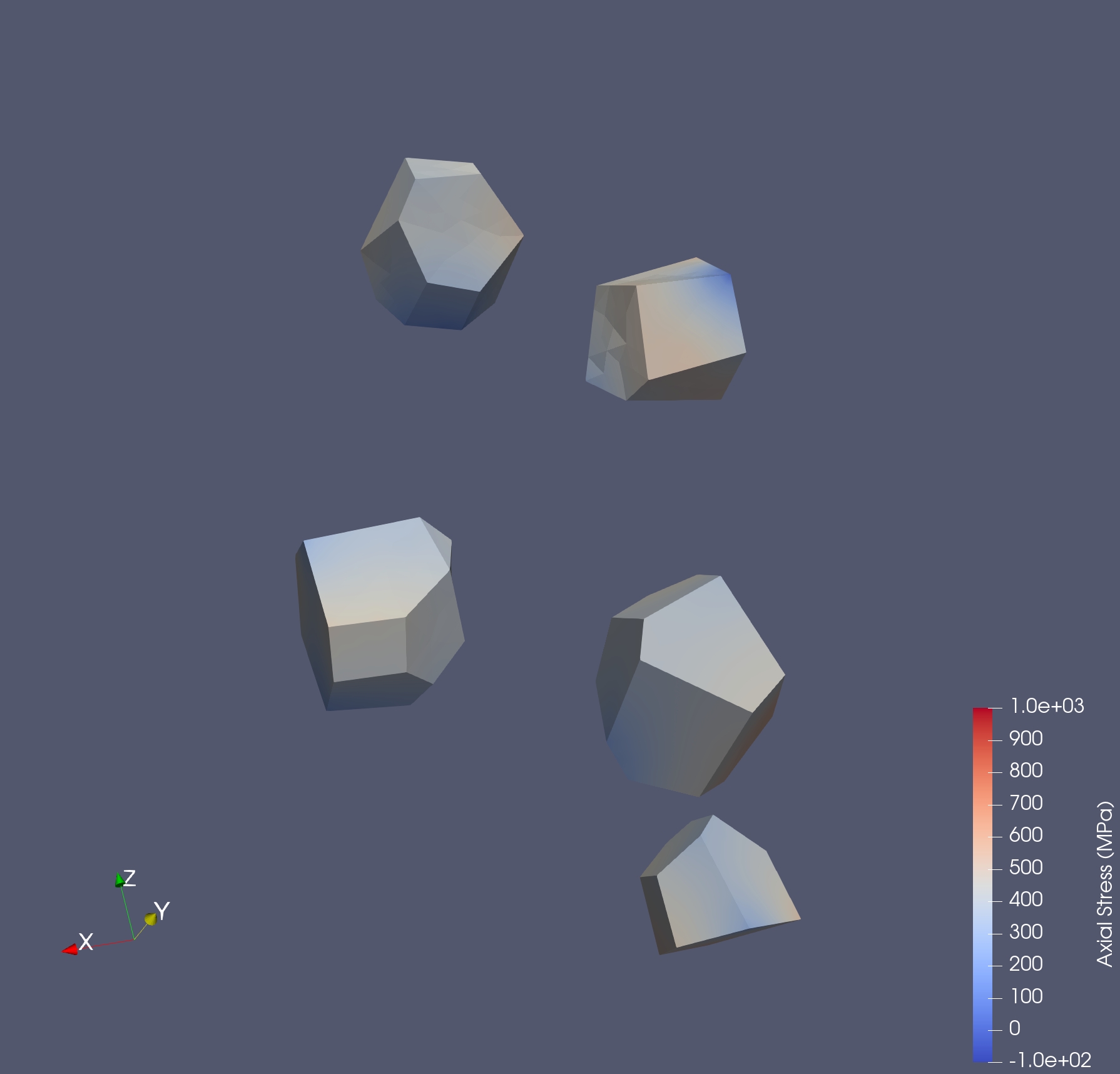}
		\caption{ }
		\label{fig:dia0p15_sph0p03_sigzz_eps0p01_5gr_10m}
	\end{subfigure}%
	\caption{Simulated axial stress distributions over Grains 48, 49, 50, 51, and 52 in the HUHS sample at nominal strains of  (a) 0.001 ; (b) 0.0025;  and (c) 0.01; 10 mode fits to the simulated stress distributions at nominal strains of (d) 0.001 ; (e) 0.0025;  and (f) 0.01.  Note that the scales are different for the different strain levels.}
		\label{fig:5gr_stress_distributions_dia0p15_sph0p03}
\end{figure}

\newpage
The stress distributions simulated with \fepx\, and represented by harmonic fits 
are shown in Figures~\ref{fig:5gr_stress_distributions_voronoi}, \ref{fig:5gr_stress_distributions_dia0p35_sph0p06}, and \ref{fig:5gr_stress_distributions_dia0p15_sph0p03}
at nominal strains of 0.1\%, 0.25\% and 1.0\% (0.001, 0.0025 and 0.01, respectively).
The harmonic fits employ only 10 harmonic modes to approximate the simulated distributions.  
The fits capture the gross trends in the data, but miss on details of the distributions.
As with any expansion, the fits generally improve with larger numbers of terms.
Note that the inclusion of more terms (modes) does not alter the weights of the modes
already employed.  
Rather, the overall approximation improves by means of
adding corrective adjustments to the approximation.
Thus we can examine the behavior of lower modes without concern that
adding more modes would alter the trends regarding the weights on the lower modes.

\subsection{Evolution of mode weights for 5 grains}
\begin{figure}[htbp]
	\centering
	\begin{subfigure}{.3\textwidth}
		\centering
		\includegraphics[width=1\linewidth]{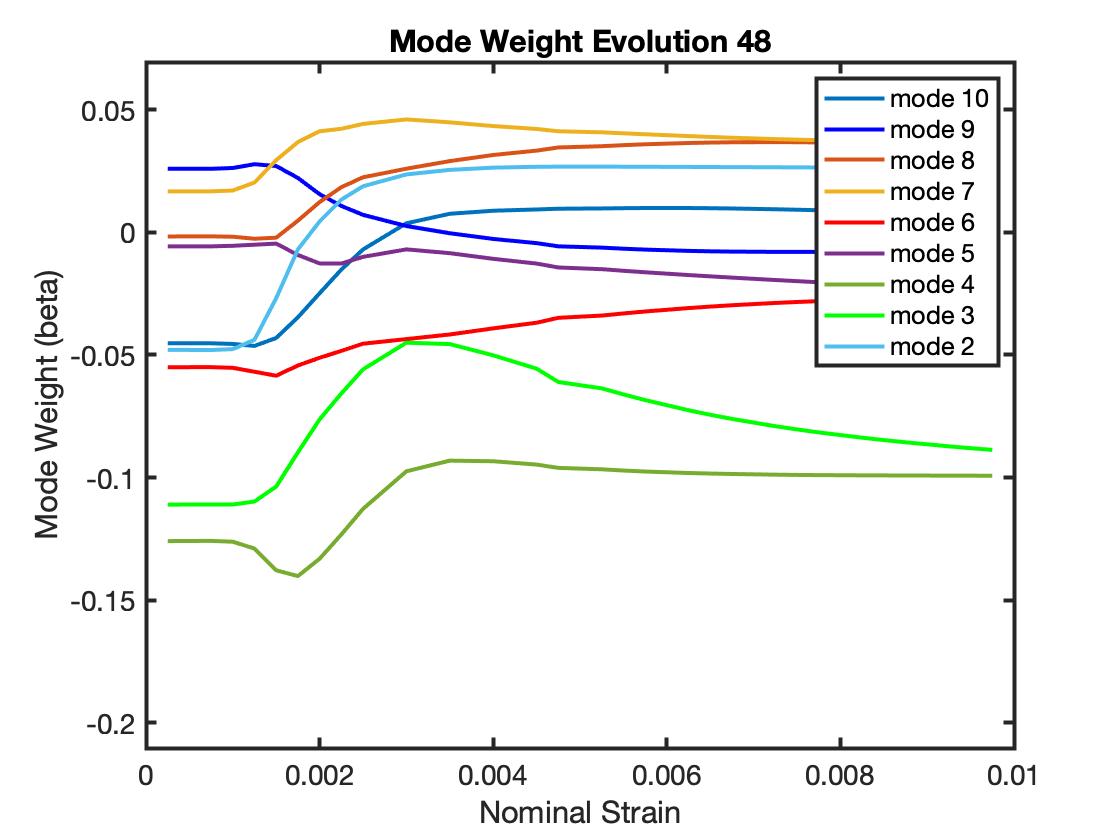}
		\caption{ }
		\label{fig:voronoi_gr48}
	\end{subfigure}%
	\quad
	\begin{subfigure}{.3\textwidth}
		\centering
		\includegraphics[width=1\linewidth]{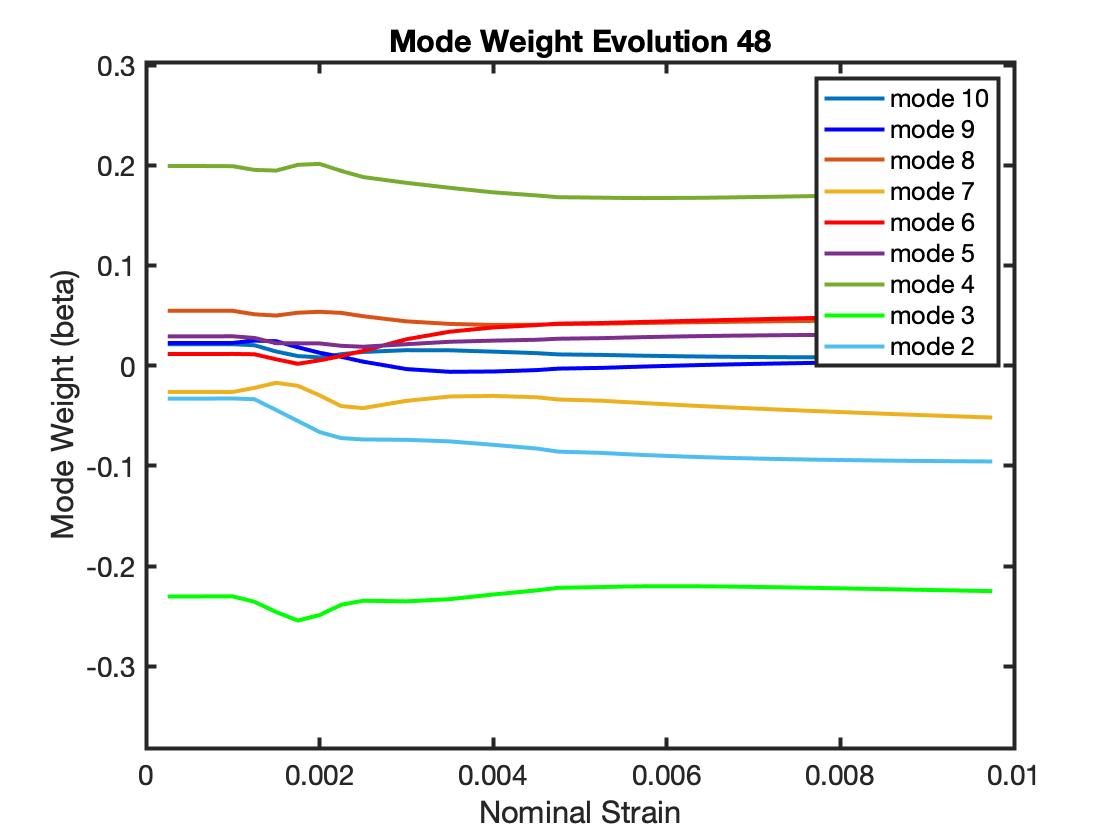}
		\caption{ }
		\label{fig:dia0p35_sph0p06_gr48}
	\end{subfigure}%
	\quad
	\begin{subfigure}{.3\textwidth}
		\centering
		\includegraphics[width=1\linewidth]{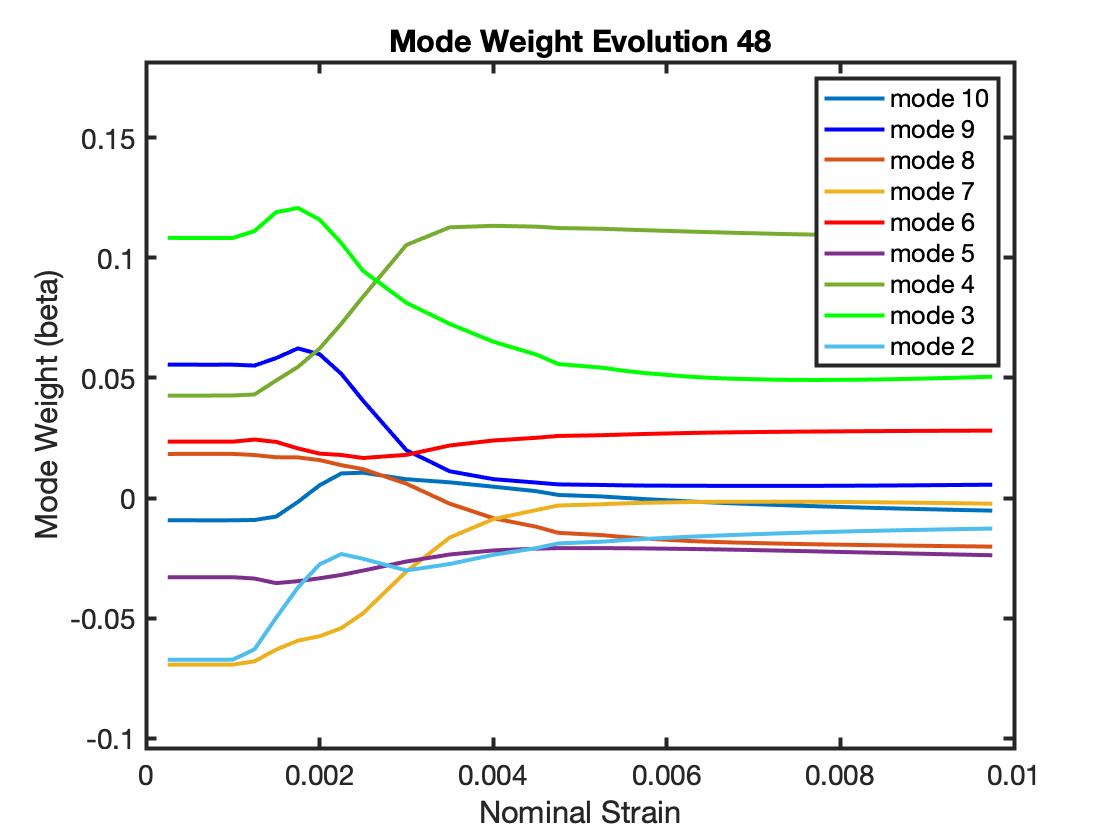}
		\caption{ }
		\label{fig:dia0p15_sph0p03_gr48}
	\end{subfigure}%
	\caption{Evolution of mode weights for Grains 48 in:  (a) Voronoi sample ; (b) LULS sample;  and (c) HUHS sample. Note that the weights are normalized by the weight of the first (constant) mode. }
		\label{fig:modeweights_gr48}
\end{figure}

\begin{figure}[htbp]
	\centering
	\begin{subfigure}{.3\textwidth}
		\centering
		\includegraphics[width=1\linewidth]{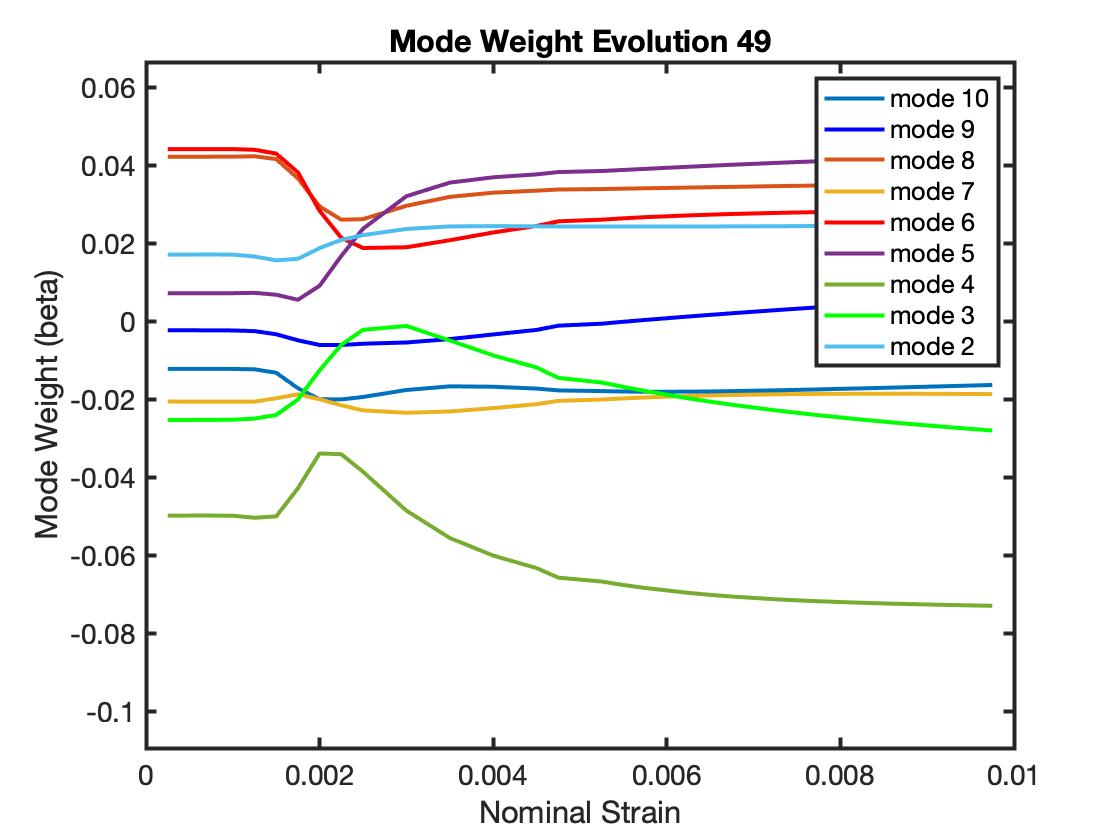}
		\caption{ }
		\label{fig:voronoi_gr49}
	\end{subfigure}%
	\quad
	\begin{subfigure}{.3\textwidth}
		\centering
		\includegraphics[width=1\linewidth]{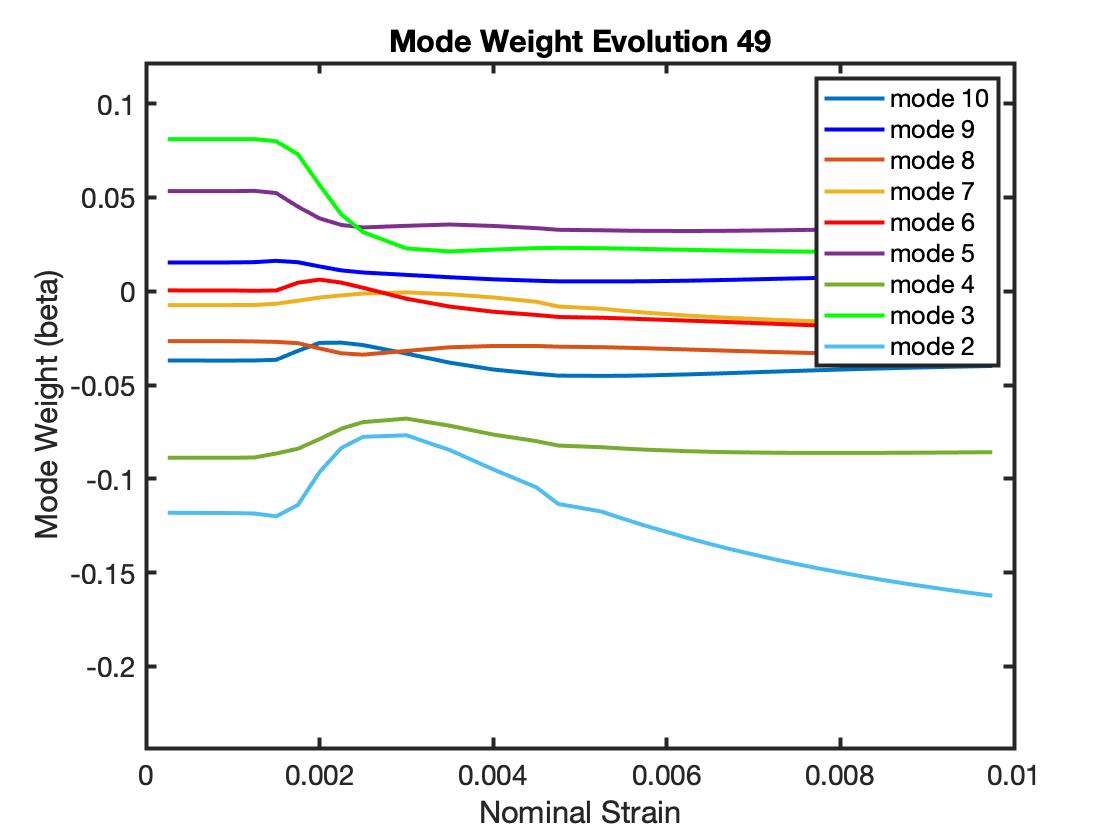}
		\caption{ }
		\label{fig:dia0p35_sph0p06_gr49}
	\end{subfigure}%
	\quad
	\begin{subfigure}{.3\textwidth}
		\centering
		\includegraphics[width=1\linewidth]{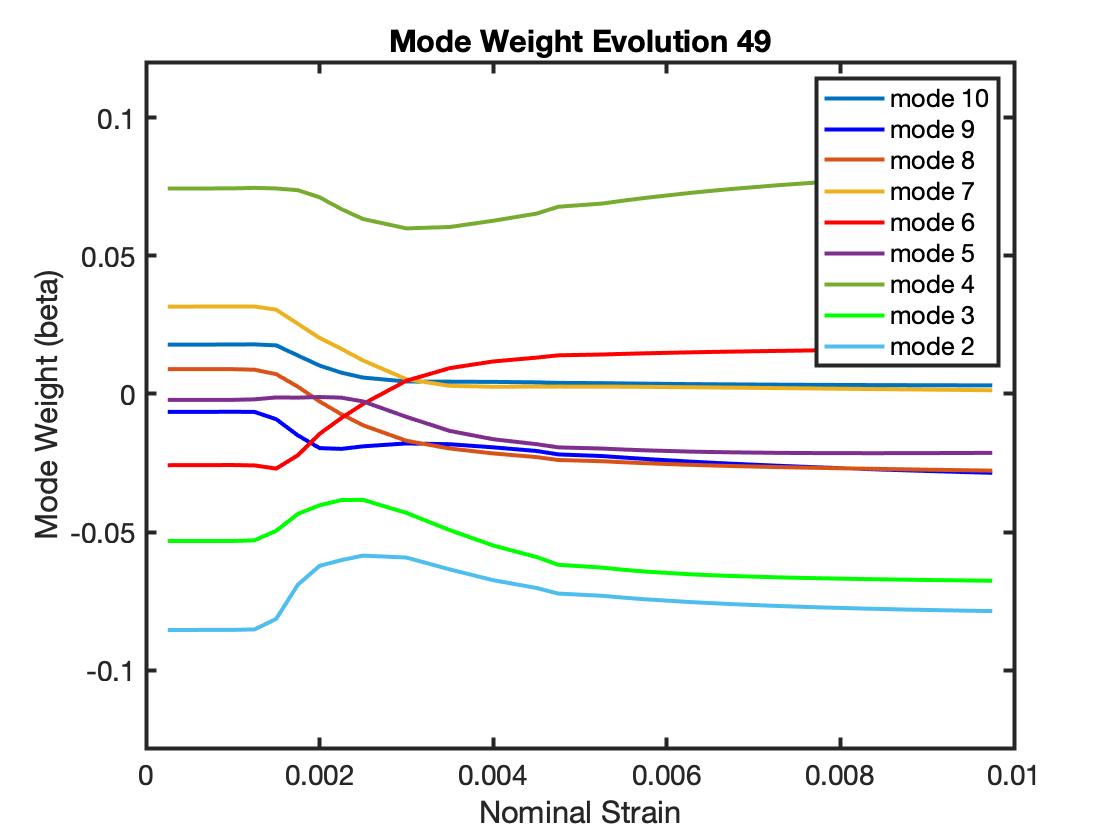}
		\caption{ }
		\label{fig:dia0p15_sph0p03_gr49}
	\end{subfigure}%
	\caption{Evolution of mode weights for Grains 49 in:  (a) Voronoi sample ; (b) LULS sample;  and (c) HUHS sample. Note that the weights are normalized by the weight of the first (constant) mode. }
		\label{fig:modeweights_gr49}
\end{figure}

\begin{figure}[htbp]
	\centering
	\begin{subfigure}{.3\textwidth}
		\centering
		\includegraphics[width=1\linewidth]{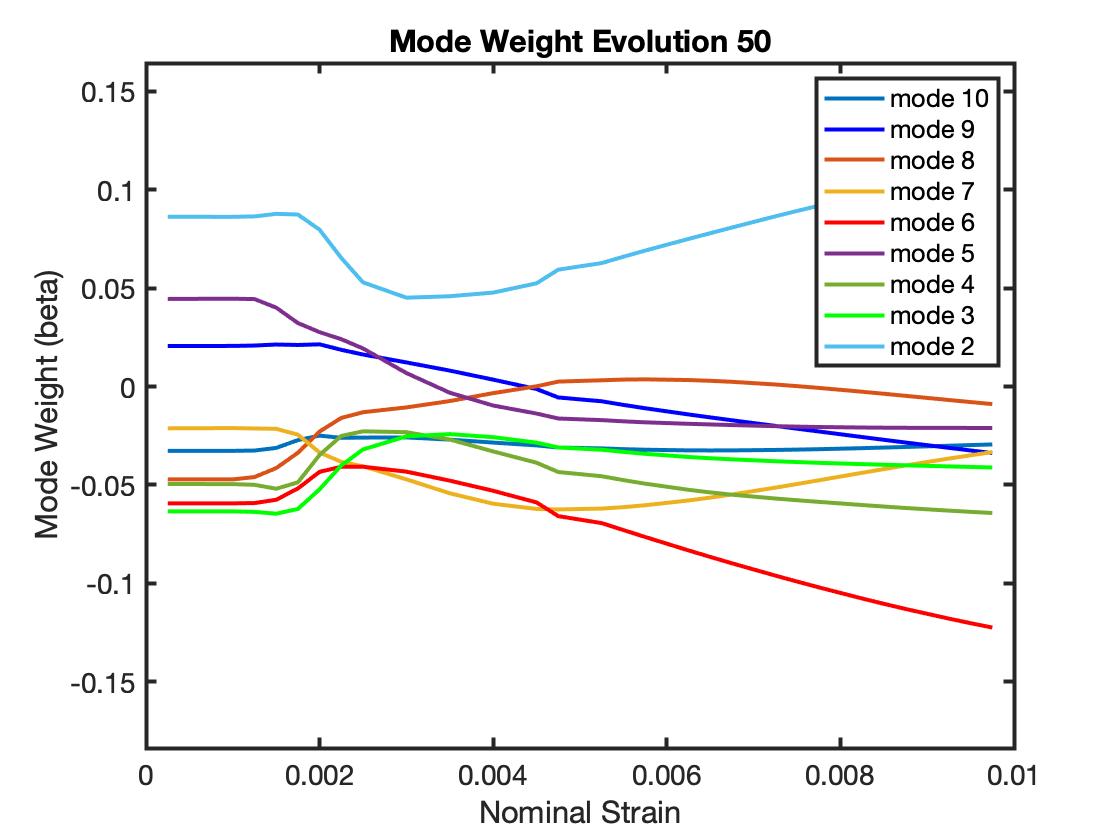}
		\caption{ }
		\label{fig:voronoi_gr50}
	\end{subfigure}%
	\quad
	\begin{subfigure}{.3\textwidth}
		\centering
		\includegraphics[width=1\linewidth]{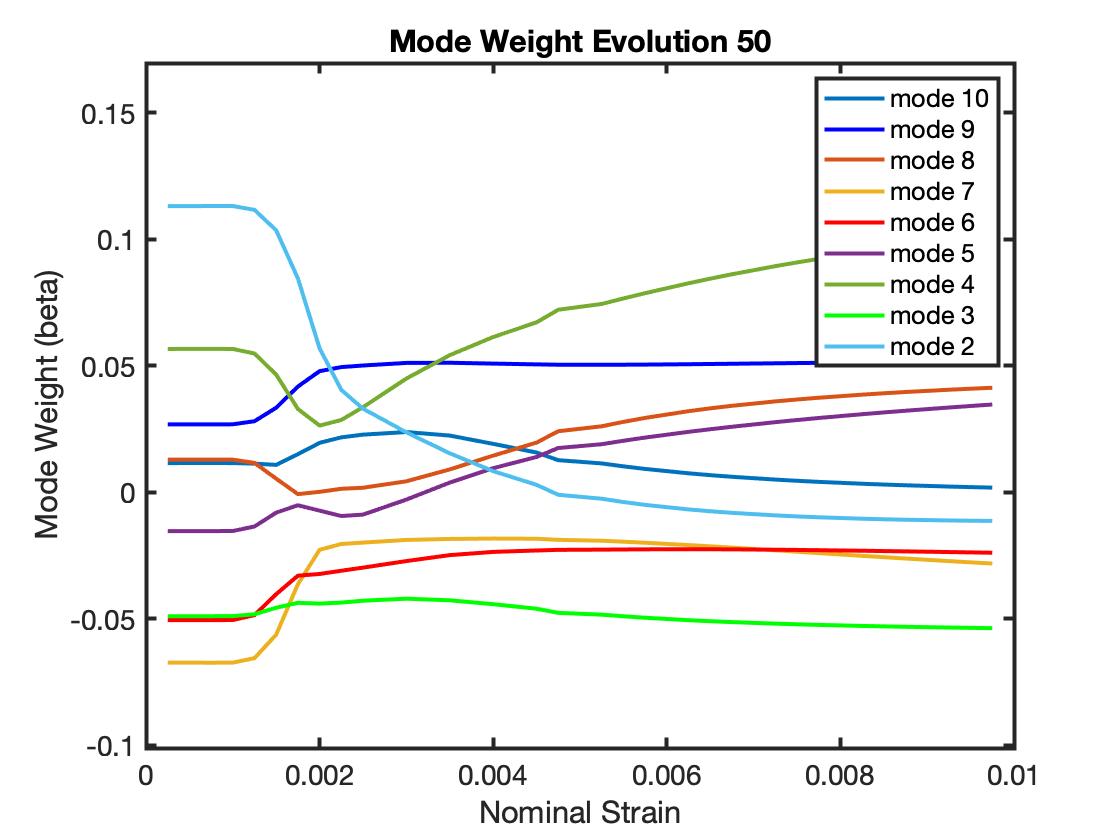}
		\caption{ }
		\label{fig:dia0p35_sph0p06_gr50}
	\end{subfigure}%
	\quad
	\begin{subfigure}{.3\textwidth}
		\centering
		\includegraphics[width=1\linewidth]{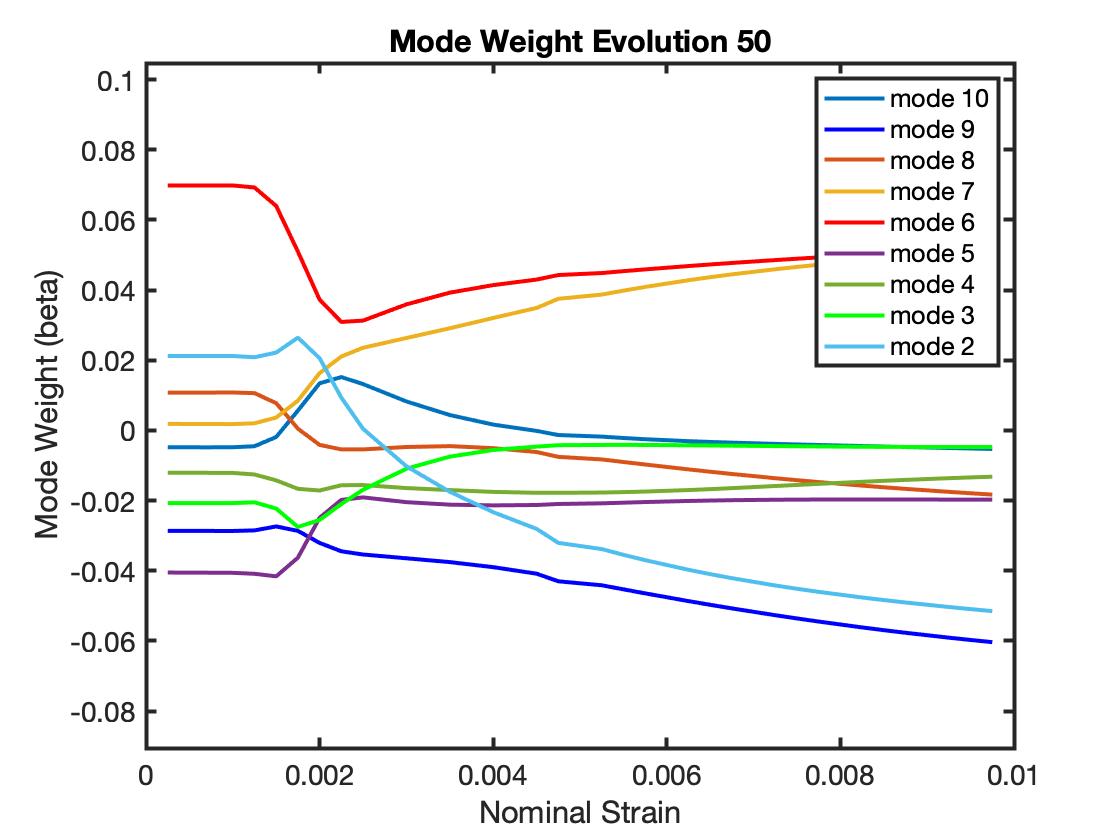}
		\caption{ }
		\label{fig:dia0p15_sph0p03_gr50}
	\end{subfigure}%
	\caption{Evolution of mode weights for Grains 50 in:  (a) Voronoi sample ; (b) LULS sample;  and (c) HUHS sample. Note that the weights are normalized by the weight of the first (constant) mode. }
		\label{fig:modeweights_gr50}
\end{figure}

\begin{figure}[htbp]
	\centering
	\begin{subfigure}{.3\textwidth}
		\centering
		\includegraphics[width=1\linewidth]{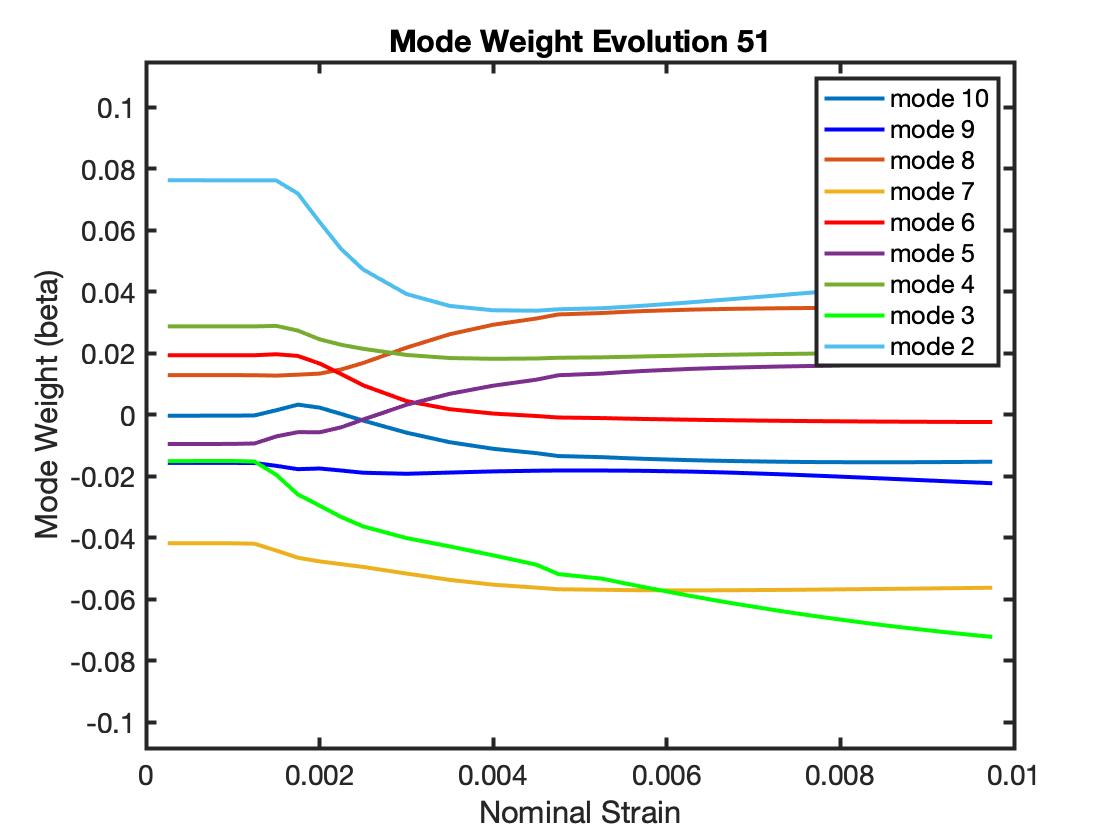}
		\caption{ }
		\label{fig:voronoi_gr51}
	\end{subfigure}%
	\quad
	\begin{subfigure}{.3\textwidth}
		\centering
		\includegraphics[width=1\linewidth]{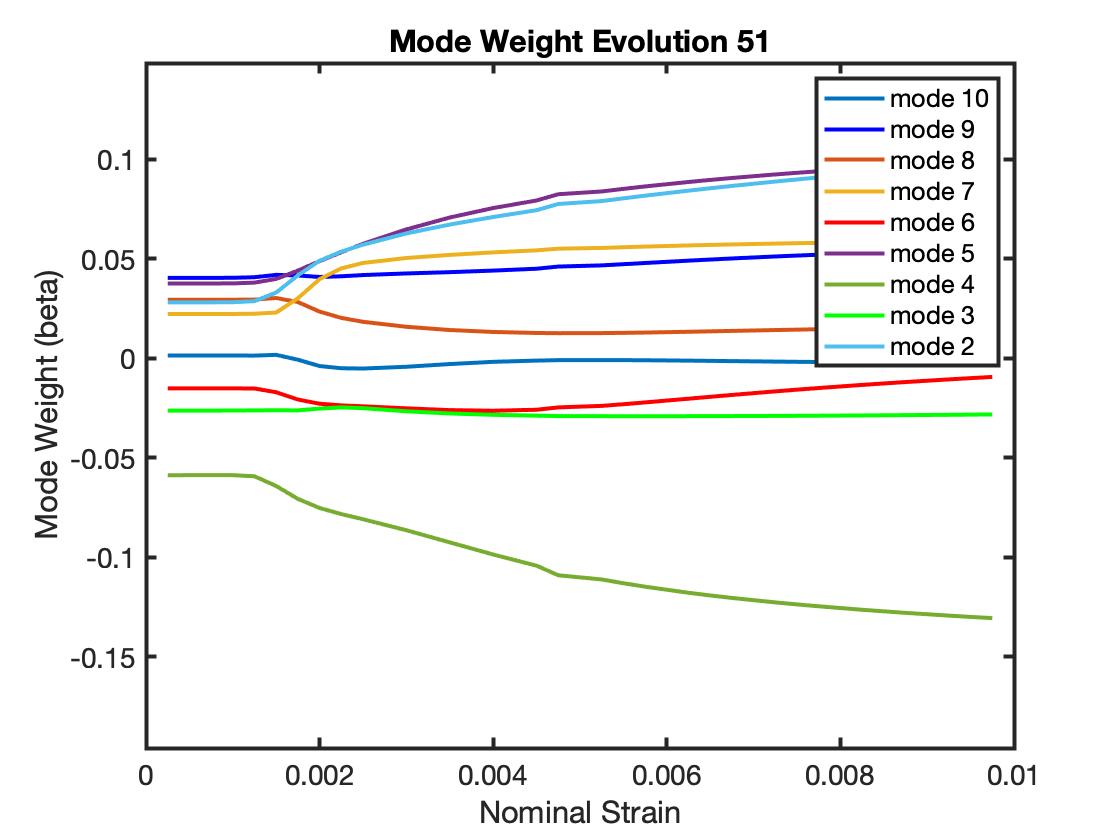}
		\caption{ }
		\label{fig:dia0p35_sph0p06_gr51}
	\end{subfigure}%
	\quad
	\begin{subfigure}{.3\textwidth}
		\centering
		\includegraphics[width=1\linewidth]{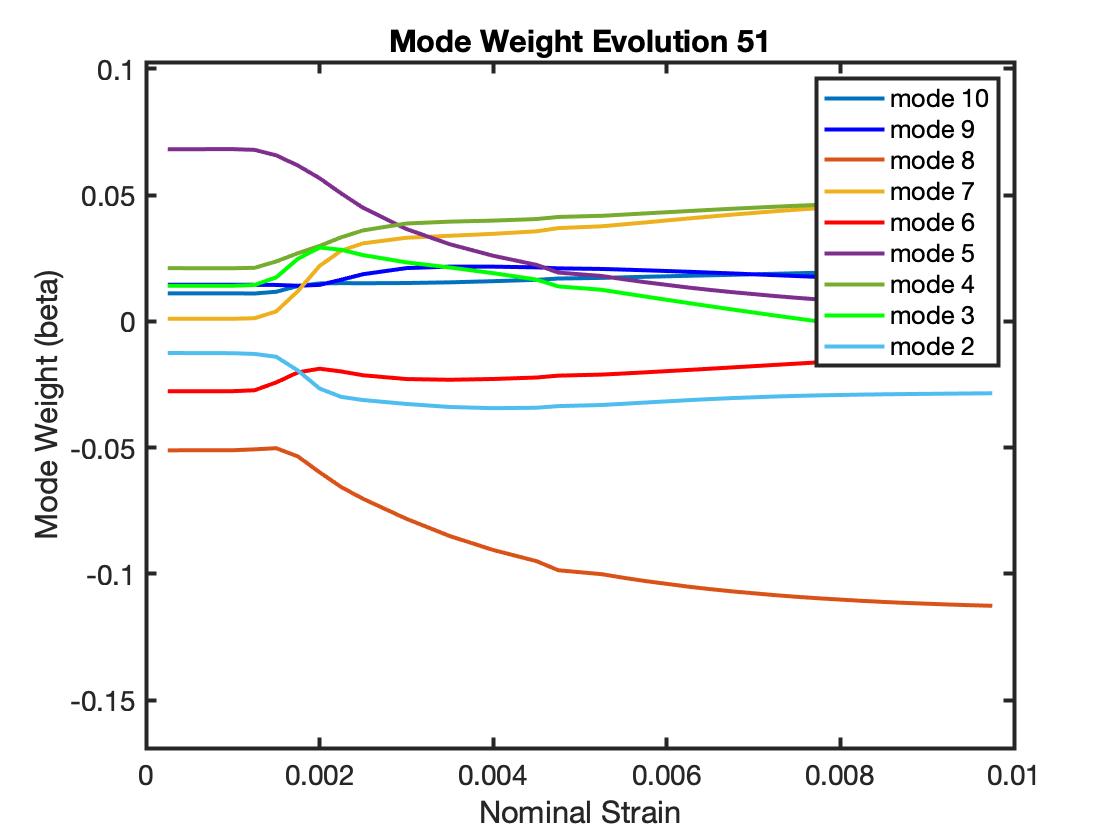}
		\caption{ }
		\label{fig:dia0p15_sph0p03_gr51}
	\end{subfigure}%
	\caption{Evolution of mode weights for Grains 51 in:  (a) Voronoi sample ; (b) LULS sample;  and (c) HUHS sample. Note that the weights are normalized by the weight of the first (constant) mode. }
		\label{fig:modeweights_gr51}
\end{figure}

\begin{figure}[htbp]
	\centering
	\begin{subfigure}{.3\textwidth}
		\centering
		\includegraphics[width=1\linewidth]{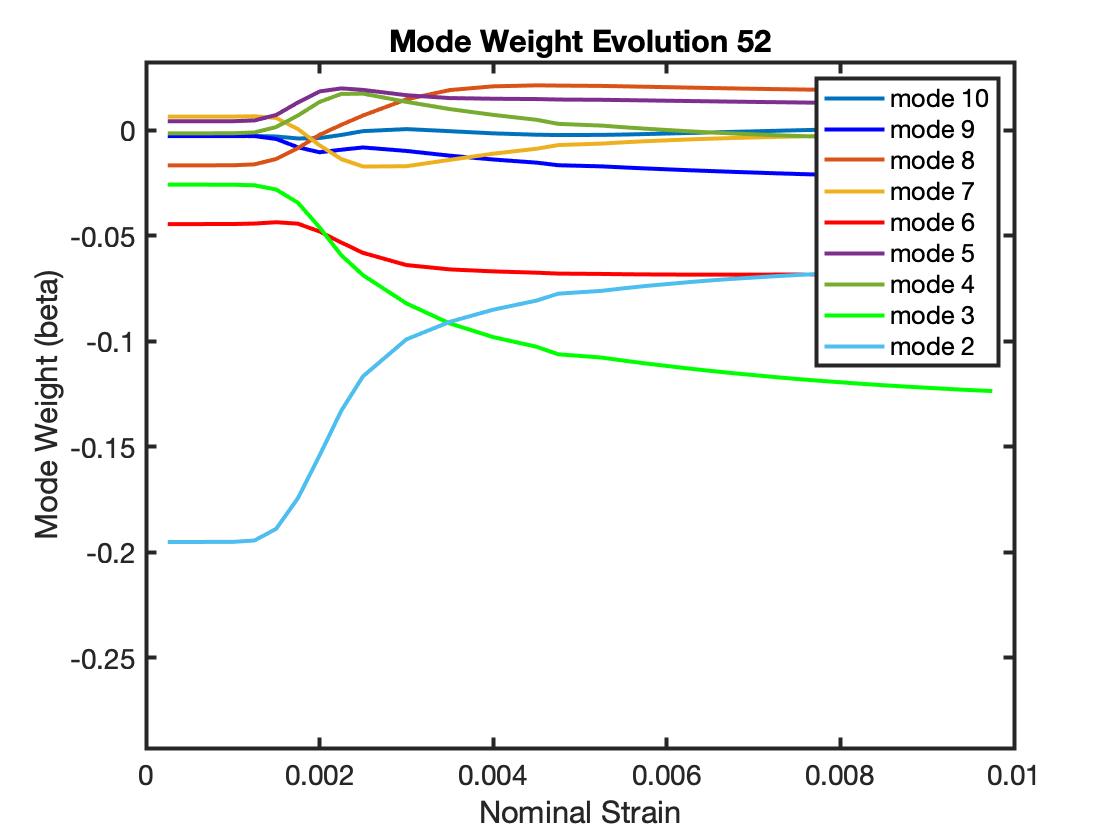}
		\caption{ }
		\label{fig:voronoi_gr52}
	\end{subfigure}%
	\quad
	\begin{subfigure}{.3\textwidth}
		\centering
		\includegraphics[width=1\linewidth]{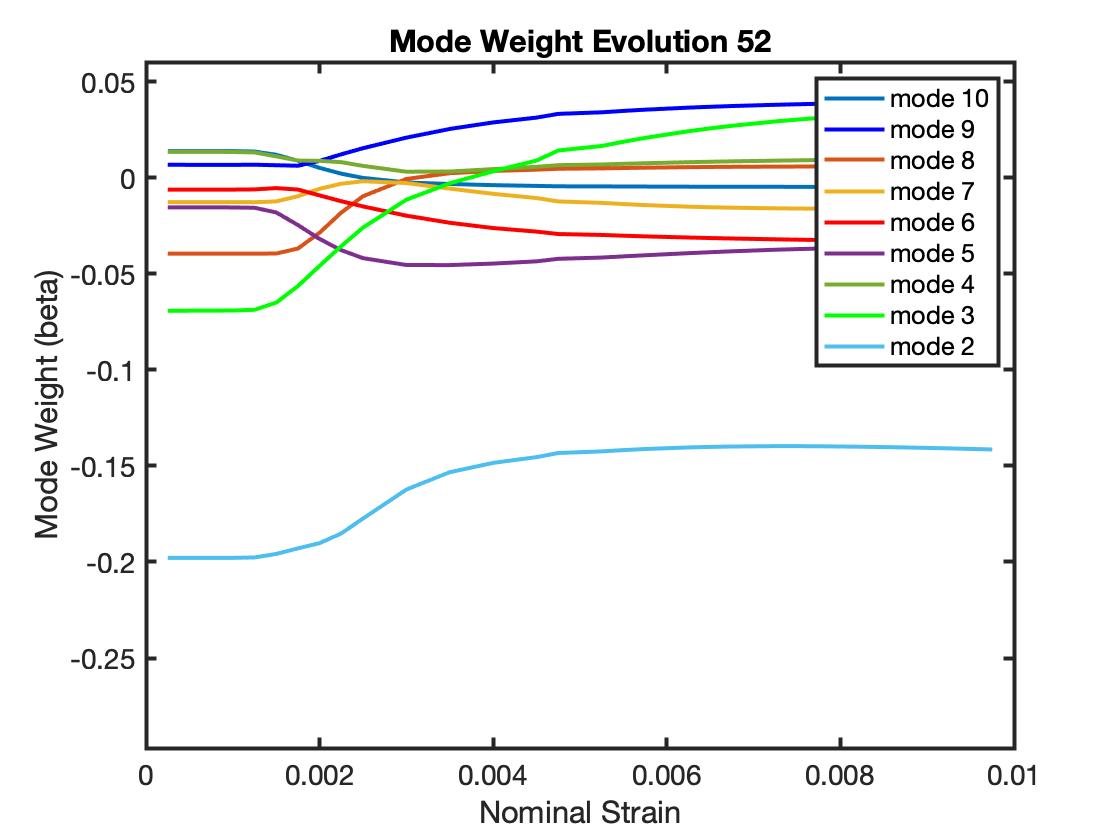}
		\caption{ }
		\label{fig:dia0p35_sph0p06_gr52}
	\end{subfigure}%
	\quad
	\begin{subfigure}{.3\textwidth}
		\centering
		\includegraphics[width=1\linewidth]{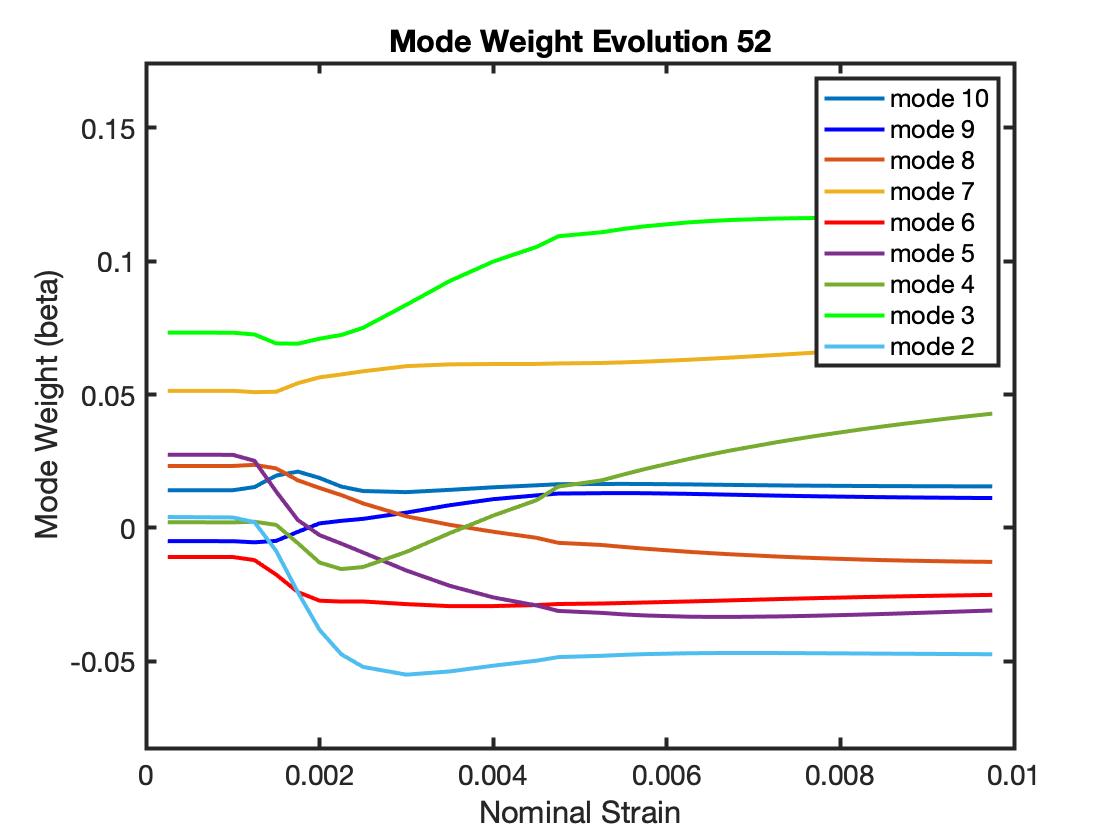}
		\caption{ }
		\label{fig:dia0p15_sph0p03_gr52}
	\end{subfigure}%
	\caption{Evolution of mode weights for Grains 52 in:  (a) Voronoi sample ; (b) LULS sample;  and (c) HUHS sample. Note that the weights are normalized by the weight of the first(constant) mode. }
		\label{fig:modeweights_gr52}
\end{figure}

The harmonic mode weights prescribe the contribution that the associated modes 
add to the complete representation of the field variable.  
If the field variable is evolving in a manner that qualitatively alters the distribution,
then the relative values of the weights change to reflect the changing distribution.
This is the case with the stress when the load increases to the point that the plastic yielding
initiates within grains.  Prior to that point,  the  response remains elastic, the stress increases
proportionally everywhere, and the harmonic mode weights maintain the same relative values.
To illustrate this behavior, the harmonic weights are plotted versus the nominal strain grain-by-grain for
the 5 grains in each of the sample variants in Figures~\ref{fig:modeweights_gr48}, \ref{fig:modeweights_gr49}, \ref{fig:modeweights_gr50}, \ref{fig:modeweights_gr51} and \ref{fig:modeweights_gr52}.  
In these plots, the evolution of the mode weights are plotted for Modes 2 to 10 after
normalization by the weight of Mode 1 (the constant mode).  Mode 1 increases monotonically over the loading sequence in proportion to the nominal load.
The relative weights remain constant over the first several steps.  The response is elastic and the stress distributions remain in the same proportion spatially.
Between a nominal strain of 0.001 and 0.002, yielding initiates and the relative weights begin an evolution that continues through the elastic-plastic transition.  
The relative weights stabilize at a nominal strain of 0.004 to 0.005.
This is more evident in some grains than others.
For all grains, it holds that the ranking of the modes is different at the 0.01 strain than
within the elastic regime.  
At 1\% strain, the stress is bounded by the yield surfac -- a constraint that alters the stress distributions within grains.  
The evolution of the  mode weights serves to quantify the stress redistribution.  
There may be some similarity in the evolution for the same grain across samples.  If true, one possible explanation lies in their  sharing the same initial lattice orientation.

\section{Future work}
The intent of this document is to describe the methodology used to compute discrete harmonic modes 
for grains of a polycrystal.  
The modes provide a set of orthonormal bases that may be use to 
represent field variables over the grains with a reduced number of parameters,
namely the weights of the modes.  
The demonstration application illustrates the process of evaluating the discrete harmonic modes,
determining the  modes weights needed to represent a field variable (in this case the axial stress
in a tensile loading of a sample), and tracking the changes of the weights through a deformation process that displays evolving stress distributions.
No attempts are made to correlate the harmonic weights, or their changes,
to attributes of the grains or the loading.  
We expect that discoveries are possible in both regards.
\mechmonics\, is available via \github\, so that researchers in addition to ourselves may pursue these.

\section{Summary}

The computational methods for computing sets of discrete harmonic modes in \mechmonics\, are presented.
One set of modes is determined for each grain of a virtual polycrystal.  
The modes may be used to represent field data over the grain domains 
by means of an expansion involving weighted contributions of the modes.  
The procedure for evaluating the mode weights 
for the expansion from the field data also is presented.
\mechmonics\, uses virtual polycrystal instantiations created and meshed by \neper.  
In addition to files containing the modes and their weights for designated data,
\mechmonics\, writes output files in VTK format for visualization using \paraview\, or \visit.

A demonstration application is provided to illustrate \mechmonics\, capabilities. 
Harmonic mode representations are computed for the axial stress computed for the 
tensile loading of a stainless steel sample with the elastoplastic code, \fepx. 
Three variants or the sample are examined that differ in their grain morphologies.
The differences are in the distributions of grain diameter and  grain sphericity.
The sample is loaded to 1\% nominal strain, which is sufficient to induce plastic flow.
The axial stress distributions change as the sample progresses from elastic to elastic-plastic behaviors.
The evolution of the mode weights captures the evolution of the stress distributions.
The sample variants show distinct histories of the mode weights stemming from the
distinct size and shape metrics for each grain.

\section*{Acknowledgements}
The research reported here was supported by the ONR under grant \# N00014-16-1-3126, Dr. William Mullins Program Manager.

%
 \section*{Conflict of interest}
 The authors declare that they have no conflict of interest.

\bibliographystyle{spmpsci}      
\bibliography{MechMonics_References.bib}  

\end{document}